\documentclass[12pt,english]{article}
\usepackage{amsmath}
\usepackage{mathptmx}
\usepackage{newtxmath}

\usepackage[T1]{fontenc}
\usepackage[latin9]{inputenc}
\usepackage[letterpaper]{geometry}
\usepackage{color}
\usepackage{babel}
\usepackage{array}
\usepackage{float}
\usepackage{rotfloat}
\usepackage{multirow}
\usepackage{graphicx}
\usepackage{setspace}
\usepackage[authoryear]{natbib}
\usepackage[unicode=true,pdfusetitle,
 bookmarks=true,bookmarksnumbered=false,bookmarksopen=false,
 breaklinks=true,pdfborder={0 0 0},pdfborderstyle={},backref=false,colorlinks=true]
 {hyperref}
\hypersetup{
 citecolor=DarkBlue,linkcolor=DarkBlue,urlcolor=DarkBlue}

\makeatletter

\providecommand{\tabularnewline}{\\}

\usepackage[svgnames]{xcolor}
\usepackage{bbm}

\@ifundefined{showcaptionsetup}{}{%
 \PassOptionsToPackage{caption=false}{subfig}}
\usepackage{subfig}
\makeatother

\begin{document}
\title{Measuring Energy-saving Technological Change:\\
{\Large{}International Trends and Differences}\thanks{We are grateful to Hirokazu Ishise, Yoshifumi Konishi, Makiko Nakano,
Filippo Pavanello, Kenji Takeuchi, anonymous referees, and seminar
participants at Keio University, Osaka University, the Annual Conference
of the European Association of Environmental and Resource Economists,
the Annual Conference of the Society for Environmental Economics and
Policy Studies, and the Japanese Economic Association Spring Meeting
for helpful comments and discussions. Inoue and Yamada acknowledge
support from the Sumitomo Foundation. Yamada acknowledges support
from JSPS KAKENHI grant numbers 17H04782 and 21H00724.}\\
\vspace{0.5cm}
}
\author{Emiko Inoue\thanks{Kyoto University. \texttt{inoue@econ.kyoto-u.ac.jp}}
\and Hiroya Taniguchi\thanks{Kyoto University. \texttt{taniguchi.hiroya.24z@st.kyoto-u.ac.jp}}
\and Ken Yamada\thanks{Kyoto University. \texttt{yamada@econ.kyoto-u.ac.jp}}\vspace{0.5cm}
}
\date{January 2022}
\maketitle
\begin{abstract}
\begin{onehalfspace}
Technological change is essential to balance economic growth and environmental
sustainability. This study documents energy-saving technological change
to understand the trends and differences therein in OECD countries.
We estimate sector-level production functions with factor-augmenting
technologies using cross-country and cross-industry panel data and
shift\textendash share instruments, thereby measuring energy-saving
technological change for each country and sector. Our results show
how the levels and growth rates of energy-saving technology vary across
countries, sectors, and time. In addition, we evaluate the extent
to which factor-augmenting technologies contribute to economic growth
and how this contribution differs across countries and sectors.\bigskip{}
\\
\textsc{Keywords}: Non-neutral technological change; capital\textendash labor\textendash energy
substitution; growth accounting; sectoral productivity.\\
\textsc{JEL classification}: E23, O33, O44, O50, Q43, Q55.
\end{onehalfspace}

\global\long\def\E{\mathbb{E}}%
\end{abstract}
\newpage{}

\captionsetup[subfigure]{font=normalsize}

\captionsetup[subtable]{font=normalsize}

\section{Introduction}

One of the greatest challenges facing society is the achievement of
economic development and environmental conservation. Technological
change has been in the past, and will be in the future, the most promising
way to balance economic growth and environmental sustainability. Society
has been developing and adopting new technology to make more efficient
use of energy and natural resources in the face of serious environmental
problems, including climate change, environmental pollution, and resource
depletion. Given the global nature of environmental problems, the
development and adoption of new technology need to expand worldwide.
However, it has been difficult to assess how energy-saving technology
has evolved across countries over time due to the lack of adequate
measurements. The aim of this study is to measure and document the
level and growth rate of energy-saving technological change for each
country and sector, thereby developing an understanding of international
trends and differences in energy-saving technological change.

Measuring energy-saving technological change is challenging. Environmentally
friendly technological change is typically measured in the literature
using data on research and development (R\&D) spending and patent
counts \citep{Popp_IRERE19}. These measures, however, have some drawbacks.
R\&D spending is a measure of input into the innovation process rather
than its outcomes. Patent counts are a measure of product innovation,
but not of process innovation. In this study, we measure energy-saving
technology in terms of output and factor inputs along the lines of
the \citet{Solow_RESTAT57} residual, also known as total factor productivity
(TFP). Similar to TFP, but different from R\&D spending and patent
counts, our measure captures the actual circumstances of national
income and technology adopted in the economy, including not only patented
product technology but also unpatented product and process technology.
At the same time, our measure differs from TFP in that it allows technological
change to be factor-augmenting.

We measure factor-augmenting technological change using sector-level
production functions and firm's optimality conditions. Our method
builds upon the seminal work of \citet{Caselli_Coleman_AERPP02,Caselli_Coleman_AER06}.
The advantage of this method is that it allows us to quantify the
level and growth rate of energy-saving technology for a given value
of the elasticity of substitution in production without assuming the
functional forms of technological change. Our analysis extends that
of \citet{Caselli_Coleman_AERPP02,Caselli_Coleman_AER06} by using
the gross-output production function at the sector level, allowing
for imperfect competition, and estimating the elasticity of substitution
between energy and non-energy inputs. Our study is the first to measure
and compare the levels and growth rates of energy-saving technology
across countries and sectors.

Although numerous studies have estimated the elasticity of substitution
between energy and non-energy inputs, this study differs from those
studies in that we take into account the cross-sectional and time
variation in the unobserved components of factor-augmenting technologies.
To do so, we use cross-country and cross-industry panel data from
12 OECD countries and construct shift\textendash share instruments.
Our results indicate that the elasticities of substitution among capital,
labor, and energy inputs are significantly less than one.

We utilize our measure of energy-saving technology in two ways. First,
we examine the nature of energy-saving technological change and its
differences across countries. Our results show that the levels and
growth rates of energy-saving technology vary substantially across
countries over time for the year 1978 to 2005 both in the goods and
service sectors. Cross-country differences in energy-saving technology
are greater than those in capital- and labor-augmenting technologies.
Moreover, in line with theoretical predictions, energy-saving technology
tends to progress in countries or years in which energy resources
are scarce. Technological change tends to be directed toward energy
as government spending on energy-related R\&D increases. Second, we
evaluate the quantitative contribution of energy-saving technology
to economic growth. Our results demonstrate that energy-saving technological
change contributes to economic growth in the goods or service sectors
of many countries.

The remainder of this paper is organized as follows. The next section
reviews the literature. Section \ref{sec: model} introduces the method
for measuring energy-saving technological change using sector-level
production functions. Section \ref{sec: estimation} considers the
identification and estimation of substitution parameters in the production
functions. Section \ref{sec: data} describes the data used in the
analysis. Section \ref{sec: results} presents the empirical results.
Section \ref{sec: extension} discusses the interpretation and implications
of the results when the model is extended. The final section summarizes
and concludes the paper.

\section{Related Literature}

This study is related to three strands of the literature. First, it
contributes to the literature that measures factor-augmenting (non-neutral)
technological change. The direction and magnitude of factor-augmenting
technological change can be measured by estimating either a production
or a cost function. \citet{Brown_DeCani_IER63} and \citet{David_VanDeLlundert}
develop an approach that uses a constant elasticity of substitution
(CES) production function (see \citealp*{LeonLedesma_McAdam_Willman_AER10}
for a list of related studies).\footnote{\citet*{Klump_McAdam_Willman_RESTAT07} measure capital- and labor-augmenting
technologies in the United States from the year 1953 to 1998. \citet*{Herrendorf_Herrington_Valentinyi_AEJ15}
measure capital- and labor-augmenting technologies in the agriculture,
manufacturing, and service sectors in the United States from the year
1947 to 2010.} \citet{VanDerWerf_EE08} adopts this type of approach to measure
energy-saving technological change in 12 OECD countries from the year
1978 to 1996. \citet{Binswanger_AER74} develops an alternative approach
that uses the factor-share equations derived from a translog cost
function (see \citealp{Jorgenson_HoE86} for a survey). \citet*{Sanstad_Roy_Sathaye_EE06}
adopt this type of approach to measure energy-saving technological
change in India from the year 1973 to 1994, the Republic of Korea
from the year 1980 to 1997, and the United States from the year 1958
to 1996. The advantages of the former approach are that it does not
require the estimation of many parameters (or dealing with many endogenous
regressors) and can estimate the key parameters in computable general
equilibrium models to analyze climate and energy policies. However,
both approaches have a common limitation in that factor-augmenting
technologies is treated as a parametric and deterministic component
in the production or cost function. Most studies assume that factor-augmenting
technologies grow at a constant rate.

\citet{Caselli_Coleman_AERPP02,Caselli_Coleman_AER06} develop an
approach that measures factor-augmenting technologies using the aggregate
production function and the firm's optimality conditions. They do
so without specifying the functional forms of technological change
for the given values of the substitution parameters in the production
function. \citet{Caselli_Coleman_AER06} measure non-neutral technologies
that augment skilled and unskilled labor in 52 countries in the year
1988, while \citet{Caselli_Coleman_AERPP02} measure non-neutral technologies
that augment capital as well as skilled and unskilled labor in the
United States from the year 1963 to 1992. \citet*{Hassler_Krusell_Olovsson_JPE21}
employ this type of approach in the first part of their analysis to
document fossil energy-saving technological change in the United States
from the year 1949 to 2018. However, as \citet{Caselli_HEG05} notes,
the intrinsic pitfall of this type of approach is that the elasticities
of substitution are not estimated in a way that takes into account
the variation in unobserved factor-augmenting technologies across
observations.

Second, this study contributes to the literature that estimates the
elasticity of substitution between energy and non-energy inputs. The
substitution parameter in the CES production function is a key parameter
in the analysis of climate and energy policies using computable general
equilibrium models \citep*{Jacoby_Reilly_McFarland_Paltsev_EE06}.
Among others, \citet*{Prywes_EE86}, \citet*{Chang_EE94}, \citet*{Kemfert_EE98},
\citet*{VanDerWerf_EE08}, and \citet*{Henningsen_Henningsen_VanDerWerf_EE19}
estimate the elasticities of substitution among capital, labor, and
energy inputs based on the CES production function,\footnote{Some studies estimate the elasticities of substitution between energy
inputs. \citet*{Papageorgiou_Saam_Schulte_RESTAT17} estimate the
elasticity of substitution between clean and dirty energy inputs in
28 industries of 19 OECD countries for the year 1995 to 2007.} while \citet{Berndt_Wood_RESTAT75} and \citet{Griffin_Gregory_AER76}
estimate them based on the translog cost function (see \citealp*{Koetse_deGroot_Florax_EE08}
for a survey). Although this literature provides various estimates
of the elasticities of substitution between energy and non-energy
inputs, it ignores the endogeneity problem associated with factor-augmenting
technological change. Consequently, there may be a bias in previous
estimates of the elasticities of substitution between energy and non-energy
inputs. Recently, \citet{Raval_RJE19} estimates the elasticity of
substitution between capital and labor in the manufacturing sector
of the United States from the year 1987 to 2007. \citet{Raval_RJE19}
addresses the endogeneity problem associated with factor-augmenting
technological change using shift\textendash share instruments. We
also adopt this type of approach to consistently estimate the elasticities
of substitution among capital, labor, and energy inputs.

Finally, this study contributes to the literature on sectoral growth
accounting. \citet*{Jorgenson_Gollop_Fraumeni_bk87} develop a growth
accounting framework to decompose the rate of growth in sectoral gross
output into the contribution of capital, labor, and intermediate inputs
and TFP in the United States for the year 1948 to 1979. \citet{OMahony_Timmer_EJ09}
apply this framework to measure the contribution of capital, labor,
and intermediate inputs and TFP to the gross output growth in business
service industries in seven OECD countries for the year 1995 to 2005.
Both studies demonstrate a significant contribution of intermediate
inputs to economic growth. While these studies focus on the growth
in gross output, many other studies focus on the growth in value-added
output (see \citealp{Caselli_HEG05} and \citealp*{Herrendorf_Rogerson_Valentinyi_HEG14}
for surveys). As far as we are aware, none of the studies evaluate
the quantitative contribution of factor-augmenting technologies to
economic growth.

\section{Model\label{sec: model}}

We assume that gross output ($y$) is produced from capital ($k$),
labor ($\ell$), and energy ($e$) in the goods and service sectors.
We denote by $r$, $w$, and $v$ the prices of capital, labor, and
energy inputs, respectively, that are normalized by the output price.
The representative firm in each sector chooses the quantities of inputs
$(k,\ell,e)$ so as to maximize its profits:
\begin{equation}
y-rk-w\ell-ve\label{eq: profit}
\end{equation}
subject to production technology:
\begin{equation}
y=f\left(k,\ell,e;a_{k},a_{\ell},a_{e}\right),\label{eq: technology}
\end{equation}
where $a_{k}$, $a_{\ell}$, and $a_{e}$ are capital-, labor-, and
energy-augmenting technologies, respectively. We interchangeably use
the terms ``energy-augmenting technology'' and ``energy-saving
technology'' throughout the paper since a rise in $a_{e}$ results
in a fall in the cost of production as well as a rise in the output
in the model presented here.

We start our analysis by considering the standard one-level CES production
function. We then extend it to the two-level nested CES production
function.

\subsection{One-level CES}

The standard one-level CES production function is in the form of:
\begin{equation}
y=\left[\left(a_{k}k\right)^{\sigma}+\left(a_{\ell}\ell\right)^{\sigma}+\left(a_{e}e\right)^{\sigma}\right]^{\frac{1}{\sigma}}\qquad\text{for}\quad\sigma<1.\label{eq: CES1}
\end{equation}
The parameter $\sigma$ governs the degree of substitution among capital,
labor, and energy inputs. The elasticity of substitution among capital,
labor, and energy inputs is $\epsilon_{\sigma}\equiv\left.1\right/\left(1-\sigma\right)>0$.
If the elasticity of substitution is one, the CES production function
reduces to the Cobb\textendash Douglas production function, in which
case the relative use of inputs is invariant to technological change.
We assume a constant returns to scale technology but confirm the robustness
of results to this assumption in the appendix.

We consider factor-augmenting technologies to be unobserved and stochastic
components in the production function. In this case, it is difficult
to estimate the parameter directly using equation \eqref{eq: CES1}
since the production function is not only non-linear in the parameters
but also non-additive in the unobserved components.

Profit maximization entails equating the ratio of input prices to
the marginal rate of technical substitution:
\begin{eqnarray}
\frac{w}{r} & = & \left(\frac{a_{\ell}}{a_{k}}\right)^{\frac{\epsilon_{\sigma}-1}{\epsilon_{\sigma}}}\left(\frac{\ell}{k}\right)^{-\frac{1}{\epsilon_{\sigma}}},\label{eq: MRTS1a}\\
\frac{w}{v} & = & \left(\frac{a_{\ell}}{a_{e}}\right)^{\frac{\epsilon_{\sigma}-1}{\epsilon_{\sigma}}}\left(\frac{\ell}{e}\right)^{-\frac{1}{\epsilon_{\sigma}}}.\label{eq: MRTS1b}
\end{eqnarray}
These equations hold irrespective of the degree of markup or the degree
of returns to scale. Equations \eqref{eq: MRTS1a} and \eqref{eq: MRTS1b}
imply that the relative use of inputs varies according to the relative
factor-augmenting technologies. When the elasticity of substitution
is less (greater) than one, the relative quantities of inputs decrease
(increase) with a rise in the relative factor-augmenting technologies.
The elasticity of substitution can be estimated using equations \eqref{eq: MRTS1a}
and \eqref{eq: MRTS1b}, as described in the next section. The ratios
of factor-augmenting technologies can then be calculated as residuals,
but their levels cannot be calculated using these two equations alone.

As noted by \citet{Caselli_Coleman_AERPP02,Caselli_Coleman_AER06},
the system of three equations \eqref{eq: CES1}\textendash \eqref{eq: MRTS1b}
contains three unknowns ($a_{k}$, $a_{\ell}$, and $a_{e}$). Factor-augmenting
technologies can be derived from these equations as:
\begin{eqnarray}
a_{k} & = & \left(\frac{rk}{rk+w\ell+ve}\right)^{\frac{\epsilon_{\sigma}}{\epsilon_{\sigma}-1}}\left(\frac{y}{k}\right),\label{eq: ak1}\\
a_{\ell} & = & \left(\frac{w\ell}{rk+w\ell+ve}\right)^{\frac{\epsilon_{\sigma}}{\epsilon_{\sigma}-1}}\left(\frac{y}{\ell}\right),\label{eq: al1}\\
a_{e} & = & \left(\frac{ve}{rk+w\ell+ve}\right)^{\frac{\epsilon_{\sigma}}{\epsilon_{\sigma}-1}}\left(\frac{y}{e}\right).\label{eq: ae1}
\end{eqnarray}
Equations \eqref{eq: ak1}\textendash \eqref{eq: ae1} imply that
factor-augmenting technology is log proportional to the factor income
share and output per factor input. Gross output is equal to the sum
of factor incomes multiplied by markup ($\omega$); that is, $y=(rk+w\ell+ve)\omega$.
Energy-saving technological change can be measured as:
\begin{equation}
\Delta\ln a_{e}=\Delta\ln\left(\frac{y}{e}\right)+\frac{\epsilon_{\sigma}}{\epsilon_{\sigma}-1}\Delta\ln\left(\frac{ve}{rk+w\ell+ve}\right).\label{eq: d_ae1}
\end{equation}
The first term is a change in output per energy input. The second
term is inversely (directly) proportional to a change in the energy
share of income when the elasticity of substitution is less (greater)
than one. As is clear from the derivation, this approach does not
require specifying the functional forms of factor-augmenting technologies.

\subsection{Two-level CES}

In the one-level CES production function, the elasticity of substitution
between energy and non-energy inputs is assumed to be identical to
the elasticity of substitution between non-energy inputs. This assumption
can be relaxed by considering the following two-level nested CES production
function:
\begin{equation}
y=\left[\left[\left(a_{k}k\right)^{\varrho}+\left(a_{\ell}\ell\right)^{\varrho}\right]^{\frac{\varsigma}{\varrho}}+\left(a_{e}e\right)^{\varsigma}\right]^{\frac{1}{\varsigma}}\qquad\text{for}\quad\varsigma,\varrho<1.\label{eq: CES2}
\end{equation}
The elasticity of substitution between capital and labor is $\epsilon_{\varrho}\equiv1/(1-\varrho)>0$,
while the elasticity of substitution between energy and non-energy
inputs is $\epsilon_{\varsigma}\equiv1/(1-\varsigma)>0$. When the
two substitution parameters $\varsigma$ and $\varrho$ are identical,
the two-level CES production function \eqref{eq: CES2} reduces to
the one-level CES production function \eqref{eq: CES1}. This nesting
structure is most commonly used in the literature and tends to fit
the data best \citep{VanDerWerf_EE08}.

Profit maximization entails equating the ratio of input prices to
the marginal rate of technical substitution:
\begin{eqnarray}
\frac{w}{r} & = & \left(\frac{a_{\ell}}{a_{k}}\right)^{\varrho}\left(\frac{\ell}{k}\right)^{\varrho-1},\label{eq: MRTS2a}\\
\frac{w}{v} & = & \left[\left(\frac{a_{k}k}{a_{\ell}\ell}\right)^{\varrho}+1\right]^{\frac{\varsigma-\varrho}{\varrho}}\left(\frac{a_{\ell}}{a_{e}}\right)^{\varsigma}\left(\frac{\ell}{e}\right)^{\varsigma-1}.\label{eq: MRTS2b}
\end{eqnarray}
The first equation retains the same form as equation \eqref{eq: MRTS1a},
while the second equation becomes more involved than equation \eqref{eq: MRTS1b}.

The system of three equations \eqref{eq: CES2}\textendash \eqref{eq: MRTS2b}
contains three unknowns ($a_{k}$, $a_{\ell}$, and $a_{e}$). Factor-augmenting
technologies can be derived from these equations as:
\begin{eqnarray}
a_{k} & = & \left(\frac{rk+w\ell}{rk+w\ell+ve}\right)^{\frac{\epsilon_{\varsigma}}{\epsilon_{\varsigma}-1}}\left(\frac{rk}{rk+w\ell}\right)^{\frac{\epsilon_{\varrho}}{\epsilon_{\varrho}-1}}\left(\frac{y}{k}\right),\label{eq: ak2}\\
a_{\ell} & = & \left(\frac{rk+w\ell}{rk+w\ell+ve}\right)^{\frac{\epsilon_{\varsigma}}{\epsilon_{\varsigma}-1}}\left(\frac{w\ell}{rk+w\ell}\right)^{\frac{\epsilon_{\varrho}}{\epsilon_{\varrho}-1}}\left(\frac{y}{\ell}\right),\label{eq: al2}\\
a_{e} & = & \left(\frac{ve}{rk+w\ell+ve}\right)^{\frac{\epsilon_{\varsigma}}{\epsilon_{\varsigma}-1}}\left(\frac{y}{e}\right).\label{eq: ae2}
\end{eqnarray}
Equations \eqref{eq: ak2}\textendash \eqref{eq: ae2} imply again
that factor-augmenting technology is log proportional to the factor
income share and output per factor input. Energy-saving technological
change can be measured as:
\begin{equation}
\Delta\ln a_{e}=\Delta\ln\left(\frac{y}{e}\right)+\frac{\epsilon_{\varsigma}}{\epsilon_{\varsigma}-1}\Delta\ln\left(\frac{ve}{rk+w\ell+ve}\right).\label{eq: d_ae2}
\end{equation}
This equation takes the same form as equation \eqref{eq: d_ae1} but
with a different parameter.

\section{Estimation\label{sec: estimation}}

We first consider how we identify and estimate the elasticity of substitution.
We then describe how we evaluate the quantitative contribution of
specific factor inputs and factor-augmenting technologies to economic
growth.

\subsection{Elasticity of substitution}

The optimality conditions \eqref{eq: MRTS1a} and \eqref{eq: MRTS1b}
form the basis for estimating the substitution parameter in the one-level
CES production function \eqref{eq: CES1}. The optimality conditions
\eqref{eq: MRTS2a} and \eqref{eq: MRTS2b} form the basis for estimating
the substitution parameters in the two-level CES production function
\eqref{eq: CES2}.

\paragraph{One-level CES}

Let $c$, $s$, and $t$ denote the indices for countries, sectors,
and years, respectively. After taking logs in equations \eqref{eq: MRTS1a}
and \eqref{eq: MRTS1b} and taking differences over time, the estimating
equations can be derived as follows:
\begin{eqnarray}
\Delta\ln\left(\frac{w_{cst}}{r_{cst}}\right) & = & -\left(1-\sigma\right)\Delta\ln\left(\frac{\ell_{cst}}{k_{cst}}\right)+\Delta v_{1cst},\label{eq: moment_ces1a}\\
\Delta\ln\left(\frac{w_{cst}}{v_{cst}}\right) & = & -\left(1-\sigma\right)\Delta\ln\left(\frac{\ell_{cst}}{e_{cst}}\right)+\Delta v_{2cst}.\label{eq: moment_ces1b}
\end{eqnarray}
where the error terms comprise the relative factor-augmenting technologies;
that is, $v_{1cst}=\sigma\ln\left(\left.a_{\ell,cst}\right/a_{k,cst}\right)$
and $v_{2cst}=\sigma\ln\left(\left.a_{\ell,cst}\right/a_{e,cst}\right)$.

Three facts about the estimating equations are worth noting. First,
the observed and unobserved terms are additively separable in equations
\eqref{eq: moment_ces1a} and \eqref{eq: moment_ces1b}, which makes
it possible to estimate the substitution parameter. Second, any time-invariant
country- and sector-specific effects are eliminated by first differencing.
Even though there are persistent and substantial differences in the
unobserved characteristics across countries and sectors, such differences
are fully controlled for. Finally, the substitution parameter can
be over-identified when using the two equations, which makes it possible
to test the validity of the equations.

One more estimating equation can be derived from the optimality conditions
with respect to capital and energy inputs as follows:

\begin{equation}
\Delta\ln\left(\frac{r_{cst}}{v_{cst}}\right)=-\left(1-\sigma\right)\Delta\ln\left(\frac{k_{cst}}{e_{cst}}\right)+\Delta v_{3cst},\label{eq: moment_ces1c}
\end{equation}
where $v_{3cst}=\sigma\ln\left(\left.a_{k,cst}\right/a_{e,cst}\right)$.
We additionally estimate this equation for the purpose of robustness
checks.

\paragraph{Two-level CES}

One of the estimating equations can be derived from equation \eqref{eq: MRTS2a}
in the same way as above:
\begin{equation}
\Delta\ln\left(\frac{w_{cst}}{r_{cst}}\right)=-\left(1-\varrho\right)\Delta\ln\left(\frac{\ell_{cst}}{k_{cst}}\right)+\Delta v_{4cst},\label{eq: moment_ces2a}
\end{equation}
where $v_{4cst}=\varrho\ln\left(\left.a_{\ell,cst}\right/a_{k,cst}\right)$.

Another estimating equation cannot be derived directly from equation
\eqref{eq: MRTS2b} since the observed capital and labor quantities
are not separated from the unobserved capital- and labor-augmenting
technologies. However, equation \eqref{eq: MRTS2a} implies that the
ratio of capital- to labor-augmenting technology is log proportional
to the relative price and relative quantity of capital to labor; that
is, $\left(a_{k}k/a_{\ell}\ell\right)^{\varrho}=rk/w\ell$. After
substituting this into equation \eqref{eq: MRTS2b}, the additional
estimating equation can be derived as follows:
\begin{equation}
\Delta\ln\left(\frac{w_{cst}}{v_{cst}}\right)=-\frac{\varsigma-\varrho}{\varrho}\Delta\ln\left(\frac{w_{cst}\ell_{cst}}{r_{cst}k_{cst}+w_{cst}\ell_{cst}}\right)-\left(1-\varsigma\right)\Delta\ln\left(\frac{\ell_{cst}}{e_{cst}}\right)+\Delta v_{5cst},\label{eq: moment_ces2b}
\end{equation}
where $v_{5cst}=\varsigma\ln\left(\left.a_{\ell,cst}\right/a_{e,cst}\right)$.

Consequently, the observed and unobserved terms are additively separable,
and any time-invariant effects are differenced out in both equations
\eqref{eq: moment_ces2a} and \eqref{eq: moment_ces2b}. By virtue
of these equations, it is possible to estimate the substitution parameters
even when factor-augmenting technologies are neither observed nor
deterministic. The substitution parameters can be over-identified
from the two equations since there are three regressors for the two
parameters in the system of equations \eqref{eq: moment_ces2a} and
\eqref{eq: moment_ces2b}.

\paragraph{Identification}

Since the relative input quantities are presumably correlated with
relative factor-augmenting technologies, the estimating equations
are likely to involve endogenous regressors. If this is not taken
into account, the estimated elasticities of substitution will be biased.\footnote{The regressor in equation \eqref{eq: moment_ces1a} or \eqref{eq: moment_ces1b}
is likely to be positively correlated with the error term. The reason
for this is that, when $0<\epsilon_{\sigma}<1$ ($\epsilon_{\sigma}>1$),
the relative input quantities should theoretically be negatively (positively)
correlated with the relative factor-augmenting technologies, and each
relative factor-augmenting technology has a negative (positive) coefficient
in the error term. The coefficient of the regressor is the negative
of the inverse of $\epsilon_{\sigma}$. The elasticity of substitution
$\epsilon_{\sigma}$ will be overestimated regardless of whether $0<\epsilon_{\sigma}<1$
or $\epsilon_{\sigma}>1$.}

We address the endogeneity problem in three ways. First, we control
for time-invariant country- and sector-specific effects, as mentioned
above. Second, we control for non-linear time trends specific to each
country-sector pair in the relative factor-augmenting technologies.
These considerations amount to decomposing each error term as:
\begin{equation}
v_{cst}=\alpha_{c}+\alpha_{s}+\sum_{q}\psi_{qcs}t^{q}+u_{cst},
\end{equation}
where $\alpha_{c}$ is a country fixed effect, $\alpha_{s}$ is a
sector fixed effect, $u_{cst}$ is an idiosyncratic shock in country
$c$, sector $s$, and year $t$, and $q$ is the order of polynomials.
We omit the first subscript of the error terms in equations \eqref{eq: moment_ces1a}\textendash \eqref{eq: moment_ces2b}
to avoid notational clutter. If time-series data from a single country
were used, it would be difficult to isolate the effect of the relative
input quantities on the relative input prices from general time trends.
However, since panel data from many countries are used in our analysis,
it is possible to identify the elasticity of substitution among inputs
by exploiting the cross-sectional and time variation in the relative
input quantities.

Finally, we use the shift\textendash share instrument, also known
as the \citet{Bartik_bk91} instrument, to allow for correlations
between the changes in the relative input quantities and idiosyncratic
shocks to the relative factor-augmenting technologies. We treat all
right-hand-side variables, except time trends, as endogenous regressors.
For each endogenous variable, we use the shift\textendash share instrument:
\begin{equation}
\Delta\ln z_{cst}=\sum_{i\in\mathcal{I}_{s^{\prime}}}\frac{z_{cs^{\prime}it_{0}}}{\sum_{i^{\prime}\in\mathcal{I}_{s^{\prime}}}z_{cs^{\prime}i^{\prime}t_{0}}}\Delta\ln\left(\sum_{c\in\mathcal{C}}z_{cs^{\prime}it}\right)\qquad\text{for }z\in\left\{ k,\ell,e,w\ell,rk+w\ell\right\} ,
\end{equation}
where $i$ is an index for subsectors (or industries), $\mathcal{C}$
is a set of countries, $\mathcal{I}_{s}$ is a set of subsectors in
sector $s$, and $t_{0}$ is the first year of observation. The shift\textendash share
instrument for the endogenous regressor in the goods (service) sector
is constructed using the data from the service (goods) sector. Basically,
we exploit demand shocks in the service (goods) sector as a source
of exogenous variation in factor supply in the goods (service) sector
\citep{Raval_RJE19,Oberfield_Raval_EM21}. The shift\textendash share
instrument is the interaction between the initial industry shares
of inputs for each country and the growth rates of inputs for each
industry. The former measures local exposure to industry shocks, while
the latter measures global shocks to industries. The identification
assumption is that either the industry shares or growth rates in one
sector are uncorrelated with idiosyncratic shocks to relative factor-augmenting
technologies in another sector (conditional on non-linear time trends
specific to each country-sector pair).\footnote{\citet*{GoldsmithPinkham_Sorkin_Swift_AER20} show that the two-stage
least squares estimator with the shift\textendash share instrument
is numerically equivalent to the generalized method of moments estimator
using industry shares as excluded instruments. \citet*{Borusyak_Hull_Jaravel_wp20}
show that it is also numerically equivalent to the two-stage least
squares estimator using industry shocks as an excluded instrument
in the exposure-weighted industry-level regression. These results
imply that the shift\textendash share instruments can be valid under
certain conditions if either initial local industry shares or global
industry growth rates are exogenous.}

\paragraph{GMM}

The elasticity of substitution in the one-level CES production function
can be estimated using equations \eqref{eq: moment_ces1a} and \eqref{eq: moment_ces1b},
while the elasticities of substitution in the two-level CES production
function can be estimated using equations \eqref{eq: moment_ces2a}
and \eqref{eq: moment_ces2b}. In both cases, the same parameter appears
in different equations, and the error terms are correlated across
equations. Hence, it is more efficient to estimate the system of equations
jointly using the generalized method of moments (GMM). In doing so,
all the right-hand-side variables except time trends are treated as
endogenous variables using the shift\textendash share instruments,
five-year differences are used, and standard errors are clustered
at the country-sector level to allow for heteroscedasticity and serial
correlation. The GMM estimator is consistent as the sample size approaches
infinity.

\subsection{Growth accounting}

We utilize our measure of factor-augmenting technologies to evaluate
the quantitative contribution of specific factor inputs and factor-augmenting
technologies to economic growth for each country and sector. Technological
change is typically measured as the Solow residual, which is the portion
of growth in output not attributable to changes in factor inputs.
The limitation of this standard approach is that it does not tell
us the type of technological change. We take the approach one step
further by leveraging our measure of factor-augmenting technologies
as follows.

We decompose the rate of growth in gross output ($y$) into changes
due to the three components in factor inputs ($k$, $\ell$, and $e$)
and three components in factor-augmenting technologies ($a_{k}$,
$a_{\ell}$, and $a_{e}$). Furthermore, we measure the contribution
of the three pairs of factor inputs and factor-augmenting technologies
($a_{k}k$, $a_{\ell}\ell$, and $a_{e}e$) to economic growth. Given
the way in which we estimate the elasticities of substitution and
measure factor-augmenting technologies, the decomposition results
are invariant to the normalization of input quantities. The issue
that arises in the implementation of the decomposition is that there
is no simple transformation to make the CES production functions additively
separable in those components. In such a case, the decomposition results
depend on the order of the decomposition. To address this issue, we
use the Shapley decomposition \citep{Shorrocks_JoEI13}. Appendix
\ref{subsec: shapley} details the decomposition procedure.

\section{Data\label{sec: data}}

The analysis described so far requires data on the prices and quantities
of capital, labor, and energy inputs. The data used for the analysis
are drawn from the EU KLEMS database and the International Energy
Agency (IEA) database. The EU KLEMS database collects information
on the quantities of and incomes from capital, labor, and energy services
in major OECD countries from the year 1970 to 2005, while the IEA
database (World Energy Prices) collects information on energy prices
inclusive of taxes since the year 1978. The wage rate can be calculated
as the ratio of labor income to hours worked. The rental price of
capital can be calculated in standard ways. The internal rate of return
approach is used as in previous analyses of data from the EU KLEMS
when markup is assumed to be absent, whereas the external rate of
return approach is used when markup is considered to be present. Appendix
\ref{subsec: rental_price} details the calculation procedure. All
the variables measured in monetary terms are converted into the value
of U.S. dollars in the year 1995.

The EU KLEMS database is created from information collected by national
statistical offices and is grounded in national accounts statistics.
The March 2008 version is used for the analysis because later versions
contain no information on energy. All countries, industries, and years,
for which the data needed for the estimation are available, are included
in the sample. The economy is composed of two sectors. The goods sector
consists of 17 industries, while the service sector consists of 12
industries.\footnote{The goods sector includes five broad categories of industries: agriculture,
hunting, forestry, and fishing; mining and quarrying; manufacturing;
electricity, gas and water supply; and construction. The service sector
includes nine broad categories of industries: wholesale and retail
trade; hotels and restaurants; transport and storage, and communication;
financial intermediation; real estate, renting, and business activities;
public administration and defense and compulsory social security;
education; health and social work; and other community, and social
and personal services.} Consequently, our sample comprises 610 country-sector-year observations
from 12 countries: Austria, the Czech Republic, Denmark, Finland,
Germany, Italy, Japan, the Netherlands, Portugal, Sweden, the United
Kingdom, and the United States.\footnote{The results remain almost unchanged if the Czech Republic is excluded
from the sample.}

When we calculate the gross output, we use the estimates of markup
by \citet{DeLoecker_Eeckhout_wp20}. When we calculate the input prices
and quantities, we adjust for the variation in input composition over
time to the extent possible. For this purpose, we make full use of
detailed information on capital, labor, and energy input components
in the EU KLEMS and IEA databases. Appendix \ref{subsec: composition}
details the adjustment procedure, including further description of
the data used.

When we compare our measure and conventional measures of energy-saving
technological change, we construct the conventional measures using
the amount of government spending on energy-related R\&D and the number
of energy-related patents (i.e., patents on climate change mitigation).
Both types of information are readily available from the OECD.Stat
database. When we examine the relationship between energy-saving technological
change and energy resource abundance, we measure the abundance of
energy resources using the self-sufficiency rate of energy supply,
defined as the ratio of the indigenous production of total primary
energy to the total primary energy supply. The self-sufficiency rate
of energy supply can be calculated from the IEA database (World Energy
Balances).

\section{Results\label{sec: results}}

We start this section by presenting the estimates of the elasticities
of substitution among capital, labor, and energy inputs. We then discuss
the international trends and differences in energy-saving technological
change. We end this section by evaluating the quantitative contribution
of energy input and energy-saving technology to economic growth.

\subsection{Production function estimates}

Table \ref{tab: elasticity1} reports the estimates of the elasticities
of substitution in the one- and two-level CES production functions.
The estimates vary slightly depending on the way in which time trends
are controlled for. The estimated elasticities of substitution ($\epsilon_{\sigma}$)
are, however, less than one in all specifications, ranging from 0.43
to 0.68, in the one-level CES production function. The same applies
to the two-level CES production function. The estimated elasticities
of substitution between energy and non-energy inputs ($\epsilon_{\varsigma}$)
range from 0.27 to 0.44, while the estimated elasticity of substitution
($\epsilon_{\varrho}$) between capital and labor ranges from 0.41
to 0.75. Since it is desirable to add extensive controls for time
trends to ensure instrument exogeneity, our preferred specification
is the one in which quadratic trends specific to each country-sector
pair are added. In this specification, the estimated elasticity of
substitution in the one-level CES production function is 0.43, while
the estimated elasticities of substitution in the two-level CES production
function are 0.41 both between energy and non-energy inputs and between
capital and labor. Thus, the estimated elasticities of substitution
between energy and non-energy inputs are approximately the same between
the one- and two-level CES production functions. Overall, our estimates
are within the range of estimates reported in \citet{VanDerWerf_EE08}
and \citet{Raval_RJE19}.\footnote{\citet{VanDerWerf_EE08} provides the estimates of the elasticities
of substitution among capital, labor, and energy inputs by country
or industry. His estimates range from 0.17 to 0.65 in the two-level
CES production function similar to equation \eqref{eq: CES2}. \citet{Raval_RJE19}
provide the estimates of the elasticity of substitution between capital
and labor by industry or year in the manufacturing sector of the United
States. Most of his estimates fall within the range between 0.15 and
0.75.}

\begin{table}[h]
\caption{Elasticities of substitution in the one- and two-level CES production
functions\label{tab: elasticity1}}

\begin{centering}
\begin{tabular}{cccc}
\hline 
 & \multicolumn{3}{c}{One-level CES}\tabularnewline
\multirow{2}{*}{$\epsilon_{\sigma}$} & 0.681 & 0.532 & 0.432\tabularnewline
 & (0.193) & (0.133) & (0.087)\tabularnewline
\cline{2-4} \cline{3-4} \cline{4-4} 
 & \multicolumn{3}{c}{Two-level CES}\tabularnewline
\multirow{2}{*}{$\epsilon_{\varsigma}$} & 0.274 & 0.441 & 0.408\tabularnewline
 & (0.200) & (0.131) & (0.108)\tabularnewline
\multirow{2}{*}{$\epsilon_{\varrho}$} & 0.748 & 0.513 & 0.414\tabularnewline
 & (0.120) & (0.511) & (0.149)\tabularnewline
\cline{2-4} \cline{3-4} \cline{4-4} 
\multirow{3}{*}{time trends} & country & country & country\tabularnewline
 & \multirow{2}{*}{linear} & sector & sector\tabularnewline
 &  & linear & quadratic\tabularnewline
\hline 
\end{tabular}
\par\end{centering}
\textit{\footnotesize{}Notes}{\footnotesize{}: Standard errors in
parentheses are clustered at the country-sector level. All specifications
use the shift\textendash share instruments. The specification in the
first column controls for linear time trends specific to each country.
The specifications in the second and third columns control for linear
and quadratic time trends specific to each country-sector pair, respectively.}{\footnotesize\par}
\end{table}

The estimated elasticities of substitution differ significantly from
one in the one-level CES production function. The Wald statistic under
the null hypothesis that the elasticity of substitution equals one
is 42.6 with a \textit{p}-value of zero in the preferred specification.
This result indicates that the Cobb-Douglas production function can
be rejected against the CES production function. At the same time,
the estimated elasticities of substitution between capital and labor
do not differ significantly from those between energy and non-energy
inputs in the two-level CES production function. The Wald statistic
under the null hypothesis that the two substitution parameters are
identical is 0.001 with a \textit{p}-value of 0.975 in the preferred
specification. This result indicates that the one-level CES production
function cannot be rejected against the two-level CES production function.

\begin{table}[h]
\caption{Elasticities of substitution estimated with different equations\label{tab: elasticity2}}

\begin{centering}
\begin{tabular}{cccc}
\hline 
 & \multicolumn{3}{c}{One-level CES}\tabularnewline
\multirow{2}{*}{$\epsilon_{\sigma}$} & 0.410 & 0.444 & 0.422\tabularnewline
 & (0.151) & (0.099) & (0.125)\tabularnewline
\multirow{1}{*}{equations} & \multirow{1}{*}{eq. \eqref{eq: moment_ces1a}} & \multirow{1}{*}{eq. \eqref{eq: moment_ces1b}} & eq. \eqref{eq: moment_ces1c}\tabularnewline
\hline 
\end{tabular}
\par\end{centering}
\textit{\footnotesize{}Notes}{\footnotesize{}: Standard errors in
parentheses are clustered at the country-sector level. All specifications
use the shift\textendash share instruments and control for quadratic
time trends specific to each country-sector pair.}{\footnotesize\par}
\end{table}

The estimated elasticities of substitution are similar irrespective
of which equation is used for estimation. Table \ref{tab: elasticity2}
reports the estimates of the elasticities of substitution in the one-level
CES production function when equations \eqref{eq: moment_ces1a},
\eqref{eq: moment_ces1b}, and \eqref{eq: moment_ces1c} are used
separately. The estimated elasticities of substitution among capital,
labor, and energy inputs fall into a tight range of 0.41 to 0.44 in
the preferred specification. The Wald statistic under the null hypothesis
that the estimated elasticities of substitution are the same between
the first (second) and second (third) columns is 0.04 (0.04) with
a \textit{p}-value of 0.85 (0.84). These results suggest the robustness
of the estimates across different CES nesting structures.

Two types of test statistics indicate that the shift\textendash share
instruments used in our analysis are valid if quadratic trends specific
to each country-sector pair are included. First, the first-stage \textit{F}
statistics under the null hypothesis that the shift\textendash share
instruments are irrelevant are 11.3 for $\ell/k$, 14.7 for $\ell/e$,
and 23.0 for $k/e$ in the one-level CES production function, while
they are 11.3 for $\ell/k$, 8.3 for $\ell/e$, and 4.6 for $w\ell/(rk+w\ell)$
in the two-level CES production function. Second, the \textit{J} statistics
under the null hypothesis that over-identifying restrictions are valid
are 0.10 with a \textit{p}-value of 0.746 in the one-level CES production
function and 0.12 with a \textit{p}-value of 0.725 in the two-level
CES production function.

We end this subsection by mentioning that the marginal rate of technical
substitution equals the ratio of input prices even if the product
market is not competitive. Accordingly, the estimating equations hold
irrespective of the presence or absence of markup. Technically, however,
many specifications assume the absence of markup so that the rental
price of capital can be calculated using the internal rate of return
approach under the assumption of competitive markets. Nevertheless,
estimating equation \eqref{eq: moment_ces1b} is robust to the presence
of markup since it does not depend on the rental price of capital.
Reassuringly, the estimated elasticities of substitution change little
irrespective of whether both equations \eqref{eq: moment_ces1a} and
\eqref{eq: moment_ces1b} are used or only equation \eqref{eq: moment_ces1b}
is used. Moreover, the first-stage \textit{F} statistic is fairly
large in the latter case. Consequently, our estimates are robust to
product market imperfections. We also report the results when estimating
the elasticity of substitution by sector in the appendix.

\subsection{Energy-saving technological change}

\subsubsection{International trends}

\begin{figure}[H]
\caption{Factor-augmenting technological change in the goods sector\label{fig: AkAlAe1_goods}}

\begin{centering}
\subfloat[Austria]{
\centering{}\includegraphics[scale=0.4]{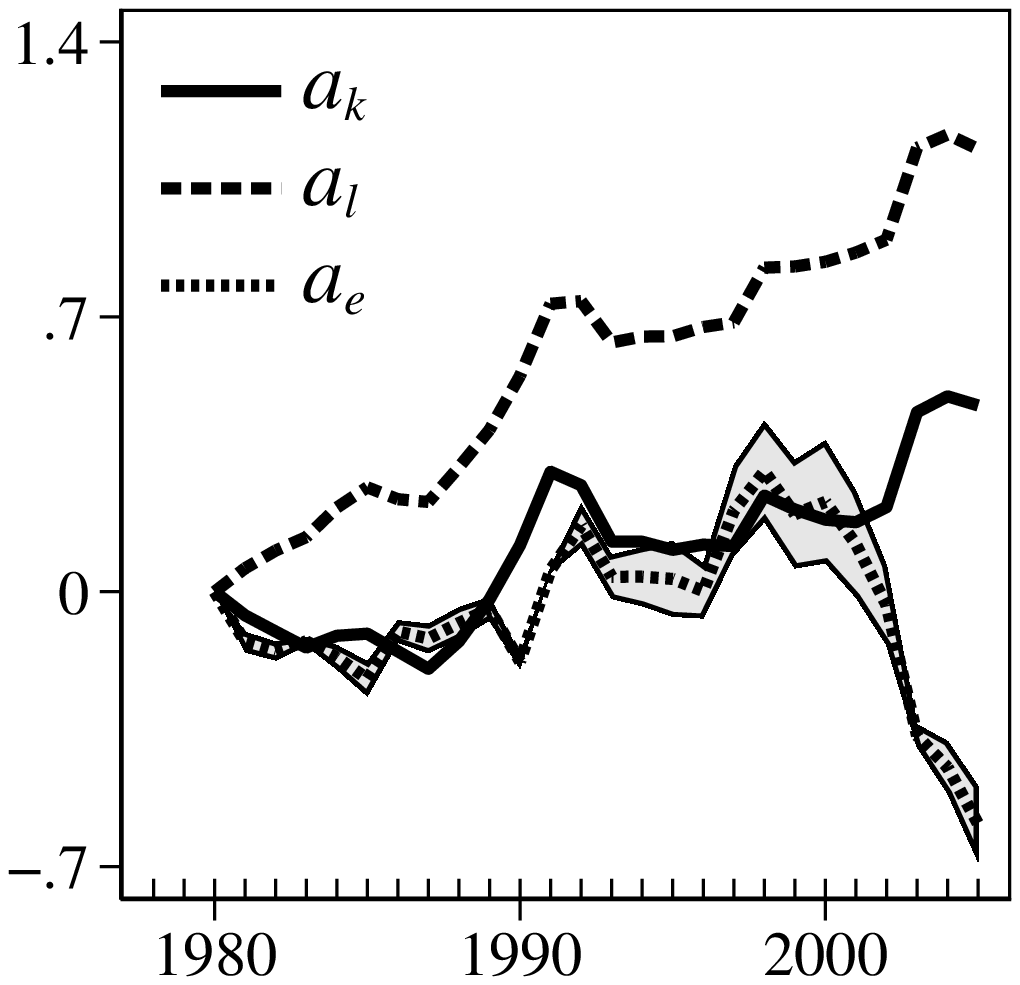}}\subfloat[Czech Republic]{
\centering{}\includegraphics[scale=0.4]{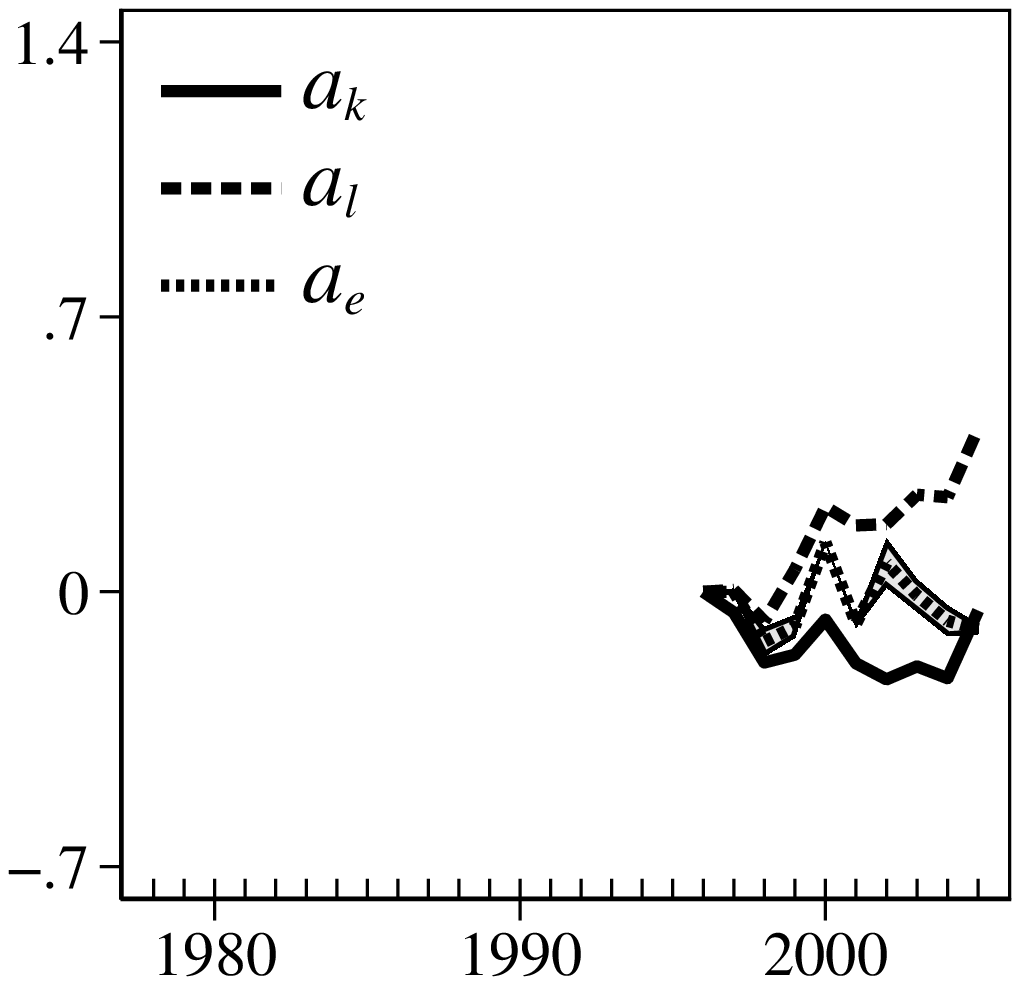}}\subfloat[Denmark]{
\centering{}\includegraphics[scale=0.4]{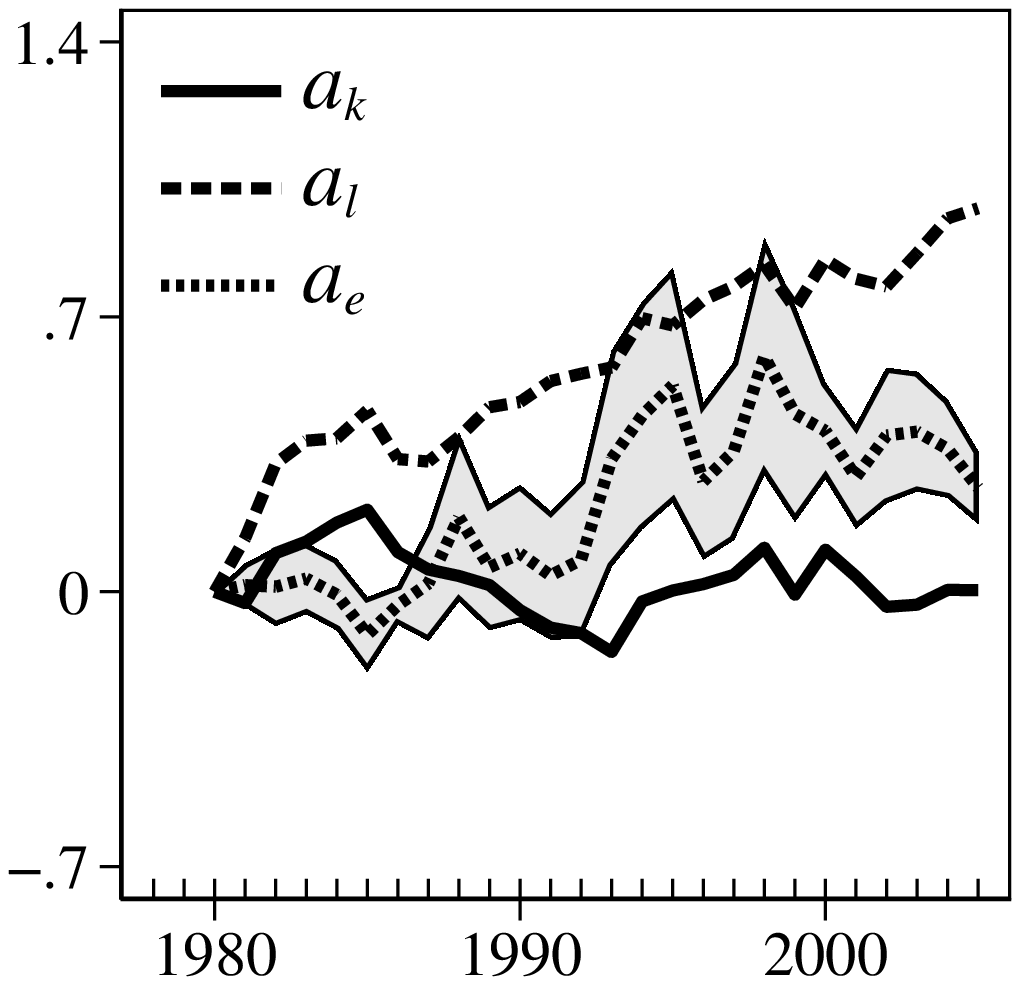}}\subfloat[Finland]{
\centering{}\includegraphics[scale=0.4]{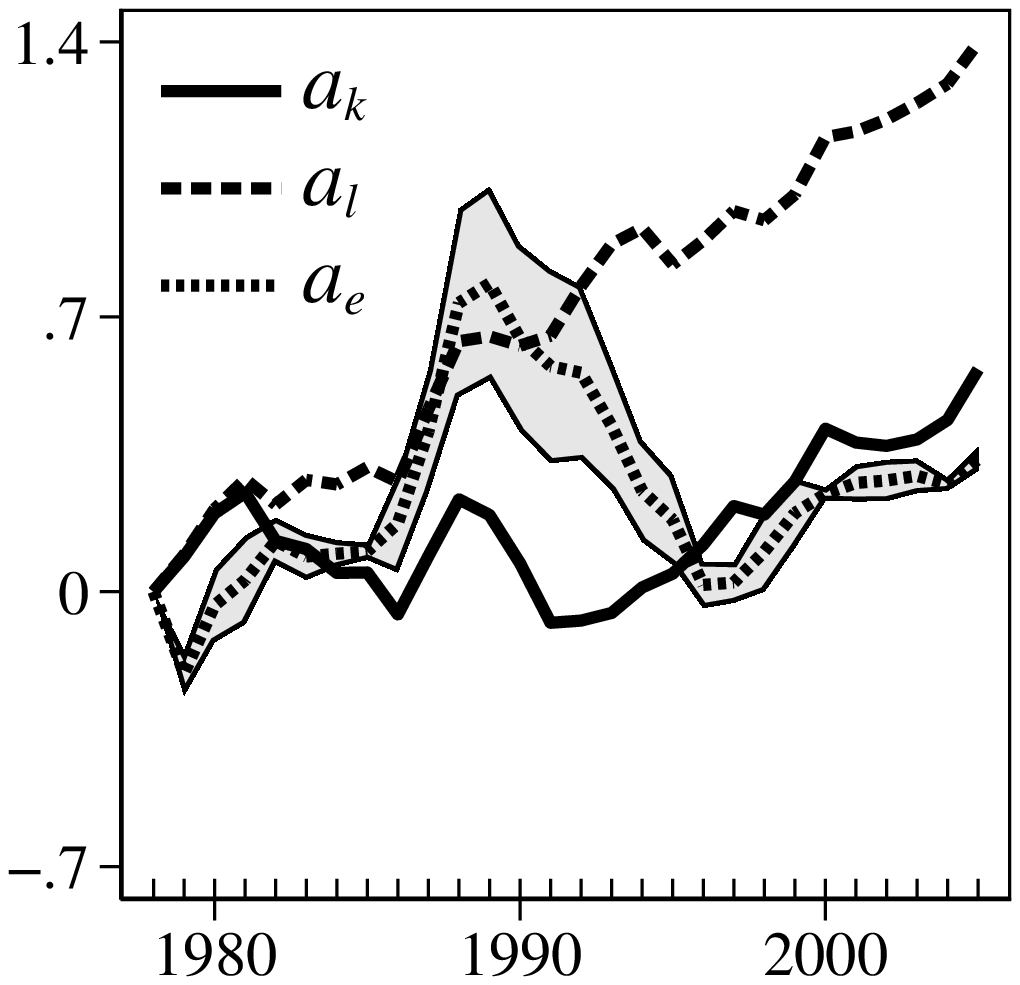}}
\par\end{centering}
\begin{centering}
\subfloat[Germany]{
\centering{}\includegraphics[scale=0.4]{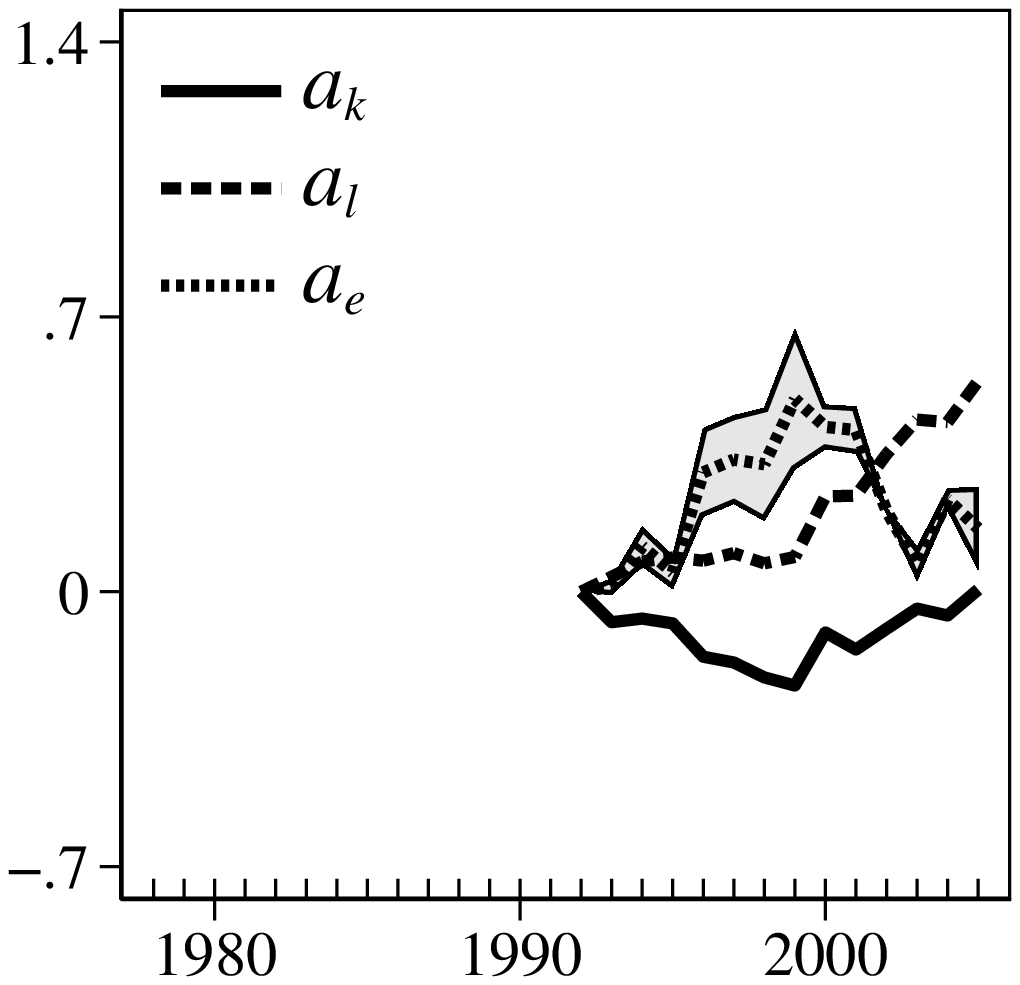}}\subfloat[Italy]{
\centering{}\includegraphics[scale=0.4]{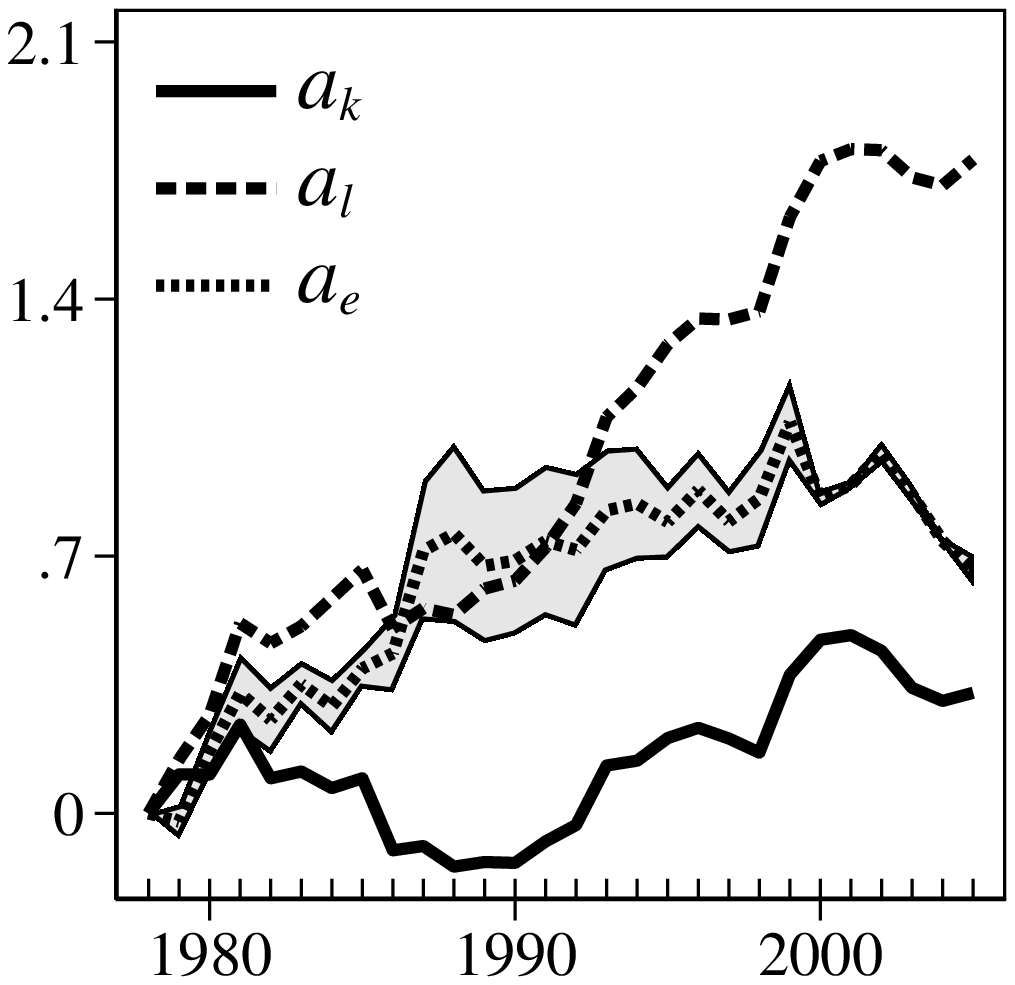}}\subfloat[Japan]{
\centering{}\includegraphics[scale=0.4]{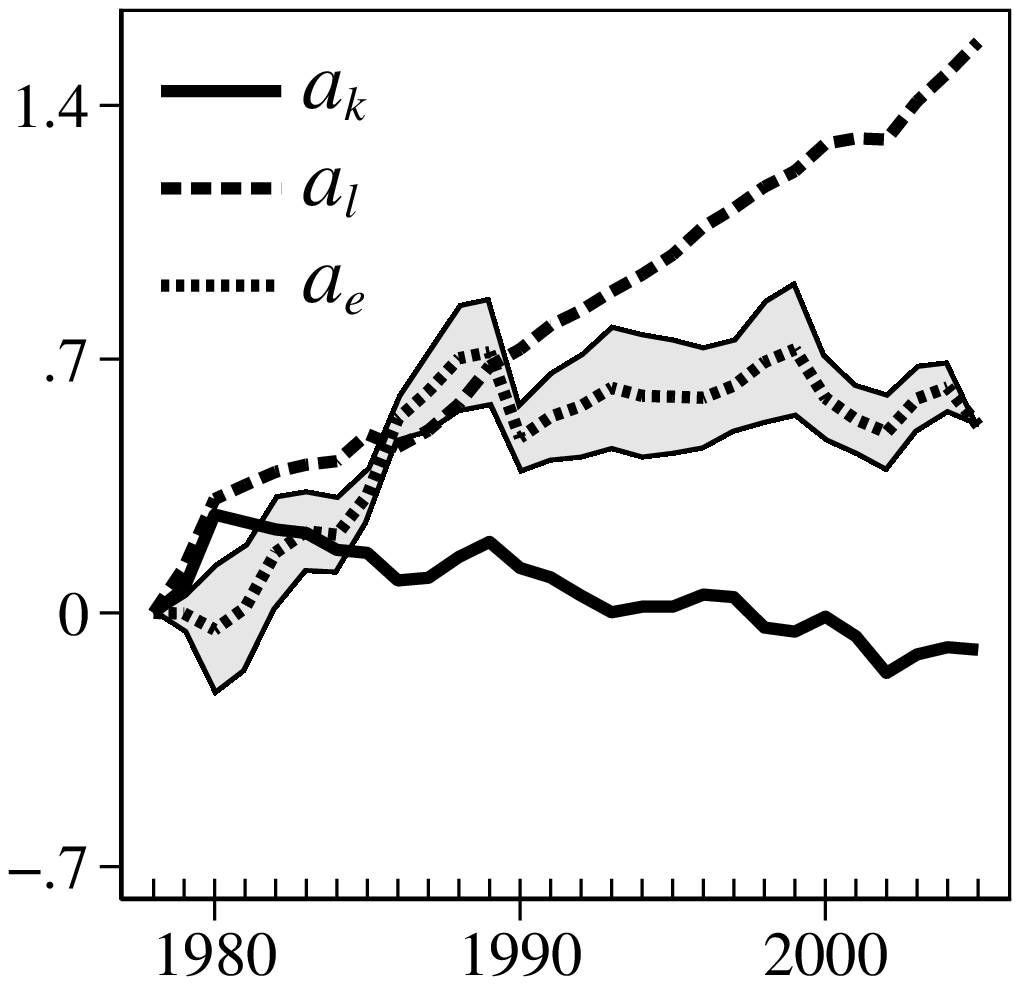}}\subfloat[Netherlands]{
\centering{}\includegraphics[scale=0.4]{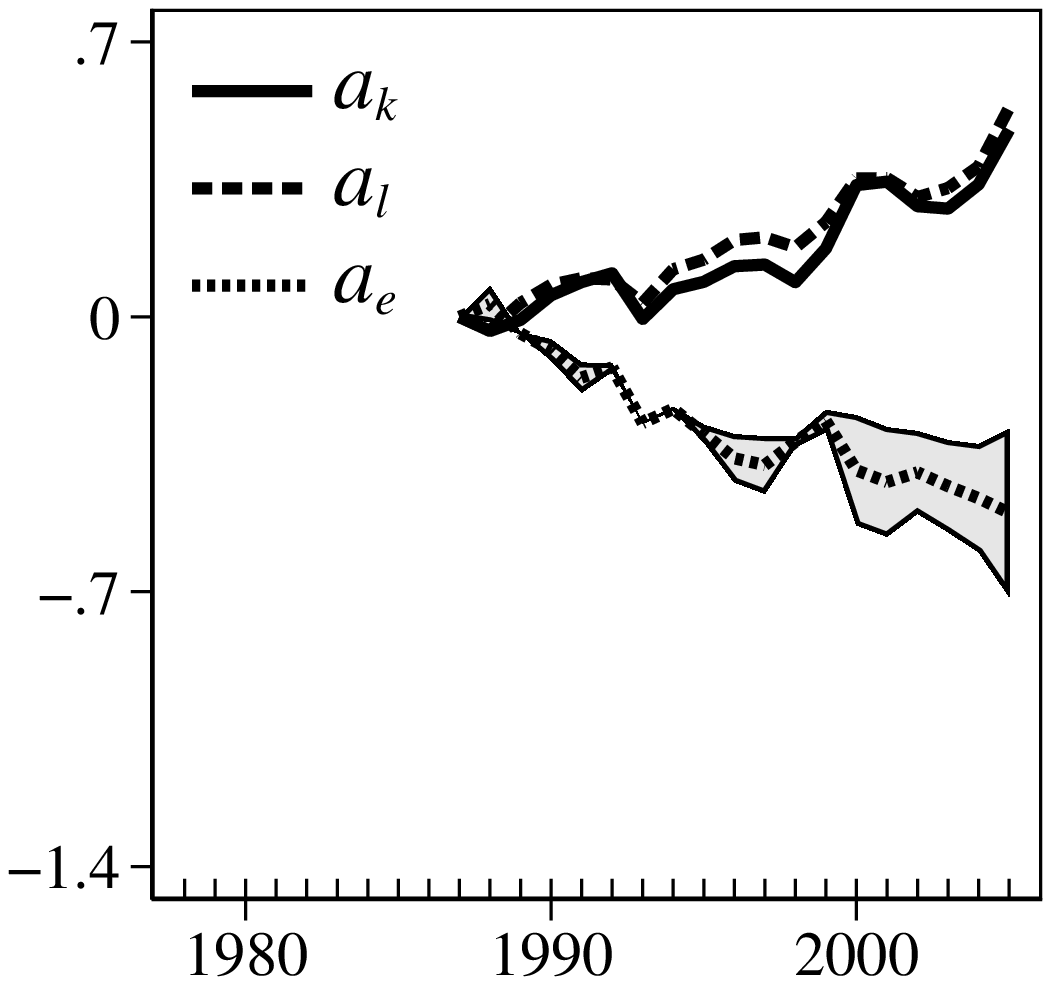}}
\par\end{centering}
\begin{centering}
\subfloat[Portugal]{
\centering{}\includegraphics[scale=0.4]{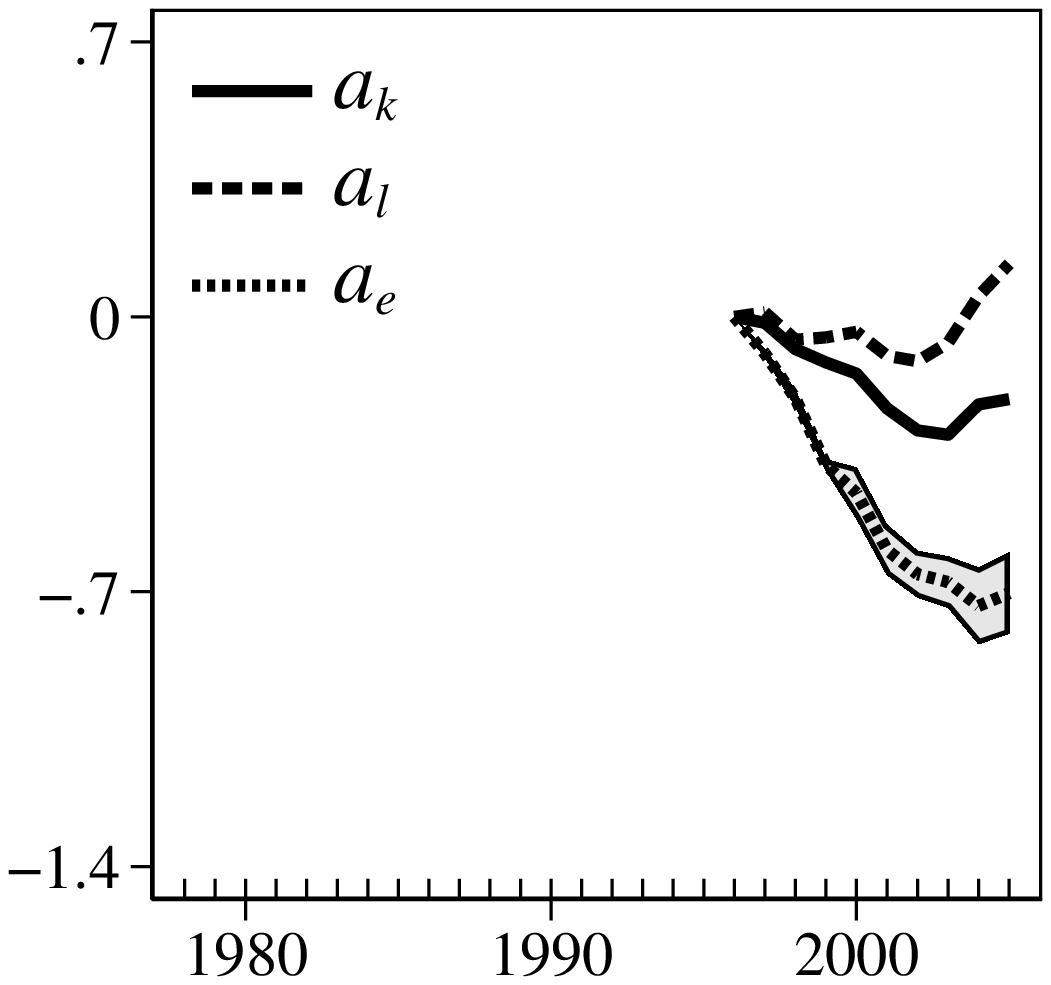}}\subfloat[Sweden]{
\centering{}\includegraphics[scale=0.4]{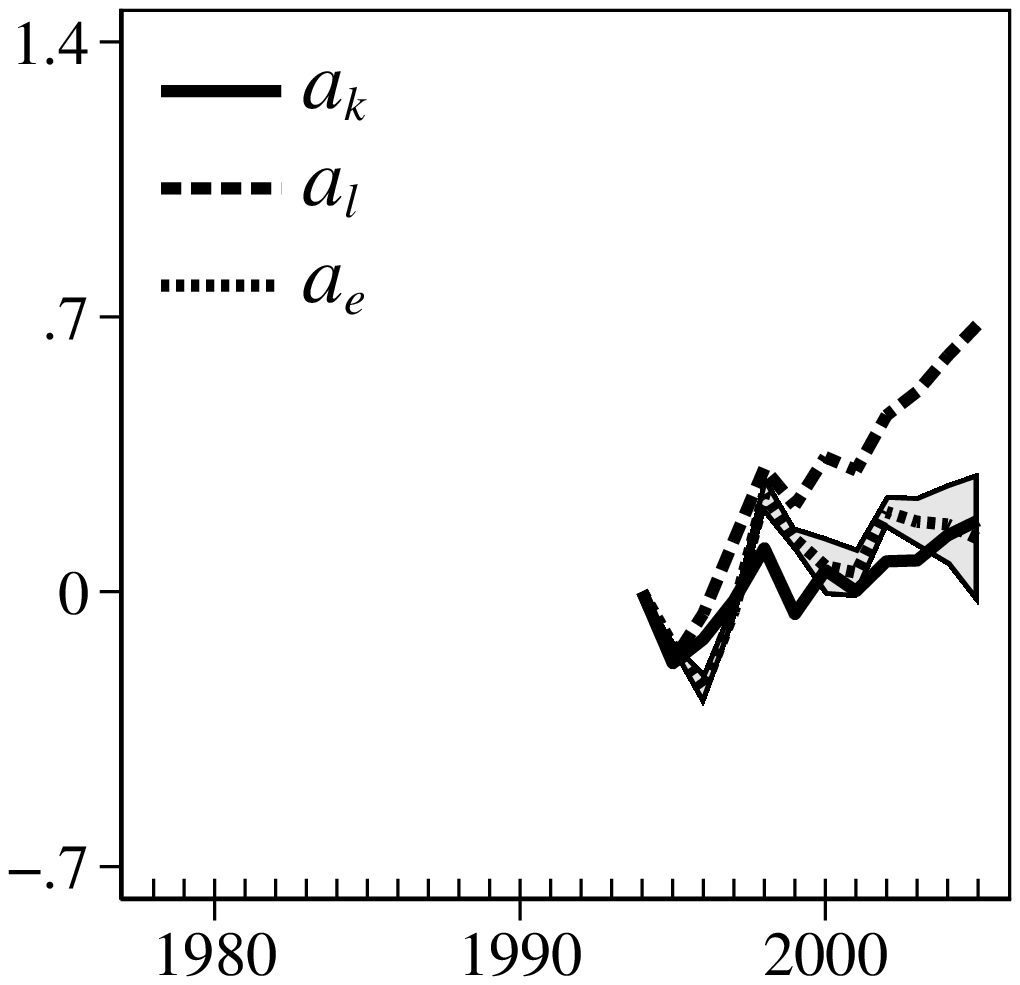}}\subfloat[United Kingdom]{
\centering{}\includegraphics[scale=0.4]{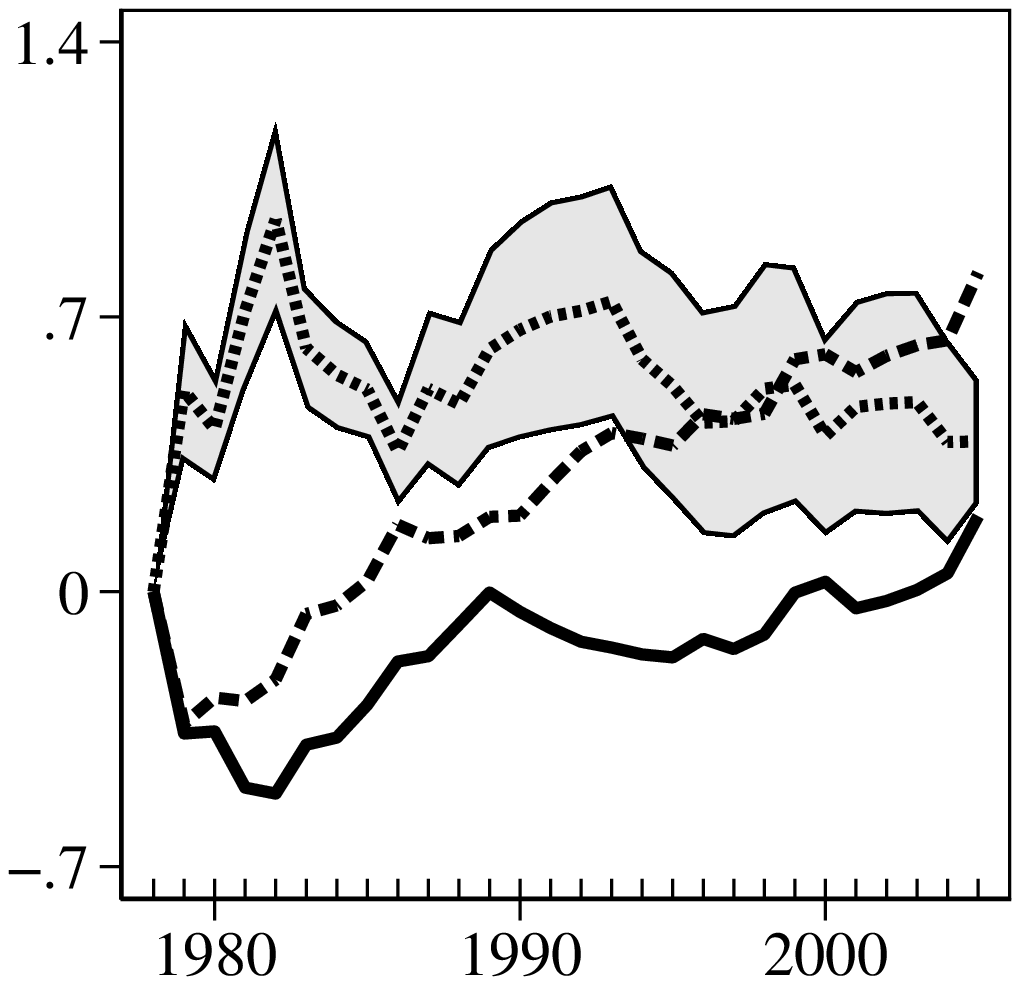}}\subfloat[United States]{
\centering{}\includegraphics[scale=0.4]{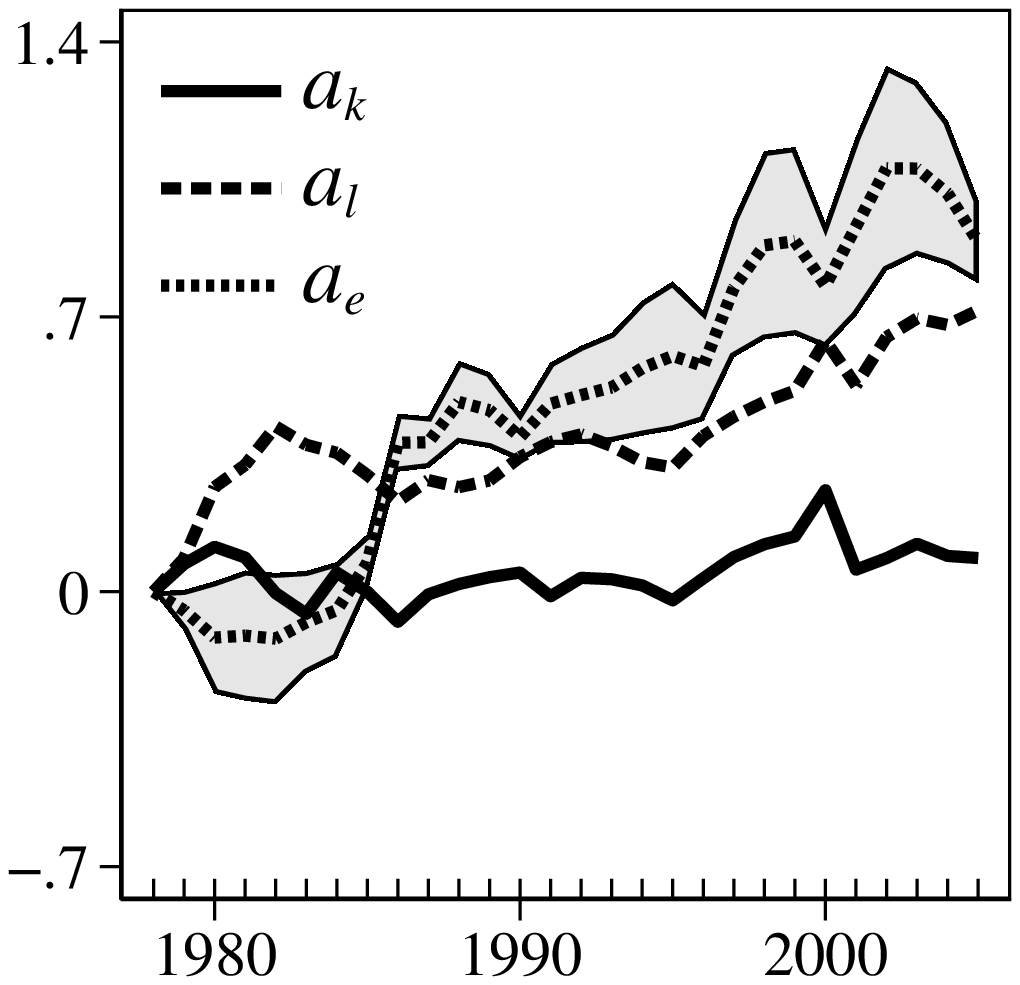}}
\par\end{centering}
\textit{\footnotesize{}Notes}{\footnotesize{}: The solid, dashed,
and dotted lines represent capital-, labor-, and energy-augmenting
technologies ($a_{k}$, $a_{\ell}$, and $a_{e}$), respectively.
The shaded area represents the 90 percent confidence interval for
$a_{e}$. All series are expressed as log differences relative to
the first year of observations.}{\footnotesize\par}
\end{figure}

\begin{figure}[H]
\caption{Factor-augmenting technological change in the service sector\label{fig: AkAlAe1_service}}

\begin{centering}
\subfloat[Austria]{
\centering{}\includegraphics[scale=0.4]{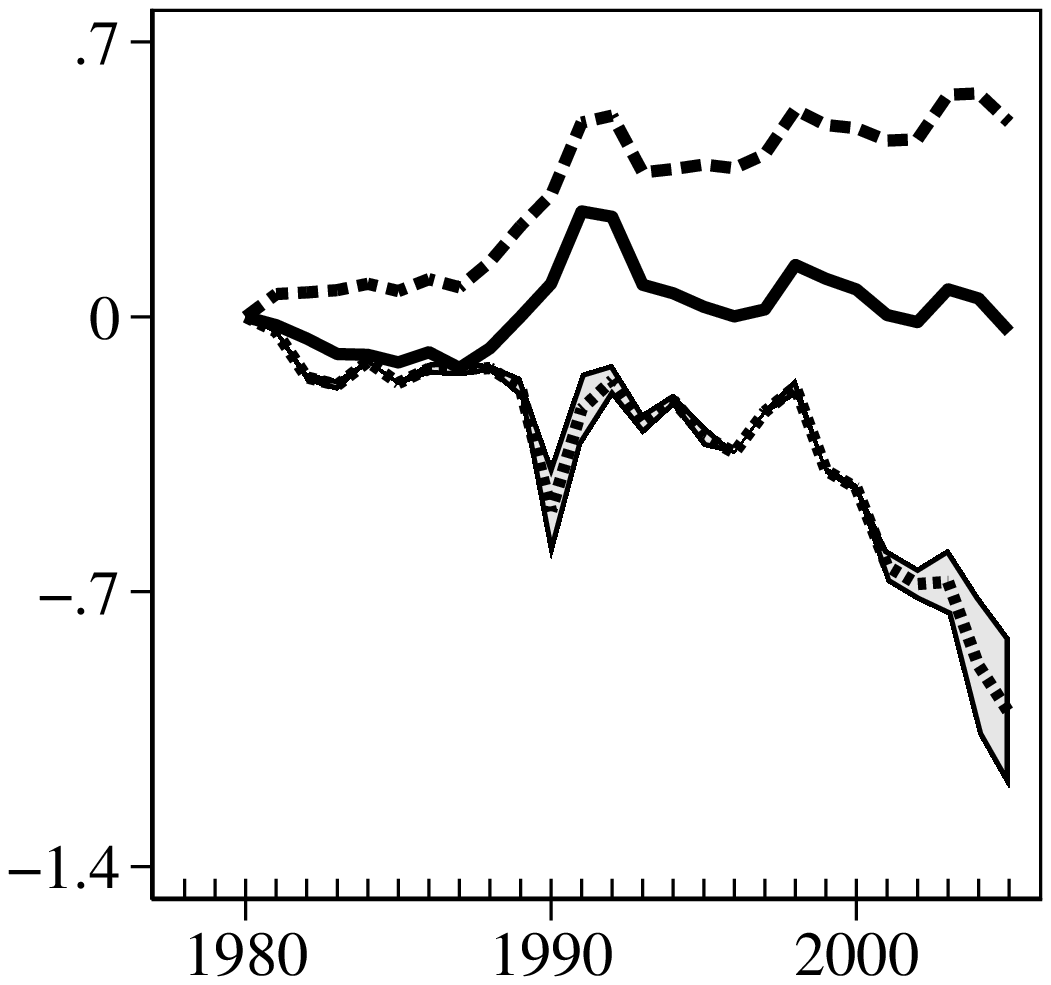}}\subfloat[Czech Republic]{
\centering{}\includegraphics[scale=0.4]{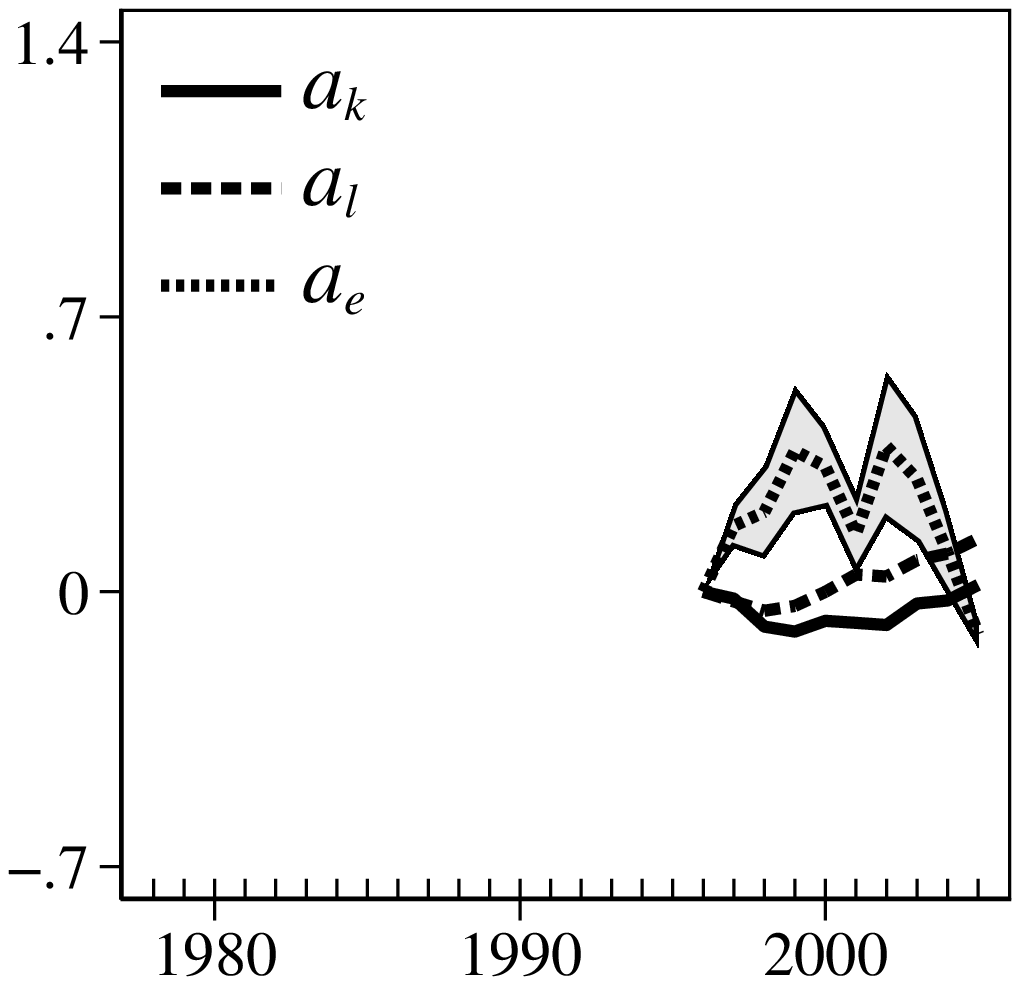}}\subfloat[Denmark]{
\centering{}\includegraphics[scale=0.4]{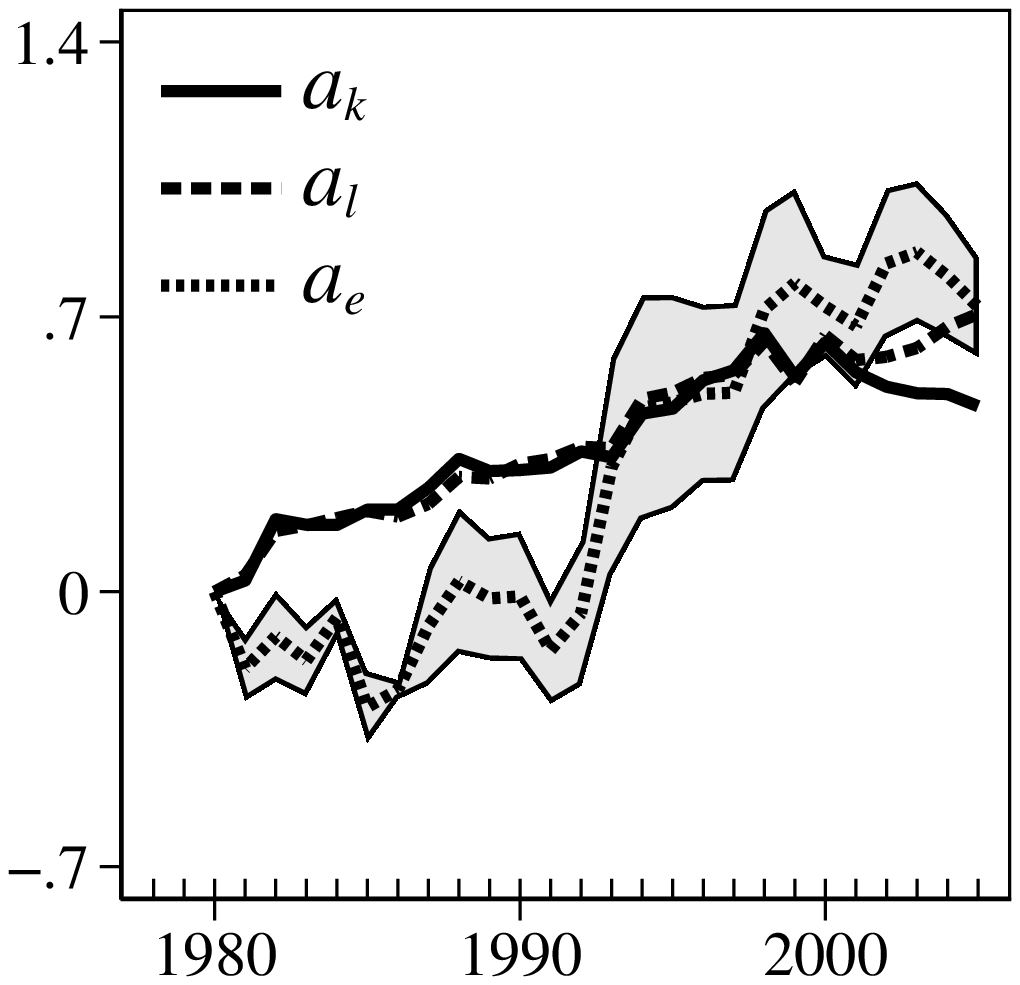}}\subfloat[Finland]{
\centering{}\includegraphics[scale=0.4]{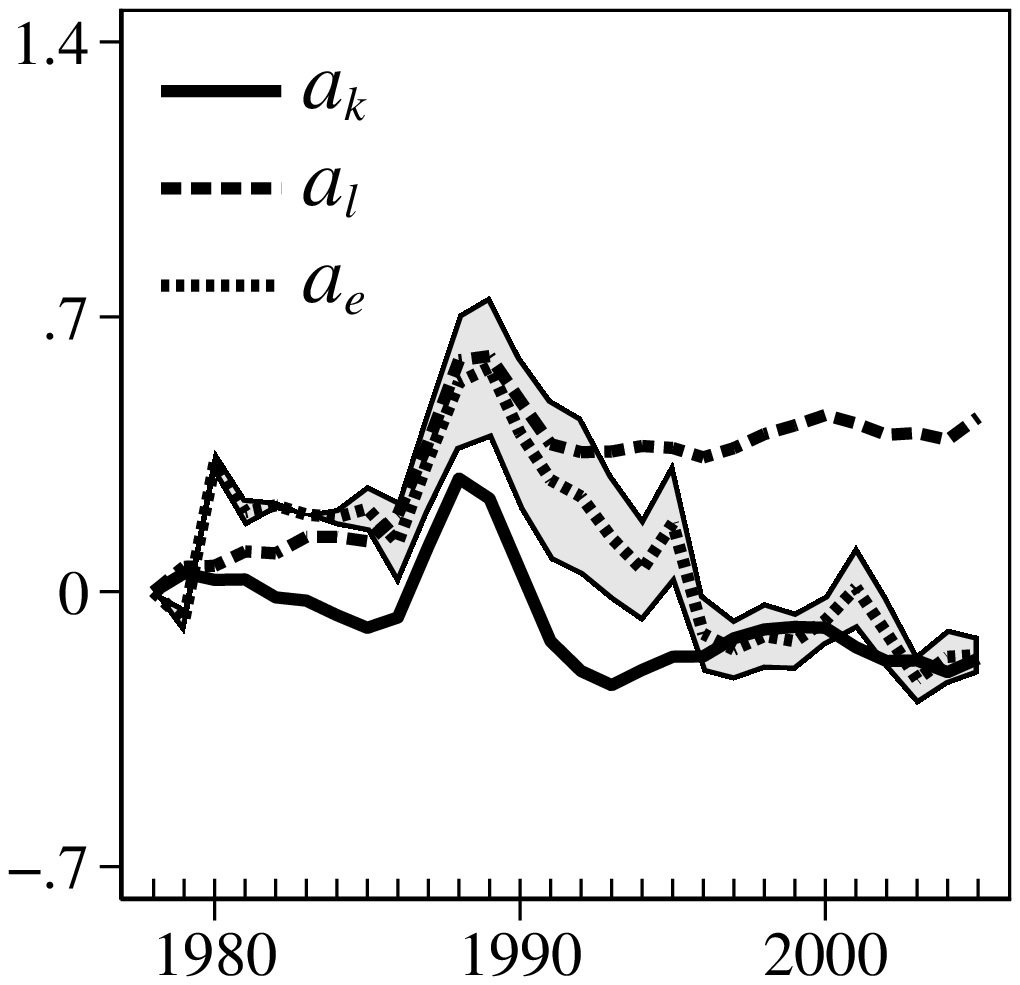}}
\par\end{centering}
\begin{centering}
\subfloat[Germany]{
\centering{}\includegraphics[scale=0.4]{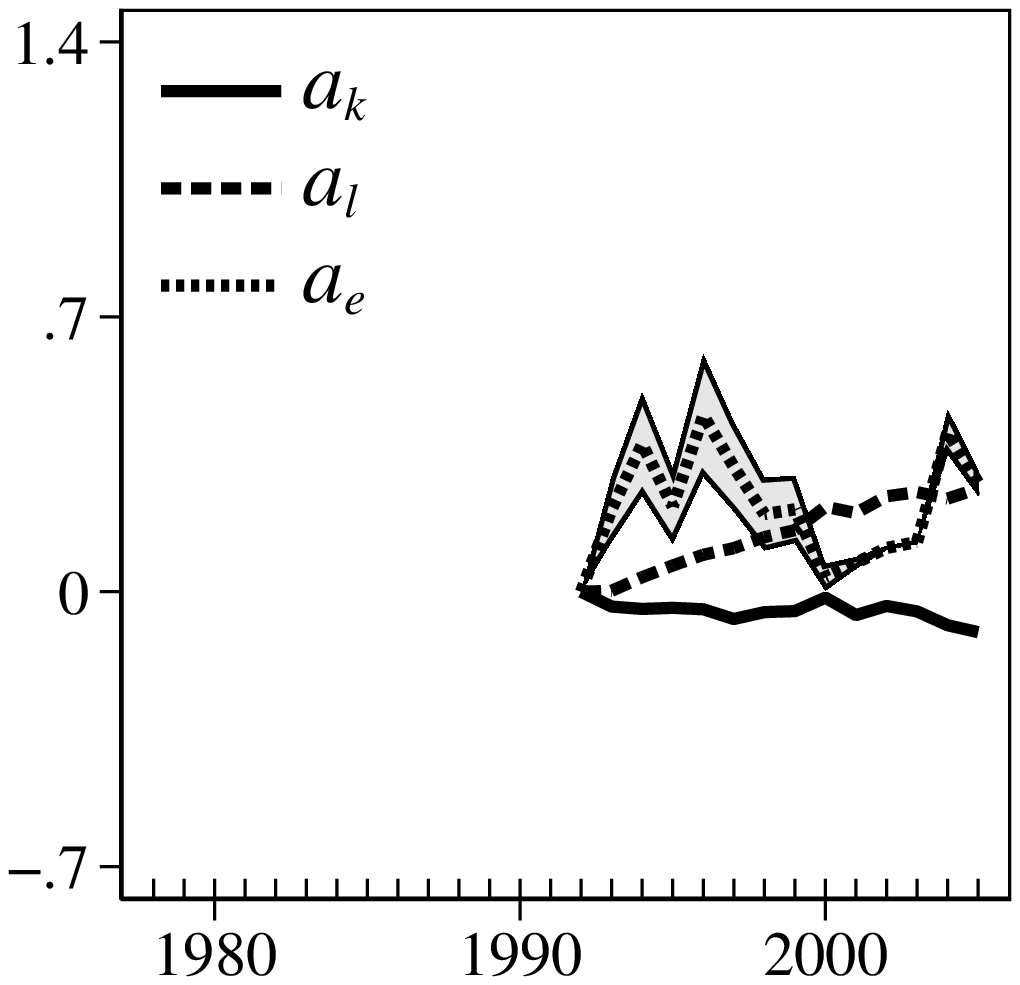}}\subfloat[Italy]{
\centering{}\includegraphics[scale=0.4]{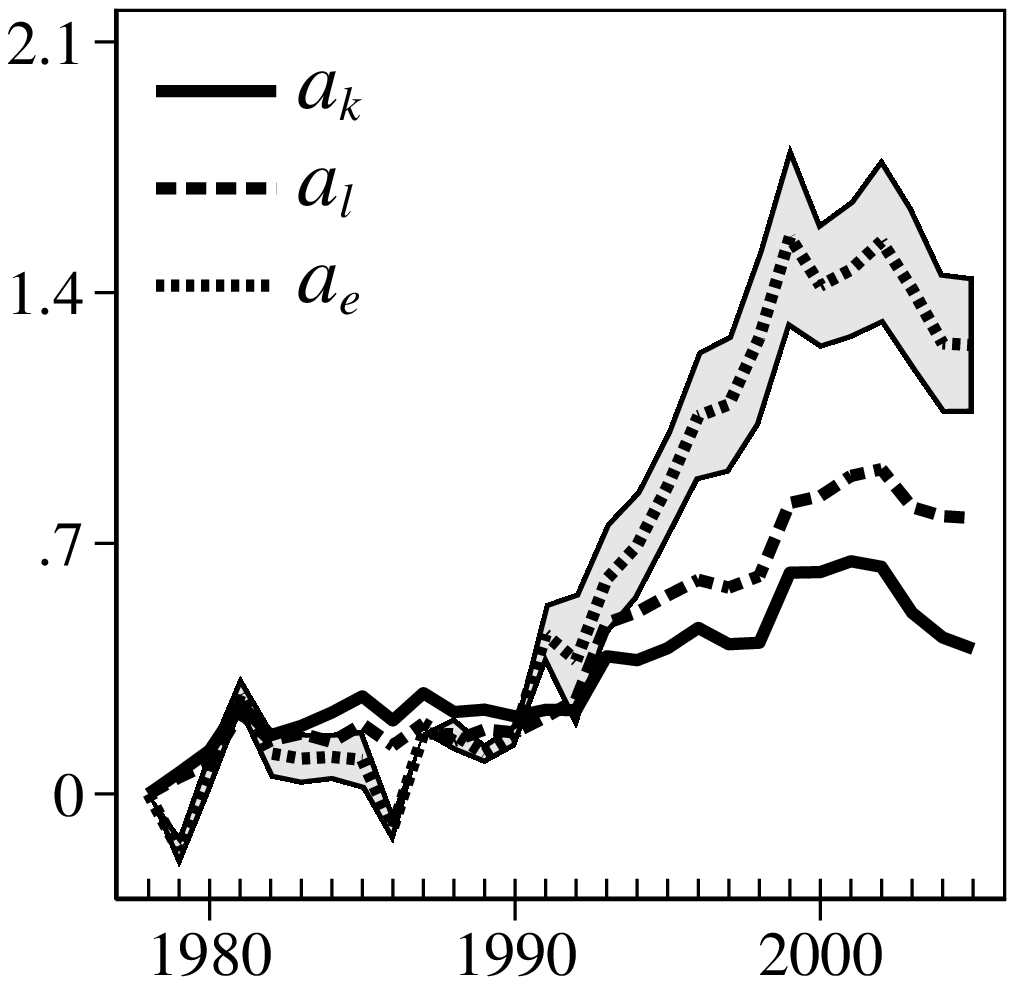}}\subfloat[Japan]{
\centering{}\includegraphics[scale=0.4]{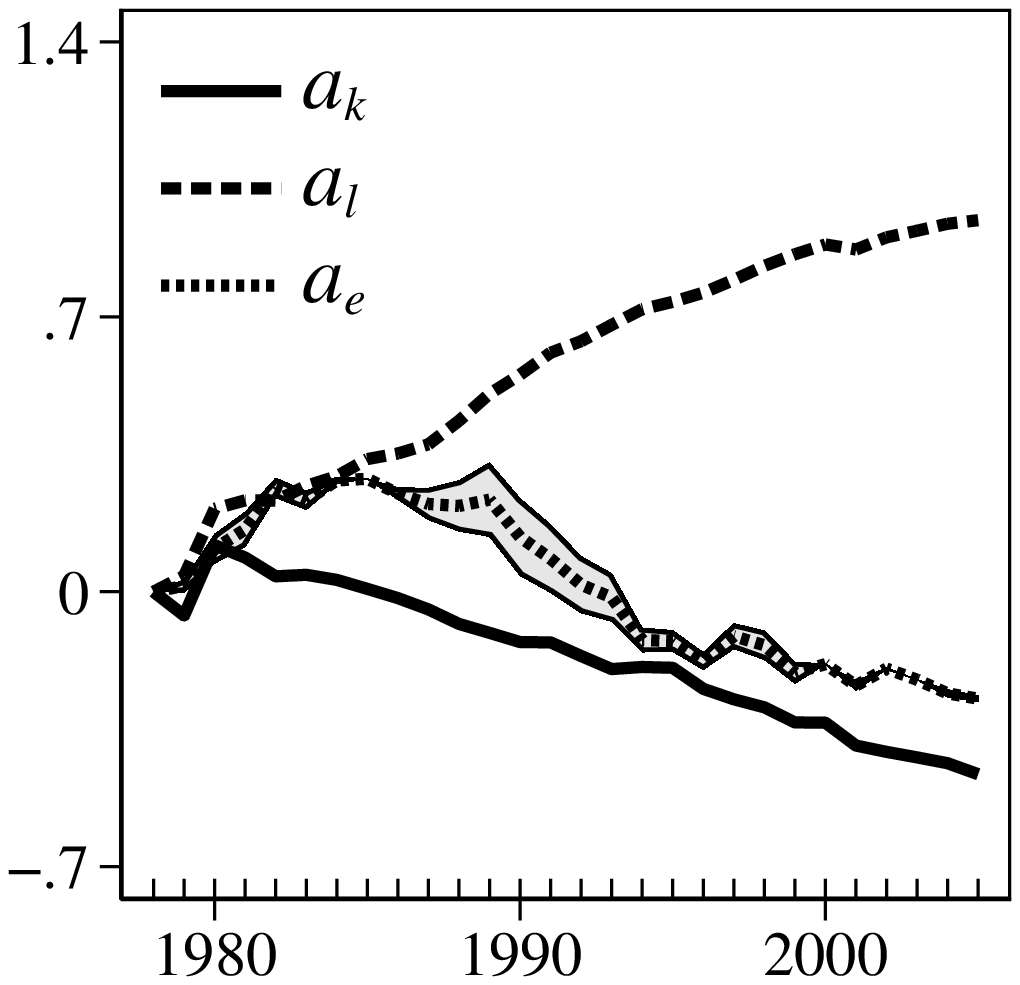}}\subfloat[Netherlands]{
\centering{}\includegraphics[scale=0.4]{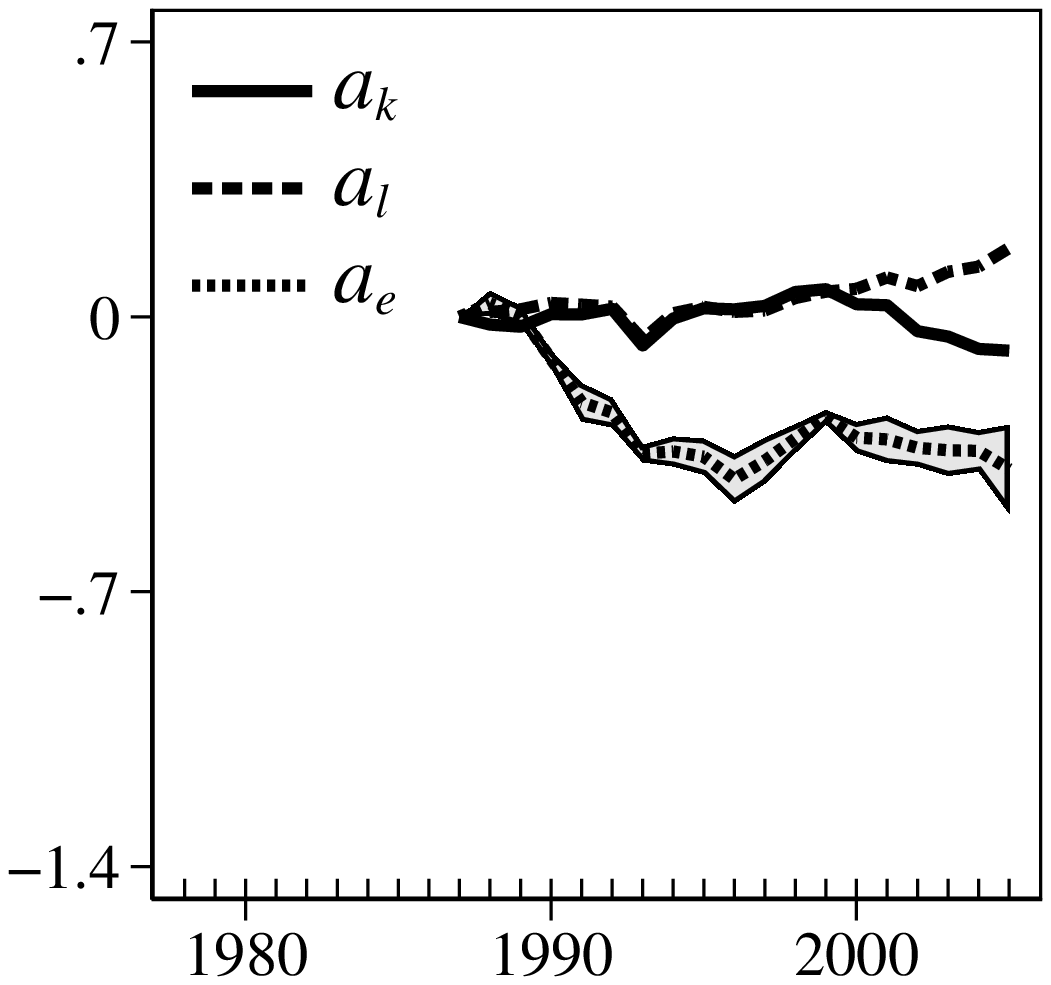}}
\par\end{centering}
\begin{centering}
\subfloat[Portugal]{
\centering{}\includegraphics[scale=0.4]{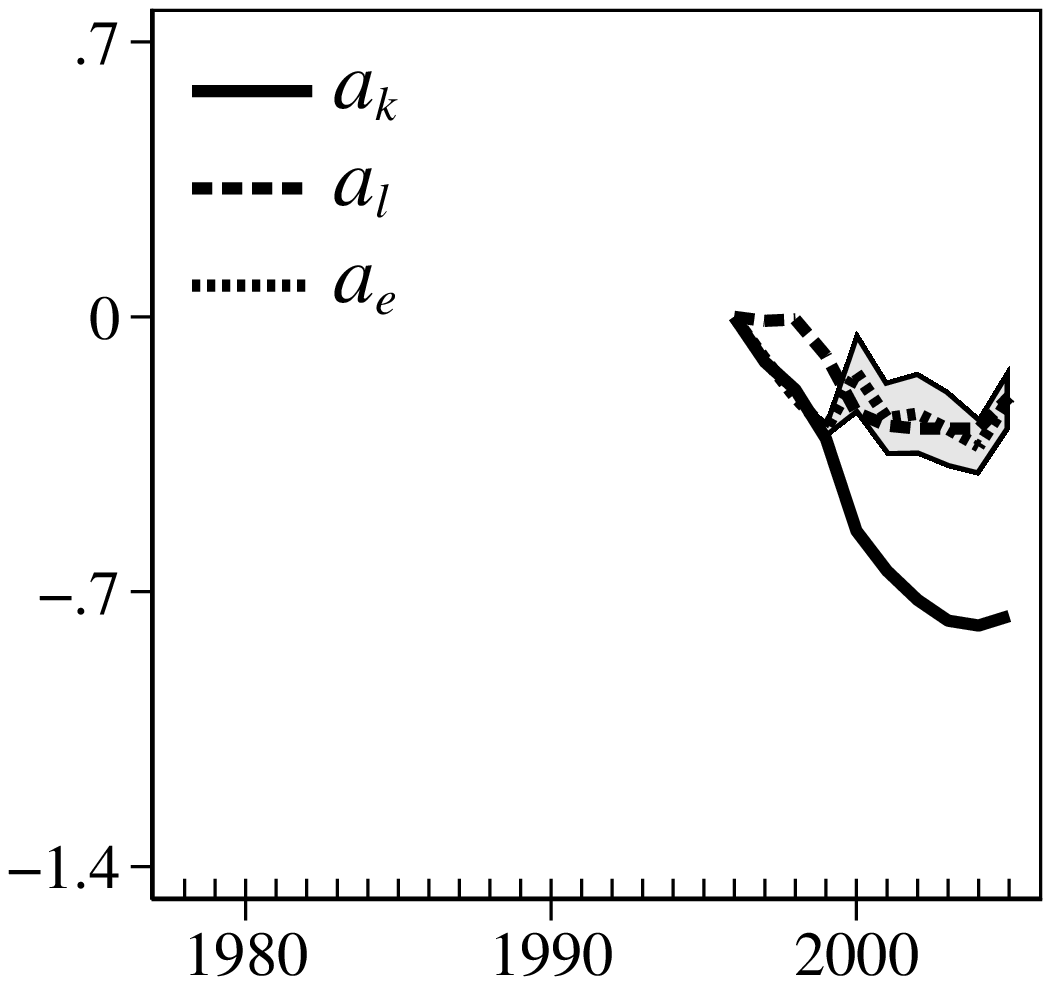}}\subfloat[Sweden]{
\centering{}\includegraphics[scale=0.4]{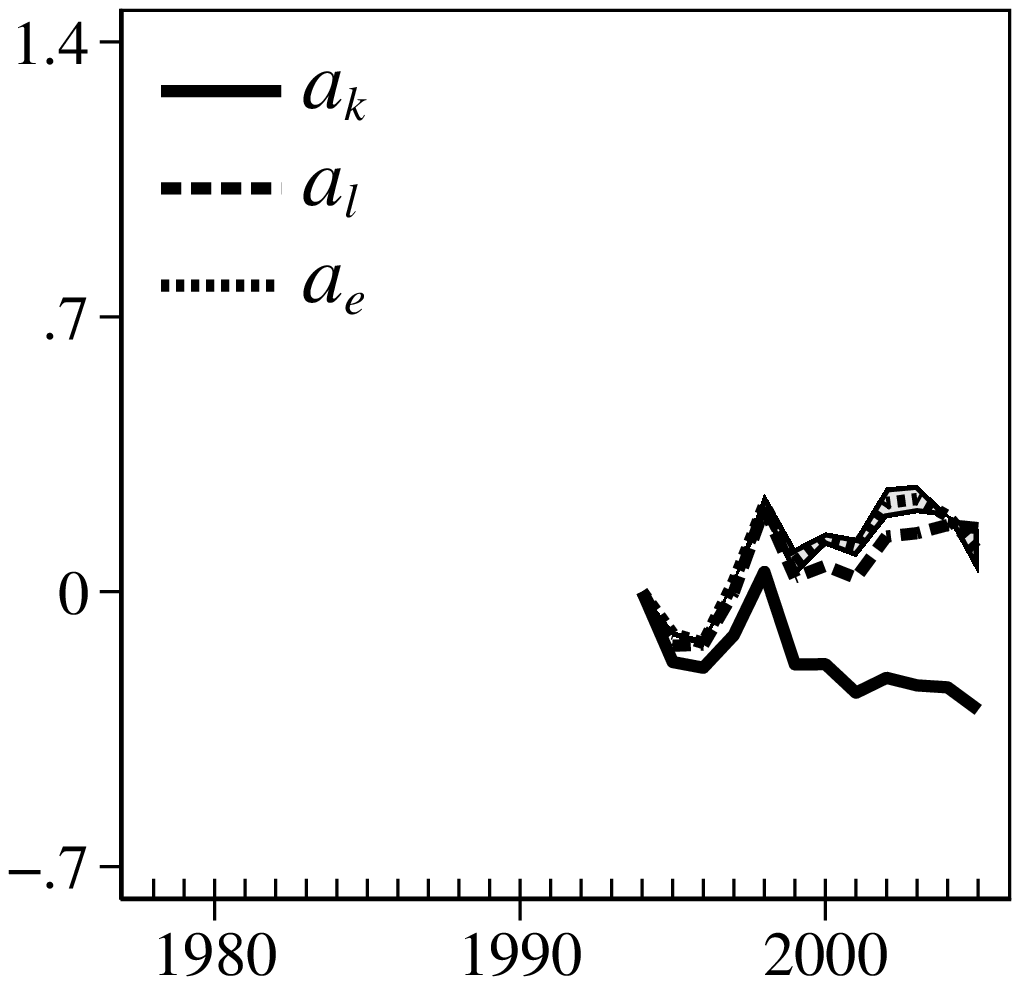}}\subfloat[United Kingdom]{
\centering{}\includegraphics[scale=0.4]{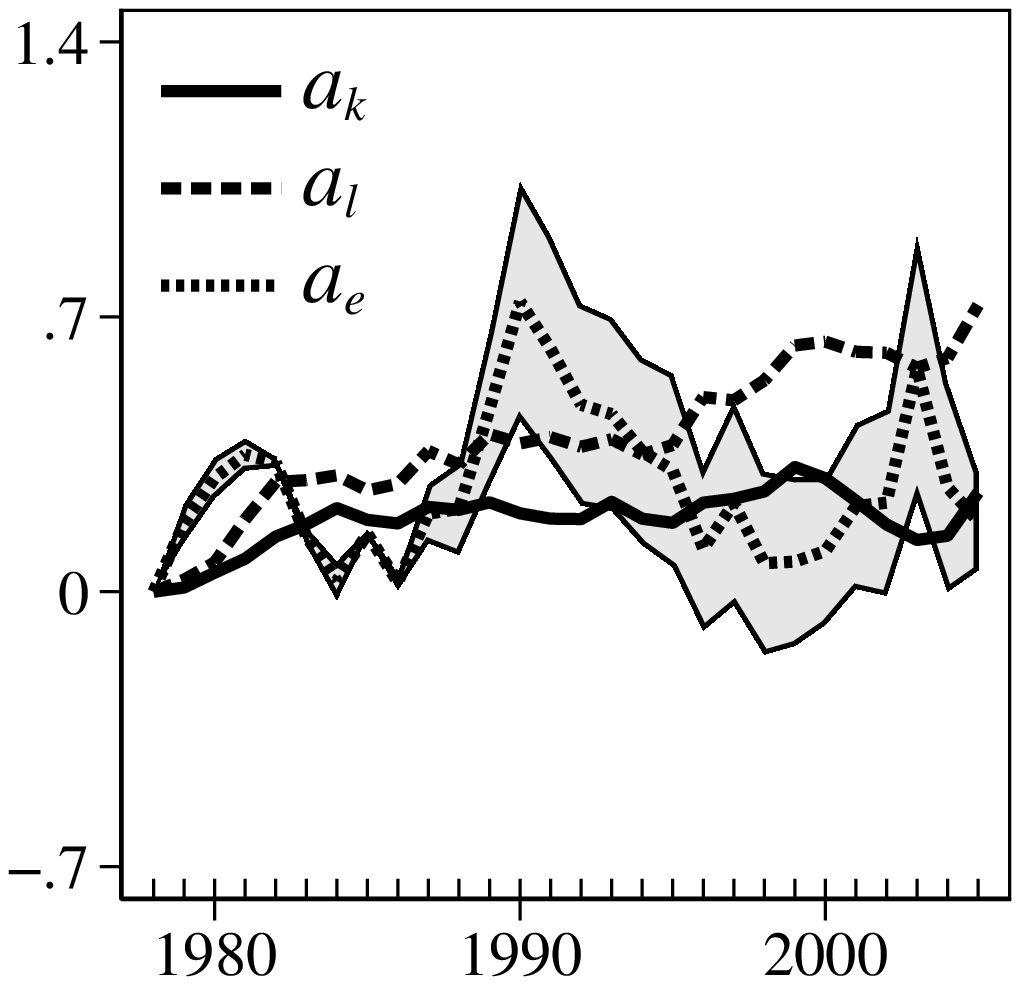}}\subfloat[United States]{
\centering{}\includegraphics[scale=0.4]{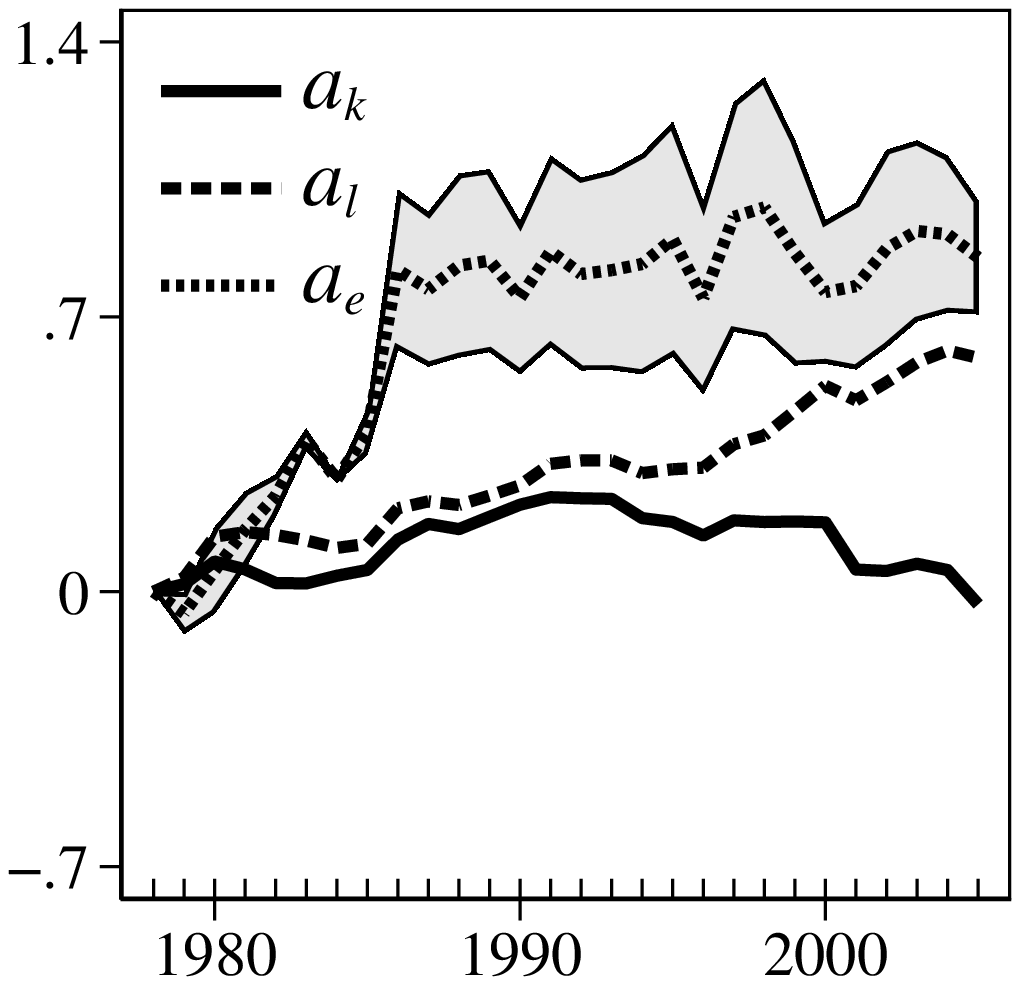}}
\par\end{centering}
\textit{\footnotesize{}Notes}{\footnotesize{}: The solid, dashed,
and dotted lines represent capital-, labor-, and energy-augmenting
technologies ($a_{k}$, $a_{\ell}$, and $a_{e}$), respectively.
The shaded area represents the 90 percent confidence interval for
$a_{e}$. All series are expressed as log differences relative to
the first year of observations.}{\footnotesize\par}
\end{figure}

We measure factor-augmenting technological change in the presence
of markup for each sector. Figures \ref{fig: AkAlAe1_goods} and \ref{fig: AkAlAe1_service}
display factor-augmenting technological change in the goods and service
sectors of 12 OECD countries. Factor-augmenting technological change
can be measured using data on output per factor input and factor income
shares for a given value of the elasticity of substitution, as seen
in equations \eqref{eq: ak1}\textendash \eqref{eq: ae1}. Given the
results above, the elasticity of substitution is set to 0.444 in the
one-level CES production function. We allow for the presence of markup
when we calculate gross output. Appendix \ref{subsec: productivity=000026income_shares}
describes the trends in output per factor input and factor income
shares.\footnote{The inverse of output per energy input (energy input per output) is
referred to as energy intensity. \citet{Mulder_deGroot_EE12} document
trends in energy intensity by industry in 18 OECD countries for the
year 1970 to 2005.}\footnote{\citet{Karabarbounis_Neiman_QJE14} document trends in the labor share
of value added in 59 countries for the year 1975 to 2012.}

Technological change is not factor-neutral. Changes in capital-, labor-,
and energy-augmenting technologies are noticeably different for each
country and sector. Capital-augmenting technology does not exhibit
a clear trend in the goods and service sectors of most countries,
but exhibits an increasing trend in the goods sector of a few countries
and a modest decreasing trend in the service sector of a few countries.
Labor-augmenting technology exhibits an increasing trend in the goods
sector of all countries and in the service sector of almost all countries.
The rate of increase in labor-augmenting technology is greater in
the goods sector than in the service sector for all countries. Labor-augmenting
technology tends to increase more than capital-augmenting technology
in the goods and service sectors of all countries. This result is
consistent with those of \citet{Klump_McAdam_Willman_RESTAT07} and
\citet{VanDerWerf_EE08}. Energy-saving technology exhibits different
trends across countries and over time for each sector. Energy-saving
technology tends to increase more than capital-augmenting technology
but less than labor-augmenting technology in the goods and service
sectors of most countries. However, the rate of increase in energy-saving
technology is similar to that in labor-augmenting technology in the
goods sector of the United Kingdom and in the service sector of Denmark,
Germany, and Sweden, and greater than that in labor-augmenting technology
in the goods and service sectors of the United States and in the service
sector of Italy. Our results indicate greater variation in energy-saving
technological change across countries and over time than those of
\citet{VanDerWerf_EE08}.

Energy-saving technological change is neither linear nor monotonic
over time. This result confirms the importance of not imposing the
functional forms of technological change. Moreover, energy-saving
technological change differs substantially across countries over time.
The rate of change in energy-saving technology was greater in the
United States than other OECD countries except Italy during the period
between the years 1978 and 2005. In the United States, energy-saving
technology started to rise after the mid-1980s and increased from
the year 1978 to 2005 by 89 and 85 log points in the goods and service
sectors, respectively. This result implies that when the years 1978
and 2005 are compared, new energy-saving technology would require
only 41 and 43 ($=100\times(a_{e,1978}/a_{e,2005})$) percent of energy
in the goods and service sectors, respectively, to produce the same
amount of output. The main reason for the difference between the United
States and other countries is the steady increase in output per energy
input in the United States (see Figures \ref{fig: factor_output_goods}
and \ref{fig: factor_output_service}). In Denmark and Italy, energy-saving
technology also increased from the 1980s to the 1990s, especially
in the service sector. In Finland, Japan, and the United Kingdom,
energy-saving technology progressed in the 1980s but stagnated in
the 1990s and 2000s. In Germany and Sweden, energy-saving technology
progressed in the 1990s but stagnated in the 2000s. The results imply
that when the first and last years of observations are compared, new
energy-saving technology would require only 77, 71, 85, 52, 60, 87,
68, and 41 percent of energy to produce the same amount of output
in the goods sector of Denmark, Finland, Germany, Italy, Japan, Sweden,
the United Kingdom, and the United States, respectively. At the same
time, new energy-saving technology would require only 48, 76, 29,
90, 84, and 43 percent of energy to produce the same amount of output
in the service sector of Denmark, Germany, Italy, Sweden, the United
Kingdom, and the United States, respectively. In Austria, energy-saving
technology modestly increased in the goods sector in the 1990s but
decreased both in the goods and service sectors in the late 1990s
and the 2000s. In the Czech Republic, energy-saving technology did
not change much during the period. In the Netherlands and Portugal,
energy-saving technology decreased during the period. Energy-saving
technology might have progressed in more countries if the technology
developed in other countries could be adopted.

Appendix \ref{subsec: results_add} provides additional results. The
magnitude and direction of energy-saving technological change can
vary significantly according to the value of the elasticity of substitution
(see Figures \ref{fig: Ae1_sigma_goods} and \ref{fig: Ae1_sigma_service}).
However, they remain almost unchanged regardless of whether we allow
the elasticity of substitution to vary across sectors (see Figures
\ref{fig: Ae1_sigma'_goods} and \ref{fig: Ae1_sigma'_service}).
The magnitude and direction of energy-saving technological change
do not vary significantly according to the degree of returns to scale
(see Figures \ref{fig: Ae1_mu_goods} and \ref{fig: Ae1_mu_service}).
Factor-augmenting technological change remains essentially unchanged
even if material inputs are taken into account (see Figures \ref{fig: AkAlAeAm_goods}
and \ref{fig: AkAlAeAm_service}).

\begin{table}[h]
\caption{Factor-augmenting technologies relative to the United States\label{tab: technology}}

\begin{centering}
\subfloat[Goods sector]{
\centering{}%
\begin{tabular}{lccr@{\extracolsep{0pt}.}lr@{\extracolsep{0pt}.}lr@{\extracolsep{0pt}.}lr@{\extracolsep{0pt}.}lr@{\extracolsep{0pt}.}lr@{\extracolsep{0pt}.}lr@{\extracolsep{0pt}.}lr@{\extracolsep{0pt}.}l}
\hline 
 &  &  & \multicolumn{4}{c}{$a_{k}$} & \multicolumn{2}{c}{} & \multicolumn{4}{c}{$a_{\ell}$} & \multicolumn{2}{c}{} & \multicolumn{4}{c}{$a_{e}$}\tabularnewline
 &  &  & \multicolumn{2}{c}{$t_{0}$} & \multicolumn{2}{c}{2005} & \multicolumn{2}{c}{} & \multicolumn{2}{c}{$t_{0}$} & \multicolumn{2}{c}{2005} & \multicolumn{2}{c}{} & \multicolumn{2}{c}{$t_{0}$} & \multicolumn{2}{c}{2005}\tabularnewline
\cline{4-7} \cline{6-7} \cline{10-13} \cline{12-13} \cline{16-19} \cline{18-19} 
Japan & 1978\textendash 2005 &  & 0&42 & 0&35 & \multicolumn{2}{c}{} & 0&17 & 0&40 & \multicolumn{2}{c}{} & 4&12 & 2&83\tabularnewline
Italy & 1978\textendash 2005 &  & 0&64 & 0&81 & \multicolumn{2}{c}{} & 0&43 & 1&26 & \multicolumn{2}{c}{} & 2&47 & 1&96\tabularnewline
Denmark & 1980\textendash 2005 &  & 0&44 & 0&45 & \multicolumn{2}{c}{} & 0&38 & 0&64 & \multicolumn{2}{c}{} & 3&51 & 1&67\tabularnewline
Sweden & 1994\textendash 2005 &  & 0&63 & 0&71 & \multicolumn{2}{c}{} & 0&65 & 0&87 & \multicolumn{2}{c}{} & 1&37 & 1&14\tabularnewline
Germany & 1992\textendash 2005 &  & 0&86 & 0&82 & \multicolumn{2}{c}{} & 0&62 & 0&77 & \multicolumn{2}{c}{} & 1&21 & 0&96\tabularnewline
Finland & 1978\textendash 2005 &  & 0&66 & 1&07 & \multicolumn{2}{c}{} & 0&37 & 0&74 & \multicolumn{2}{c}{} & 1&47 & 0&84\tabularnewline
United Kingdom & 1978\textendash 2005 &  & 0&95 & 1&06 & \multicolumn{2}{c}{} & 0&70 & 0&77 & \multicolumn{2}{c}{} & 0&97 & 0&58\tabularnewline
Austria & 1980\textendash 2005 &  & 0&33 & 0&55 & \multicolumn{2}{c}{} & 0&32 & 0&63 & \multicolumn{2}{c}{} & 2&06 & 0&42\tabularnewline
Czech Republic & 1996\textendash 2005 &  & 0&47 & 0&43 & \multicolumn{2}{c}{} & 0&63 & 0&69 & \multicolumn{2}{c}{} & 0&64 & 0&42\tabularnewline
Netherlands & 1987\textendash 2005 &  & 0&54 & 0&79 & \multicolumn{2}{c}{} & 0&83 & 0&91 & \multicolumn{2}{c}{} & 1&05 & 0&38\tabularnewline
Portugal & 1996\textendash 2005 &  & 0&76 & 0&59 & \multicolumn{2}{c}{} & 0&55 & 0&45 & \multicolumn{2}{c}{} & 0&94 & 0&34\tabularnewline
\hline 
\end{tabular}}
\par\end{centering}
\begin{centering}
\subfloat[Service sector]{
\centering{}%
\begin{tabular}{lccr@{\extracolsep{0pt}.}lr@{\extracolsep{0pt}.}lr@{\extracolsep{0pt}.}lr@{\extracolsep{0pt}.}lr@{\extracolsep{0pt}.}lr@{\extracolsep{0pt}.}lr@{\extracolsep{0pt}.}lr@{\extracolsep{0pt}.}l}
\hline 
 &  &  & \multicolumn{4}{c}{$a_{k}$} & \multicolumn{2}{c}{} & \multicolumn{4}{c}{$a_{\ell}$} & \multicolumn{2}{c}{} & \multicolumn{4}{c}{$a_{e}$}\tabularnewline
 &  &  & \multicolumn{2}{c}{$t_{0}$} & \multicolumn{2}{c}{2005} & \multicolumn{2}{c}{} & \multicolumn{2}{c}{$t_{0}$} & \multicolumn{2}{c}{2005} & \multicolumn{2}{c}{} & \multicolumn{2}{c}{$t_{0}$} & \multicolumn{2}{c}{2005}\tabularnewline
\cline{4-7} \cline{6-7} \cline{10-13} \cline{12-13} \cline{16-19} \cline{18-19} 
Netherlands & 1987\textendash 2005 &  & 0&42 & 0&47 & \multicolumn{2}{c}{} & 1&09 & 0&90 & \multicolumn{2}{c}{} & 5&93 & 3&71\tabularnewline
Denmark & 1980\textendash 2005 &  & 0&29 & 0&51 & \multicolumn{2}{c}{} & 0&86 & 1&10 & \multicolumn{2}{c}{} & 1&62 & 1&51\tabularnewline
Italy & 1978\textendash 2005 &  & 0&83 & 1&28 & \multicolumn{2}{c}{} & 0&92 & 1&10 & \multicolumn{2}{c}{} & 0&81 & 1&21\tabularnewline
Germany & 1992\textendash 2005 &  & 0&33 & 0&39 & \multicolumn{2}{c}{} & 0&92 & 0&92 & \multicolumn{2}{c}{} & 0&81 & 1&02\tabularnewline
Japan & 1978\textendash 2005 &  & 0&74 & 0&48 & \multicolumn{2}{c}{} & 0&33 & 0&47 & \multicolumn{2}{c}{} & 2&86 & 0&93\tabularnewline
United Kingdom & 1978\textendash 2005 &  & 0&63 & 0&84 & \multicolumn{2}{c}{} & 0&62 & 0&71 & \multicolumn{2}{c}{} & 1&34 & 0&68\tabularnewline
Sweden & 1994\textendash 2005 &  & 0&84 & 0&77 & \multicolumn{2}{c}{} & 0&79 & 0&69 & \multicolumn{2}{c}{} & 0&44 & 0&48\tabularnewline
Portugal & 1996\textendash 2005 &  & 0&60 & 0&33 & \multicolumn{2}{c}{} & 1&10 & 0&68 & \multicolumn{2}{c}{} & 0&35 & 0&26\tabularnewline
Austria & 1980\textendash 2005 &  & 0&42 & 0&45 & \multicolumn{2}{c}{} & 0&64 & 0&66 & \multicolumn{2}{c}{} & 1&54 & 0&25\tabularnewline
Czech Republic & 1996\textendash 2005 &  & 0&21 & 0&25 & \multicolumn{2}{c}{} & 0&90 & 0&78 & \multicolumn{2}{c}{} & 0&31 & 0&25\tabularnewline
Finland & 1978\textendash 2005 &  & 0&80 & 0&70 & \multicolumn{2}{c}{} & 0&72 & 0&61 & \multicolumn{2}{c}{} & 0&35 & 0&13\tabularnewline
\hline 
\end{tabular}}
\par\end{centering}
\textit{\footnotesize{}Notes}{\footnotesize{}: Countries are arranged
in descending order of energy-saving technology in the year 2005 by
sector. This table reports the levels of capital-, labor-, and energy-augmenting
technologies relative to the United States ($a_{k,c}/a_{k,US}$, $a_{\ell,c}/a_{\ell,US}$,
and $a_{e,c}/a_{e,US}$) in the first and last years of observations.}{\footnotesize\par}
\end{table}

\subsubsection{International differences}

The levels of factor-augmenting technologies differ substantially
across countries. Table \ref{tab: technology} reports the levels
of capital-, labor-, and energy-augmenting technologies relative to
the United States in the first and last years of observations for
each country and sector. The level of capital-augmenting technology
relative to the United States in the year 2005 ranges from 0.35 (0.25)
to 1.07 (1.28) in the goods (service) sector. The level of labor-augmenting
technology relative to the United States in the year 2005 ranges from
0.40 (0.47) to 1.26 (1.10) in the goods (service) sector. Although
the levels of capital- and labor-augmenting technologies were higher
in the United States than in most other countries for each sector
and year, the level of energy-saving technology was lower in the United
States than many other countries especially in the 1970s and 1980s.
Looking at the first year of observations, eight (five) countries
had a higher level of energy-saving technology than the United States
in the goods (service) sector. The United States achieved progress
in energy-saving technology after the mid-1980s, as shown above. Looking
at the year 2005, the United States was placed fifth behind Japan,
Italy, Denmark, and Sweden in the goods sector, and fifth behind the
Netherlands, Denmark, Italy, and Germany in the service sector. The
level of energy-saving technology relative to the United States in
the year 2005 ranges from 0.34 (0.13) to 2.83 (3.71) in the goods
(service) sector.

There was greater variation across countries in the level of energy-saving
technology than in the levels of capital- and labor-augmenting technologies
for each sector and year. In the year 2005, the ratios of the maximum
of energy-saving technology to the minimum were 8.4 and 29.0 in the
goods and service sectors, respectively, while the ratios of the maximum
of capital- (labor-) augmenting technology to the minimum were 3.1
(3.1) in the goods sector and 5.0 (2.3) in the service sector. One
reason for this may be that the cross-country differences in energy-saving
technology were more persistent than those in capital- and labor-augmenting
technologies.

\begin{figure}[H]
\caption{Convergence of factor-augmenting technologies\label{fig: convergence}}

\begin{centering}
\subfloat[Goods sector]{
\centering{}\includegraphics[scale=0.55]{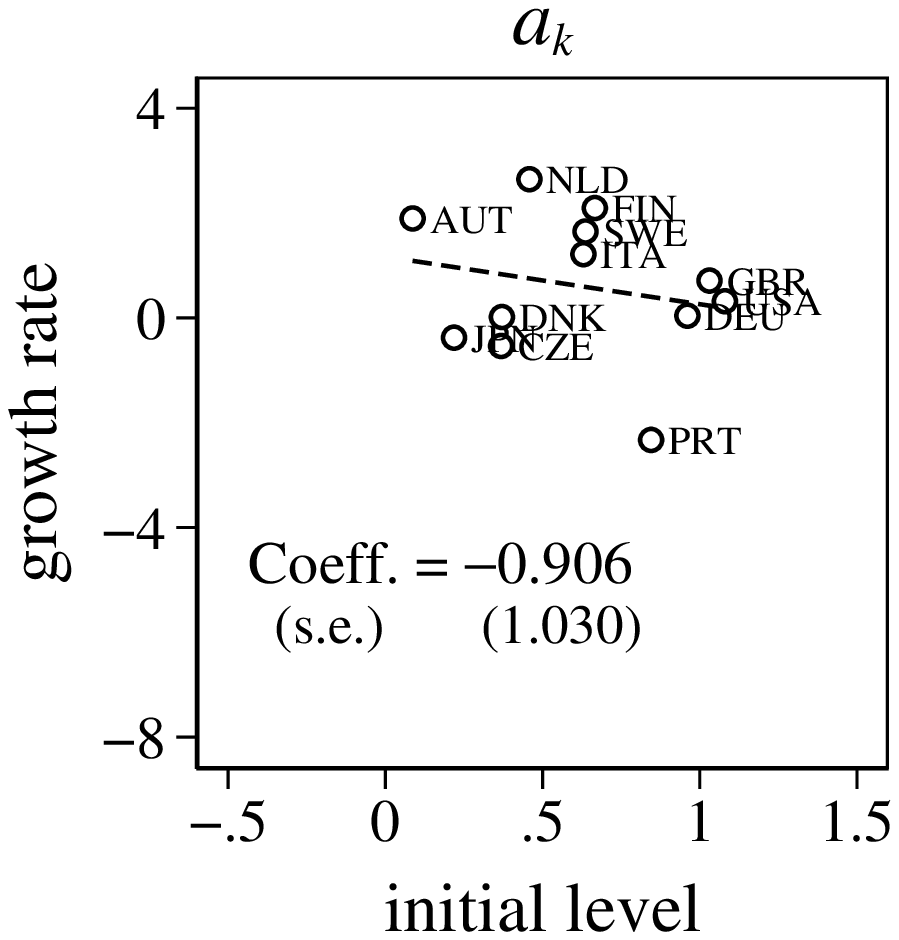}\enskip{}\includegraphics[scale=0.55]{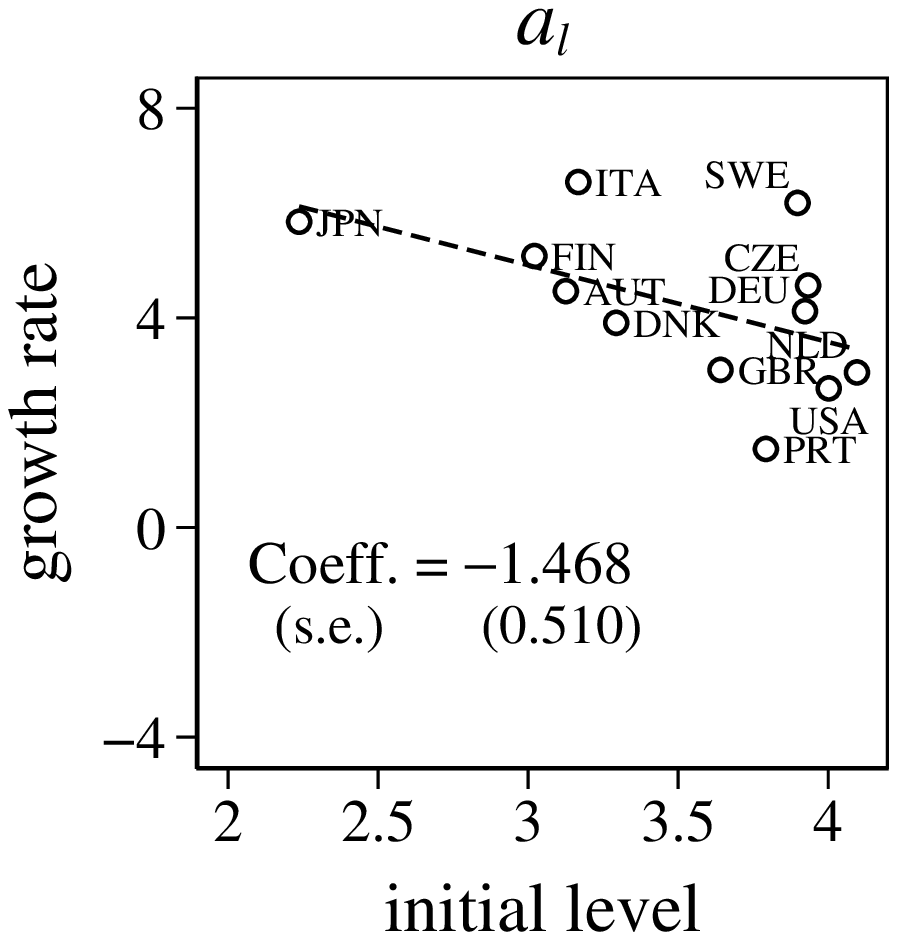}\enskip{}\includegraphics[scale=0.55]{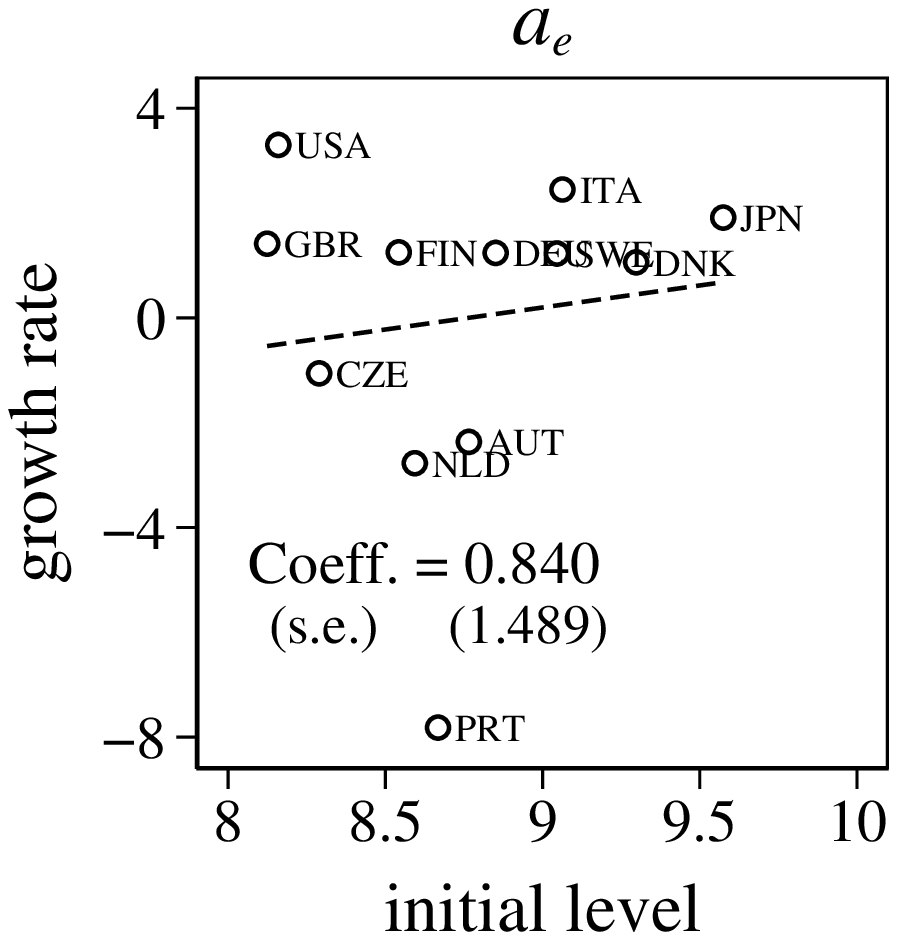}}
\par\end{centering}
\begin{centering}
\subfloat[Service sector]{
\centering{}\includegraphics[scale=0.55]{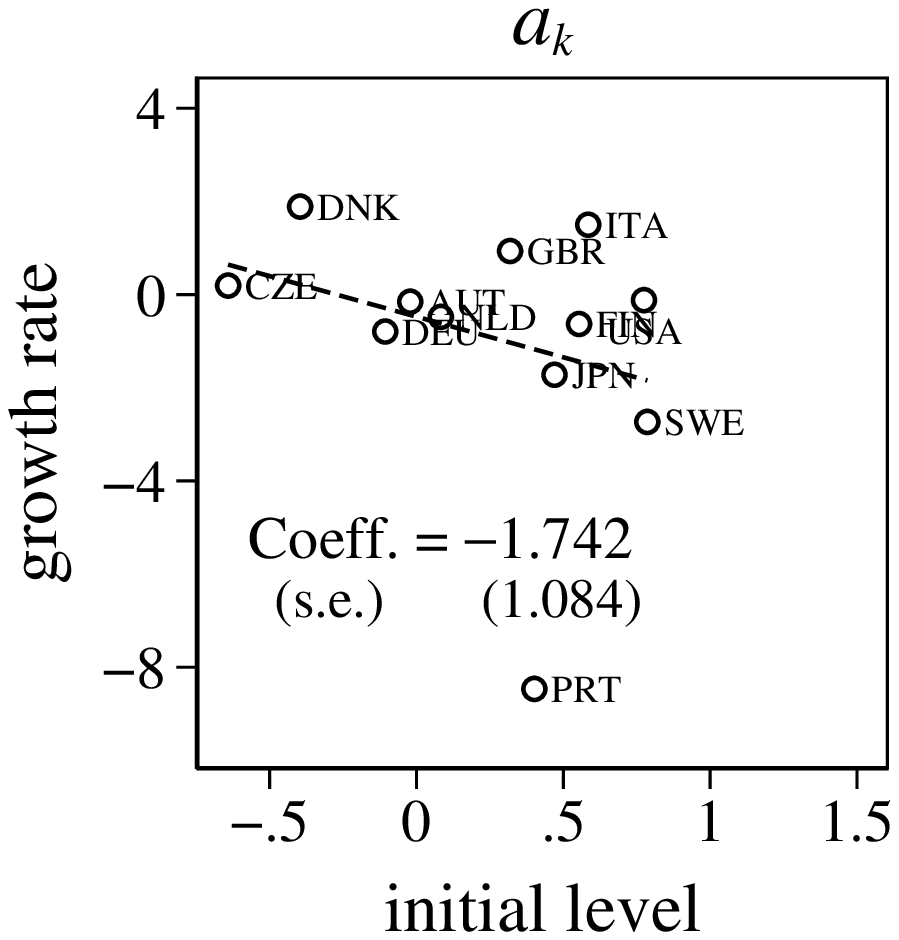}\enskip{}\includegraphics[scale=0.55]{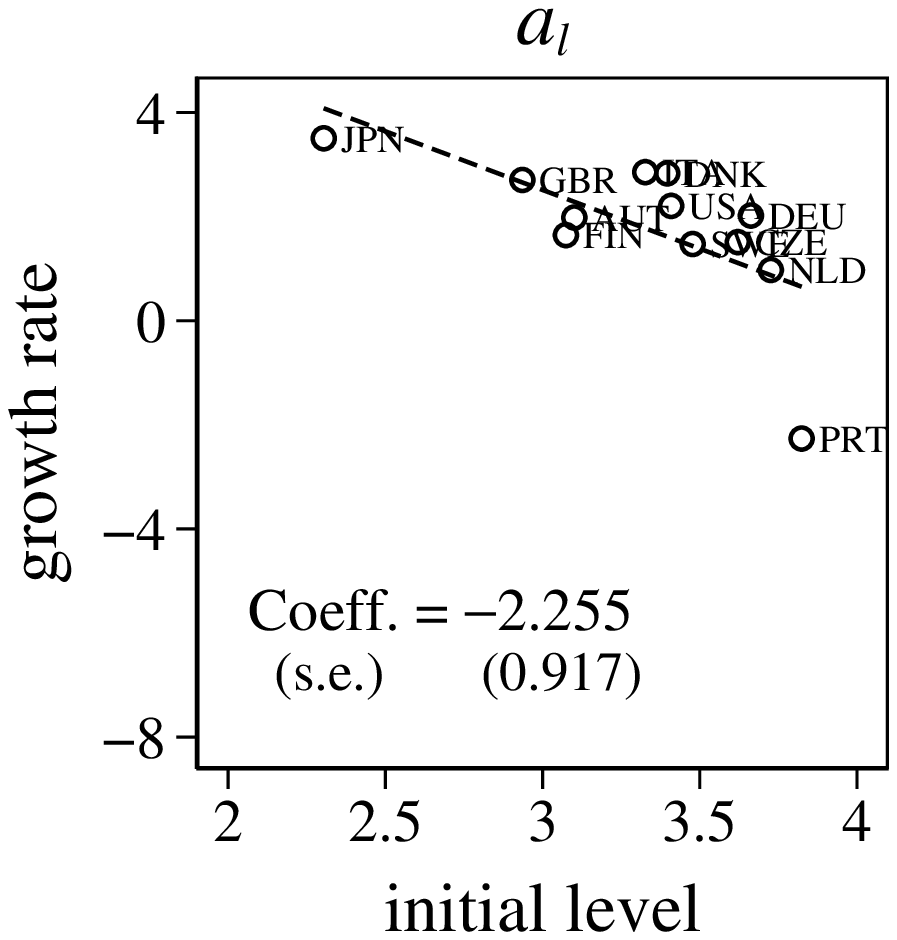}\enskip{}\includegraphics[scale=0.55]{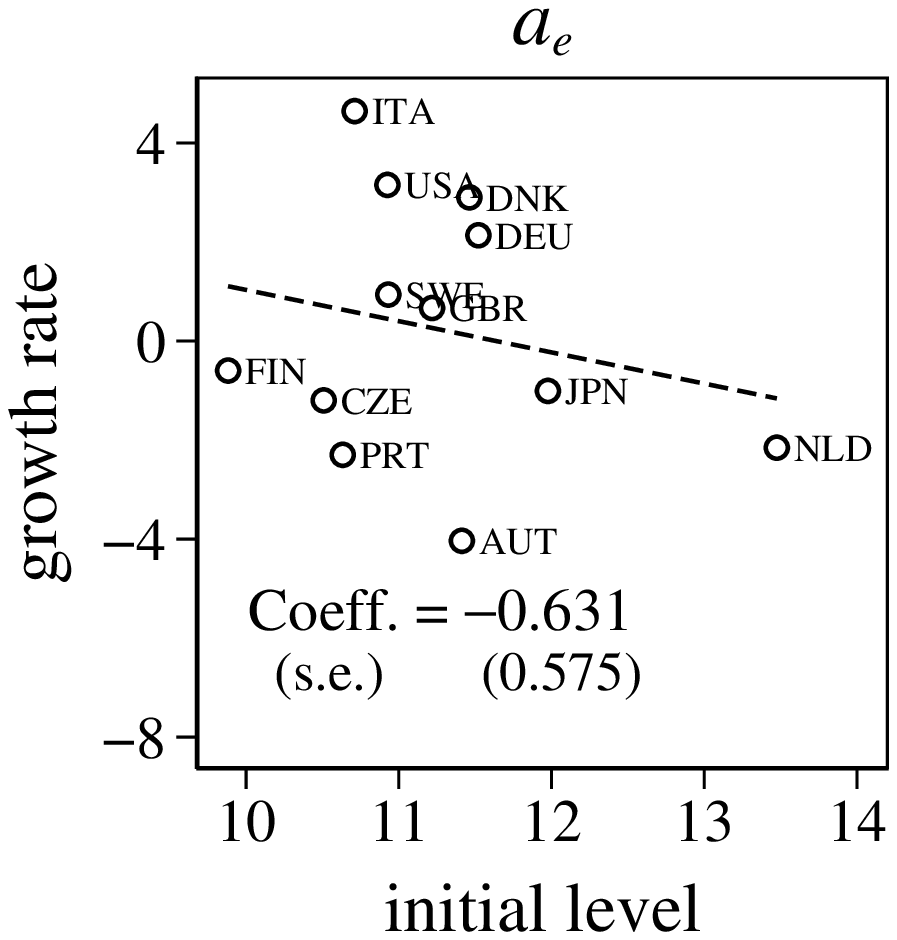}}
\par\end{centering}
\textit{\footnotesize{}Notes}{\footnotesize{}: The vertical axis indicates
the annual rates of growth in factor-augmenting technologies ($a_{k}$,
$a_{\ell}$, or $a_{e}$) from the first year of observation ($t_{0}$)
to 2005 (e.g., $100\times(\ln a_{k,2005}-\ln a_{k,t_{0}})/(2005-t_{0})$).
The horizontal axis indicates the logs of factor-augmenting technologies
in the first year of observation (e.g., $\ln a_{k,t_{0}}$). The estimated
coefficients in the regressions of the annual rates of growth in factor-augmenting
technologies on the initial levels of factor-augmenting technologies
are reported at the lower left of each figure. The heteroscedasticity-robust
standard errors are also reported in parentheses at the lower left
of each figure.}{\footnotesize\par}
\end{figure}

Figure \ref{fig: convergence} illustrates the convergence of factor-augmenting
technologies by plotting the annual rates of growth in factor-augmenting
technologies against their initial levels for each sector. Capital-
and labor-augmenting technologies were more likely to grow faster
in countries where they were initially low. Thus, they had a tendency
to converge among countries for each sector. Moreover, the rate of
convergence was faster in the service sector, where more capital and
labor were used in the process of production, than in the goods sector.
However, energy-saving technology did not converge among countries
in the goods sector, where more energy was used in the process of
production. In the year 2005, energy was used on average 2.4 times
more in the goods sector than in the service sector, while capital
and labor were used on average 3.2 and 2.6 times more in the service
sector than in the goods sector.

The differences in the tendency to converge among factor-augmenting
technologies may be related to difficulties in technology adoption.
Energy-saving technology tends to be specific to the country for at
least three reasons. First, there is a difference in the availability
of fossil fuels across countries. Second, there is a difference in
available renewable energy sources across countries. The type of renewable
energy available depends on the climate and geography of the country.
Finally, there is a difference in public attitudes across countries
toward energy sources such as nuclear power. Consequently, if new
technologies that can save capital, labor, and energy are developed
in one country, the energy-saving technology is presumably less likely
to be adopted in other countries than the capital- and labor-saving
technologies. Nonetheless, more research may be needed to determine
the cause and mechanism of the slower rate of convergence in energy-saving
technology.

\subsubsection{R\&D spending and patent counts}

We correlate our measure of energy-saving technological change with
conventional measures of energy-saving technological change. The first
row of Table \ref{tab: correlation} reports the correlation coefficients
between our measure and conventional measures of energy-saving technological
change for each sector. Our measure of energy-saving technology is
positively and significantly correlated with energy-related R\&D spending
by the government and the number of energy-related patents in logs
(odd-numbered columns). The correlation coefficients are greater than
0.3 for both R\&D spending and patent counts. Meanwhile, our measure
of energy-saving technology is not significantly correlated with energy-related
R\&D spending or energy-related patent counts in 10-year growth rates
(even-numbered columns). The correlation may be obscured by some common
factors in factor-augmenting technological change. The second and
third rows report the correlation coefficients calculated using the
ratio of energy- to labor- or capital-augmenting technology ($a_{e}/a_{\ell}$
and $a_{e}/a_{k}$). These ratios represent the direction of technological
change and are not subject to the influence of any common factors.
The relative energy-saving technology is positively and significantly
correlated in 10-year growth rates with energy-related R\&D spending
but not with energy-related patent counts (even-numbered columns).
The former result indicates that technological change tends to be
directed toward energy as government spending on energy-related R\&D
increases. The size of the correlation is greater in the goods sector
than in the service sector, reflecting the fact that more energy is
used in the goods sector than in the service sector. The latter result
suggests that our measure of energy-saving technological change contains
complementary information on unpatented innovation and/or no unnecessary
information on useless patents.

\begin{table}[h]
\caption{Correlations with alternative measures\label{tab: correlation}}

\begin{centering}
\begin{tabular}{lr@{\extracolsep{0pt}.}lr@{\extracolsep{0pt}.}lr@{\extracolsep{0pt}.}lr@{\extracolsep{0pt}.}lr@{\extracolsep{0pt}.}lr@{\extracolsep{0pt}.}lr@{\extracolsep{0pt}.}lr@{\extracolsep{0pt}.}lr@{\extracolsep{0pt}.}l}
\hline 
 & \multicolumn{8}{c}{Goods sector} & \multicolumn{2}{c}{} & \multicolumn{8}{c}{Service sector}\tabularnewline
 & \multicolumn{4}{c}{R\&D} & \multicolumn{4}{c}{Patents} & \multicolumn{2}{c}{} & \multicolumn{4}{c}{R\&D} & \multicolumn{4}{c}{Patents}\tabularnewline
 & \multicolumn{2}{c}{log} & \multicolumn{2}{c}{change} & \multicolumn{2}{c}{log} & \multicolumn{2}{c}{change} & \multicolumn{2}{c}{} & \multicolumn{2}{c}{log} & \multicolumn{2}{c}{change} & \multicolumn{2}{c}{log} & \multicolumn{2}{c}{change}\tabularnewline
\cline{2-9} \cline{4-9} \cline{6-9} \cline{8-9} \cline{12-19} \cline{14-19} \cline{16-19} \cline{18-19} 
$a_{e}$ & 0&328 & 0&120 & 0&376 & \textendash 0&064 & \multicolumn{2}{c}{} & 0&304 & 0&042 & 0&380 & \textendash 0&195\tabularnewline
 & {[}0&000{]} & {[}0&186{]} & {[}0&000{]} & {[}0&646{]} & \multicolumn{2}{c}{} & {[}0&000{]} & {[}0&645{]} & {[}0&000{]} & {[}0&159{]}\tabularnewline
$a_{e}/a_{\ell}$ & 0&196 & 0&310 & 0&255 & 0&139 & \multicolumn{2}{c}{} & 0&364 & 0&187 & 0&439 & \textendash 0&047\tabularnewline
 & {[}0&002{]} & {[}0&001{]} & {[}0&001{]} & {[}0&316{]} & \multicolumn{2}{c}{} & {[}0&000{]} & {[}0&040{]} & {[}0&000{]} & {[}0&737{]}\tabularnewline
$a_{e}/a_{k}$ & 0&098 & 0&306 & 0&194 & 0&168 & \multicolumn{2}{c}{} & 0&048 & 0&170 & 0&213 & \textendash 0&040\tabularnewline
 & {[}0&131{]} & {[}0&001{]} & {[}0&010{]} & {[}0&224{]} & \multicolumn{2}{c}{} & {[}0&459{]} & {[}0&062{]} & {[}0&005{]} & {[}0&772{]}\tabularnewline
\hline 
\end{tabular}
\par\end{centering}
\textit{\footnotesize{}Notes}{\footnotesize{}: Correlation coefficients
are reported. All variables are taken in logs. The numbers in square
brackets are }\textit{\footnotesize{}p}{\footnotesize{}-values under
the null hypothesis of no correlation.}{\footnotesize\par}
\end{table}

\subsection{Growth accounting}

We measure the contribution of factor inputs and factor-augmenting
technologies to economic growth for each country and sector. The first
column of Table \ref{tab: growth} reports the rate of growth in gross
output. We allow for the presence of markup in the growth decomposition.
The second column reports the rate of growth in markup based on the
results of \citet{DeLoecker_Eeckhout_wp20}. When countries are arranged
in descending order of growth rate of gross output by sector, Italy
and Portugal are placed quite differently depending on whether changes
in markup are taken into account. Most of the countries are, however,
placed roughly the same irrespective of the presence or absence of
markup. The third to fifth columns report the portions attributable
to the pairs of factor inputs and factor-augmenting technologies ($a_{k}k$,
$a_{\ell}\ell$, and $a_{e}e$). The sixth to eleventh columns report
the portions attributable to each factor input ($k$, $\ell$, and
$e$) and each factor-augmenting technology ($a_{k}$, $a_{\ell}$,
and $a_{e}$). The last column reports the portions attributable to
TFP, calculated as the sum of the portions attributable to factor-augmenting
technologies for each country. Given the results above, the decomposition
results are calculated based on the one-level CES production function
in which the elasticity of substitution is set to 0.444.

\begin{sidewaystable}[ph]
\caption{Sources of economic growth\label{tab: growth}}

\begin{centering}
\subfloat[Goods sector]{
\centering{}%
\begin{tabular}{llr@{\extracolsep{0pt}.}lr@{\extracolsep{0pt}.}lr@{\extracolsep{0pt}.}lr@{\extracolsep{0pt}.}lr@{\extracolsep{0pt}.}lr@{\extracolsep{0pt}.}lr@{\extracolsep{0pt}.}lr@{\extracolsep{0pt}.}lr@{\extracolsep{0pt}.}lr@{\extracolsep{0pt}.}lr@{\extracolsep{0pt}.}lr@{\extracolsep{0pt}.}lr@{\extracolsep{0pt}.}lr@{\extracolsep{0pt}.}lr@{\extracolsep{0pt}.}l}
\hline 
 &  & \multicolumn{2}{c}{$y$} & \multicolumn{2}{c}{$\omega$} & \multicolumn{2}{c}{} & \multicolumn{2}{c}{$a_{k}k$} & \multicolumn{2}{c}{$a_{\ell}\ell$} & \multicolumn{2}{c}{$a_{e}e$} & \multicolumn{2}{c}{} & \multicolumn{2}{c}{$k$} & \multicolumn{2}{c}{$\ell$} & \multicolumn{2}{c}{$e$} & \multicolumn{2}{c}{$a_{k}$} & \multicolumn{2}{c}{$a_{\ell}$} & \multicolumn{2}{c}{$a_{e}$} & \multicolumn{2}{c}{} & \multicolumn{2}{c}{$a$}\tabularnewline
\cline{3-6} \cline{5-6} \cline{9-14} \cline{11-14} \cline{13-14} \cline{17-28} \cline{19-28} \cline{21-28} \cline{23-28} \cline{25-28} \cline{27-28} \cline{31-32} 
Sweden & 1994\textendash 2005 & 5&47 & \textendash 0&07 & \multicolumn{2}{c}{} & 1&36 & 3&42 & 0&70 & \multicolumn{2}{c}{} & 0&96 & 0&08 & 0&43 & 0&40 & 3&34 & 0&27 & \multicolumn{2}{c}{} & 4&01\tabularnewline
Italy & 1978\textendash 2005 & 4&76 & 3&22 & \multicolumn{2}{c}{} & 0&85 & 2&73 & 1&18 & \multicolumn{2}{c}{} & 0&58 & \textendash 0&70 & 0&54 & 0&27 & 3&43 & 0&64 & \multicolumn{2}{c}{} & 4&34\tabularnewline
Japan & 1978\textendash 2005 & 3&98 & 0&89 & \multicolumn{2}{c}{} & 0&83 & 2&59 & 0&56 & \multicolumn{2}{c}{} & 0&93 & \textendash 0&83 & 0&29 & \textendash 0&10 & 3&41 & 0&27 & \multicolumn{2}{c}{} & 3&58\tabularnewline
Finland & 1978\textendash 2005 & 3&78 & 1&02 & \multicolumn{2}{c}{} & 0&72 & 1&99 & 1&07 & \multicolumn{2}{c}{} & 0&34 & \textendash 0&72 & 0&70 & 0&38 & 2&71 & 0&37 & \multicolumn{2}{c}{} & 3&46\tabularnewline
Czech Republic & 1996\textendash 2005 & 2&87 & 1&16 & \multicolumn{2}{c}{} & 0&84 & 0&98 & 1&05 & \multicolumn{2}{c}{} & 0&98 & \textendash 0&63 & 1&47 & \textendash 0&14 & 1&61 & \textendash 0&42 & \multicolumn{2}{c}{} & 1&05\tabularnewline
Unites States & 1978\textendash 2005 & 2&74 & 1&26 & \multicolumn{2}{c}{} & 0&37 & 1&47 & 0&90 & \multicolumn{2}{c}{} & 0&32 & \textendash 0&01 & 0&01 & 0&05 & 1&47 & 0&89 & \multicolumn{2}{c}{} & 2&42\tabularnewline
Denmark & 1980\textendash 2005 & 2&70 & 2&07 & \multicolumn{2}{c}{} & 0&49 & 1&47 & 0&74 & \multicolumn{2}{c}{} & 0&49 & \textendash 0&66 & 0&50 & 0&00 & 2&12 & 0&24 & \multicolumn{2}{c}{} & 2&37\tabularnewline
Austria & 1980\textendash 2005 & 2&66 & 1&06 & \multicolumn{2}{c}{} & 0&60 & 1&45 & 0&61 & \multicolumn{2}{c}{} & 0&17 & \textendash 0&74 & 1&28 & 0&43 & 2&19 & \textendash 0&67 & \multicolumn{2}{c}{} & 1&95\tabularnewline
Netherlands & 1987\textendash 2005 & 2&40 & 0&30 & \multicolumn{2}{c}{} & 0&80 & 1&39 & 0&21 & \multicolumn{2}{c}{} & 0&26 & 0&00 & 1&10 & 0&54 & 1&38 & \textendash 0&89 & \multicolumn{2}{c}{} & 1&04\tabularnewline
United Kingdom & 1978\textendash 2005 & 1&77 & 1&72 & \multicolumn{2}{c}{} & 0&29 & 0&46 & 1&03 & \multicolumn{2}{c}{} & 0&17 & \textendash 0&91 & 0&48 & 0&11 & 1&37 & 0&55 & \multicolumn{2}{c}{} & 2&03\tabularnewline
Germany & 1992\textendash 2005 & 0&95 & 0&90 & \multicolumn{2}{c}{} & 0&19 & 0&82 & \textendash 0&07 & \multicolumn{2}{c}{} & 0&19 & \textendash 1&46 & \textendash 0&39 & 0&01 & 2&28 & 0&32 & \multicolumn{2}{c}{} & 2&61\tabularnewline
Portugal & 1996\textendash 2005 & \textendash 0&06 & \textendash 2&71 & \multicolumn{2}{c}{} & 0&27 & 0&39 & \textendash 0&73 & \multicolumn{2}{c}{} & 0&81 & \textendash 0&14 & 2&48 & \textendash 0&54 & 0&54 & \textendash 3&21 & \multicolumn{2}{c}{} & \textendash 3&21\tabularnewline
\hline 
\end{tabular}}
\par\end{centering}
\begin{centering}
\subfloat[Service sector]{
\centering{}%
\begin{tabular}{llr@{\extracolsep{0pt}.}lr@{\extracolsep{0pt}.}lr@{\extracolsep{0pt}.}lr@{\extracolsep{0pt}.}lr@{\extracolsep{0pt}.}lr@{\extracolsep{0pt}.}lr@{\extracolsep{0pt}.}lr@{\extracolsep{0pt}.}lr@{\extracolsep{0pt}.}lr@{\extracolsep{0pt}.}lr@{\extracolsep{0pt}.}lr@{\extracolsep{0pt}.}lr@{\extracolsep{0pt}.}lr@{\extracolsep{0pt}.}lr@{\extracolsep{0pt}.}l}
\hline 
 &  & \multicolumn{2}{c}{$y$} & \multicolumn{2}{c}{$\omega$} & \multicolumn{2}{c}{} & \multicolumn{2}{c}{$a_{k}k$} & \multicolumn{2}{c}{$a_{\ell}\ell$} & \multicolumn{2}{c}{$a_{e}e$} & \multicolumn{2}{c}{} & \multicolumn{2}{c}{$k$} & \multicolumn{2}{c}{$\ell$} & \multicolumn{2}{c}{$e$} & \multicolumn{2}{c}{$a_{k}$} & \multicolumn{2}{c}{$a_{\ell}$} & \multicolumn{2}{c}{$a_{e}$} & \multicolumn{2}{c}{} & \multicolumn{2}{c}{$a$}\tabularnewline
\cline{3-6} \cline{5-6} \cline{9-14} \cline{11-14} \cline{13-14} \cline{17-28} \cline{19-28} \cline{21-28} \cline{23-28} \cline{25-28} \cline{27-28} \cline{31-32} 
Italy & 1978\textendash 2005 & 4&68 & 3&22 & \multicolumn{2}{c}{} & 0&95 & 3&23 & 0&50 & \multicolumn{2}{c}{} & 0&62 & 1&25 & 0&10 & 0&33 & 1&98 & 0&40 & \multicolumn{2}{c}{} & 2&71\tabularnewline
United Kingdom & 1978\textendash 2005 & 4&69 & 1&72 & \multicolumn{2}{c}{} & 1&08 & 3&14 & 0&47 & \multicolumn{2}{c}{} & 0&88 & 1&27 & 0&41 & 0&20 & 1&87 & 0&06 & \multicolumn{2}{c}{} & 2&13\tabularnewline
Unites States & 1978\textendash 2005 & 4&37 & 1&26 & \multicolumn{2}{c}{} & 0&83 & 3&21 & 0&33 & \multicolumn{2}{c}{} & 0&85 & 1&58 & 0&13 & \textendash 0&02 & 1&63 & 0&20 & \multicolumn{2}{c}{} & 1&81\tabularnewline
Japan & 1978\textendash 2005 & 4&05 & 0&89 & \multicolumn{2}{c}{} & 0&84 & 3&02 & 0&19 & \multicolumn{2}{c}{} & 1&27 & 0&57 & 0&24 & \textendash 0&43 & 2&45 & \textendash 0&05 & \multicolumn{2}{c}{} & 1&97\tabularnewline
Denmark & 1980\textendash 2005 & 3&84 & 2&07 & \multicolumn{2}{c}{} & 1&12 & 2&38 & 0&35 & \multicolumn{2}{c}{} & 0&61 & 0&53 & 0&13 & 0&51 & 1&85 & 0&22 & \multicolumn{2}{c}{} & 2&57\tabularnewline
Austria & 1980\textendash 2005 & 3&27 & 1&06 & \multicolumn{2}{c}{} & 0&73 & 2&29 & 0&25 & \multicolumn{2}{c}{} & 0&77 & 0&98 & 0&66 & \textendash 0&04 & 1&31 & \textendash 0&41 & \multicolumn{2}{c}{} & 0&87\tabularnewline
Finland & 1978\textendash 2005 & 3&08 & 1&02 & \multicolumn{2}{c}{} & 0&53 & 2&06 & 0&49 & \multicolumn{2}{c}{} & 0&65 & 0&96 & 0&57 & \textendash 0&12 & 1&10 & \textendash 0&09 & \multicolumn{2}{c}{} & 0&89\tabularnewline
Netherlands & 1987\textendash 2005 & 3&04 & 0&30 & \multicolumn{2}{c}{} & 0&75 & 2&24 & 0&05 & \multicolumn{2}{c}{} & 0&87 & 1&54 & 0&10 & \textendash 0&12 & 0&71 & \textendash 0&05 & \multicolumn{2}{c}{} & 0&54\tabularnewline
Czech Republic & 1996\textendash 2005 & 2&92 & 1&16 & \multicolumn{2}{c}{} & 1&38 & 1&11 & 0&43 & \multicolumn{2}{c}{} & 1&31 & 0&36 & 0&58 & 0&07 & 0&76 & \textendash 0&15 & \multicolumn{2}{c}{} & 0&68\tabularnewline
Germany & 1992\textendash 2005 & 2&69 & 0&90 & \multicolumn{2}{c}{} & 0&68 & 1&88 & 0&13 & \multicolumn{2}{c}{} & 0&88 & 0&51 & 0&01 & \textendash 0&21 & 1&37 & 0&12 & \multicolumn{2}{c}{} & 1&28\tabularnewline
Sweden & 1994\textendash 2005 & 2&41 & \textendash 0&07 & \multicolumn{2}{c}{} & 0&41 & 1&89 & 0&11 & \multicolumn{2}{c}{} & 0&95 & 0&81 & 0&05 & \textendash 0&54 & 1&08 & 0&06 & \multicolumn{2}{c}{} & 0&60\tabularnewline
Portugal & 1996\textendash 2005 & \textendash 0&16 & \textendash 2&71 & \multicolumn{2}{c}{} & \textendash 0&29 & \textendash 0&03 & 0&16 & \multicolumn{2}{c}{} & 2&02 & 1&30 & 0&47 & \textendash 2&31 & \textendash 1&34 & \textendash 0&32 & \multicolumn{2}{c}{} & \textendash 3&96\tabularnewline
\hline 
\end{tabular}}
\par\end{centering}
\textit{\footnotesize{}Notes}{\footnotesize{}: The first and second
columns report the annual rates of growth in gross output ($y$) and
markup ($\omega$) from the first year of observation ($t_{0}$) to
2005 (i.e., $100\times(\ln y_{2005}-\ln y_{t_{0}})/(2005-t_{0})$).
The sixth to eleventh columns report the results of the Shapley decomposition.
The third, fourth, and fifth columns report the sum of the numbers
in the sixth and ninth columns, the seventh and tenth columns, and
eighth and eleventh columns, respectively. The last column reports
the sum of the numbers from the ninth to eleventh columns. Countries
are arranged in descending order of growth rate of gross output by
sector.}{\footnotesize\par}
\end{sidewaystable}

All the pairs of factor inputs and factor-augmenting technologies
contribute to economic growth in the goods and service sectors of
almost all countries. The contribution of factor inputs and factor-augmenting
technologies to the gross output growth is proportional to the factor
share of income and the rate of increase in factor inputs and factor-augmenting
technologies, as can be readily shown. The average capital, labor,
and energy shares of income in the year 2005 are, respectively, 22
(24), 49 (69), and 29 (7) percent in the goods (service) sector. Since
the labor share of income is much greater than the capital share of
income and the energy share of income, the pair of labor input and
labor-augmenting technology contributes most to economic growth. Since
the energy share of income is greater in the goods sector than in
the service sector, the pair of energy input and energy-saving technology
tends to contribute to economic growth more in the goods sector than
in the service sector. Consequently, the pair of energy input and
energy-saving technology contributes to economic growth more than
the pair of capital input and capital-augmenting technology in the
goods sector of half of the countries but less than the pair of capital
input and capital-augmenting technology in the service sector of almost
all countries.

Capital input contributes to economic growth in the goods and service
sectors of all countries. The contribution of capital input is greater
than that of capital-augmenting technology in the goods sector of
most countries and in the service sector of all countries. Labor input
does not contribute to economic growth in the goods sector but contributes
to economic growth in the service sector, reflecting the fact that
labor input decreased in the goods sector but increased in the service
sector. In contrast, the contribution of labor-augmenting technology
is greater in the goods sector than in the service sector for most
of the countries. Among the six elements of factor inputs and factor-augmenting
technologies, labor-augmenting technology contributes the most to
economic growth in the goods and service sectors of most countries.
Energy input contributes to economic growth in the goods and service
sectors of almost all countries, while energy-saving technology contributes
to economic growth in the goods or service sector of the majority
of the countries, including Denmark, Finland, Germany, Italy, Japan,
Sweden, the United Kingdom, and the United States. This result holds
even when material inputs are taken into account (see Table \ref{tab: growth_material}
in Appendix \ref{subsec: results_add}). The contribution of energy-saving
technology is greater than that of energy input in the goods and service
sectors of several countries. The TFP has a great deal of contribution
to economic growth or stagnation in all countries. The rate of increase
in the TFP is greater in the goods sector than in the service sector
for almost all countries, which is consistent with existing studies
of value-added growth accounting \citep{Herrendorf_Rogerson_Valentinyi_HEG14}.

\begin{sidewaystable}[ph]
\caption{Differences in economic growth compared to the United States\label{tab: growth_difference}}

\begin{centering}
\subfloat[Goods sector]{
\centering{}%
\begin{tabular}{llr@{\extracolsep{0pt}.}lr@{\extracolsep{0pt}.}lr@{\extracolsep{0pt}.}lr@{\extracolsep{0pt}.}lr@{\extracolsep{0pt}.}lr@{\extracolsep{0pt}.}lr@{\extracolsep{0pt}.}lr@{\extracolsep{0pt}.}lr@{\extracolsep{0pt}.}lr@{\extracolsep{0pt}.}lr@{\extracolsep{0pt}.}lr@{\extracolsep{0pt}.}lr@{\extracolsep{0pt}.}lr@{\extracolsep{0pt}.}lr@{\extracolsep{0pt}.}l}
\hline 
 &  & \multicolumn{2}{c}{$y$} & \multicolumn{2}{c}{$\omega$} & \multicolumn{2}{c}{} & \multicolumn{2}{c}{$a_{k}k$} & \multicolumn{2}{c}{$a_{\ell}\ell$} & \multicolumn{2}{c}{$a_{e}e$} & \multicolumn{2}{c}{} & \multicolumn{2}{c}{$k$} & \multicolumn{2}{c}{$\ell$} & \multicolumn{2}{c}{$e$} & \multicolumn{2}{c}{$a_{k}$} & \multicolumn{2}{c}{$a_{\ell}$} & \multicolumn{2}{c}{$a_{e}$} & \multicolumn{2}{c}{} & \multicolumn{2}{c}{$a$}\tabularnewline
\cline{3-6} \cline{5-6} \cline{9-14} \cline{11-14} \cline{13-14} \cline{17-28} \cline{19-28} \cline{21-28} \cline{23-28} \cline{25-28} \cline{27-28} \cline{31-32} 
Sweden & 1994\textendash 2005 & 2&54 & \textendash 0&89 & \multicolumn{2}{c}{} & 0&81 & 1&50 & 0&24 & \multicolumn{2}{c}{} & 0&53 & 0&22 & 0&63 & 0&28 & 1&28 & \textendash 0&39 & \multicolumn{2}{c}{} & 1&17\tabularnewline
Italy & 1978\textendash 2005 & 2&02 & 1&96 & \multicolumn{2}{c}{} & 0&48 & 1&27 & 0&28 & \multicolumn{2}{c}{} & 0&26 & \textendash 0&69 & 0&53 & 0&21 & 1&96 & \textendash 0&26 & \multicolumn{2}{c}{} & 1&92\tabularnewline
Japan & 1978\textendash 2005 & 1&23 & \textendash 0&36 & \multicolumn{2}{c}{} & 0&45 & 1&12 & \textendash 0&34 & \multicolumn{2}{c}{} & 0&61 & \textendash 0&82 & 0&28 & \textendash 0&16 & 1&94 & \textendash 0&63 & \multicolumn{2}{c}{} & 1&16\tabularnewline
Finland & 1978\textendash 2005 & 1&04 & \textendash 0&23 & \multicolumn{2}{c}{} & 0&35 & 0&52 & 0&16 & \multicolumn{2}{c}{} & 0&02 & \textendash 0&71 & 0&69 & 0&33 & 1&23 & \textendash 0&53 & \multicolumn{2}{c}{} & 1&04\tabularnewline
Denmark & 1980\textendash 2005 & 0&30 & 0&83 & \multicolumn{2}{c}{} & 0&23 & 0&55 & \textendash 0&47 & \multicolumn{2}{c}{} & 0&20 & \textendash 0&65 & 0&54 & 0&02 & 1&19 & \textendash 1&01 & \multicolumn{2}{c}{} & 0&20\tabularnewline
Austria & 1980\textendash 2005 & 0&27 & \textendash 0&18 & \multicolumn{2}{c}{} & 0&34 & 0&53 & \textendash 0&60 & \multicolumn{2}{c}{} & \textendash 0&12 & \textendash 0&73 & 1&32 & 0&45 & 1&26 & \textendash 1&93 & \multicolumn{2}{c}{} & \textendash 0&21\tabularnewline
Czech Republic & 1996\textendash 2005 & 0&11 & 0&36 & \multicolumn{2}{c}{} & 0&32 & \textendash 0&75 & 0&55 & \multicolumn{2}{c}{} & 0&56 & \textendash 0&29 & 1&78 & \textendash 0&24 & \textendash 0&45 & \textendash 1&24 & \multicolumn{2}{c}{} & \textendash 1&93\tabularnewline
Netherlands & 1987\textendash 2005 & \textendash 0&09 & \textendash 0&61 & \multicolumn{2}{c}{} & 0&37 & 0&04 & \textendash 0&50 & \multicolumn{2}{c}{} & \textendash 0&08 & 0&01 & 1&11 & 0&45 & 0&03 & \textendash 1&60 & \multicolumn{2}{c}{} & \textendash 1&13\tabularnewline
United Kingdom & 1978\textendash 2005 & \textendash 0&97 & 0&46 & \multicolumn{2}{c}{} & \textendash 0&09 & \textendash 1&01 & 0&12 & \multicolumn{2}{c}{} & \textendash 0&14 & \textendash 0&90 & 0&47 & 0&06 & \textendash 0&11 & \textendash 0&35 & \multicolumn{2}{c}{} & \textendash 0&40\tabularnewline
Germany & 1992\textendash 2005 & \textendash 1&55 & 0&34 & \multicolumn{2}{c}{} & \textendash 0&29 & \textendash 0&65 & \textendash 0&60 & \multicolumn{2}{c}{} & \textendash 0&23 & \textendash 1&53 & \textendash 0&21 & \textendash 0&06 & 0&88 & \textendash 0&39 & \multicolumn{2}{c}{} & 0&42\tabularnewline
Portugal & 1996\textendash 2005 & \textendash 2&81 & \textendash 3&51 & \multicolumn{2}{c}{} & \textendash 0&25 & \textendash 1&33 & \textendash 1&23 & \multicolumn{2}{c}{} & 0&39 & 0&19 & 2&79 & \textendash 0&64 & \textendash 1&53 & \textendash 4&03 & \multicolumn{2}{c}{} & \textendash 6&20\tabularnewline
\hline 
\end{tabular}}
\par\end{centering}
\begin{centering}
\subfloat[Service sector]{
\centering{}%
\begin{tabular}{llr@{\extracolsep{0pt}.}lr@{\extracolsep{0pt}.}lr@{\extracolsep{0pt}.}lr@{\extracolsep{0pt}.}lr@{\extracolsep{0pt}.}lr@{\extracolsep{0pt}.}lr@{\extracolsep{0pt}.}lr@{\extracolsep{0pt}.}lr@{\extracolsep{0pt}.}lr@{\extracolsep{0pt}.}lr@{\extracolsep{0pt}.}lr@{\extracolsep{0pt}.}lr@{\extracolsep{0pt}.}lr@{\extracolsep{0pt}.}lr@{\extracolsep{0pt}.}l}
\hline 
 &  & \multicolumn{2}{c}{$y$} & \multicolumn{2}{c}{$\omega$} & \multicolumn{2}{c}{} & \multicolumn{2}{c}{$a_{k}k$} & \multicolumn{2}{c}{$a_{\ell}\ell$} & \multicolumn{2}{c}{$a_{e}e$} & \multicolumn{2}{c}{} & \multicolumn{2}{c}{$k$} & \multicolumn{2}{c}{$\ell$} & \multicolumn{2}{c}{$e$} & \multicolumn{2}{c}{$a_{k}$} & \multicolumn{2}{c}{$a_{\ell}$} & \multicolumn{2}{c}{$a_{e}$} & \multicolumn{2}{c}{} & \multicolumn{2}{c}{$a$}\tabularnewline
\cline{3-6} \cline{5-6} \cline{9-14} \cline{11-14} \cline{13-14} \cline{17-28} \cline{19-28} \cline{21-28} \cline{23-28} \cline{25-28} \cline{27-28} \cline{31-32} 
Italy & 1978\textendash 2005 & 0&31 & 1&96 & \multicolumn{2}{c}{} & 0&12 & 0&02 & 0&18 & \multicolumn{2}{c}{} & \textendash 0&23 & \textendash 0&33 & \textendash 0&03 & 0&35 & 0&34 & 0&21 & \multicolumn{2}{c}{} & 0&90\tabularnewline
United Kingdom & 1978\textendash 2005 & 0&33 & 0&46 & \multicolumn{2}{c}{} & 0&25 & \textendash 0&07 & 0&14 & \multicolumn{2}{c}{} & 0&03 & \textendash 0&31 & 0&28 & 0&23 & 0&24 & \textendash 0&14 & \multicolumn{2}{c}{} & 0&33\tabularnewline
Denmark & 1980\textendash 2005 & \textendash 0&18 & 0&83 & \multicolumn{2}{c}{} & 0&35 & \textendash 0&50 & \textendash 0&04 & \multicolumn{2}{c}{} & \textendash 0&24 & \textendash 1&01 & \textendash 0&04 & 0&59 & 0&51 & 0&00 & \multicolumn{2}{c}{} & 1&10\tabularnewline
Japan & 1978\textendash 2005 & \textendash 0&32 & \textendash 0&36 & \multicolumn{2}{c}{} & 0&01 & \textendash 0&19 & \textendash 0&14 & \multicolumn{2}{c}{} & 0&42 & \textendash 1&01 & 0&11 & \textendash 0&41 & 0&82 & \textendash 0&25 & \multicolumn{2}{c}{} & 0&17\tabularnewline
Netherlands & 1987\textendash 2005 & \textendash 0&74 & \textendash 0&61 & \multicolumn{2}{c}{} & 0&09 & \textendash 0&72 & \textendash 0&11 & \multicolumn{2}{c}{} & 0&00 & 0&13 & \textendash 0&04 & 0&09 & \textendash 0&85 & \textendash 0&07 & \multicolumn{2}{c}{} & \textendash 0&83\tabularnewline
Austria & 1980\textendash 2005 & \textendash 0&75 & \textendash 0&18 & \multicolumn{2}{c}{} & \textendash 0&03 & \textendash 0&59 & \textendash 0&13 & \multicolumn{2}{c}{} & \textendash 0&08 & \textendash 0&56 & 0&49 & 0&05 & \textendash 0&03 & \textendash 0&63 & \multicolumn{2}{c}{} & \textendash 0&61\tabularnewline
Germany & 1992\textendash 2005 & \textendash 0&99 & 0&34 & \multicolumn{2}{c}{} & 0&12 & \textendash 1&13 & 0&02 & \multicolumn{2}{c}{} & \textendash 0&04 & \textendash 0&95 & \textendash 0&09 & 0&16 & \textendash 0&19 & 0&11 & \multicolumn{2}{c}{} & 0&08\tabularnewline
Finland & 1978\textendash 2005 & \textendash 1&29 & \textendash 0&23 & \multicolumn{2}{c}{} & \textendash 0&30 & \textendash 1&15 & 0&16 & \multicolumn{2}{c}{} & \textendash 0&21 & \textendash 0&61 & 0&45 & \textendash 0&09 & \textendash 0&53 & \textendash 0&29 & \multicolumn{2}{c}{} & \textendash 0&91\tabularnewline
Czech Republic & 1996\textendash 2005 & \textendash 1&50 & 0&36 & \multicolumn{2}{c}{} & 0&73 & \textendash 2&53 & 0&30 & \multicolumn{2}{c}{} & 0&31 & \textendash 0&87 & 0&50 & 0&42 & \textendash 1&66 & \textendash 0&20 & \multicolumn{2}{c}{} & \textendash 1&44\tabularnewline
Sweden & 1994\textendash 2005 & \textendash 1&73 & \textendash 0&89 & \multicolumn{2}{c}{} & \textendash 0&20 & \textendash 1&54 & 0&01 & \multicolumn{2}{c}{} & \textendash 0&01 & \textendash 0&54 & \textendash 0&04 & \textendash 0&19 & \textendash 1&00 & 0&06 & \multicolumn{2}{c}{} & \textendash 1&13\tabularnewline
Portugal & 1996\textendash 2005 & \textendash 4&59 & \textendash 3&51 & \multicolumn{2}{c}{} & \textendash 0&94 & \textendash 3&68 & 0&04 & \multicolumn{2}{c}{} & 1&02 & 0&08 & 0&40 & \textendash 1&96 & \textendash 3&75 & \textendash 0&37 & \multicolumn{2}{c}{} & \textendash 6&09\tabularnewline
\hline 
\end{tabular}}
\par\end{centering}
\textit{\footnotesize{}Notes}{\footnotesize{}: The first and second
columns report the annual rates of growth in gross output and markup
of each country relative to the United States. The sixth to eleventh
columns report the results of the Shapley decomposition. The third,
fourth, and fifth columns report the sum of the numbers in the sixth
and ninth columns, the seventh and tenth columns, and eighth and eleventh
columns, respectively. The last column reports the sum of the numbers
from the ninth to eleventh columns. Countries are arranged in descending
order of growth rate of gross output by sector.}{\footnotesize\par}
\end{sidewaystable}

Table \ref{tab: growth_difference} presents the quantitative contribution
of factor inputs and factor-augmenting technologies to international
differences in economic growth compared to the United States. The
results in Table \ref{tab: growth_difference} are based on those
in Table \ref{tab: growth} and obtained by subtracting each value
in the United States from the corresponding values in other countries
for the same period. The differences in the rates of growth in gross
output compared to the United States are attributed mainly to the
pair of labor input and labor-augmenting technology in the goods and
service sectors of most countries. However, they are also attributed
to the pair of capital input and capital-augmenting technology in
the goods sector of most countries and in the service sector of half
of the countries and to the pair of energy input and energy-saving
technology in the goods and service sectors of several countries.
The differences from the United States tend to be attributed to energy
input more than energy-saving technology in the goods sector, but
to energy-saving technology more than energy input in the service
sector. A significant fraction of the differences from the United
States can be attributed to the TFP in the goods and service sectors
of most countries.

\section{Extension\label{sec: extension}}

We end our analysis by discussing the interpretation of our results
and testing the main implication of the model when factor-augmenting
technologies are endogenous.

\subsection{A model of technology choice}

Following \citet{Caselli_Coleman_AER06}, we consider a model of technology
choice, in which the representative firm chooses factor-augmenting
technologies $(a_{k},a_{\ell},a_{e})$ as well as the quantities of
inputs $(k,\ell,e)$, so as to maximize its profits subject to production
technology and a technology frontier. The technology frontier, from
which the firm chooses the optimal mix of technologies, is given by:
\begin{equation}
\left[\left(\frac{a_{k}}{A_{k}}\right)^{\eta}+\left(\frac{a_{\ell}}{A_{\ell}}\right)^{\eta}+\left(\frac{a_{e}}{A_{e}}\right)^{\eta}\right]^{\frac{1}{\eta}}\le B,\qquad\text{for }\eta>\frac{\sigma}{1-\sigma}\label{eq: frontier}
\end{equation}
where $A_{k}$, $A_{\ell}$, and $A_{e}$ govern the trade-offs between
capital-, labor-, and energy-augmenting technologies, and $B$ governs
the technology frontier. We assume $\eta>\left.\sigma\right/(1-\sigma)$
as well as $\sigma<1$ to satisfy the second-order condition of this
problem. This assumption implies $\eta>\sigma$ and $\eta\sigma/(\eta-\sigma)<1$.

Suppose that the production technology is represented by the one-level
CES production function \eqref{eq: CES1}. For a given choice of input
quantities $(k,\ell,e)$, the optimal choice of technologies can be
written as:
\begin{eqnarray}
a_{k} & = & B\left[\left(A_{k}k\right)^{\frac{\eta\sigma}{\eta-\sigma}}+\left(A_{\ell}\ell\right)^{\frac{\eta\sigma}{\eta-\sigma}}+\left(A_{e}e\right)^{\frac{\eta\sigma}{\eta-\sigma}}\right]^{-\frac{1}{\eta}}A_{k}^{\frac{\eta}{\eta-\sigma}}k^{\frac{\sigma}{\eta-\sigma}},\\
a_{\ell} & = & B\left[\left(A_{k}k\right)^{\frac{\eta\sigma}{\eta-\sigma}}+\left(A_{\ell}\ell\right)^{\frac{\eta\sigma}{\eta-\sigma}}+\left(A_{e}e\right)^{\frac{\eta\sigma}{\eta-\sigma}}\right]^{-\frac{1}{\eta}}A_{\ell}^{\frac{\eta}{\eta-\sigma}}\ell^{\frac{\sigma}{\eta-\sigma}},\\
a_{e} & = & B\left[\left(A_{k}k\right)^{\frac{\eta\sigma}{\eta-\sigma}}+\left(A_{\ell}\ell\right)^{\frac{\eta\sigma}{\eta-\sigma}}+\left(A_{e}e\right)^{\frac{\eta\sigma}{\eta-\sigma}}\right]^{-\frac{1}{\eta}}A_{e}^{\frac{\eta}{\eta-\sigma}}e^{\frac{\sigma}{\eta-\sigma}}.
\end{eqnarray}
The production function \eqref{eq: CES1} can then be rewritten as:
\begin{equation}
y=\left[\left(A_{k}Bk\right)^{\frac{\eta\sigma}{\eta-\sigma}}+\left(A_{\ell}B\ell\right)^{\frac{\eta\sigma}{\eta-\sigma}}+\left(A_{e}Be\right)^{\frac{\eta\sigma}{\eta-\sigma}}\right]^{\frac{\eta-\sigma}{\eta\sigma}}\qquad\text{for }\frac{\eta\sigma}{\eta-\sigma}<1,\label{eq: CES3}
\end{equation}
where $\left.\eta\sigma\right/\left(\eta-\sigma\right)$ represents
the degree of substitution among capital, labor, and energy inputs
when the firm can adjust the mix of factor-augmenting technologies
as well as the mix of factor inputs in response to changes in factor
prices. In this case, the elasticity of substitution among capital,
labor, and energy inputs is $\epsilon_{\eta\sigma}\equiv\left.\left(\eta-\sigma\right)\right/\left(\eta-\sigma-\eta\sigma\right)>0$.
Therefore, the elasticity of substitution is greater when both factor
inputs and factor-augmenting technologies are endogenous than when
only factor inputs are endogenous (i.e., $\epsilon_{\eta\sigma}>\epsilon_{\sigma}$).

The first-order conditions with respect to $k$, $\ell$, and $e$
imply that
\begin{eqnarray}
\frac{w}{r} & = & \left(\frac{A_{\ell}}{A_{k}}\right)^{\frac{\epsilon_{\eta\sigma}-1}{\epsilon_{\eta\sigma}}}\left(\frac{\ell}{k}\right)^{-\frac{1}{\epsilon_{\eta\sigma}}},\label{eq: MRTS3a}\\
\frac{w}{v} & = & \left(\frac{A_{\ell}}{A_{e}}\right)^{\frac{\epsilon_{\eta\sigma}-1}{\epsilon_{\eta\sigma}}}\left(\frac{\ell}{e}\right)^{-\frac{1}{\epsilon_{\eta\sigma}}}.\label{eq: MRTS3b}
\end{eqnarray}
These equations are of the same form as equations \eqref{eq: MRTS1a}
and \eqref{eq: MRTS1b} with a different parameter. Hence, the elasticity
of substitution ($\epsilon_{\eta\sigma}$ or $\epsilon_{\sigma}$)
can be estimated using the same equations, irrespective of whether
technologies are endogenous or exogenous. Whether we estimate the
elasticity $\epsilon_{\eta\sigma}$ (or $\epsilon_{\sigma}$) from
equations \eqref{eq: moment_ces1a} and \eqref{eq: moment_ces1b}
depends on whether we use differences over a long (or short) period
during which technologies are endogenous (or exogenous) in the estimation.

The coefficients of capital, labor, and energy inputs in the production
function \eqref{eq: CES3} can be derived from equations \eqref{eq: CES3}\textendash \eqref{eq: MRTS3b}
as follows:
\begin{eqnarray}
A_{k}B & = & \left(\frac{rk}{rk+w\ell+ve}\right)^{\frac{\epsilon_{\eta\sigma}}{\epsilon_{\eta\sigma}-1}}\left(\frac{y}{k}\right),\label{eq: AkB3}\\
A_{\ell}B & = & \left(\frac{w\ell}{rk+w\ell+ve}\right)^{\frac{\epsilon_{\eta\sigma}}{\epsilon_{\eta\sigma}-1}}\left(\frac{y}{\ell}\right),\label{eq: AlB3}\\
A_{e}B & = & \left(\frac{ve}{rk+w\ell+ve}\right)^{\frac{\epsilon_{\eta\sigma}}{\epsilon_{\eta\sigma}-1}}\left(\frac{y}{e}\right).\label{eq: AeB3}
\end{eqnarray}
These equations are of the same form as equations \eqref{eq: ak1}\textendash \eqref{eq: ae1},
yet with a different parameter. Moreover, it can be readily shown
that factor-augmenting technologies $(a_{k},a_{\ell},a_{e})$ retain
the same form as equations \eqref{eq: ak1}\textendash \eqref{eq: ae1}
with the same parameter, irrespective of whether technologies are
endogenous or exogenous. The same results apply to the case in which
the production technology is represented by the two-level nested CES
production function \eqref{eq: CES2}.

Whether we measure $(A_{k}B,A_{\ell}B,A_{e}B)$ or $(a_{k},a_{\ell},a_{e})$
as a result of calculating the right-hand side of equations \eqref{eq: AkB3}\textendash \eqref{eq: AeB3}
depends on whether we use the elasticity $\epsilon_{\eta\sigma}$
or $\epsilon_{\sigma}$ in the calculation. We are concerned with
the possibility that our estimates of the elasticities of substitution
may be closer to those of $\epsilon_{\eta\sigma}$ than those of $\epsilon_{\sigma}$.
However, we obtain similar estimates of the elasticities of substitution
if we look at the differences over a shorter period. This result implies
that our estimates would be closer to those of $\epsilon_{\sigma}$,
and a longer span of data would be needed to estimate $\epsilon_{\eta\sigma}$.

Equations \eqref{eq: MRTS1a} and \eqref{eq: MRTS1b} imply that technological
change is directed toward scarce (abundant) factors when the elasticity
of substitution $\epsilon_{\sigma}$ is less (greater) than one; in
other words, factor inputs are complementary (substitutable) to some
extent \citep{Acemoglu_RES02,Caselli_Coleman_AER06}. Consequently,
energy-saving technology tends to progress in countries with scarce
(abundant) energy resources when the elasticity of substitution is
less (greater) than one, as can be seen from equation \eqref{eq: ae1}.
Our estimates of the elasticity of substitution imply that energy-saving
technology should progress in countries or years in which energy resources
are scarce.

\subsection{Energy resources and development}

\begin{figure}[H]
\caption{Energy-saving technological change and energy resource abundance\label{fig: self_sufficiency}}

\begin{centering}
\subfloat[Goods sector]{
\centering{}\includegraphics[scale=0.6]{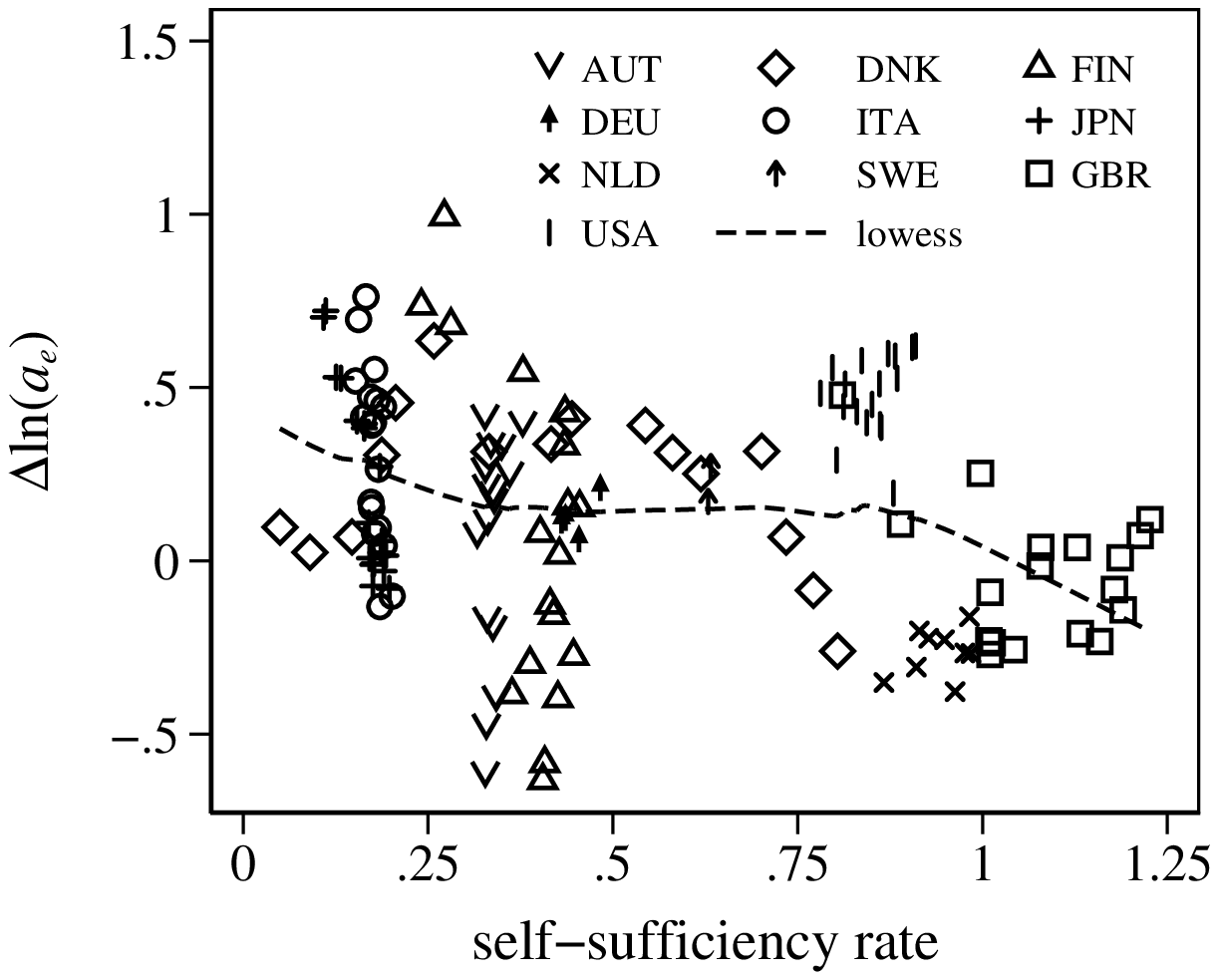}}\subfloat[Service sector]{
\centering{}\includegraphics[scale=0.6]{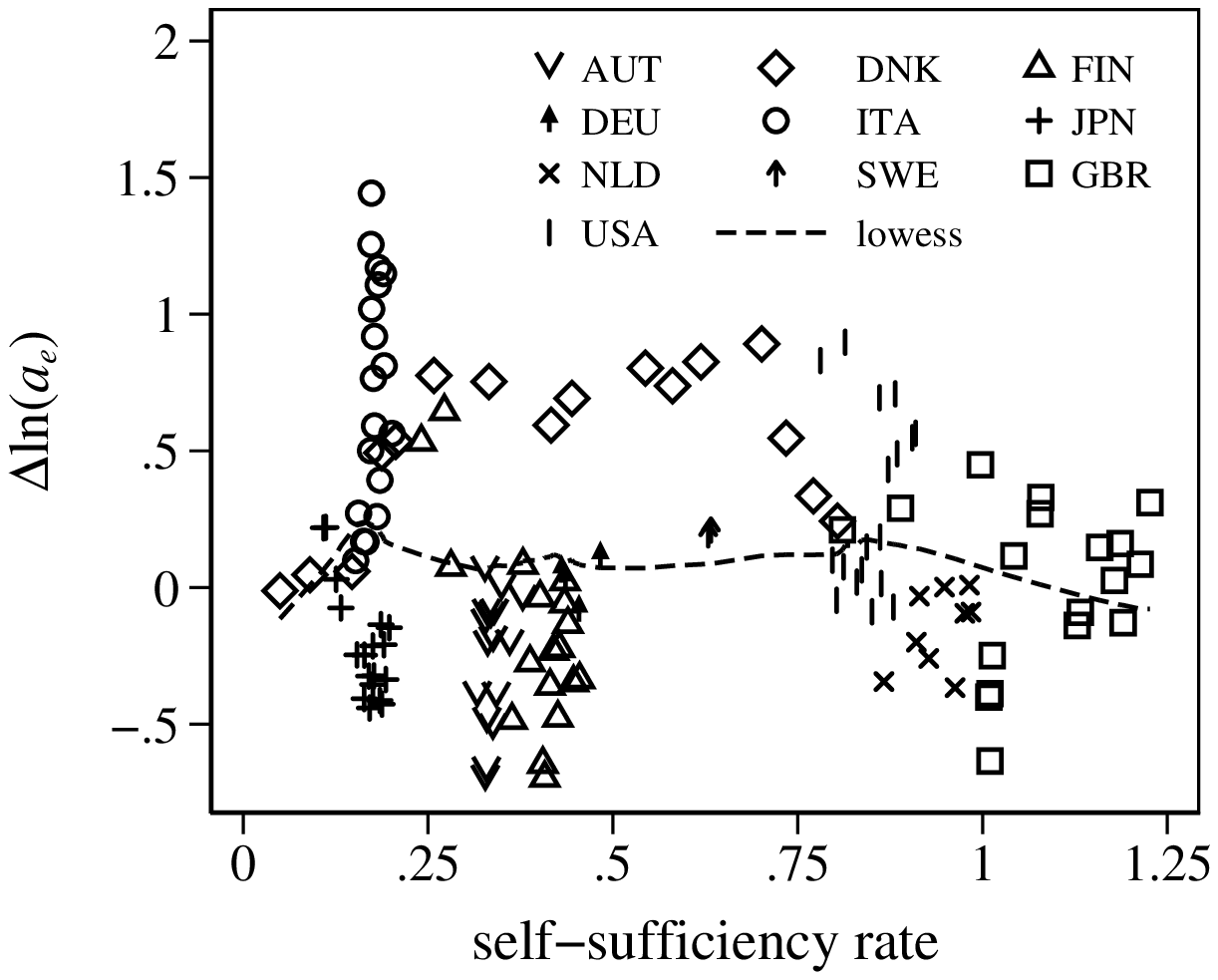}}
\par\end{centering}
\textit{\footnotesize{}Notes}{\footnotesize{}: The dashed curve is
the lowess (locally weighted scatterplot smoother) curve. Country
names are abbreviated as follows: AUT, Austria; DNK, Denmark; FIN,
Finland; DEU, Germany; ITA, Italy; JPN, Japan; NLD, the Netherlands;
SWE, Sweden; GBR, the United Kingdom; and USA, the United States.}{\footnotesize\par}
\end{figure}

We examine whether energy-saving technology tends to progress in countries
or years in which energy resources are scarce. Figure \ref{fig: self_sufficiency}
illustrates the relationship between energy-saving technological change
and energy resource abundance for each sector. Energy-saving technological
change is measured using the 10-year change in the log of energy-saving
technology. The abundance of energy resources is measured using the
self-sufficiency rate of energy supply. The Czech Republic and Portugal
are not included in the figure due to an insufficient number of periods.

As of the year 1978, the self-sufficiency rates were low at 4.4 percent
in Denmark, 10.9 percent in Japan, 16.6 percent in Italy, and 24.1
percent in Finland; medium at 35.8 percent in Austria and Sweden and
49.9 percent in Germany; and high at 78.1 percent in the United States,
81.1 percent in the United Kingdom, and 107.6 percent in the Netherlands.
Looking at the goods sector, where more energy is used in the process
of production, energy-saving technological change occurred in Denmark,
Finland, Italy, and Japan, which were not endowed with fossil fuels.
At the same time, energy-saving technological change did not occur
in the Netherlands, which was endowed with abundant natural gas. In
the United Kingdom, the self-sufficiency rate used to be lower than
50 percent in the early 1970s but increased up to 100 percent in the
early 1980s, in part due to the development of oil and natural gas
in the North Sea. In the United States, the self-sufficiency rate
gradually declined from 91 to 80 (70) percent between the years 1982
and 1995 (2005). In the two countries, energy-saving technological
change occurred in the period during which energy resources were less
abundant.

Overall, energy-saving technological change is negatively associated
with the self-sufficiency rate at a correlation coefficient of \textendash 0.216
(\textendash 0.038) and a \textit{p}-value of 0.011 (0.662) in the
goods (service) sector. The non-parametric regression curves indicate
that energy-saving technological change is negatively associated with
the self-sufficiency rate in the range where the self-sufficiency
rate is lower than 30 percent and higher than 80 percent for both
sectors. The observed negative association between energy-saving technological
change and energy resource abundance is consistent with the prediction
of the model when energy and non-energy inputs are complementary to
some extent.

We close this section by touching upon energy development and policies
in countries, where energy-saving technology has made progress. The
two oil crises in the 1970s were turning points in energy development
and policies. Many countries subsequently promoted energy conservation
and sought alternative energy sources. On the one hand, countries
without fossil fuels advanced the use of renewable energy and nuclear
energy. On the basis of climatic and geographical conditions, Denmark,
Japan, and Finland achieved the development of wind, solar, and woody
biomass energy, respectively. In addition, Japan and Finland achieved
the development of nuclear energy. Italy did not win public support
for nuclear power but managed to import electricity generated by nuclear
power and hydro power from France and Switzerland. On the other hand,
countries with fossil fuels advanced the expansion of indigenous production.
Against a backdrop of stable energy supply, the United Kingdom and
the United States deregulated the energy market from the 1980s to
the 1990s. Therefore, energy-saving technological change might be
associated with the development of alternative energy and the deregulation
of energy markets. We leave it for future research to determine the
impact of specific energy policies and regulations.

\section{Conclusion\label{sec: conclusion}}

This study aims to provide an understanding of the trends and differences
in energy-saving technological change among OECD countries. When we
measure the level and growth rate of energy-saving technology for
each country and sector, we use the theoretical result that technological
change can be measured using data on output per factor input and factor
income shares for a given value of the elasticity of substitution
in production. The main challenge that arises here is to estimate
the elasticity of substitution between energy and non-energy inputs
after taking into account the cross-sectional and time variation in
the unobserved components of factor-augmenting technologies. We address
this issue by using cross-country and cross-industry panel data and
shift\textendash share instruments. Our results indicate that capital,
labor, and energy inputs are complementary to some extent in production.

Our study is the first to measure and compare the levels and growth
rates of energy-saving technology across countries and sectors. This
makes it possible to conduct two kinds of analysis. First, we examine
the nature of energy-saving technological change and its differences
across countries. Second, we evaluate the quantitative contribution
of energy-saving technology to economic growth.

Our analysis yields the following findings. First, the levels and
growth rates of energy-saving technology vary substantially across
countries over time both in the goods and service sectors. Cross-country
differences in energy-saving technology are more persistent than those
in capital- and labor-augmenting technologies. Second, energy-saving
technology has progressed in countries or years in which energy resources
are scarce, as theory predicts when factor inputs are complementary
to some extent in production. Energy-saving technology relative to
other technologies has also progressed in countries in which government
spending on energy-related R\&D is large. Finally, energy-saving technological
change has a positive contribution to economic growth in the goods
or service sector of many countries, although not to the extent of
labor-augmenting technological change.

Our findings suggest that global energy efficiency can be improved
by developing general-purpose energy-saving technology and accelerating
international transfers of such technology. Such efforts would subsequently
lead to an increase in the national income for each country.

\clearpage{}

\bibliographystyle{econ}
\bibliography{references}

\clearpage{}

\appendix

\section{Appendix}

\captionsetup[subfigure]{font=normalsize}

\captionsetup[subtable]{font=normalsize}

\setcounter{table}{0} \renewcommand{\thetable}{A\arabic{table}}

\setcounter{figure}{0} \renewcommand{\thefigure}{A\arabic{figure}}

\setcounter{equation}{0}

\renewcommand\theequation{A.\arabic{equation}}

\subsection{Shapley decomposition\label{subsec: shapley}}

We describe the procedure to decompose the rate of growth in gross
output into the contribution of specific factor inputs and factor-augmenting
technologies. Let $\mathcal{Y}$ denote output and $d_{m}$ for $m\in\{1,2,\ldots,M\}=\mathcal{\mathcal{M}}$
denote its determinant factors, including factor inputs ($k$, $\ell$,
and $e$) and factor-augmenting technologies ($a_{k}$, $a_{\ell}$,
and $a_{e}$). For a given country, sector, and year, the natural
log of output is given by
\begin{equation}
\ln\mathcal{Y}=\mathcal{F}\left(d_{1},d_{2},\ldots,d_{M}\right).
\end{equation}
To quantify the contribution of each factor, we consider counterfactual
situations in which some or all of the factors are fixed at the initial
level. Let $\Gamma\left(\mathcal{G}\right)$ denote the value that
$\ln\mathcal{Y}$ takes if the factors $d_{m}$ for $m\notin\mathcal{G}\subseteq\mathcal{\mathcal{M}}$
are fixed at the initial level, $o=\left(o_{1},o_{2},\ldots,o_{M}\right)\in\mathcal{O}$
denote the order in which the factors are fixed, and $\mathcal{G}\left(o_{\tau},o\right)=\left\{ \left.o_{\tau^{\prime}}\right|\tau^{\prime}>\tau\right\} $
denote the set of factors that remain unfixed after the $\tau$-th
factor is fixed. The marginal contribution of the $m$-th factor to
the log changes in output, $\Delta\ln\mathcal{Y}$, can be measured
as:
\begin{equation}
\Lambda_{d_{m}}^{o}=\Gamma\left(\mathcal{G}\left(d_{m},o\right)\cup\left\{ d_{m}\right\} \right)-\Gamma\left(\mathcal{G}\left(d_{m},o\right)\right).
\end{equation}
The marginal contribution, $\Lambda_{d_{m}}^{o}$, depends on the
order in which the factors are fixed, but the average of the marginal
contributions over all possible sequences, $\Lambda_{d_{m}}$, does
not. The Shapley decomposition is
\begin{equation}
\Delta\ln\mathcal{Y}=\sum_{m\mathcal{\in\mathcal{M}}}\Lambda_{d_{m}},
\end{equation}
where
\begin{equation}
\Lambda_{d_{m}}=\frac{1}{M!}\sum_{o\in\mathcal{O}}\Lambda_{d_{m}}^{o}.
\end{equation}
This decomposition is not only path independent but also exact \citep{Shorrocks_JoEI13}.
The results of the decomposition are expressed in terms of growth
rates by dividing by the number of years between the first and last
years.

\subsection{Rental price of capital\label{subsec: rental_price}}

We describe the procedure to calculate the rental price of capital.
Capital is divided into capital equipment and structure. Capital equipment
is composed of computing equipment, communications equipment, software,
transport equipment, and other machinery and equipment, while capital
structure is composed of non-residential structures and infrastructures,
residential structures, and other assets. Let $j$ $\in$ \{equipment,
structure\} denote an index for the capital components. The rental
price of capital ($r_{jt}$) is determined by the price of investment
($q_{jt}$), the depreciation rate ($\delta_{j}$), and the interest
rate ($\iota_{t}$). The price of investment is calculated by dividing
the nominal value by the real value of investment for each component.
The depreciation rate is the time average of those obtained from the
law of motion of capital.

We calculate the rental price of capital in two ways. When we assume
the absence of markup, we adopt the internal rate of return approach
of \citet{OMahony_Timmer_EJ09}, who calculate the rental price of
capital as:
\begin{equation}
r_{jt}=\delta_{j}q_{jt}+\left(\iota_{t}-\frac{q_{jt}-q_{j,t-1}}{q_{j,t-1}}\right)q_{j,t-1},
\end{equation}
where the interest rate is the internal rate of return:
\begin{equation}
\iota_{t}=\frac{\sum_{j}r_{jt}k_{jt}-\sum_{j}\delta_{j}q_{jt}k_{jt}+\sum_{j}\left(q_{jt}-q_{j,t-1}\right)k_{jt}}{\sum_{j}q_{j,t-1}k_{jt}}.
\end{equation}
The advantage of this approach is to maintain consistency between
national income and production accounts. When we allow for the presence
of markup, we adopt the external rate of return approach of \citet*{Harper_Berndt_Wood_bc89},
who calculate the rental price of capital as:
\begin{equation}
r_{jt}=\delta_{j}q_{jt}+\widetilde{\iota}_{t}q_{j,t-1},
\end{equation}
where the real rate of return ($\tilde{\iota}$) is set to a constant
3.5 percent. The advantage of this approach is not to require the
assumption of competitive markets.

\subsection{Adjustment for input composition\label{subsec: composition}}

We describe the procedure used to adjust for the variation in the
composition of capital, labor, and energy inputs over time when calculating
the prices and quantities of capital, labor, and energy inputs. The
procedure is similar to that used by \citet*{Autor_Katz_Kearney_RESTAT08},
who adjust for the compositional changes in labor inputs when estimating
the aggregate production function with two types of labor in the United
States. In our case, capital is divided into capital equipment and
structure; labor is divided into skilled and unskilled labor; and
energy is divided into sulfur fuel oil, light fuel oil, natural gas,
electricity, automotive diesel, steam coal, and coking coal. The procedure
requires the assumption that the input components are perfect substitutes
within each type of input.

Here, we denote the price of input by $p\in\left\{ r,w,v\right\} $
and the quantity of input by $x\in\{k,\ell,e\}$ $\in$ \{(equipment,
structure), (skilled, unskilled), (sulfur fuel oil, light fuel oil,
natural gas, electricity, automotive diesel, steam coal, coking coal)\}.
We use squiggles to represent unadjusted prices and quantities. If
there were no need to make adjustments to input prices and quantities,
we could calculate the price of input in country $c$, sector $s$,
and year $t$ as $\widetilde{p}_{cst}=\sum_{j}\theta_{jcst}^{x}\widetilde{p}_{jcst}$,
where $\theta_{jcst}^{x}$ is the share of component $j$ in input
$x$ (i.e., $\theta_{jcst}^{x}=\widetilde{x}_{jcst}/\sum_{j}\widetilde{x}_{jcst}$),
and the quantity of input in country $c$, sector $s$, and year $t$
as $\widetilde{x}_{cst}=\sum_{j}\widetilde{x}_{jcst}$.

We adjust for the variation in input composition over time by holding
the shares of input components constant when calculating input prices
and by using time-invariant efficiency units as weights when calculating
input quantities. Let $T_{c}$ denote the number of years observed
for country $c$ and $J_{x}$ denote the number of components in input
$x$. We can calculate the composition-adjusted price of input in
country $c$, sector $s$, and year $t$ as $p_{cst}=\sum_{j}\overline{\theta}_{jcs}^{x}\widetilde{p}_{jcst}$,
where $\overline{\theta}_{jcs}^{x}$ is the country- and sector-specific
mean of $\theta_{jcst}^{x}$ (i.e., $\overline{\theta}_{jcs}^{x}=\sum_{t=1}^{T_{c}}\theta_{jcst}^{x}/T_{c}$),
and the composition-adjusted quantity of input in country $c$, sector
$s$, and year $t$ as $x_{cst}=\sum_{j}(\overline{p}_{jcs}/\overline{p}_{cs})\widetilde{x}_{jcst}$,
where the weight is the country- and sector-specific mean of $\widetilde{p}_{jcst}$
(i.e., $\overline{p}_{jcs}=\sum_{t=1}^{T_{c}}\widetilde{p}_{jcst}/T_{c}$)
normalized by its mean across components (i.e., $\overline{p}_{cs}=\sum_{j}\overline{p}_{jcs}/J_{x}$).

We construct the data on the prices of input components $\widetilde{r}_{j}$
and $\widetilde{w}_{j}$, the shares of input components $\theta_{j}^{k}$
and $\theta_{j}^{\ell}$, and the quantities of inputs $\widetilde{k}$,
$\widetilde{\ell}$, and $\widetilde{e}$ from the EU KLEMS database
and obtain the data on the energy price $\widetilde{v}_{j}$ and the
share of energy components $\theta_{j}^{e}$ from the IEA database
(World Energy Prices and World Energy Balances).

\subsection{Factor productivity and factor income shares\label{subsec: productivity=000026income_shares}}

\begin{figure}[H]
\caption{Output per factor input in the goods sector\label{fig: factor_output_goods}}

\begin{centering}
\subfloat[Austria]{
\centering{}\includegraphics[scale=0.4]{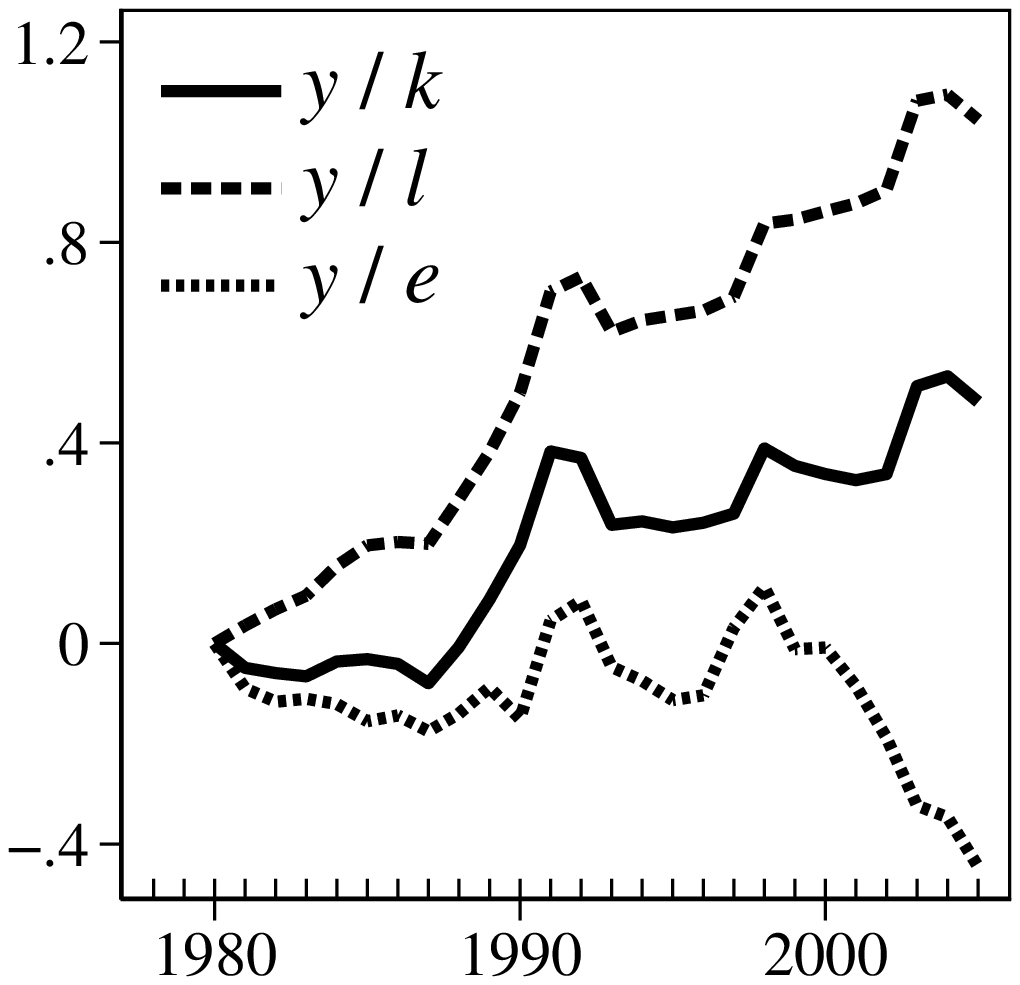}}\subfloat[Czech Republic]{
\centering{}\includegraphics[scale=0.4]{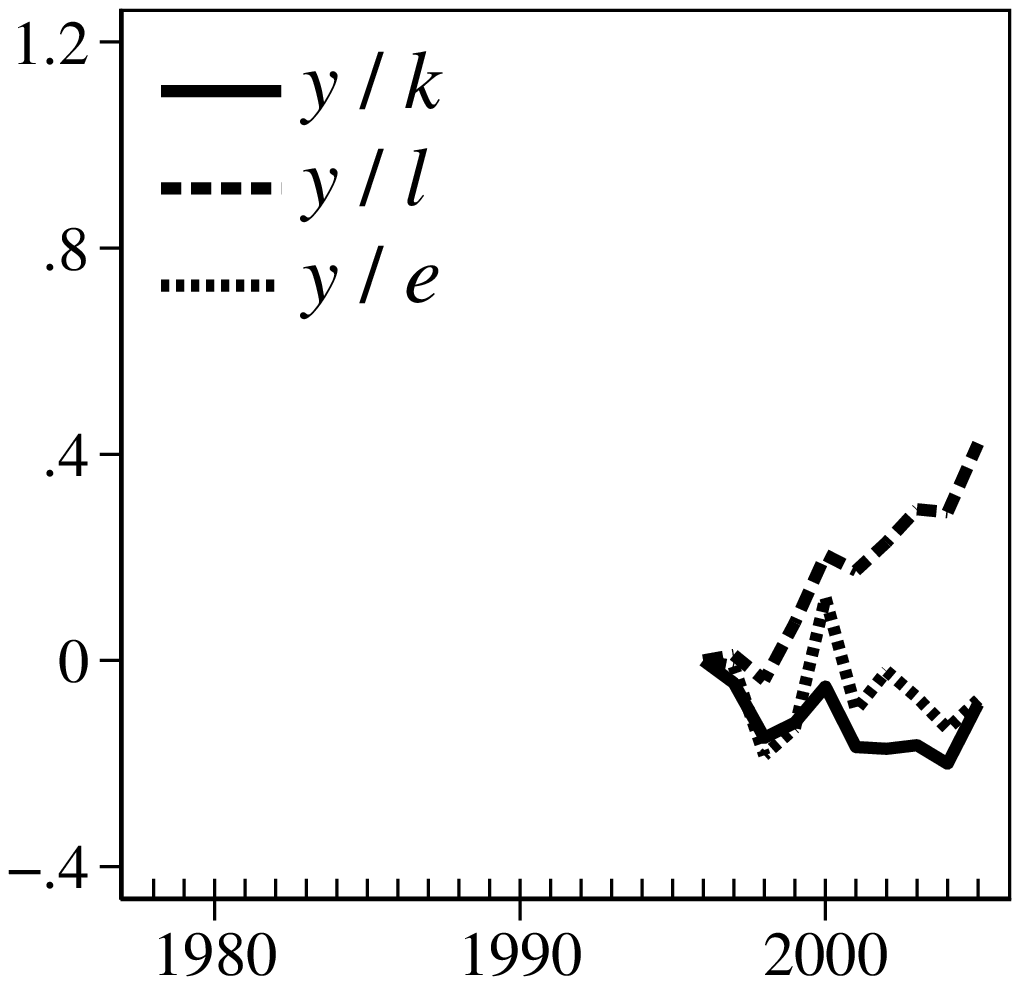}}\subfloat[Denmark]{
\centering{}\includegraphics[scale=0.4]{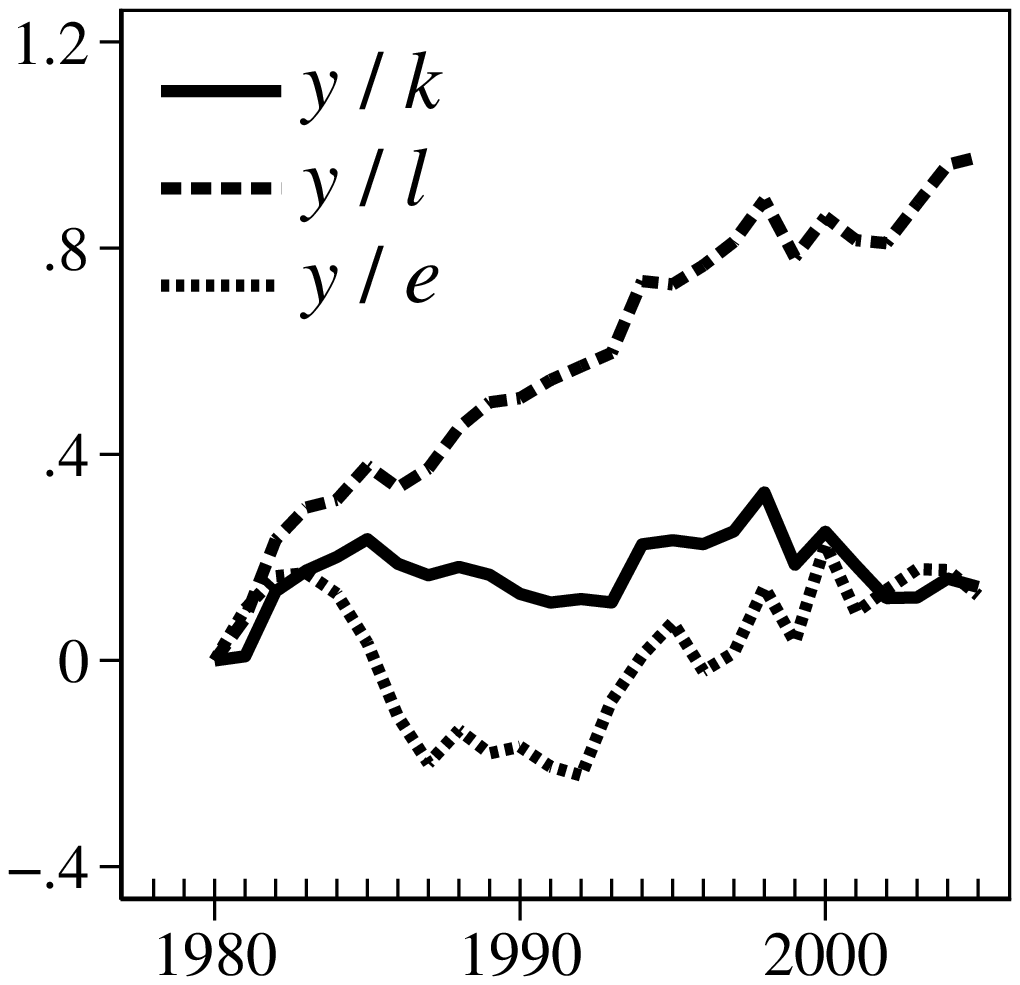}}\subfloat[Finland]{
\centering{}\includegraphics[scale=0.4]{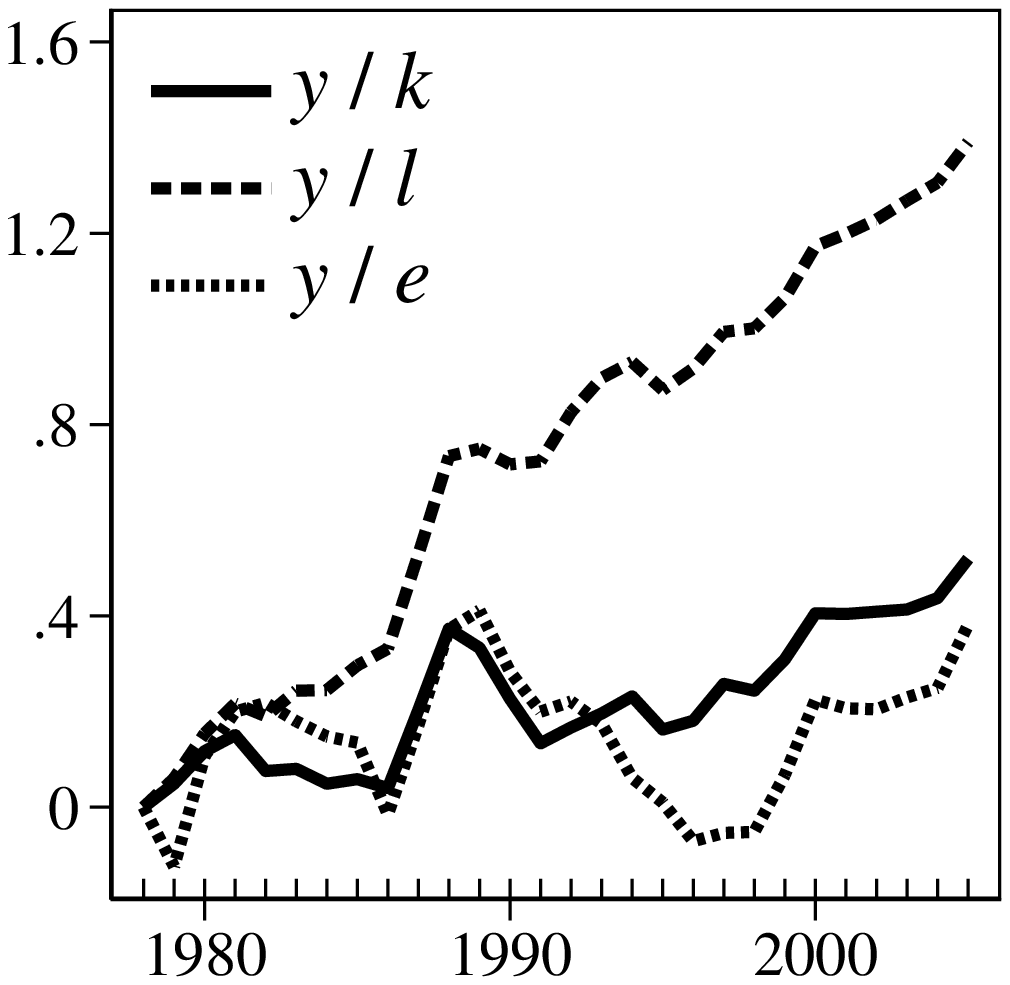}}
\par\end{centering}
\begin{centering}
\subfloat[Germany]{
\centering{}\includegraphics[scale=0.4]{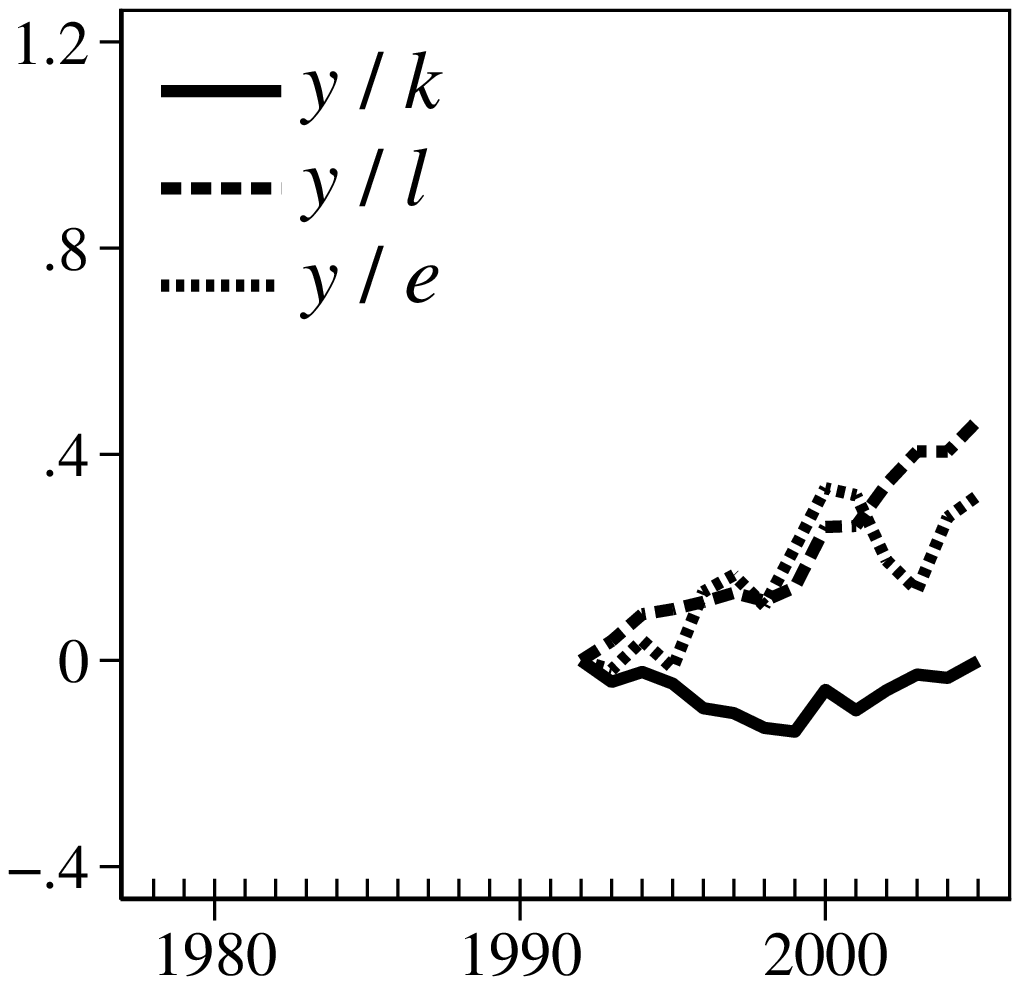}}\subfloat[Italy]{
\centering{}\includegraphics[scale=0.4]{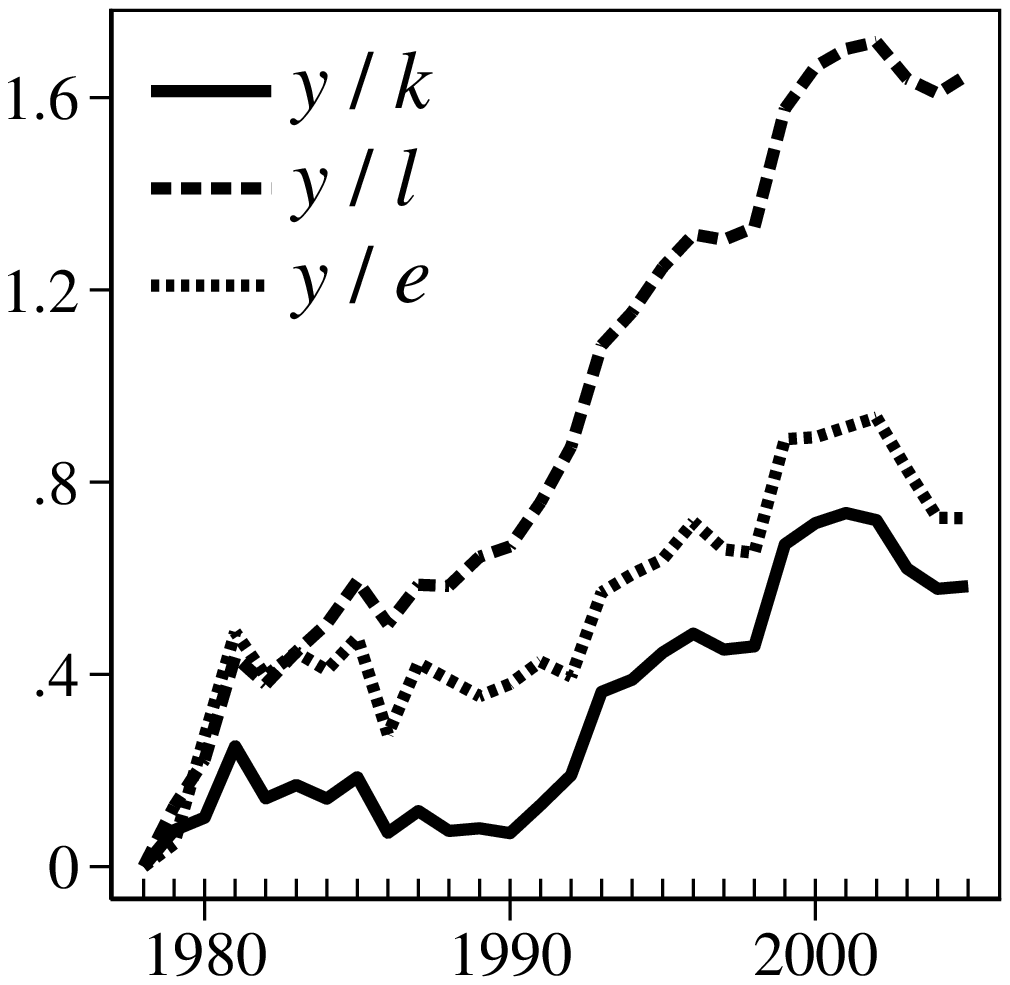}}\subfloat[Japan]{
\centering{}\includegraphics[scale=0.4]{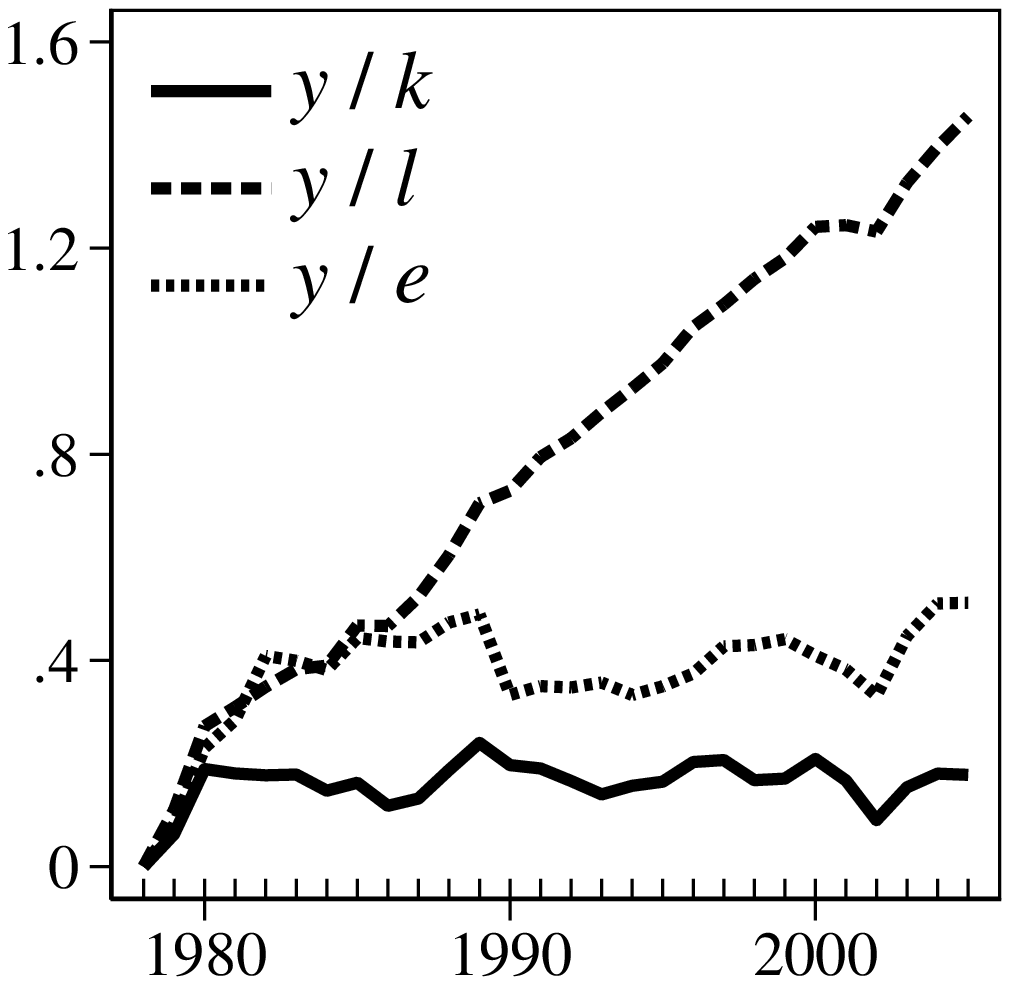}}\subfloat[Netherlands]{
\centering{}\includegraphics[scale=0.4]{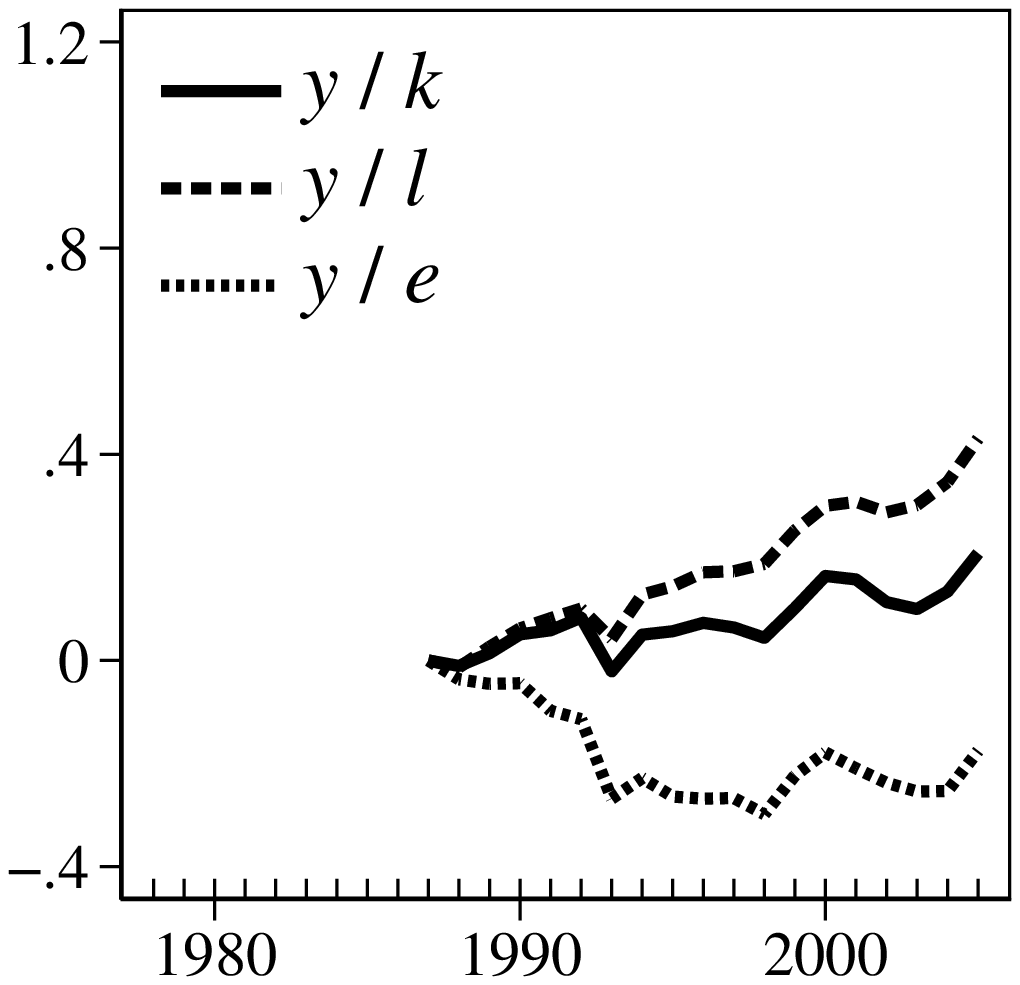}}
\par\end{centering}
\begin{centering}
\subfloat[Portugal]{
\centering{}\includegraphics[scale=0.4]{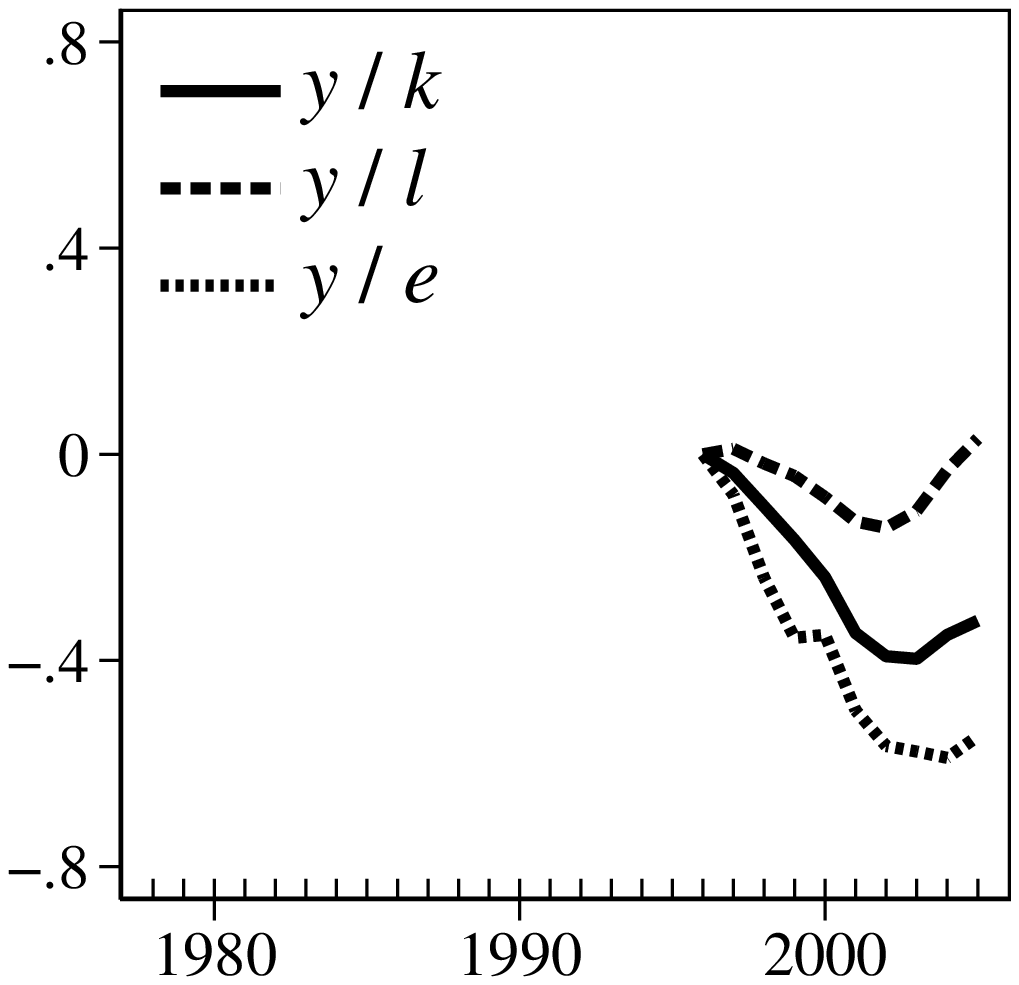}}\subfloat[Sweden]{
\centering{}\includegraphics[scale=0.4]{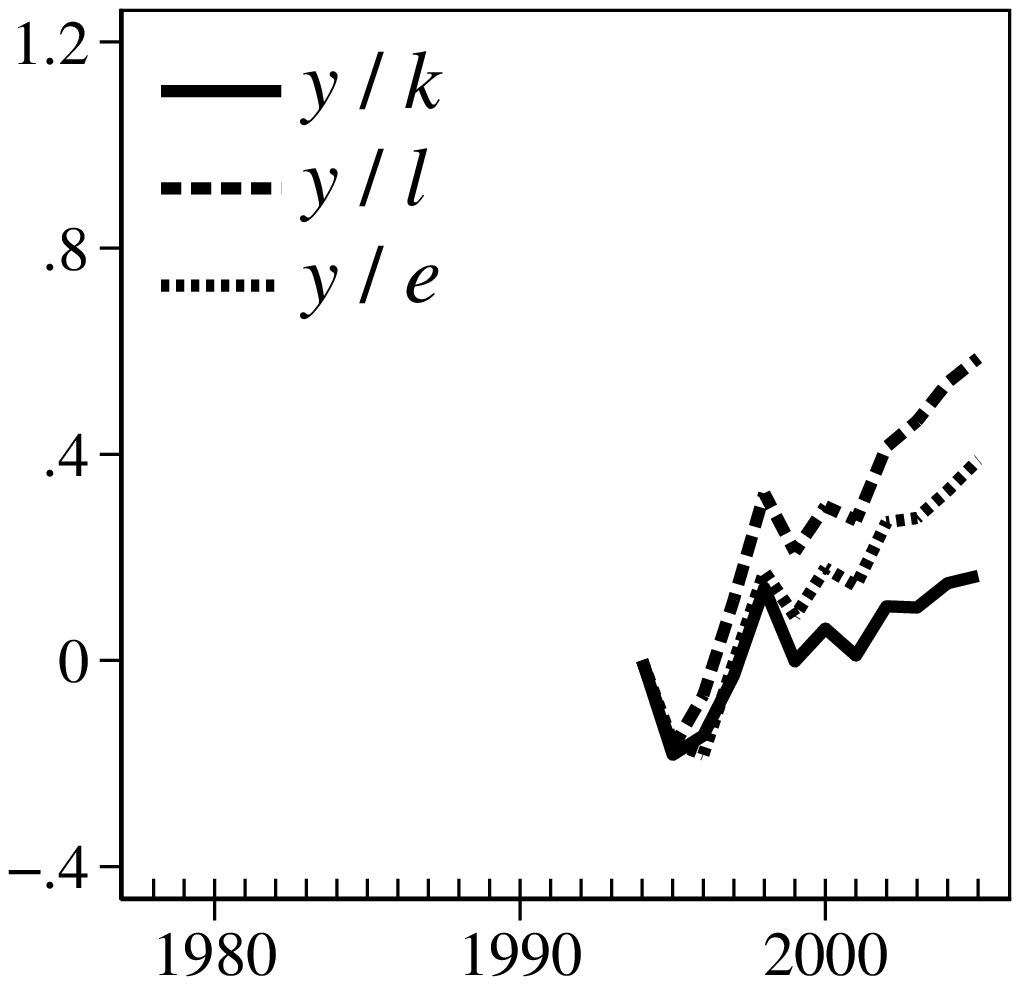}}\subfloat[United Kingdom]{
\centering{}\includegraphics[scale=0.4]{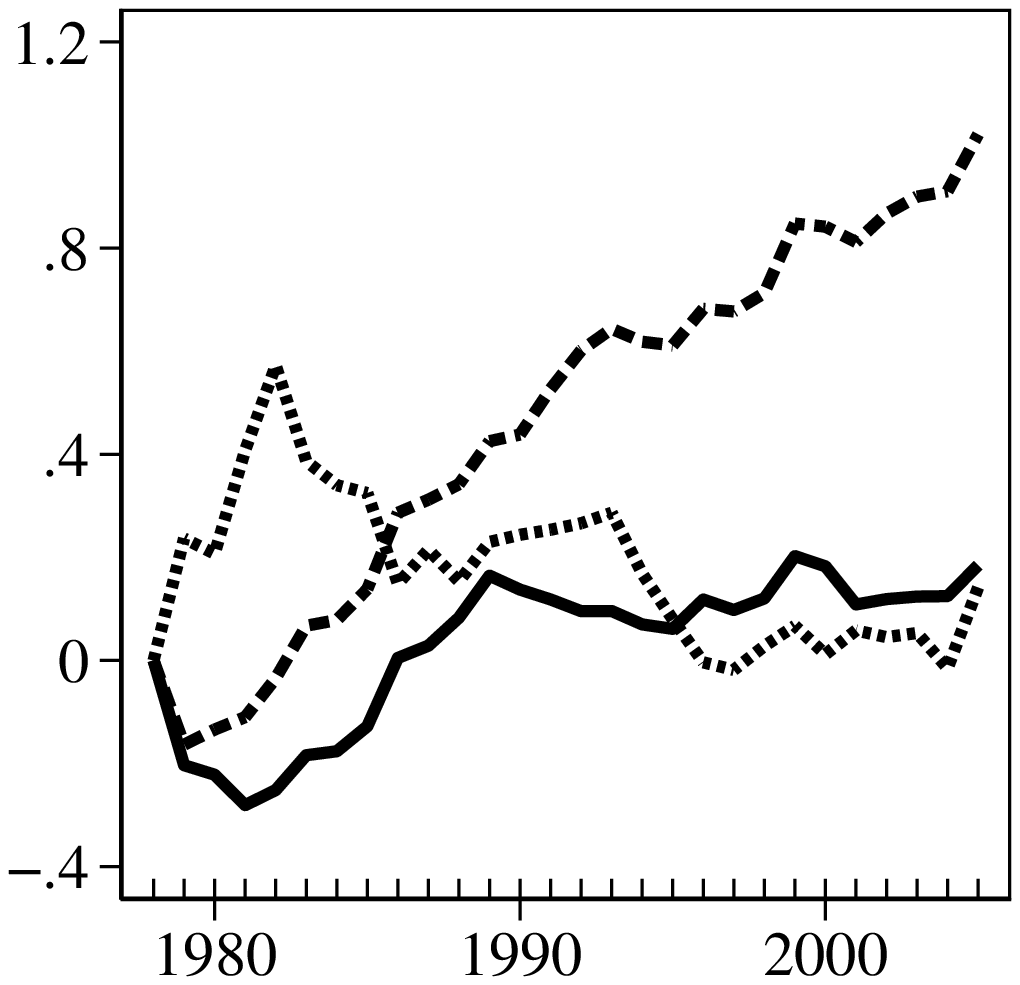}}\subfloat[United States]{
\centering{}\includegraphics[scale=0.4]{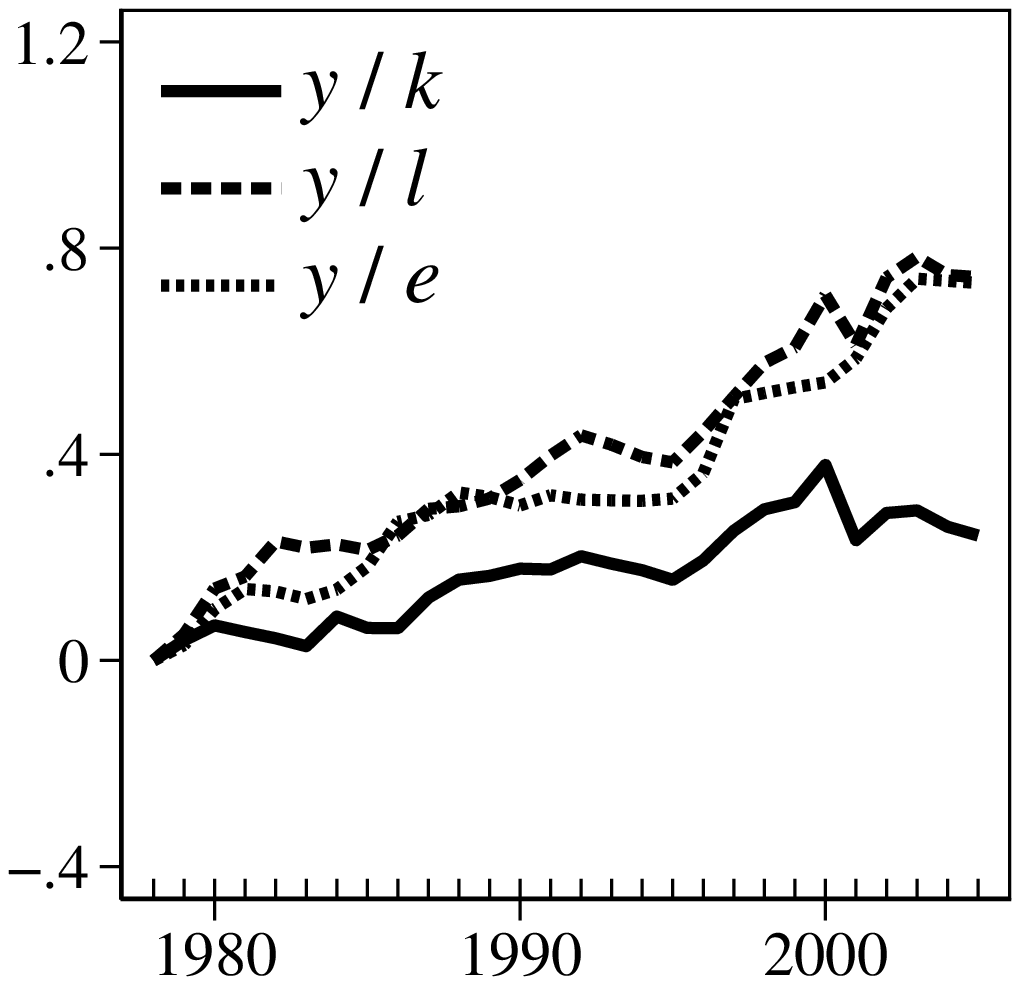}}
\par\end{centering}
\textit{\footnotesize{}Notes}{\footnotesize{}: The solid, dashed,
and dotted lines represent output per capital input ($y/k$), output
per labor input ($y/\ell$), and output per energy input ($y/e$),
respectively. All series are expressed as log differences relative
to the first year of observations.}{\footnotesize\par}
\end{figure}

\begin{figure}[H]
\caption{Output per factor input in the service sector\label{fig: factor_output_service}}

\begin{centering}
\subfloat[Austria]{
\centering{}\includegraphics[scale=0.4]{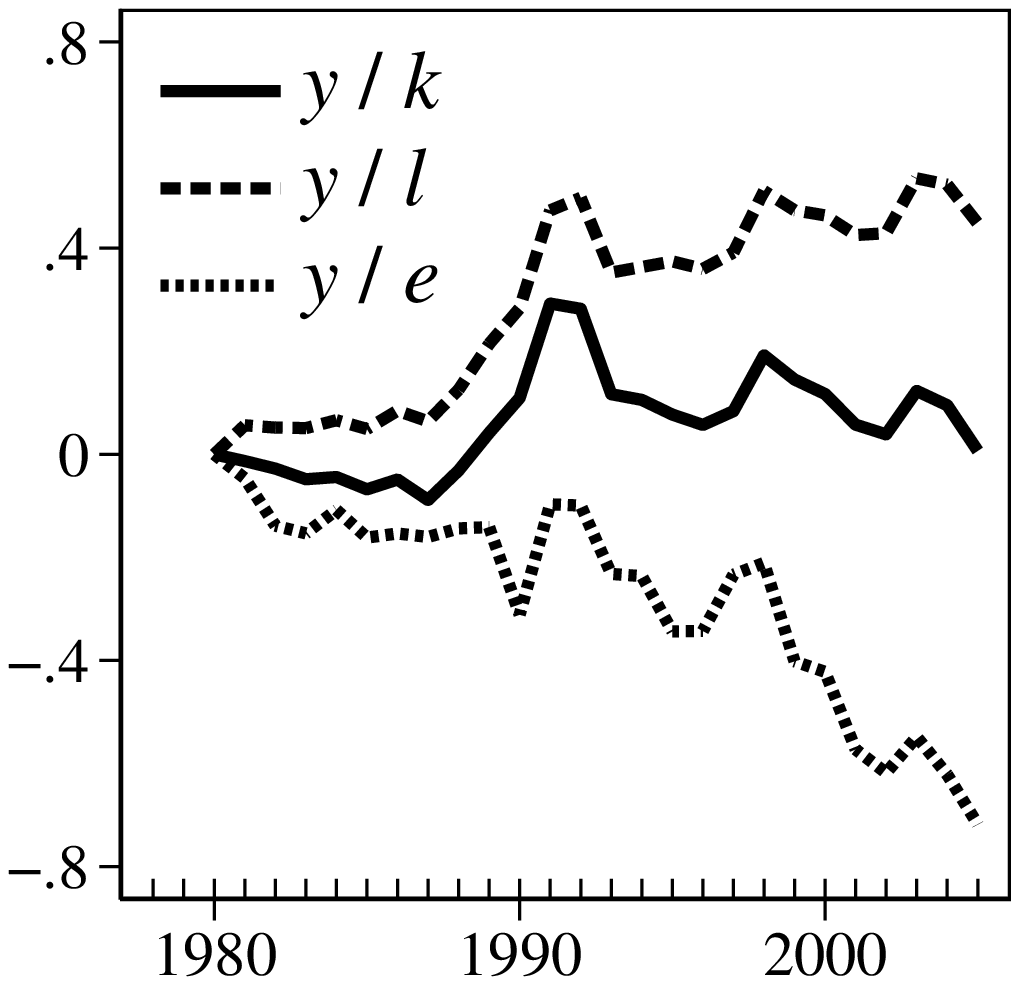}}\subfloat[Czech Republic]{
\centering{}\includegraphics[scale=0.4]{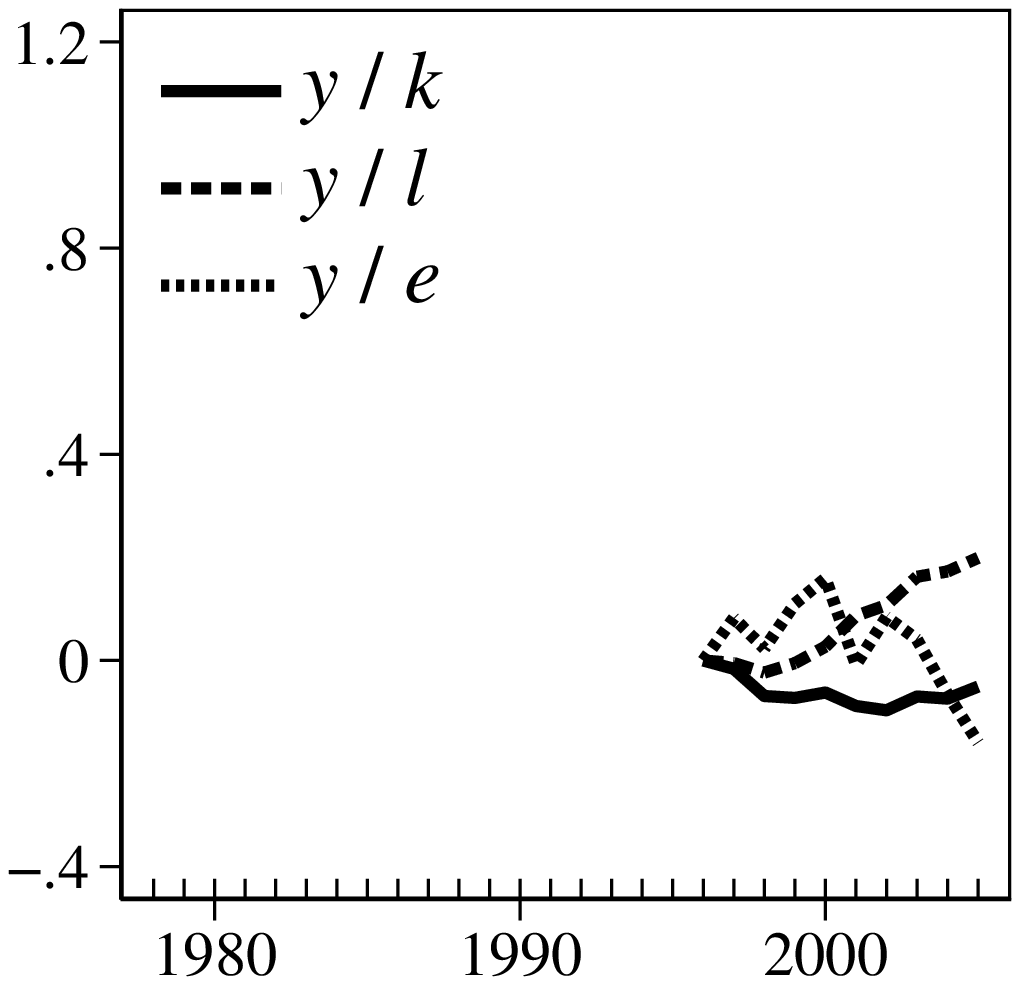}}\subfloat[Denmark]{
\centering{}\includegraphics[scale=0.4]{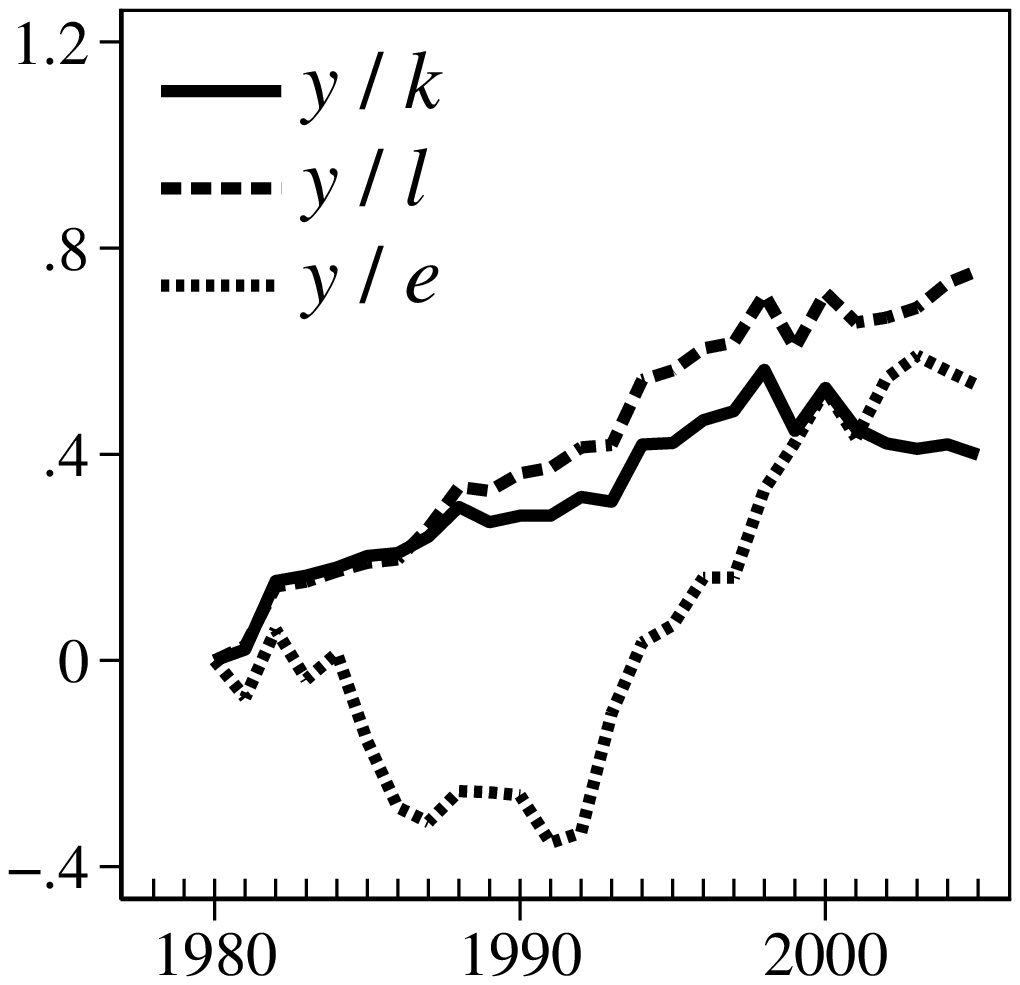}}\subfloat[Finland]{
\centering{}\includegraphics[scale=0.4]{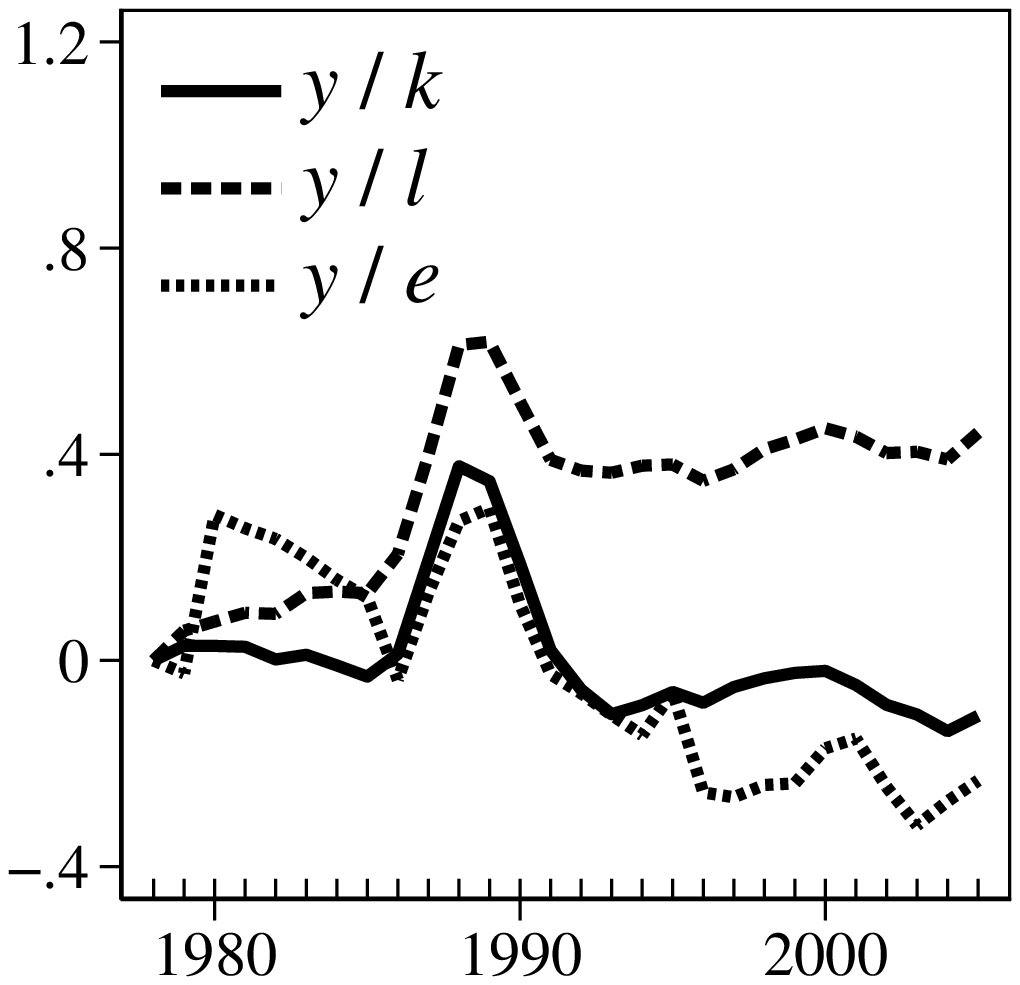}}
\par\end{centering}
\begin{centering}
\subfloat[Germany]{
\centering{}\includegraphics[scale=0.4]{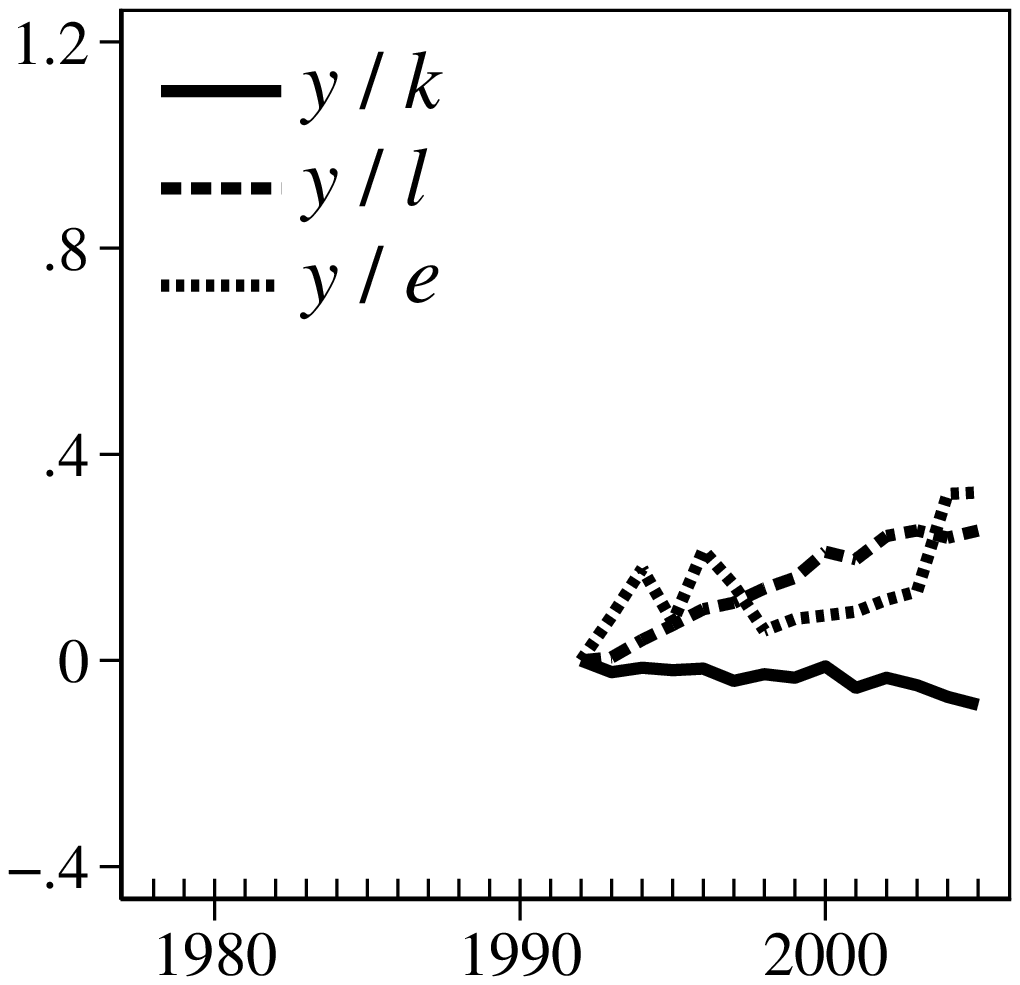}}\subfloat[Italy]{
\centering{}\includegraphics[scale=0.4]{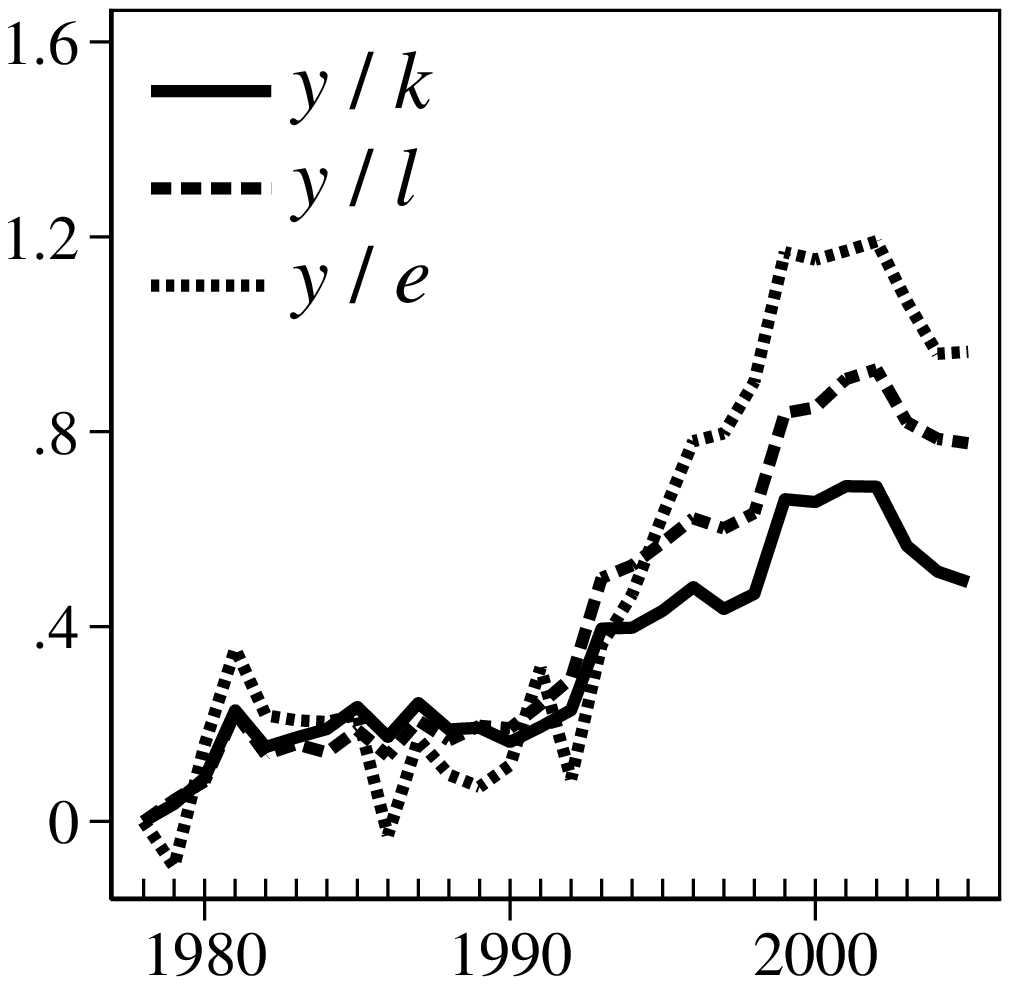}}\subfloat[Japan]{
\centering{}\includegraphics[scale=0.4]{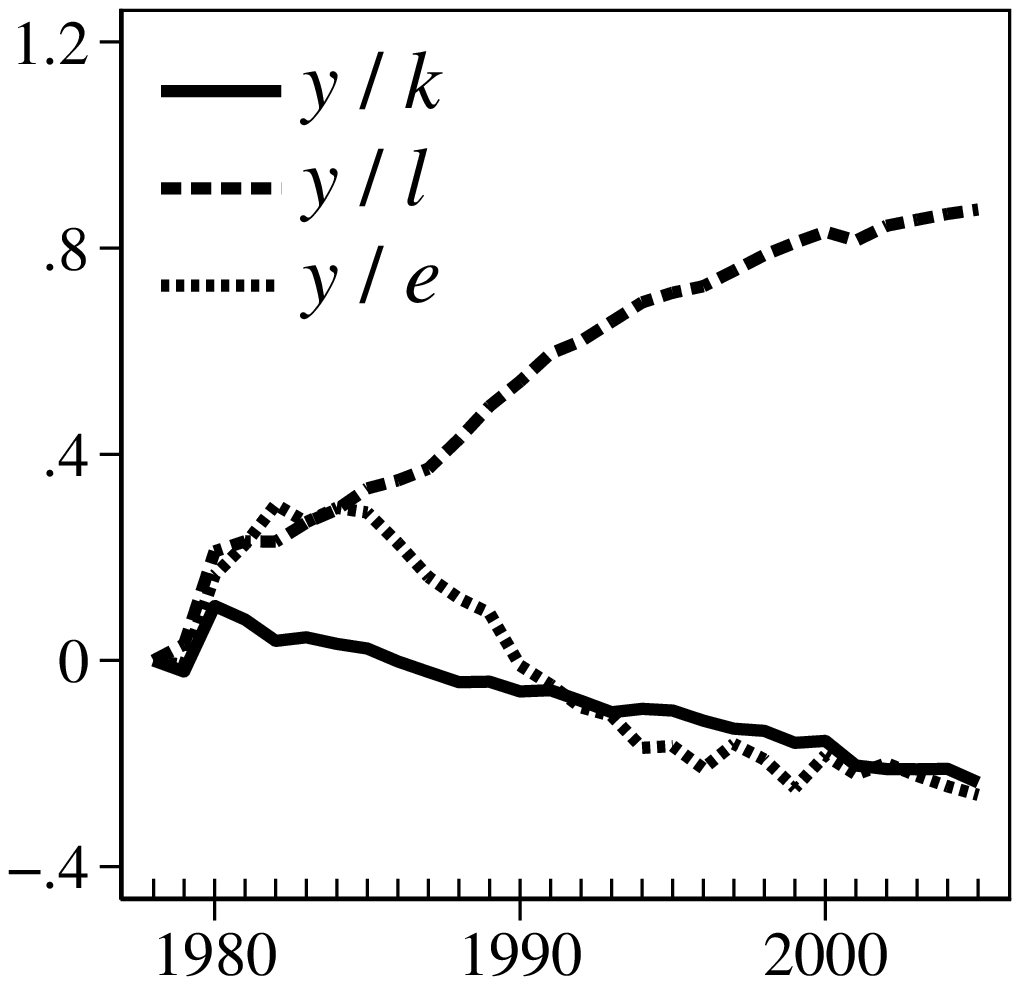}}\subfloat[Netherlands]{
\centering{}\includegraphics[scale=0.4]{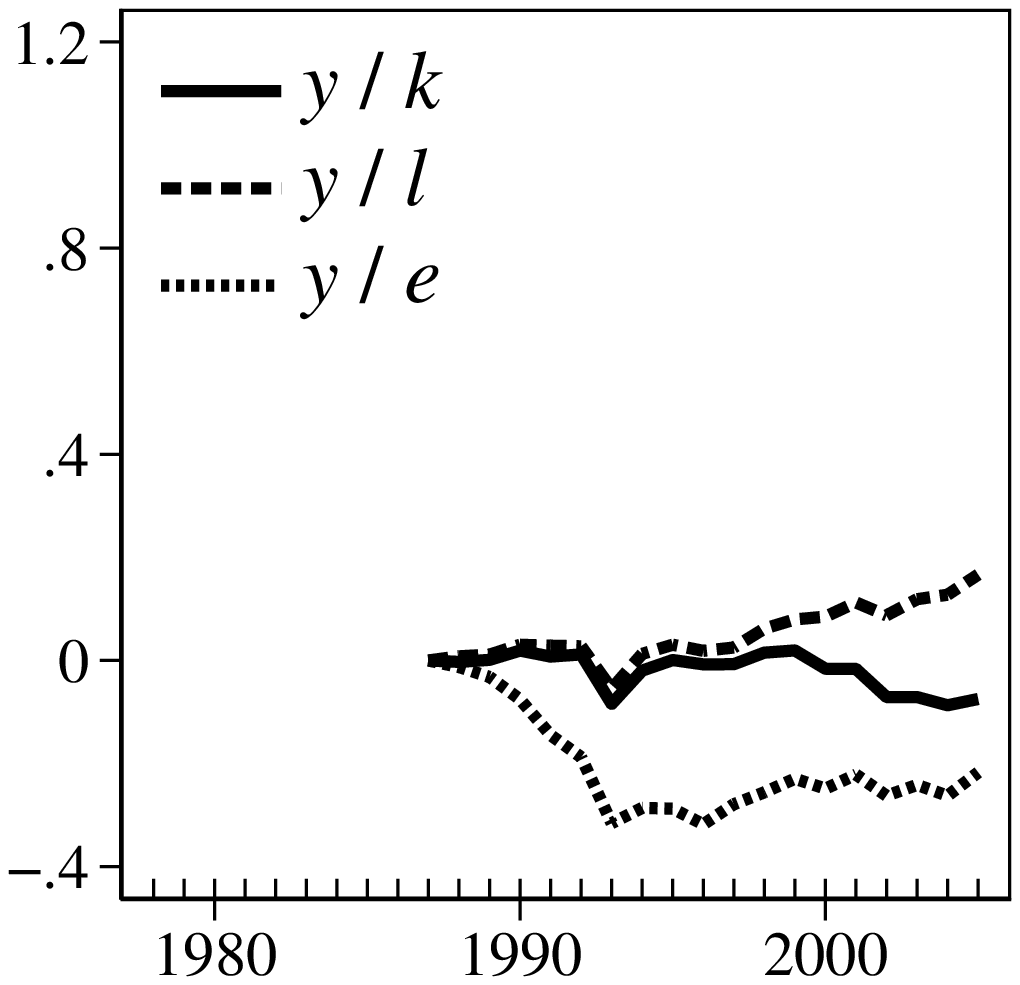}}
\par\end{centering}
\begin{centering}
\subfloat[Portugal]{
\centering{}\includegraphics[scale=0.4]{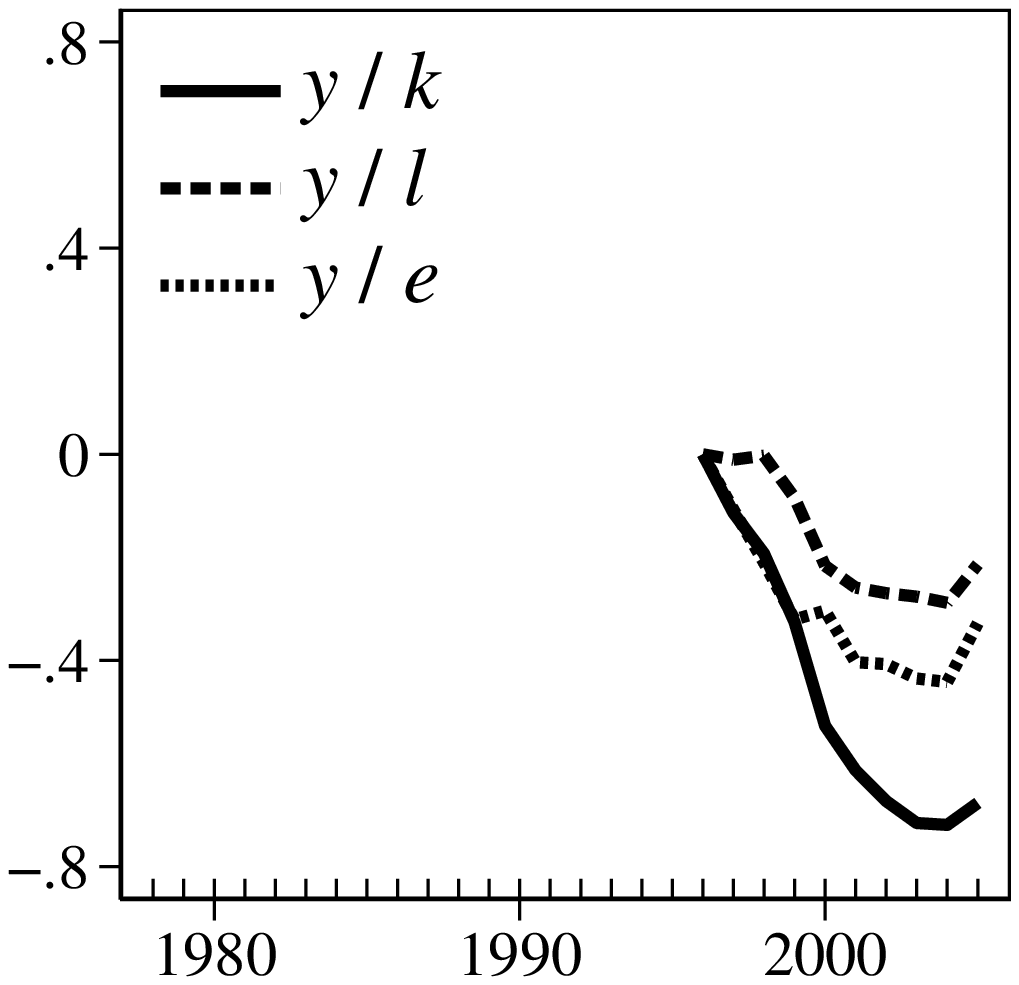}}\subfloat[Sweden]{
\centering{}\includegraphics[scale=0.4]{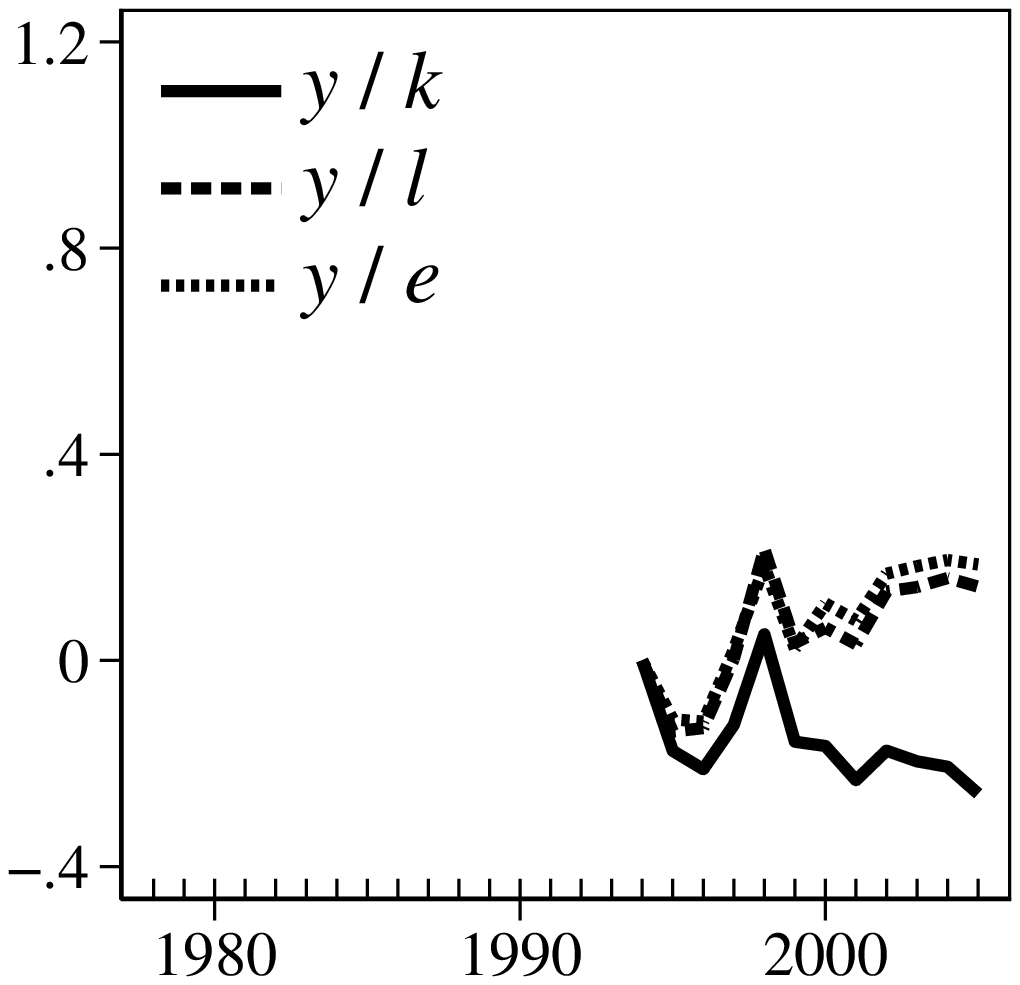}}\subfloat[United Kingdom]{
\centering{}\includegraphics[scale=0.4]{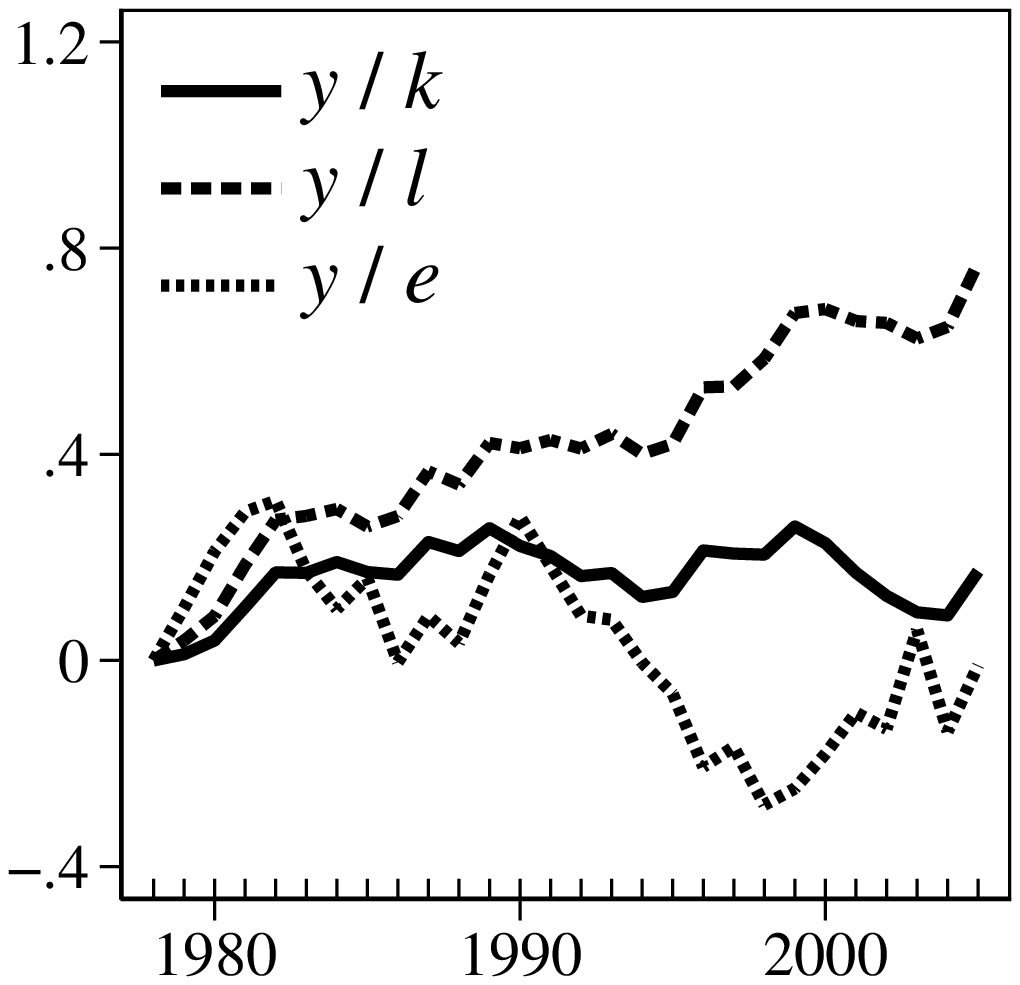}}\subfloat[United States]{
\centering{}\includegraphics[scale=0.4]{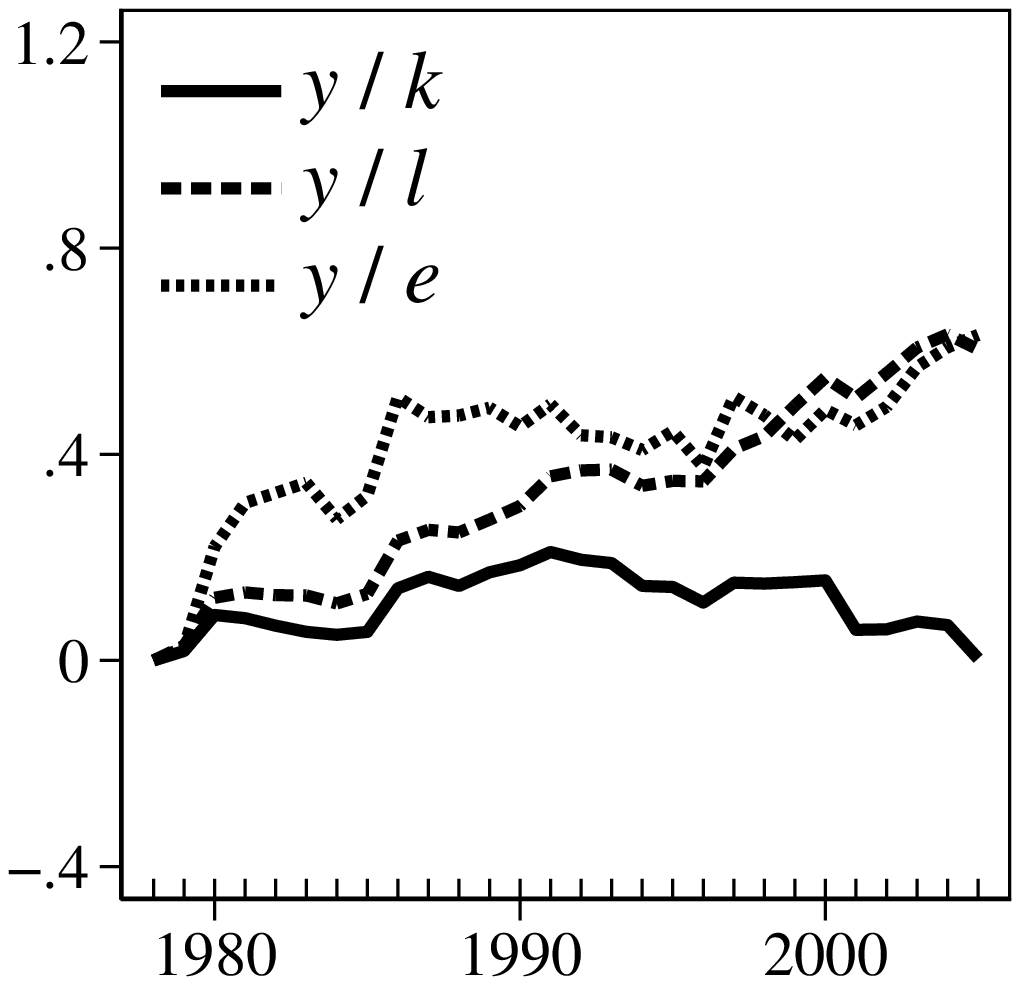}}
\par\end{centering}
\textit{\footnotesize{}Notes}{\footnotesize{}: The solid, dashed,
and dotted lines represent output per capital input ($y/k$), output
per labor input ($y/\ell$), and output per energy input ($y/e$),
respectively. All series are expressed as log differences relative
to the first year of observations.}{\footnotesize\par}
\end{figure}

Figures \ref{fig: factor_output_goods} and \ref{fig: factor_output_service}
show the trends in output per factor input in the goods and service
sectors of 12 OECD countries. Output per factor input is often used
as the measure of factor productivity. Output per capital input does
not exhibit a clear trend in many countries but exhibits a moderate
increasing trend in the goods sector of several countries and a modest
decreasing trend in the service sector of several countries. Output
per labor input exhibits an increasing trend in the goods and service
sectors of almost all countries. The rate of increase in output per
labor input is greater in the goods sector than in the service sector
for each country. Output per energy input exhibits different trends
across countries over time both in the goods and service sectors.
When these trends are compared, output per labor input tends to increase
more or decrease less than output per capital (energy) input in the
goods and service sectors of all (most) countries. However, the rate
of increase in output per energy input is similar to that in output
per labor input in the goods and service sectors of the United States
and the service sector of Germany and Sweden, and greater than that
in output per labor input in the service sector of Italy. The increase
in output per energy input is noticeable in Italy and the United States.

\begin{figure}[H]
\caption{Factor income shares in the goods sector\label{fig: factor_income_goods}}

\begin{centering}
\subfloat[Austria]{
\centering{}\includegraphics[scale=0.4]{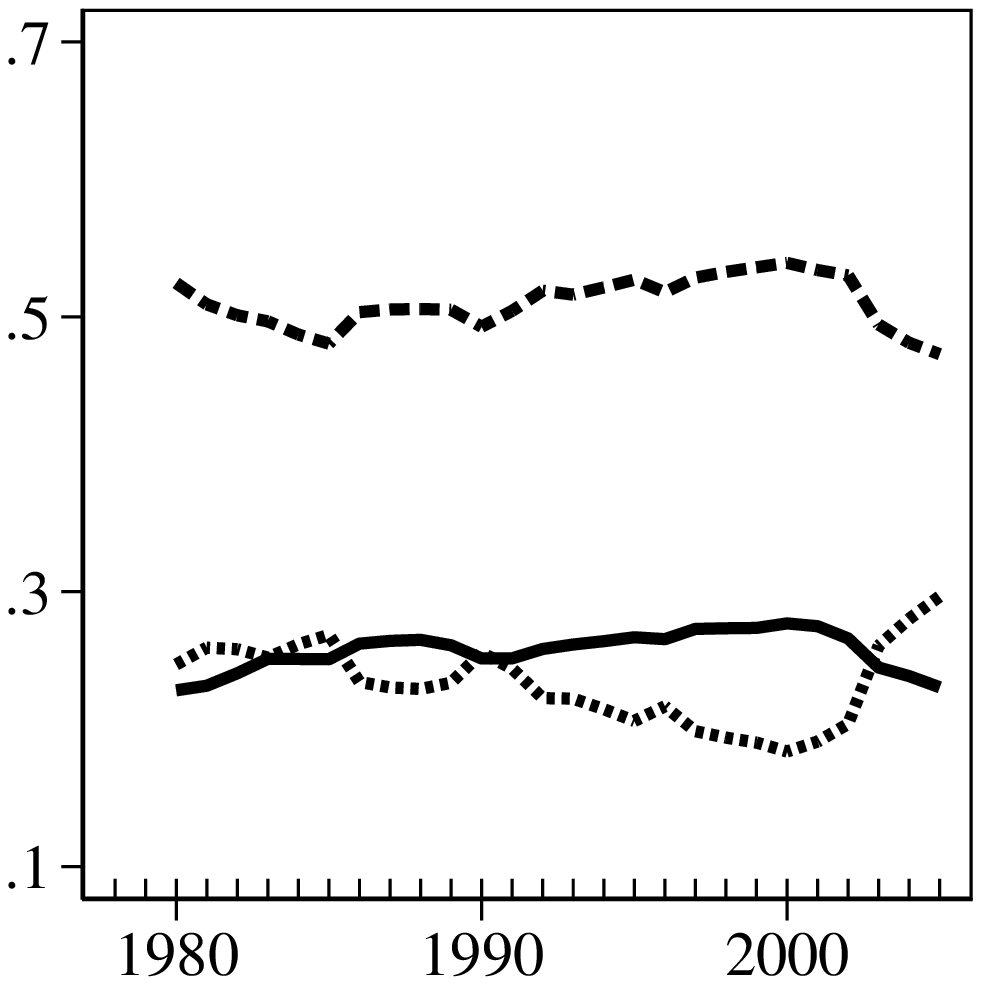}}\subfloat[Czech Republic]{
\centering{}\includegraphics[scale=0.4]{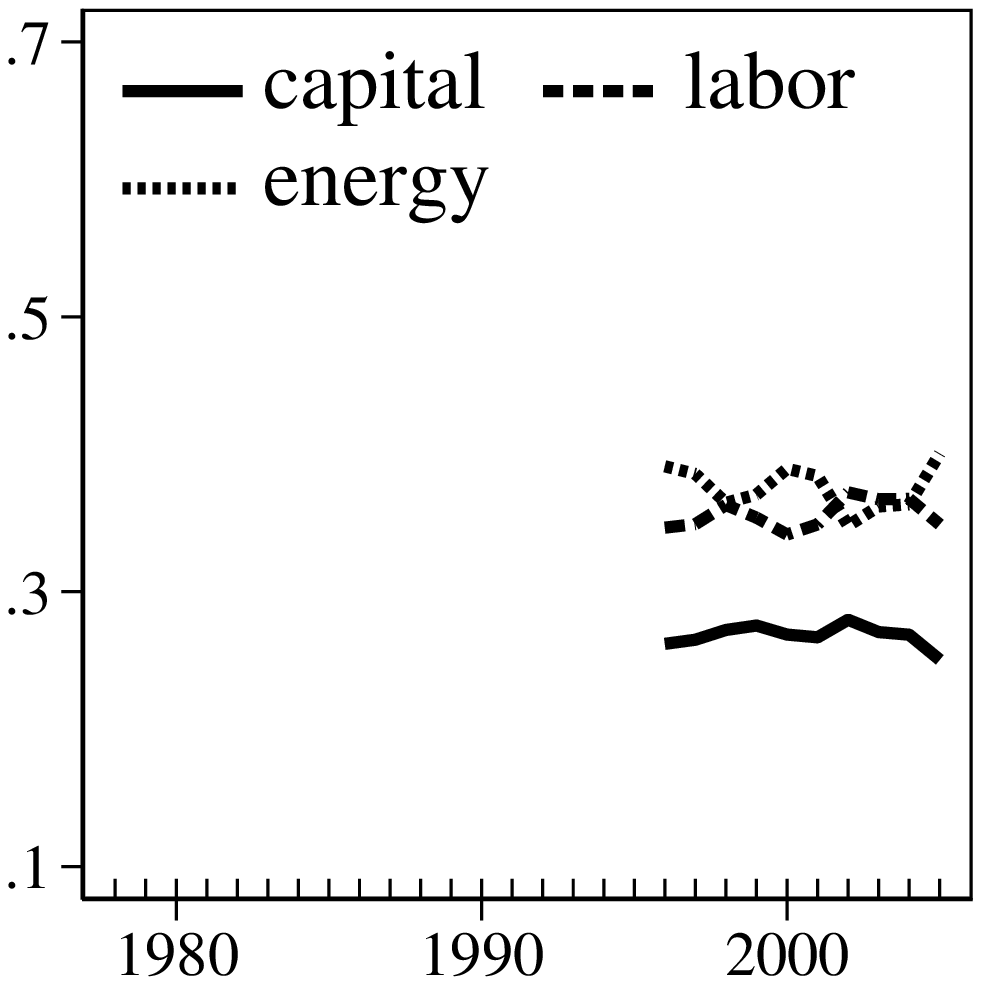}}\subfloat[Denmark]{
\centering{}\includegraphics[scale=0.4]{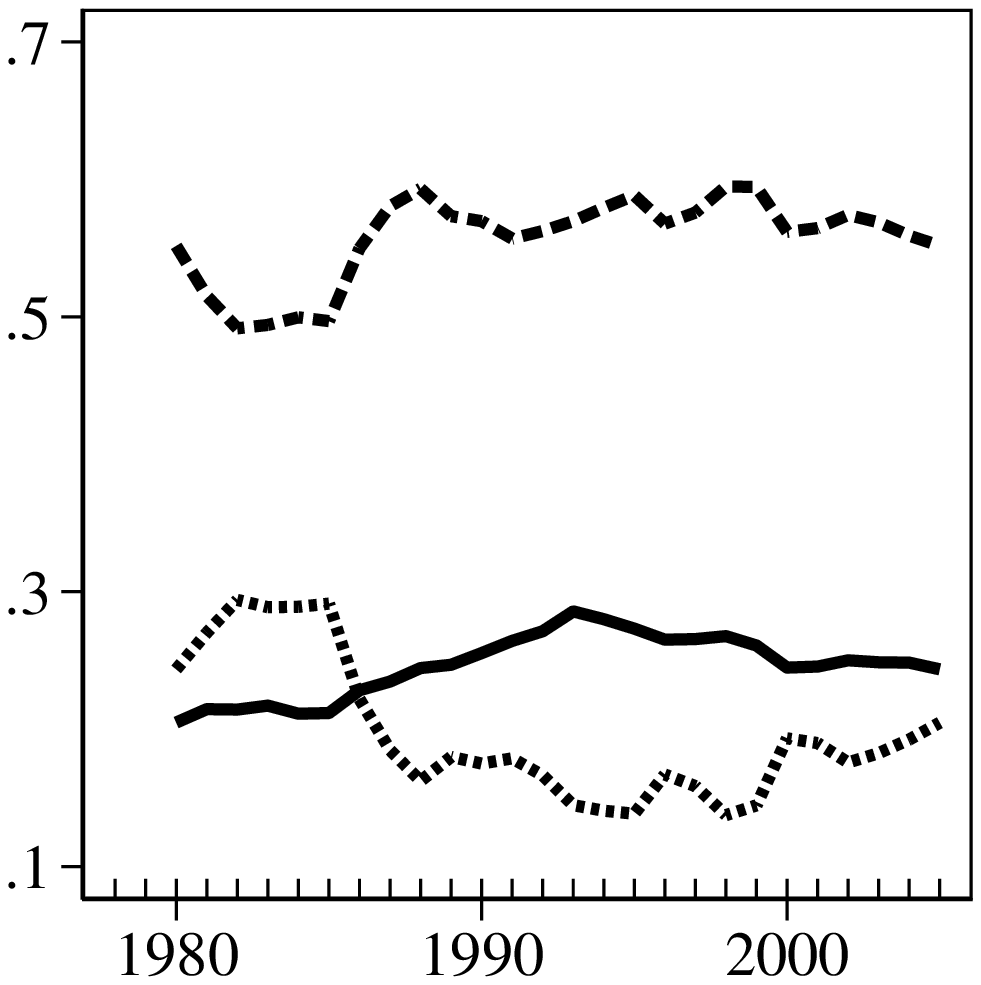}}\subfloat[Finland]{
\centering{}\includegraphics[scale=0.4]{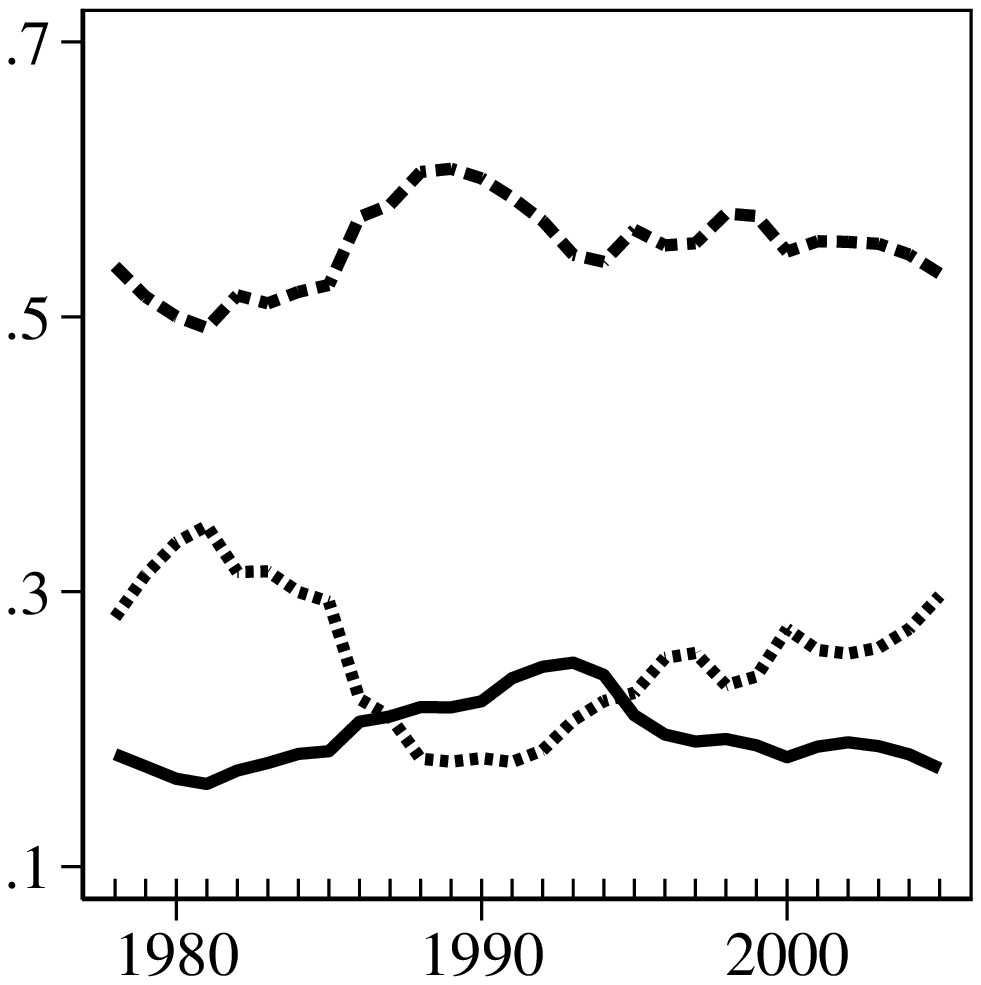}}
\par\end{centering}
\begin{centering}
\subfloat[Germany]{
\centering{}\includegraphics[scale=0.4]{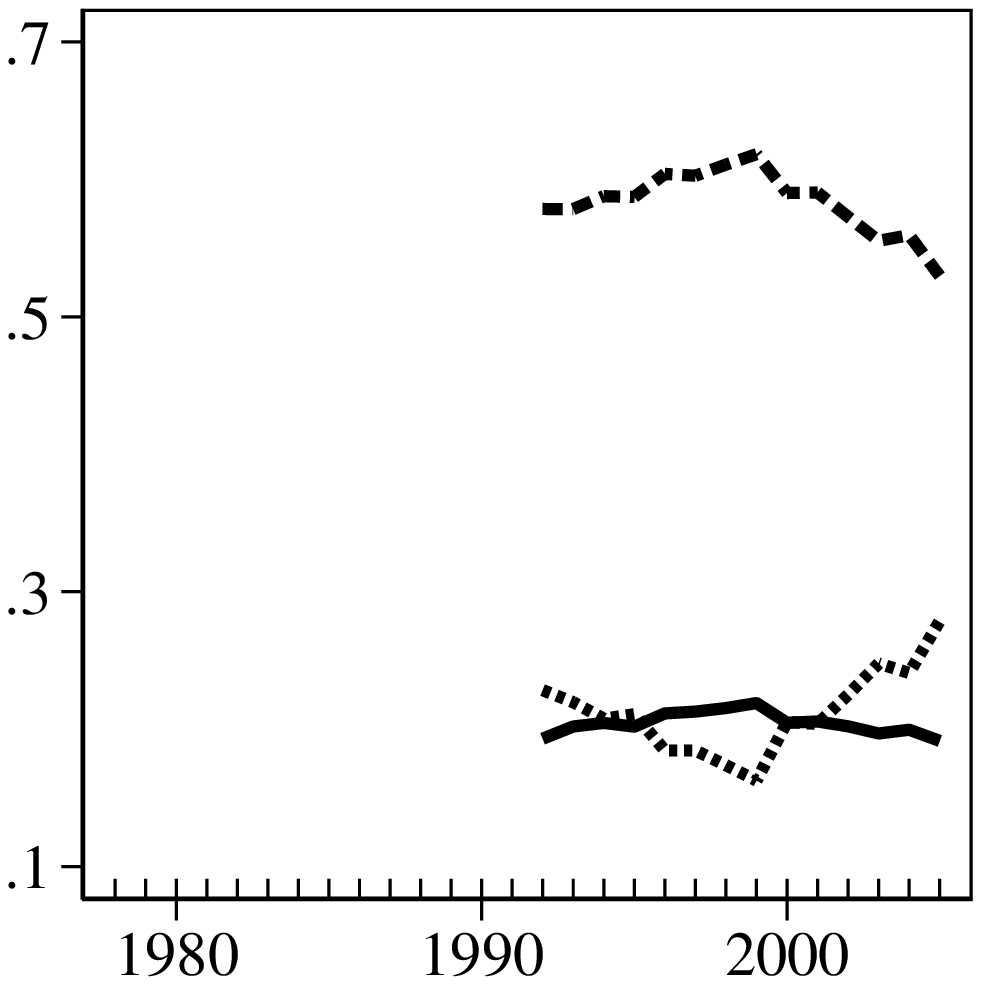}}\subfloat[Italy]{
\centering{}\includegraphics[scale=0.4]{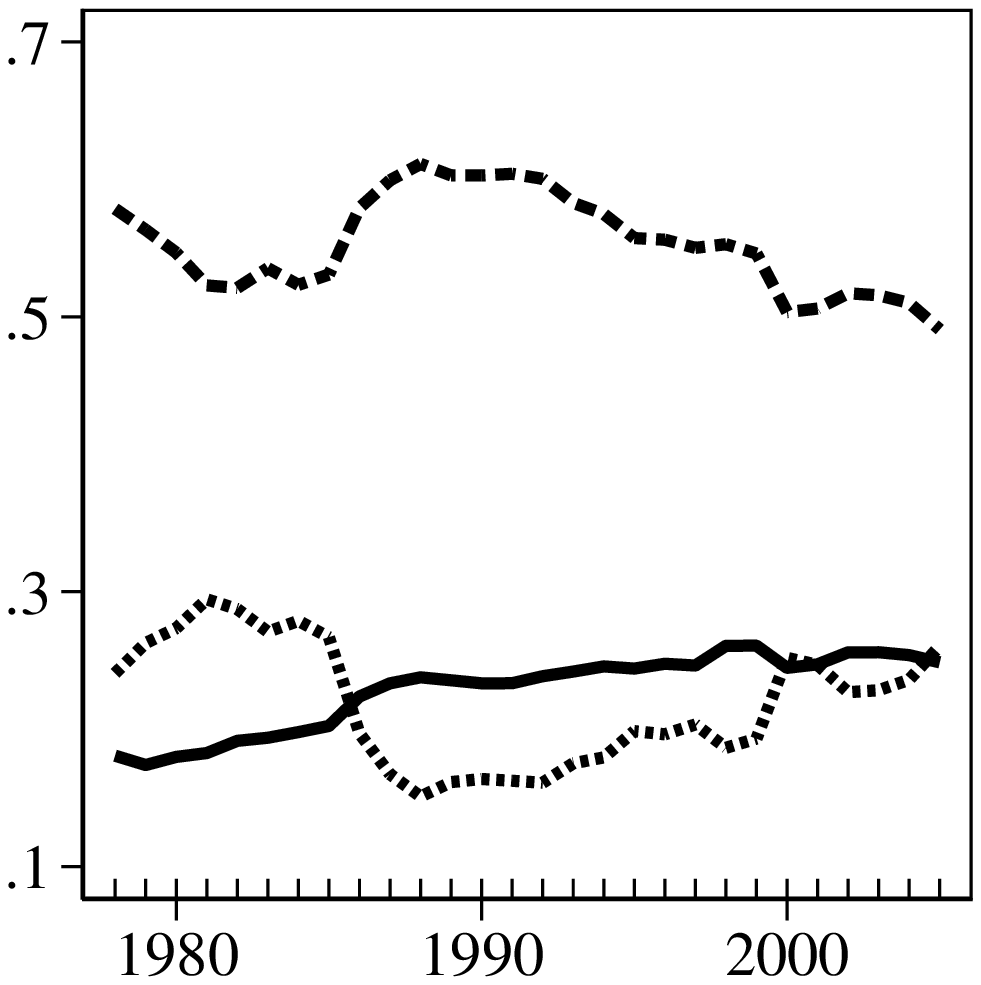}}\subfloat[Japan]{
\centering{}\includegraphics[scale=0.4]{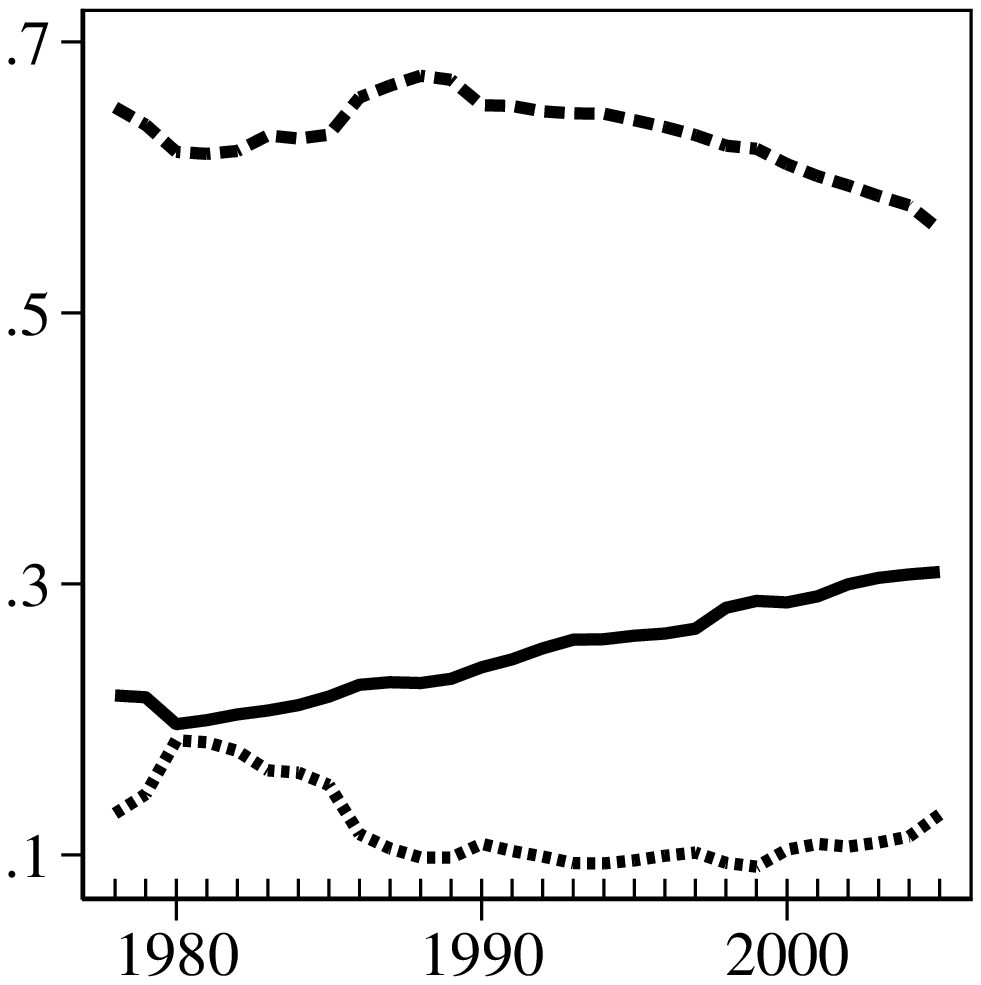}}\subfloat[Netherlands]{
\centering{}\includegraphics[scale=0.42]{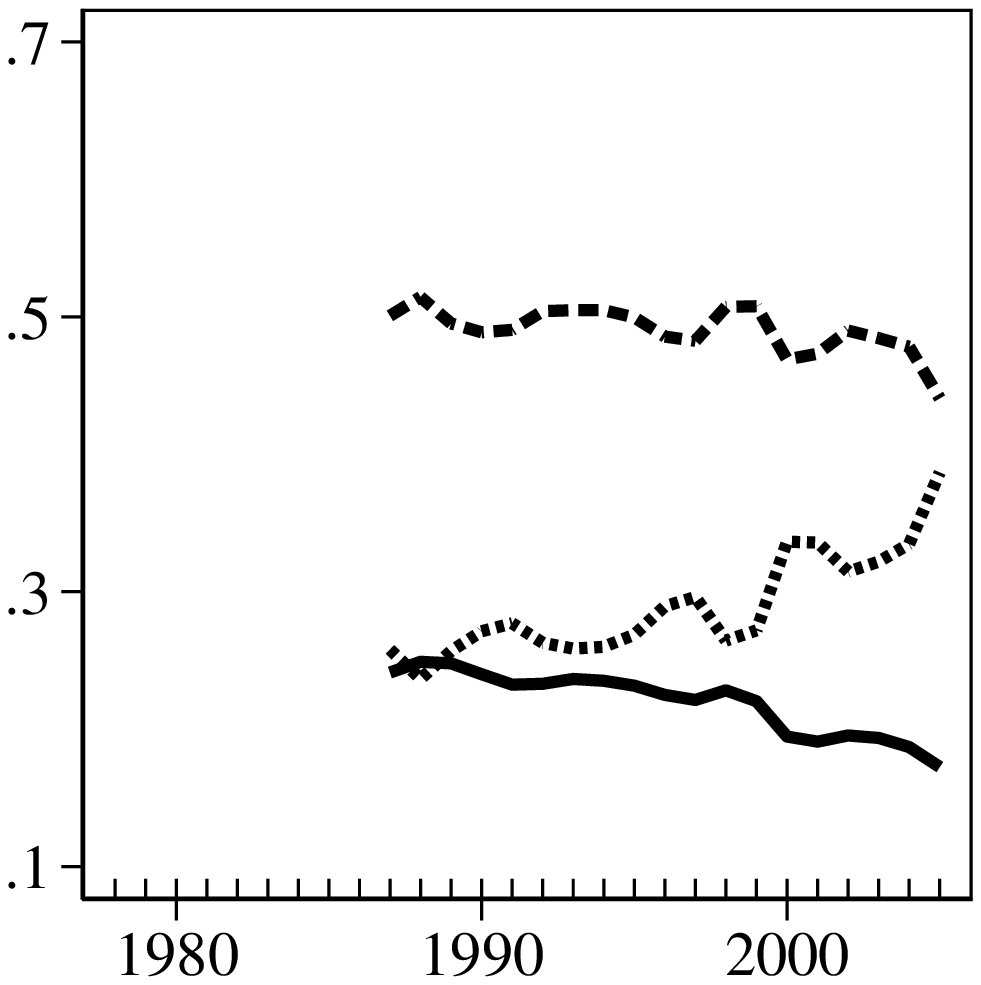}}
\par\end{centering}
\begin{centering}
\subfloat[Portugal]{
\centering{}\includegraphics[scale=0.4]{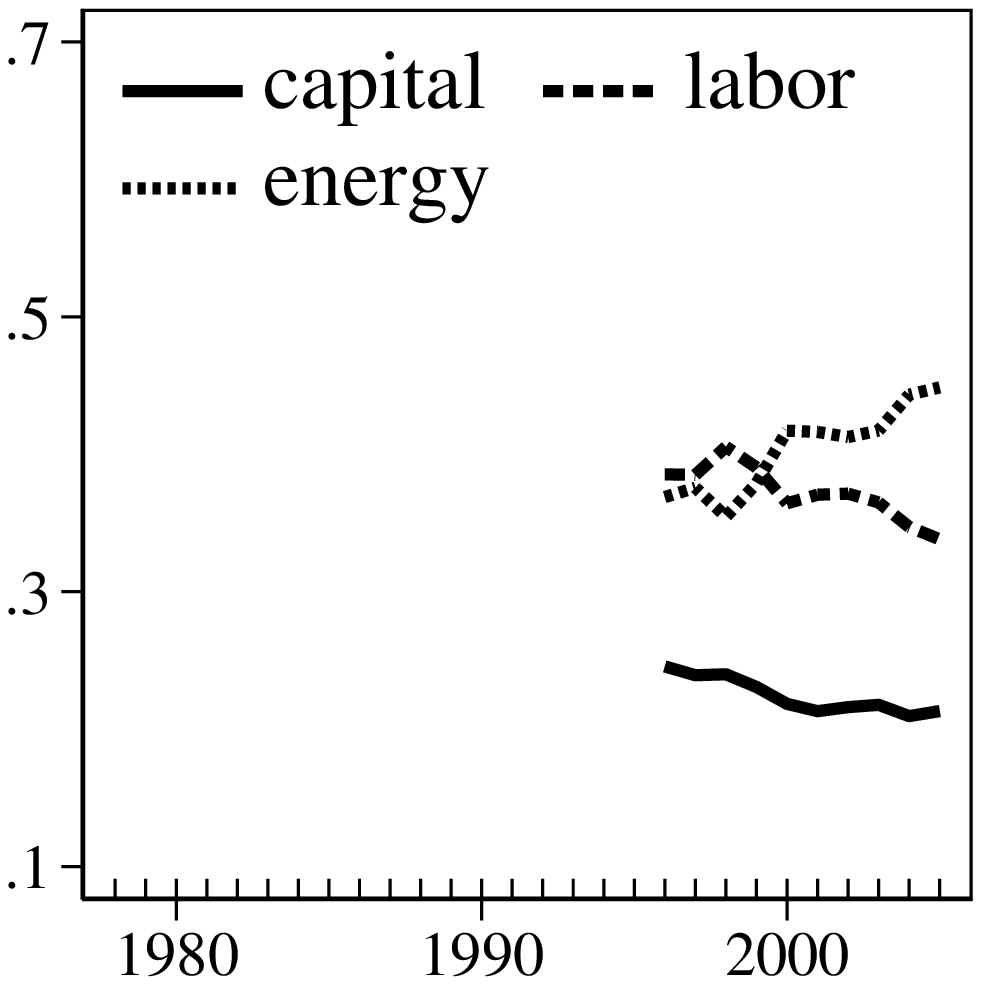}}\subfloat[Sweden]{
\centering{}\includegraphics[scale=0.4]{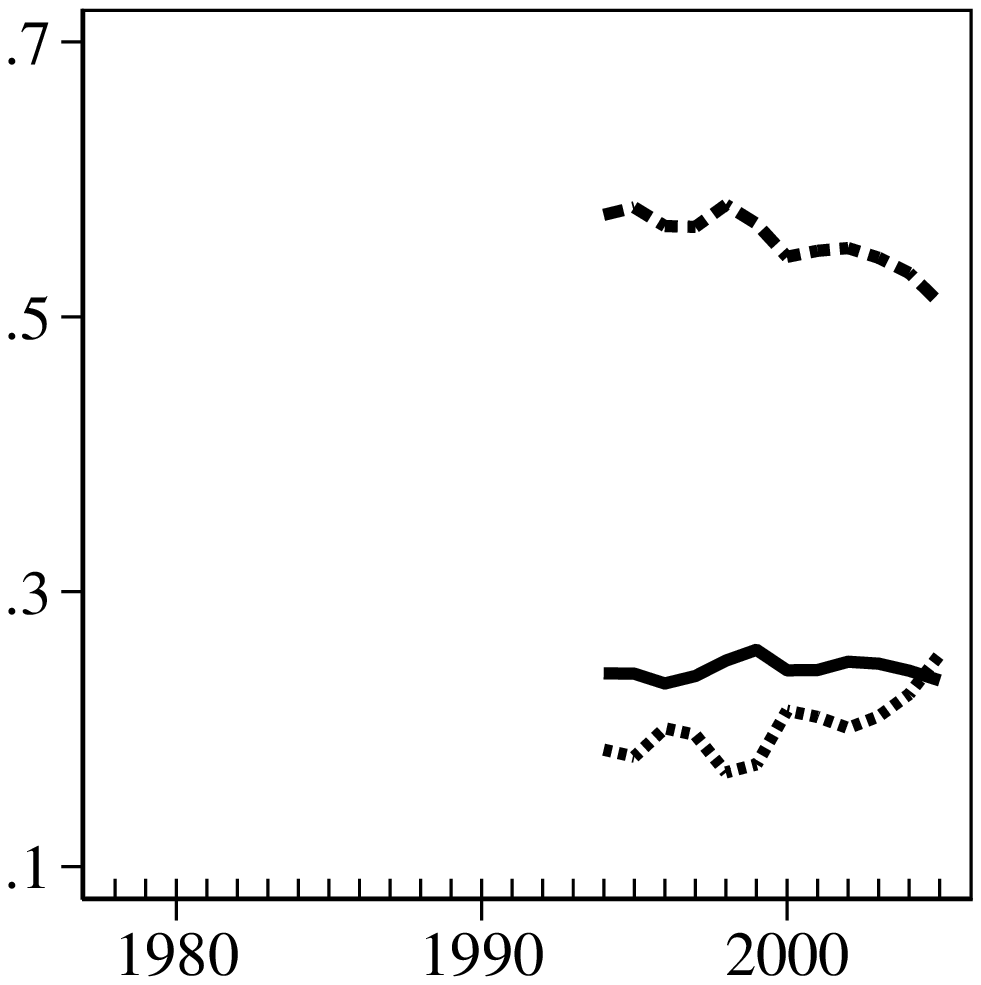}}\subfloat[United Kingdom]{
\centering{}\includegraphics[scale=0.4]{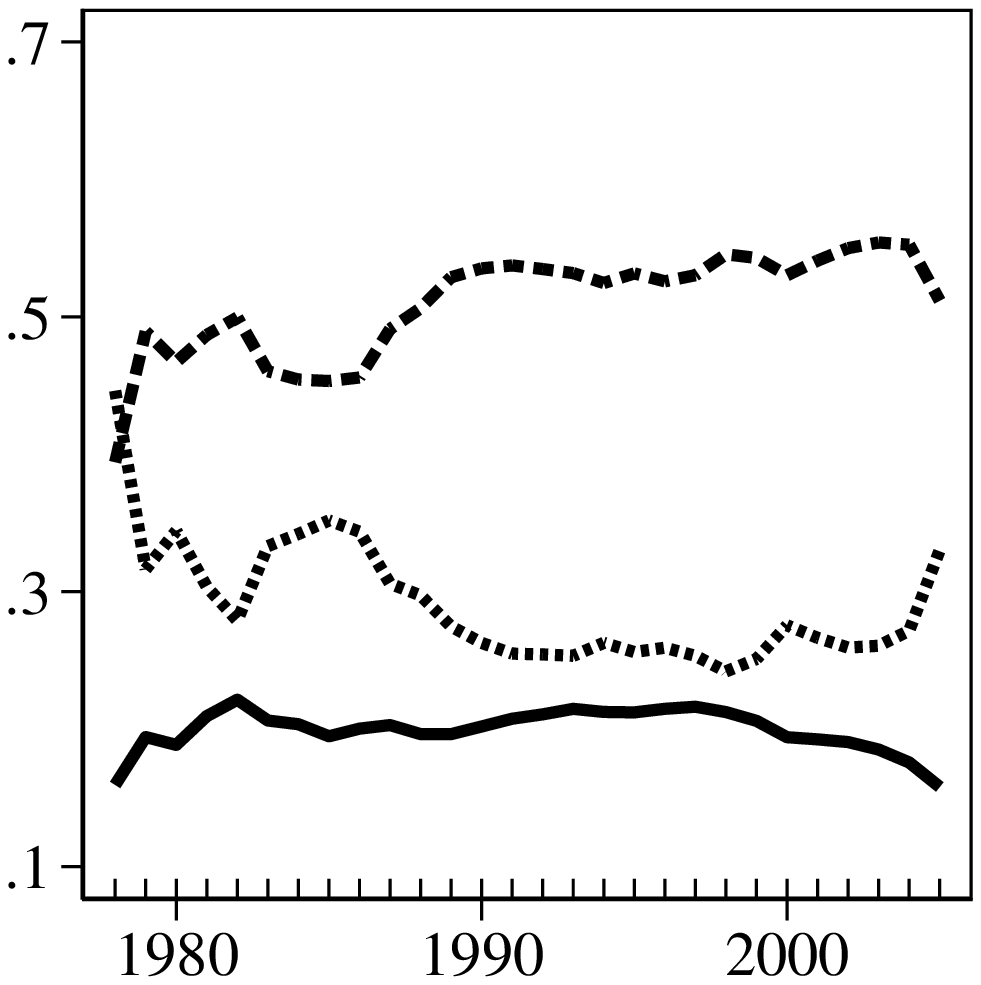}}\subfloat[United States]{
\centering{}\includegraphics[scale=0.4]{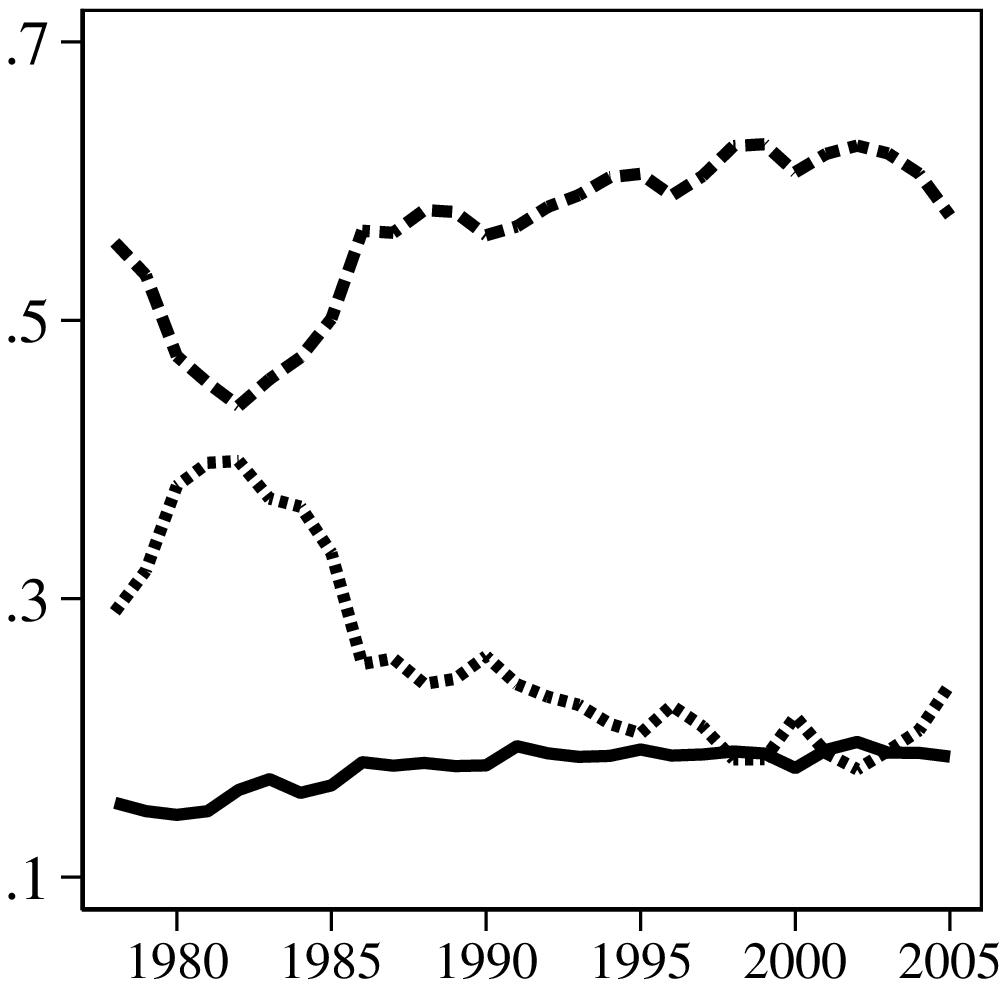}}
\par\end{centering}
\textit{\footnotesize{}Notes}{\footnotesize{}: The solid, dashed,
and dotted lines represent the capital, labor, and energy shares of
income (i.e., $rk/(rk+w\ell+ve)$, $w\ell/(rk+w\ell+ve)$, and $ve/(rk+w\ell+ve)$),
respectively.}{\footnotesize\par}
\end{figure}

\begin{figure}[H]
\caption{Factor income shares in the service sector\label{fig: factor_income_service}}

\begin{centering}
\subfloat[Austria]{
\centering{}\includegraphics[scale=0.4]{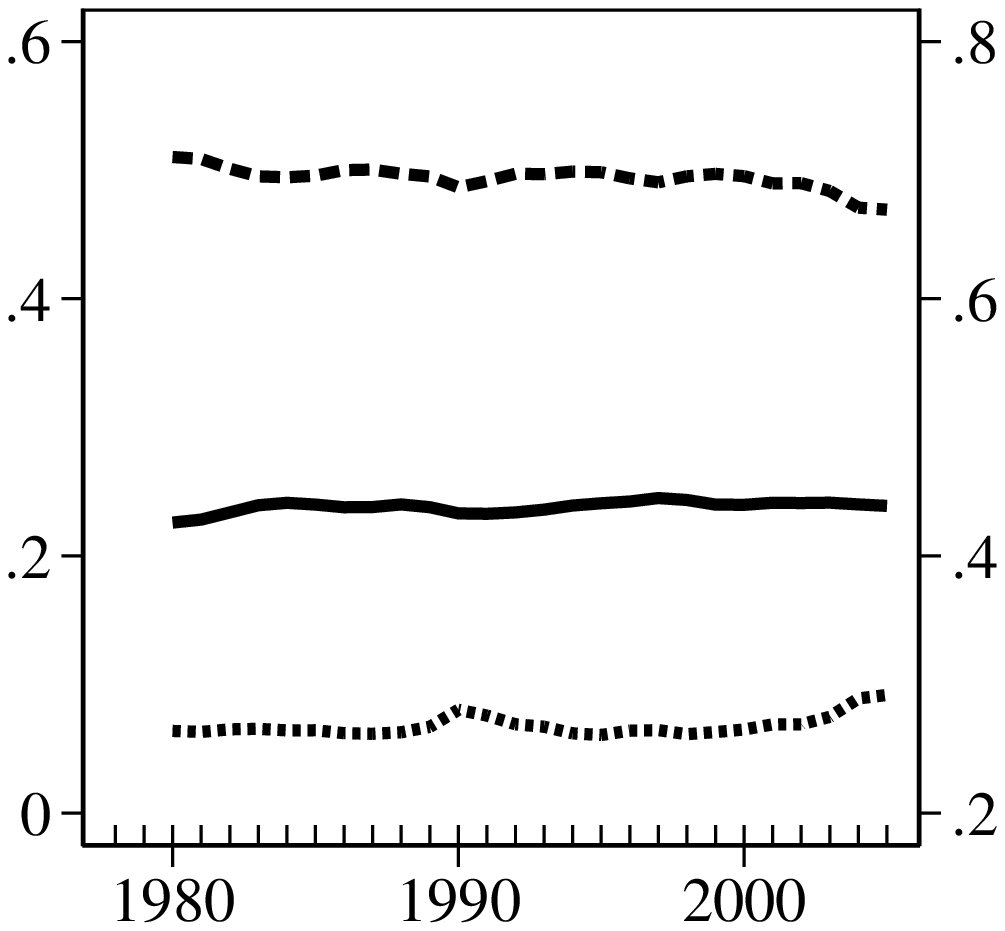}}\subfloat[Czech Republic]{
\centering{}\includegraphics[scale=0.4]{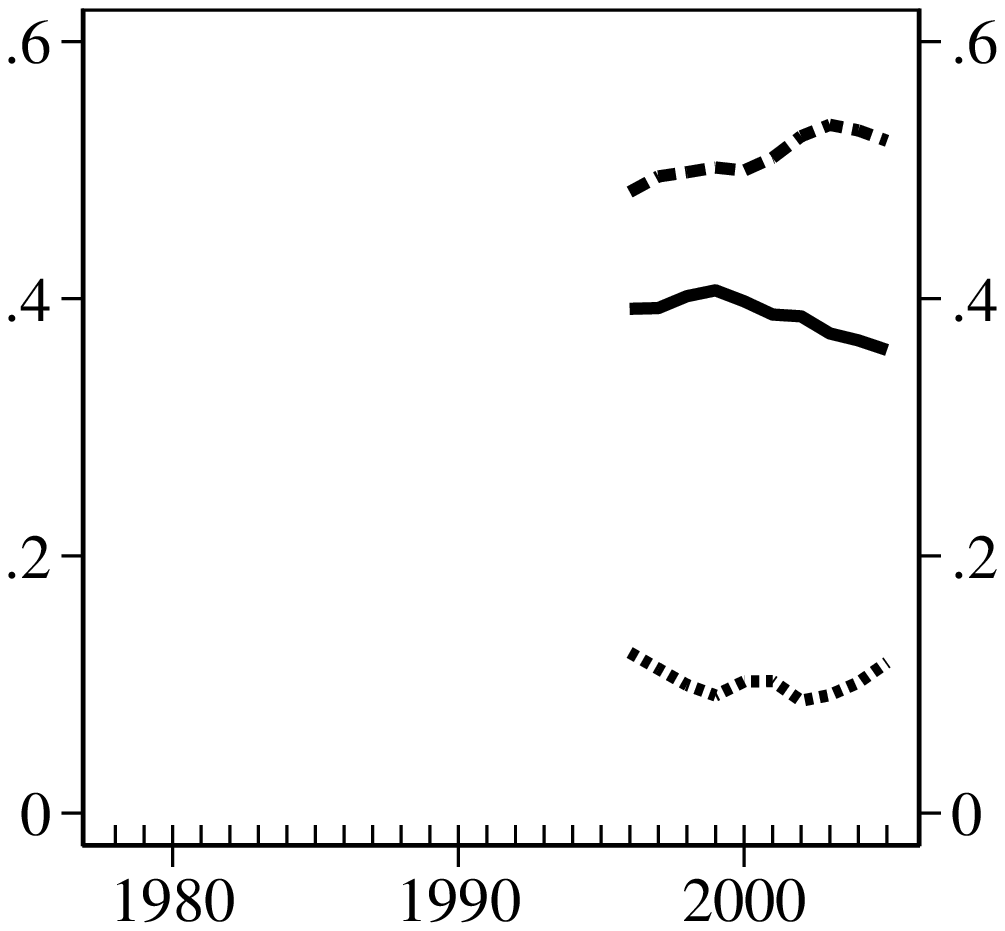}}\subfloat[Denmark]{
\centering{}\includegraphics[scale=0.4]{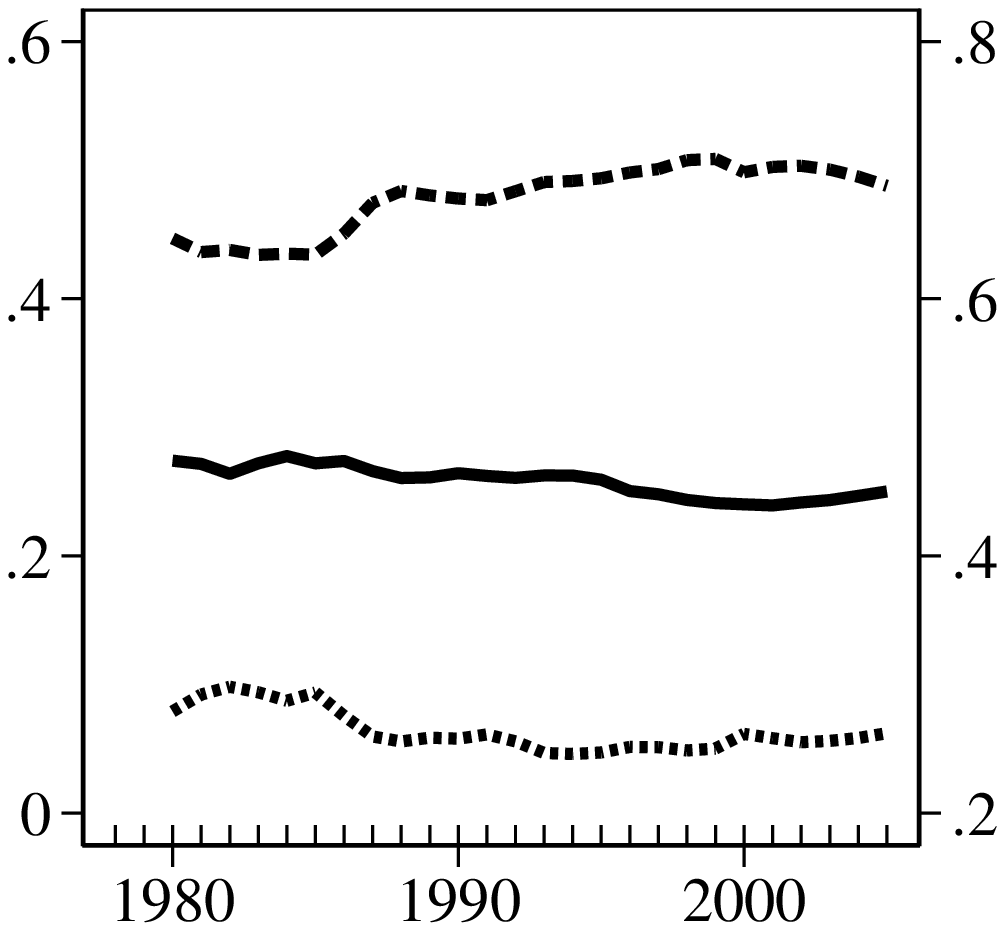}}\subfloat[Finland]{
\centering{}\includegraphics[scale=0.4]{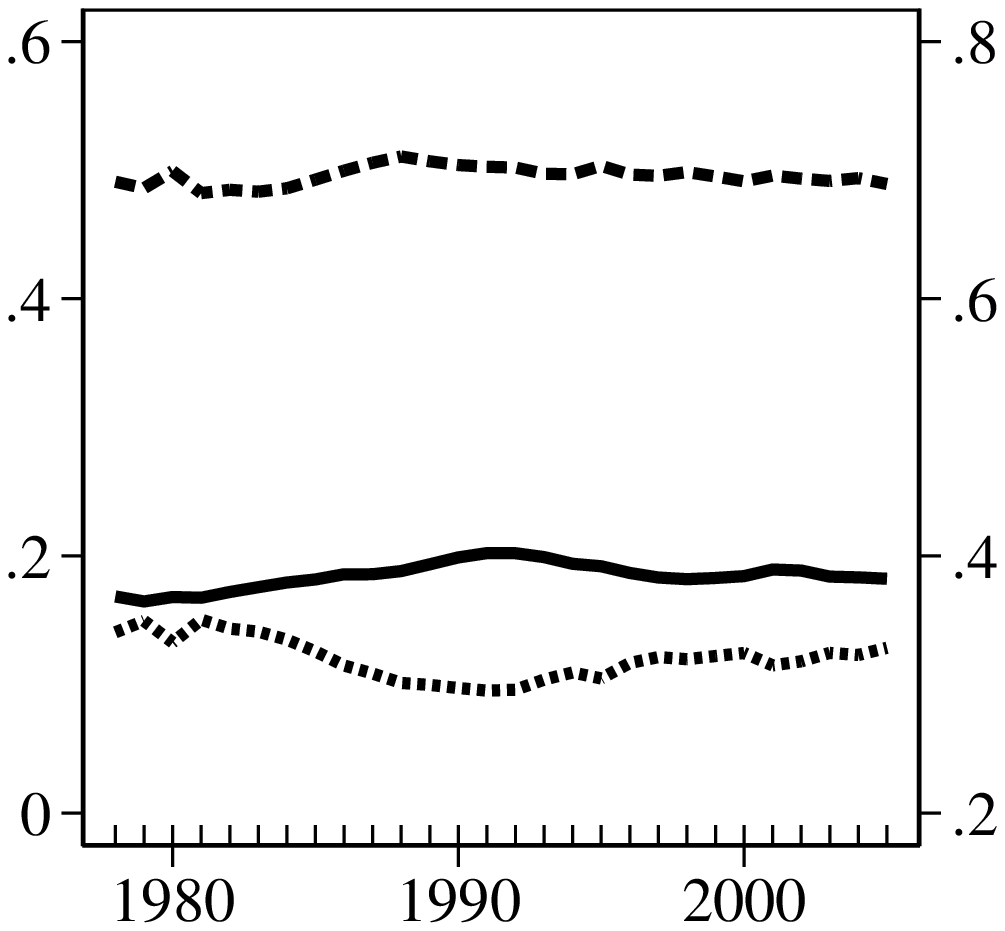}}
\par\end{centering}
\begin{centering}
\subfloat[Germany]{
\centering{}\includegraphics[scale=0.4]{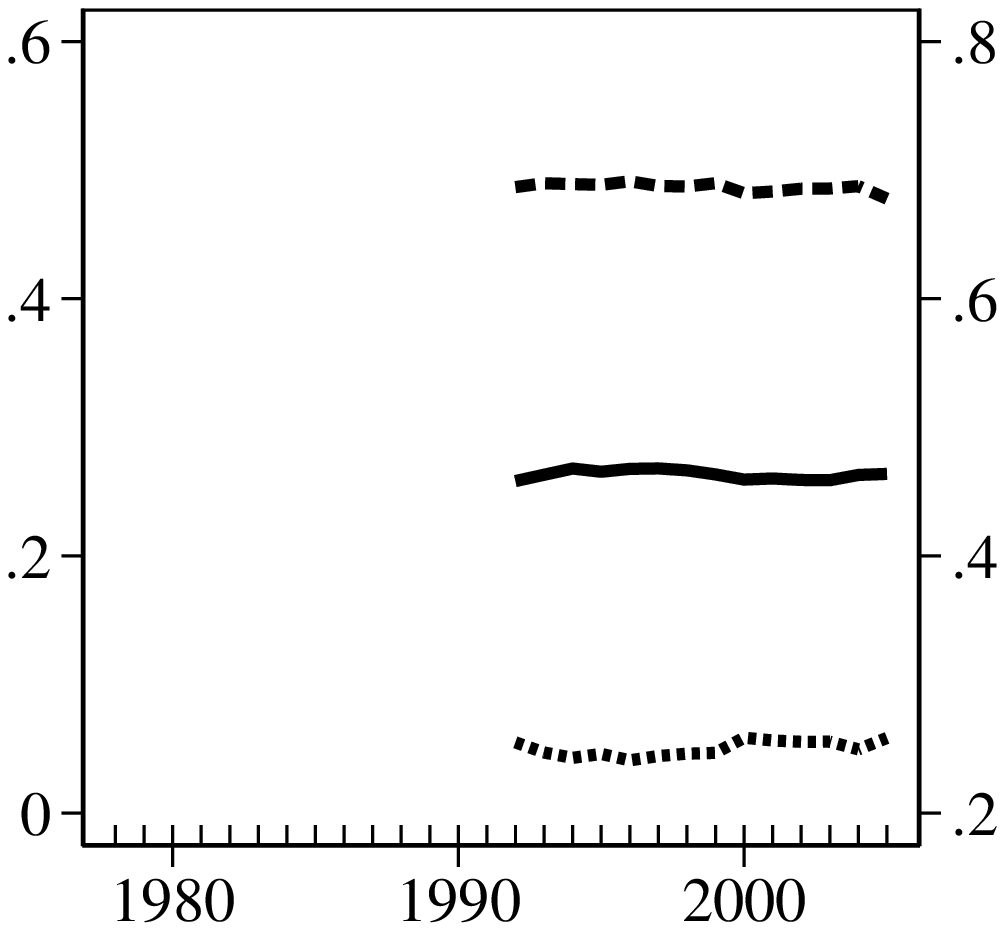}}\subfloat[Italy]{
\centering{}\includegraphics[scale=0.4]{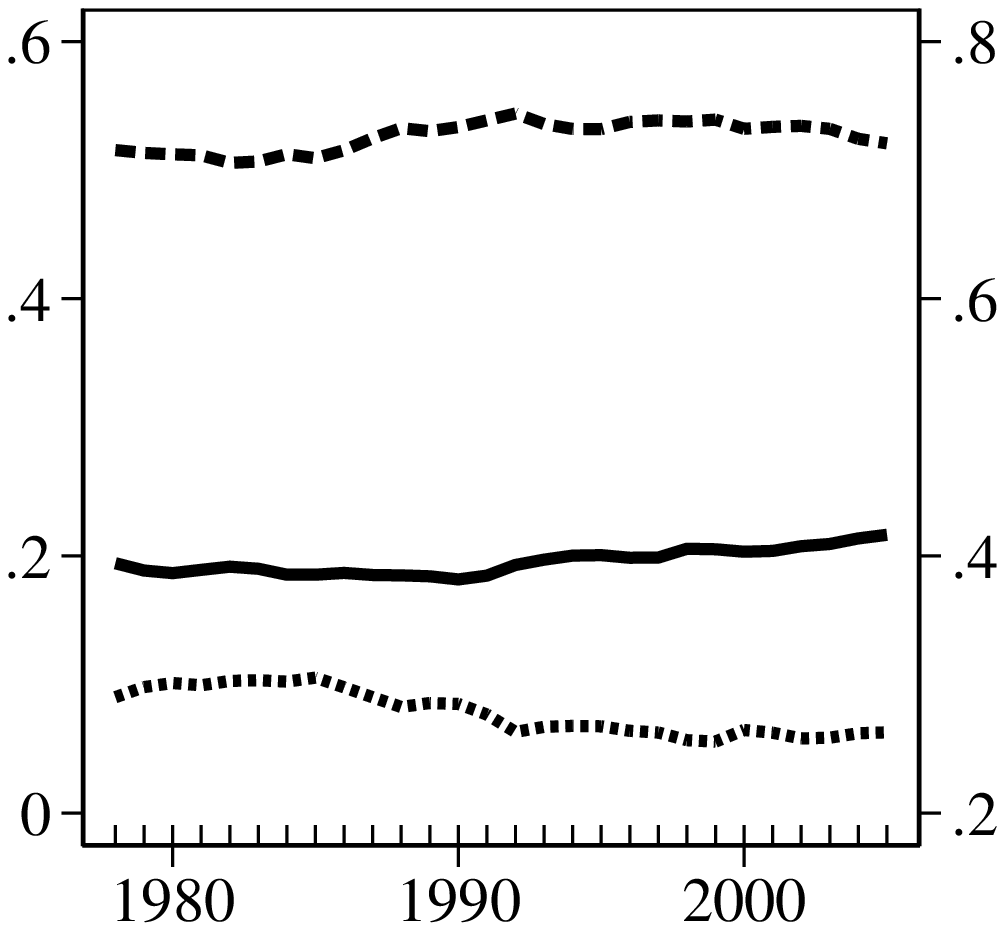}}\subfloat[Japan]{
\centering{}\includegraphics[scale=0.4]{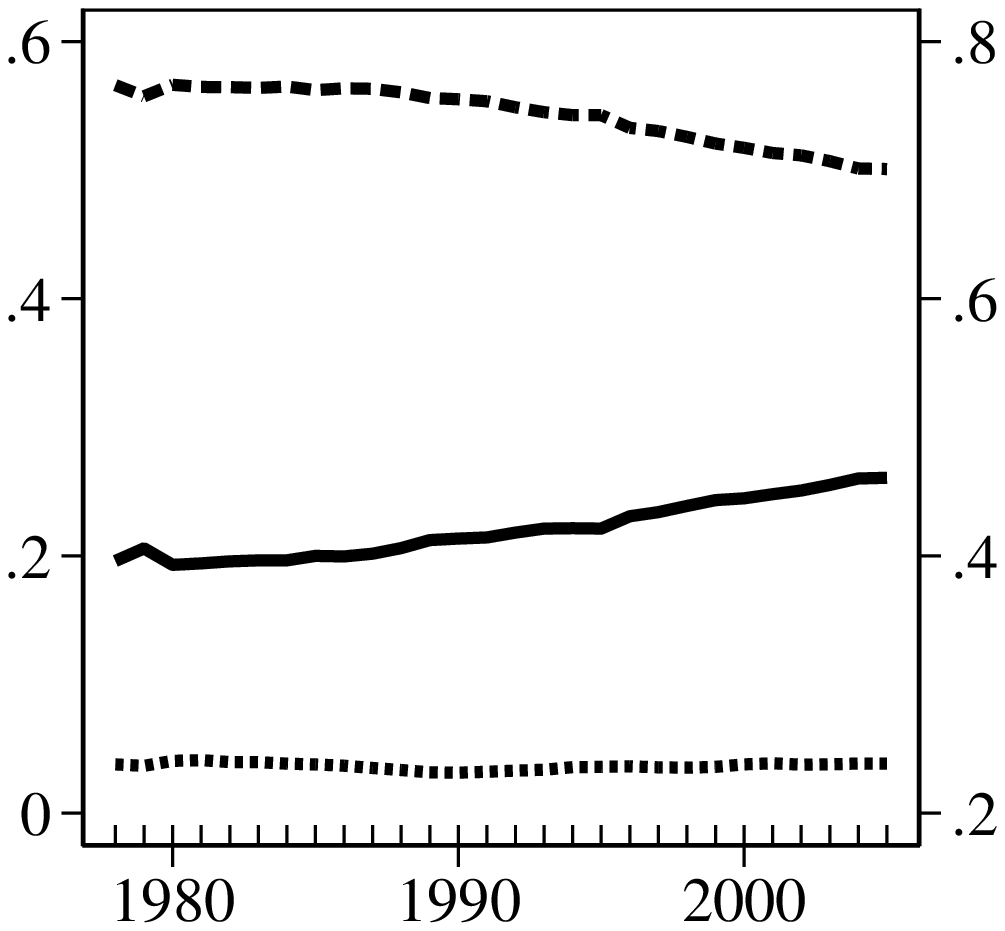}}\subfloat[Netherlands]{
\centering{}\includegraphics[scale=0.42]{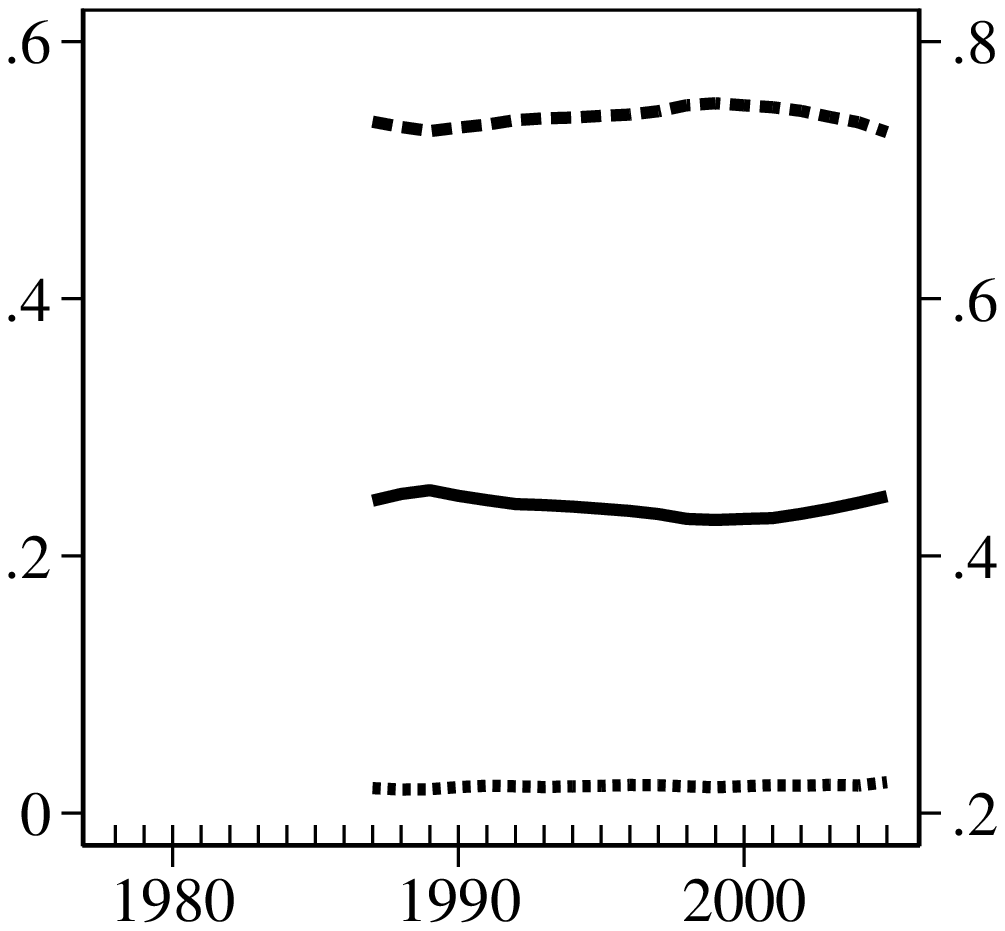}}
\par\end{centering}
\begin{centering}
\subfloat[Portugal]{
\centering{}\includegraphics[scale=0.4]{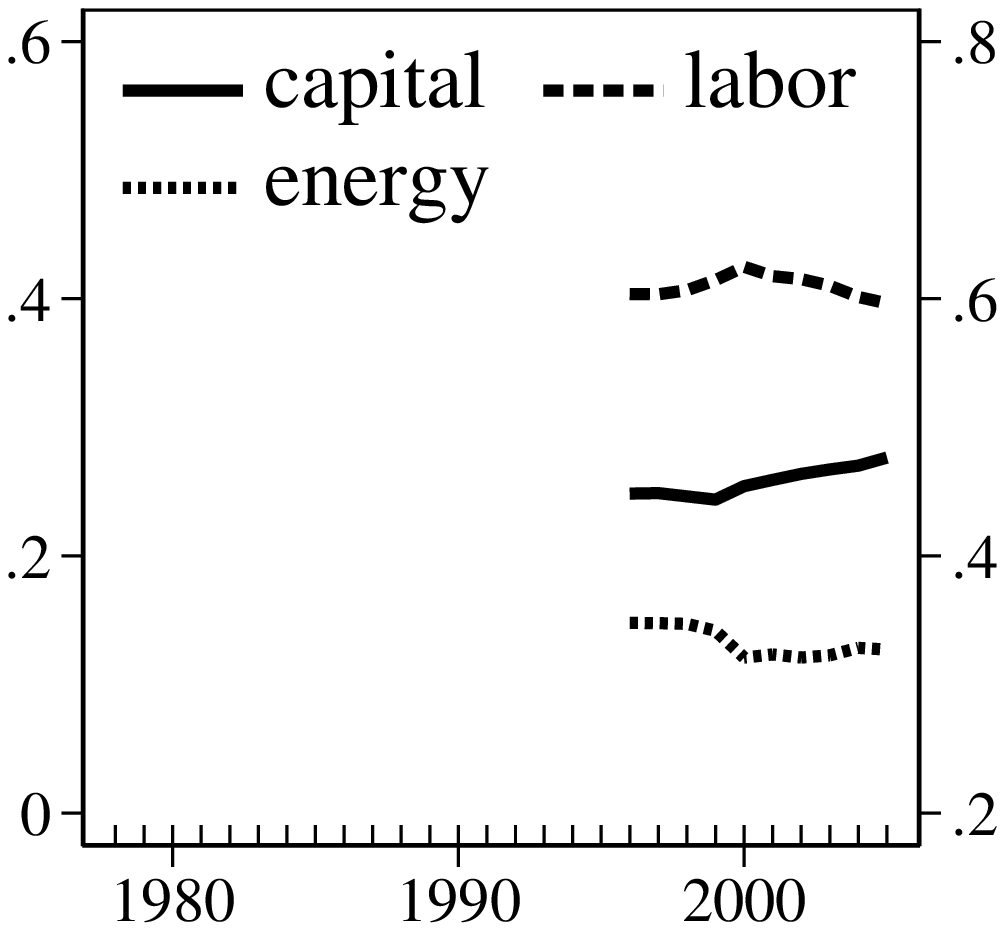}}\subfloat[Sweden]{
\centering{}\includegraphics[scale=0.4]{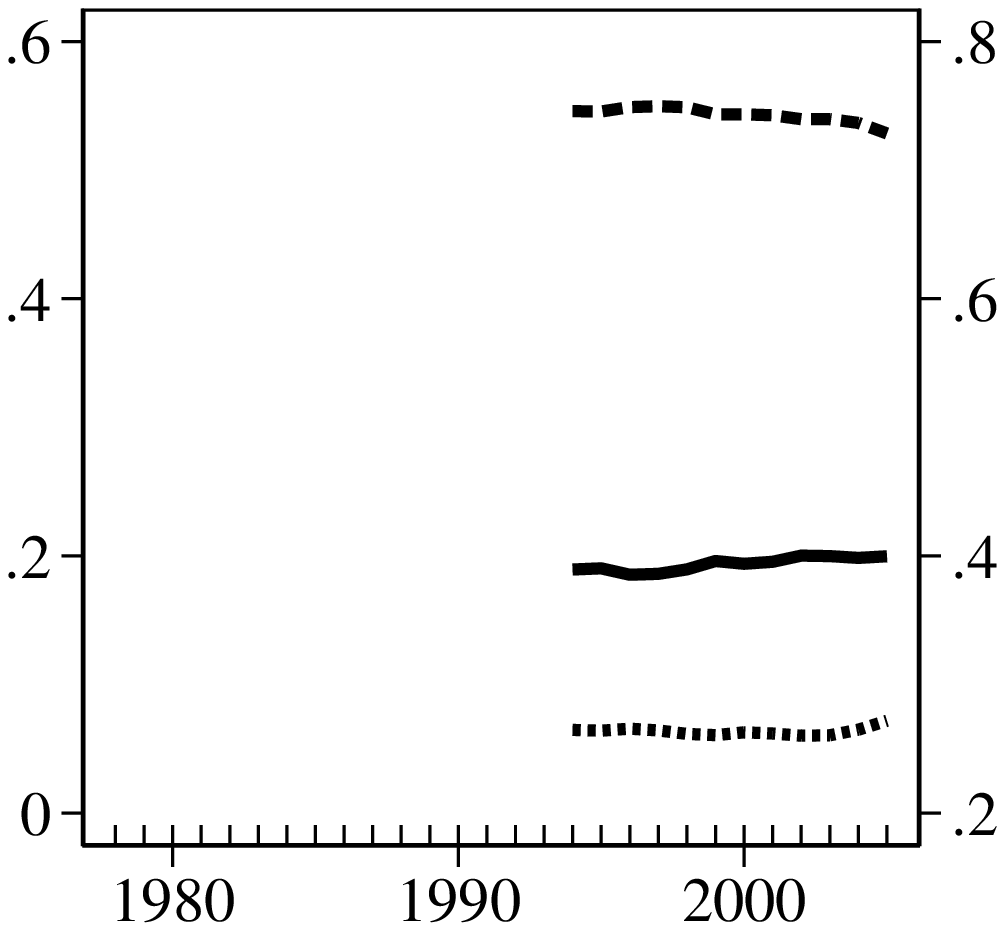}}\subfloat[United Kingdom]{
\centering{}\includegraphics[scale=0.4]{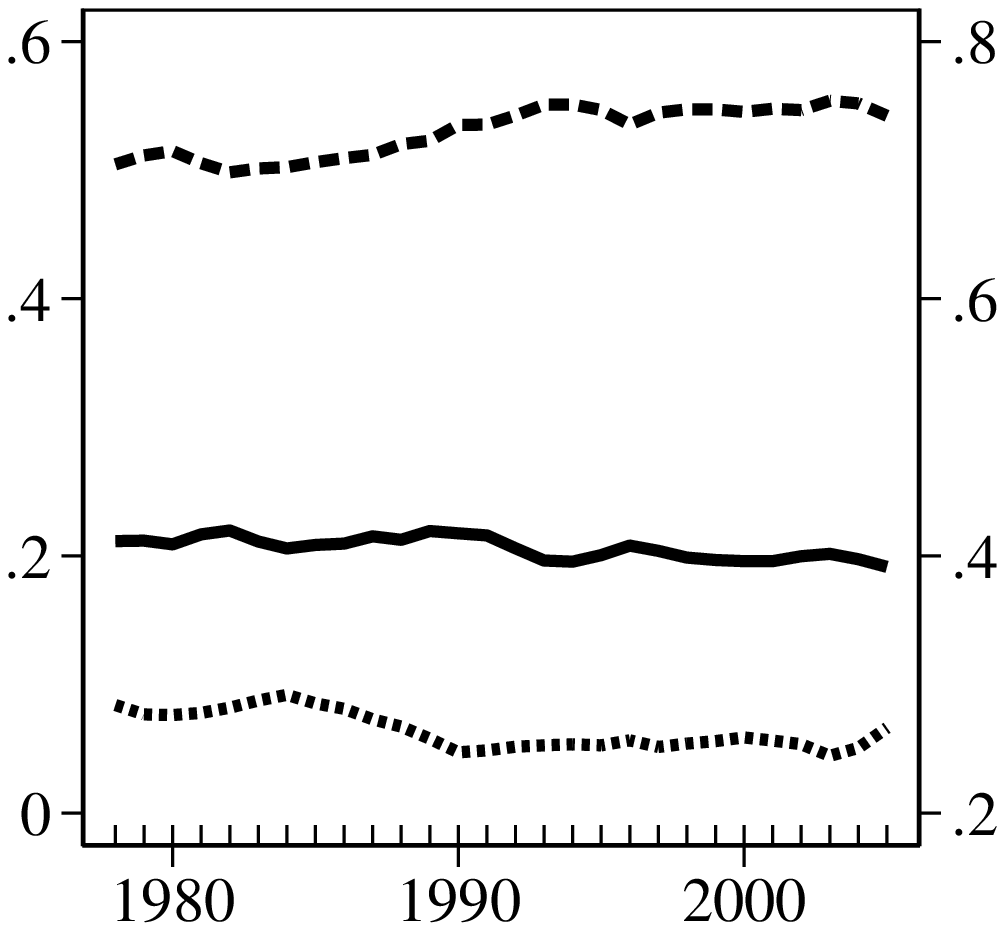}}\subfloat[United States]{
\centering{}\includegraphics[scale=0.4]{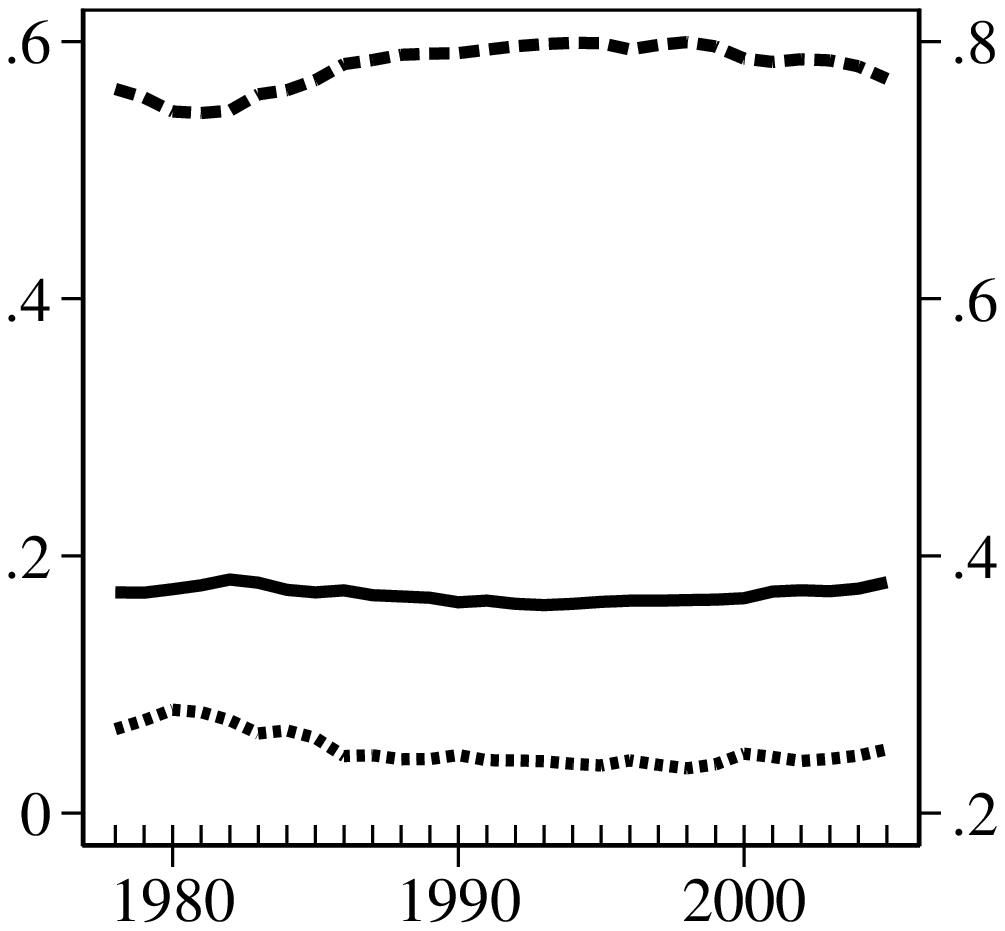}}
\par\end{centering}
\textit{\footnotesize{}Notes}{\footnotesize{}: The solid, dashed,
and dotted lines represent the capital, labor, and energy shares of
income (i.e., $rk/(rk+w\ell+ve)$, $w\ell/(rk+w\ell+ve)$, and $ve/(rk+w\ell+ve)$),
respectively. The left vertical axis indicates the capital and energy
shares, while the right axis indicates the labor share.}{\footnotesize\par}
\end{figure}

Figures \ref{fig: factor_income_goods} and \ref{fig: factor_income_service}
show the trends in the capital, labor, and energy shares of income
in the goods and service sectors of 12 OECD countries. The capital
share of income tends to increase modestly in the goods sector of
many countries, while the labor share of income tends to decrease
modestly in the goods sector of many countries. The rates of change
in the capital and labor shares of income tend to be smaller in the
service sector than in the goods sector. The energy share of income
decreased in the 1980s or 1990s and increased in the late 1990s or
2000s in the goods and service sectors of many countries. The rate
of decrease is large especially in Denmark, Italy, the United Kingdom,
and the United States.

\subsection{Additional results\label{subsec: results_add}}

\begin{figure}[H]
\caption{Energy-saving technological change for different values of the elasticity
of substitution in the goods sector\label{fig: Ae1_sigma_goods}}

\begin{centering}
\subfloat[Austria]{
\centering{}\includegraphics[scale=0.4]{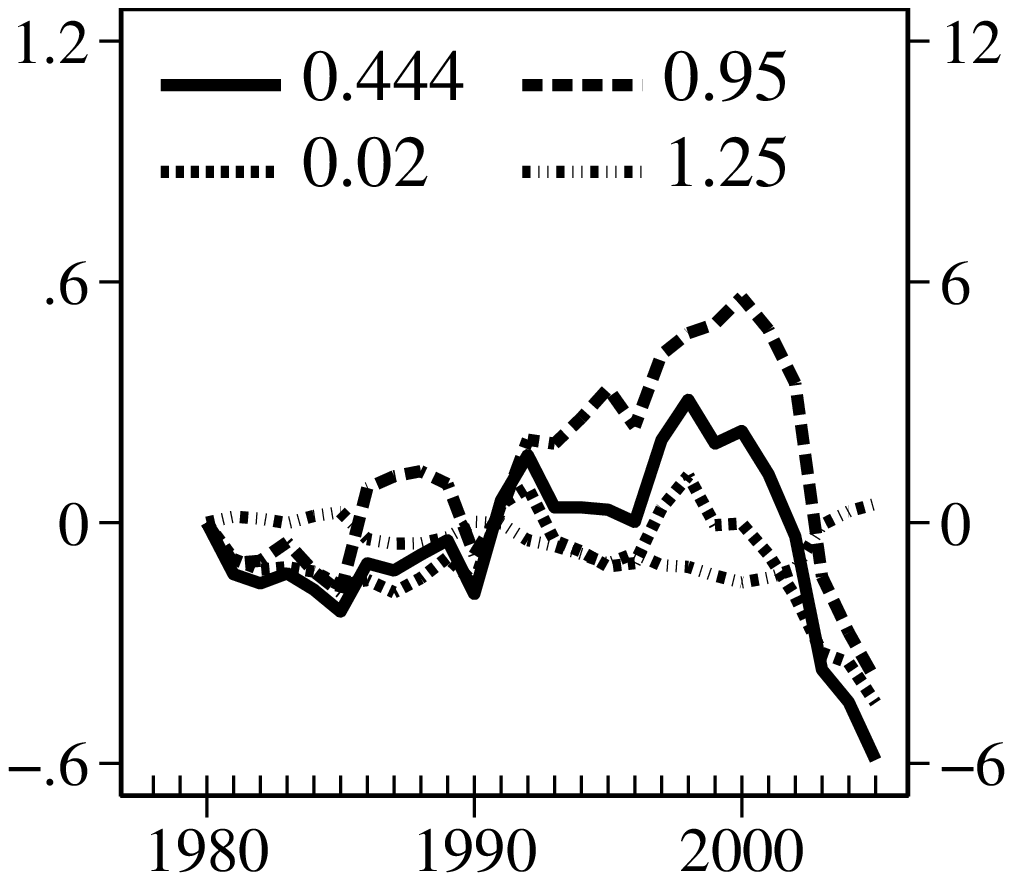}}\subfloat[Czech Republic]{
\centering{}\includegraphics[scale=0.4]{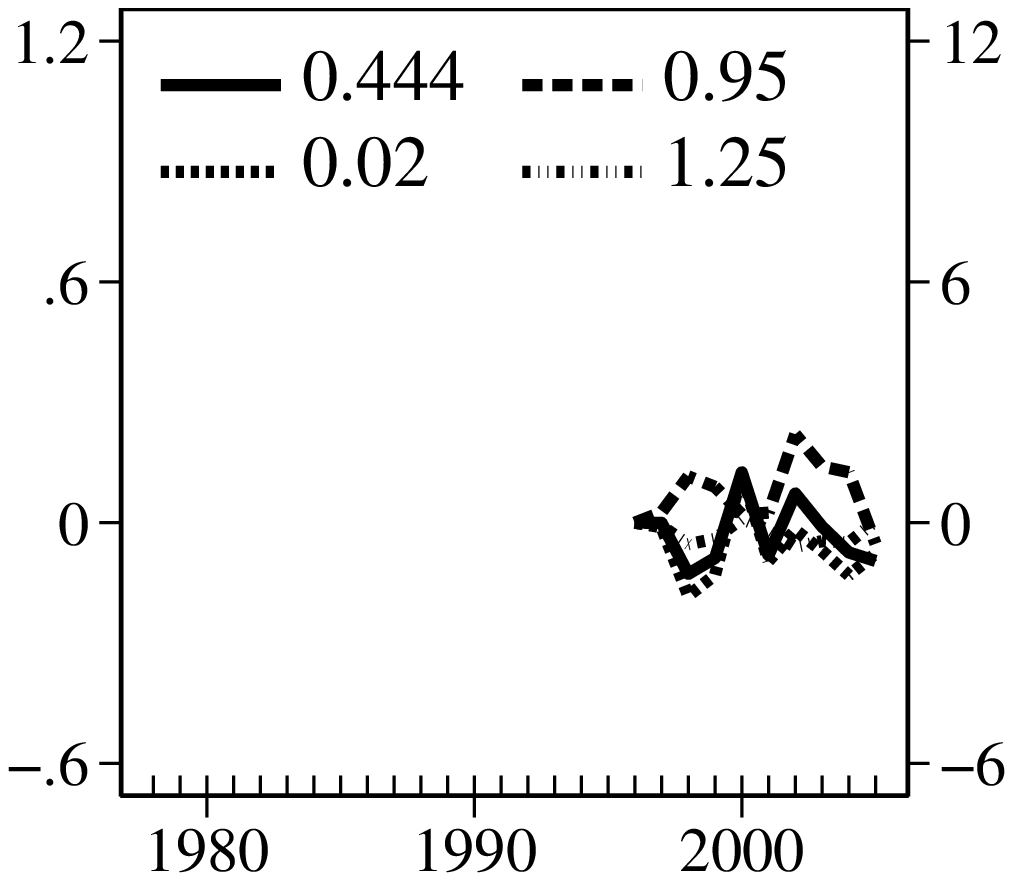}}\subfloat[Denmark]{
\centering{}\includegraphics[scale=0.4]{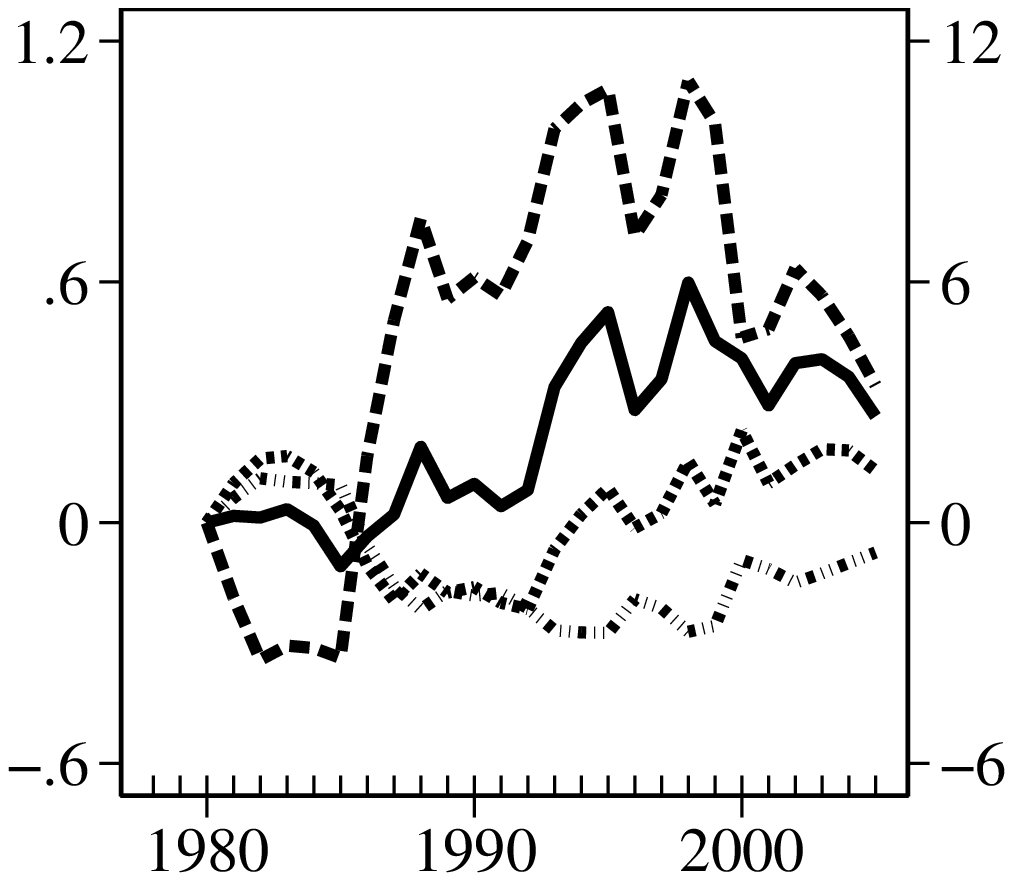}}\subfloat[Finland]{
\centering{}\includegraphics[scale=0.4]{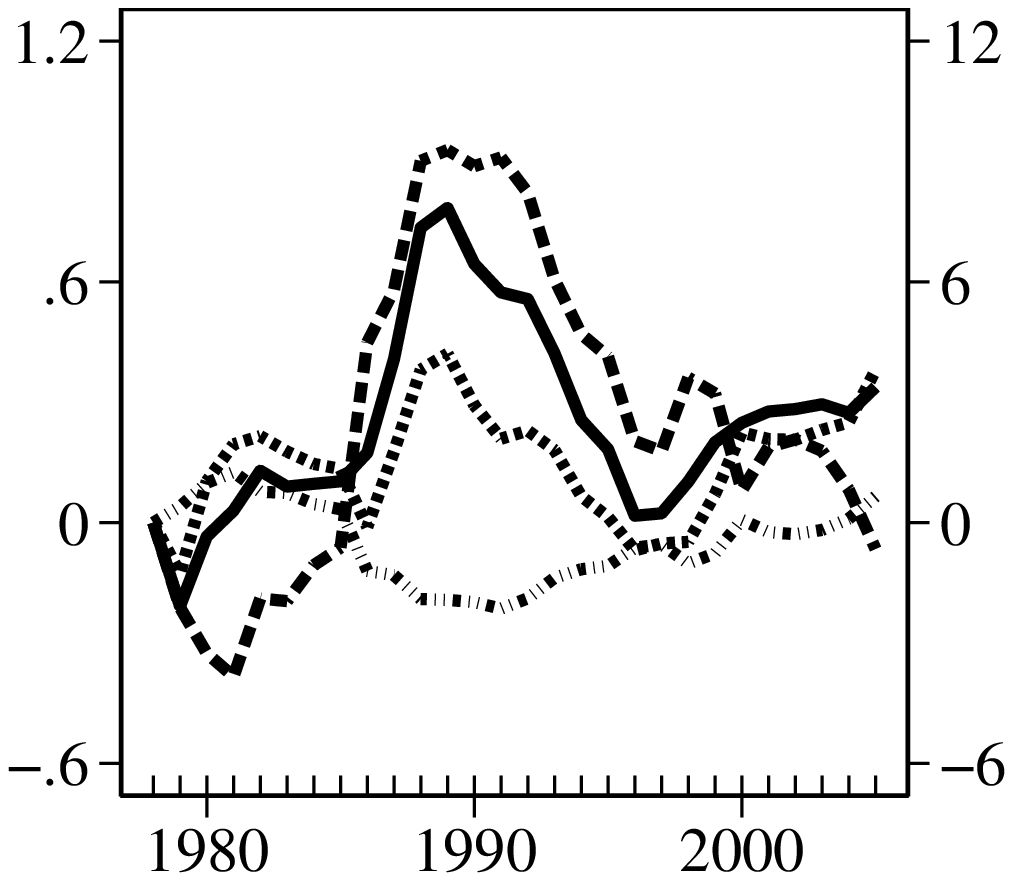}}
\par\end{centering}
\begin{centering}
\subfloat[Germany]{
\centering{}\includegraphics[scale=0.4]{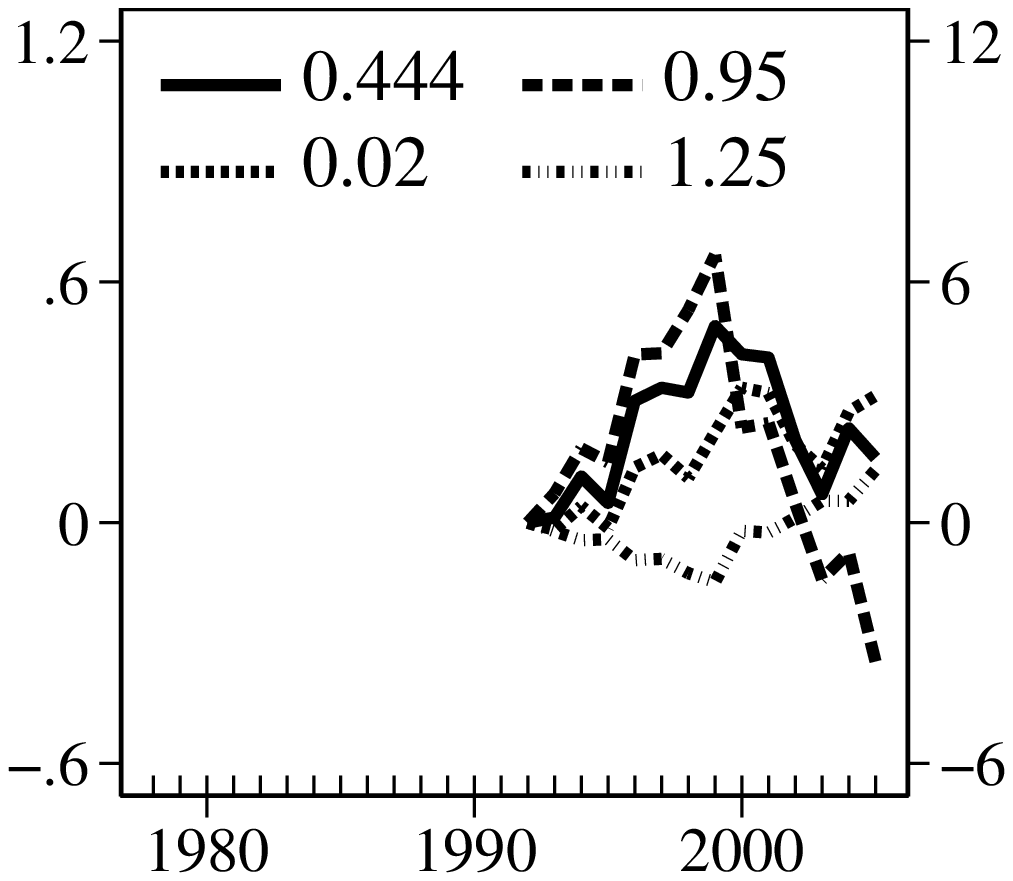}}\subfloat[Italy]{
\centering{}\includegraphics[scale=0.4]{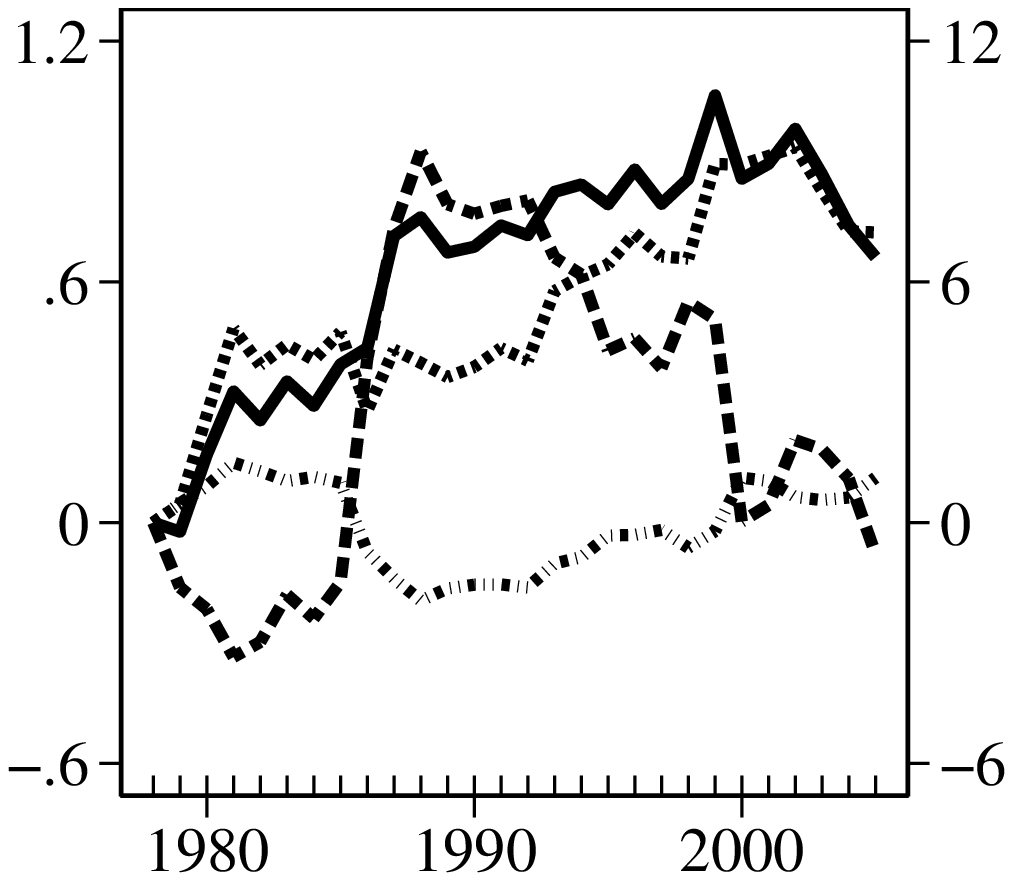}}\subfloat[Japan]{
\centering{}\includegraphics[scale=0.4]{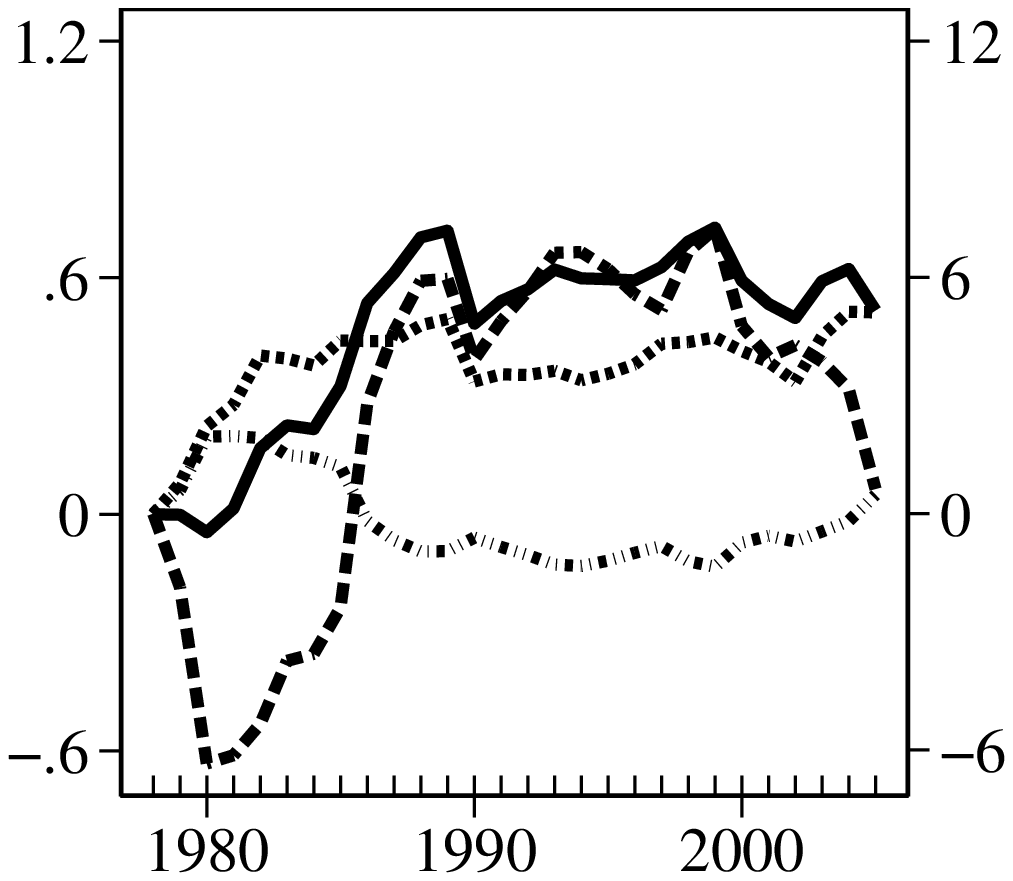}}\subfloat[Netherlands]{
\centering{}\includegraphics[scale=0.4]{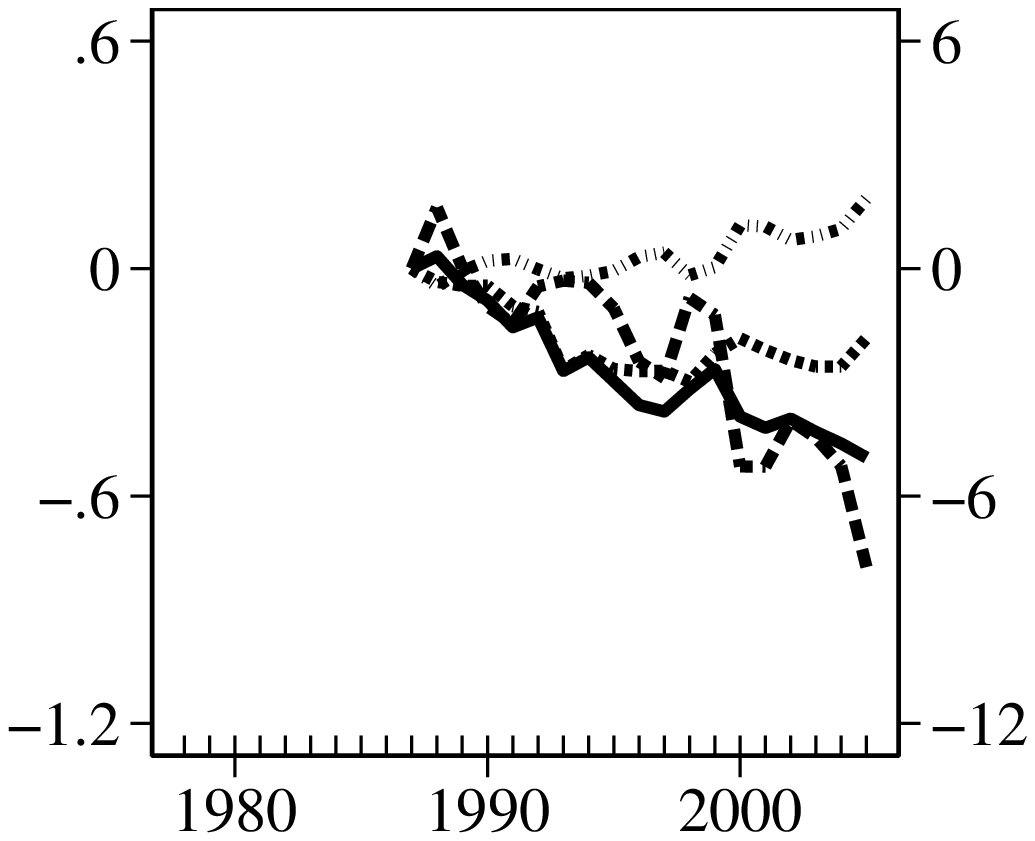}}
\par\end{centering}
\begin{centering}
\subfloat[Portugal]{
\centering{}\includegraphics[scale=0.4]{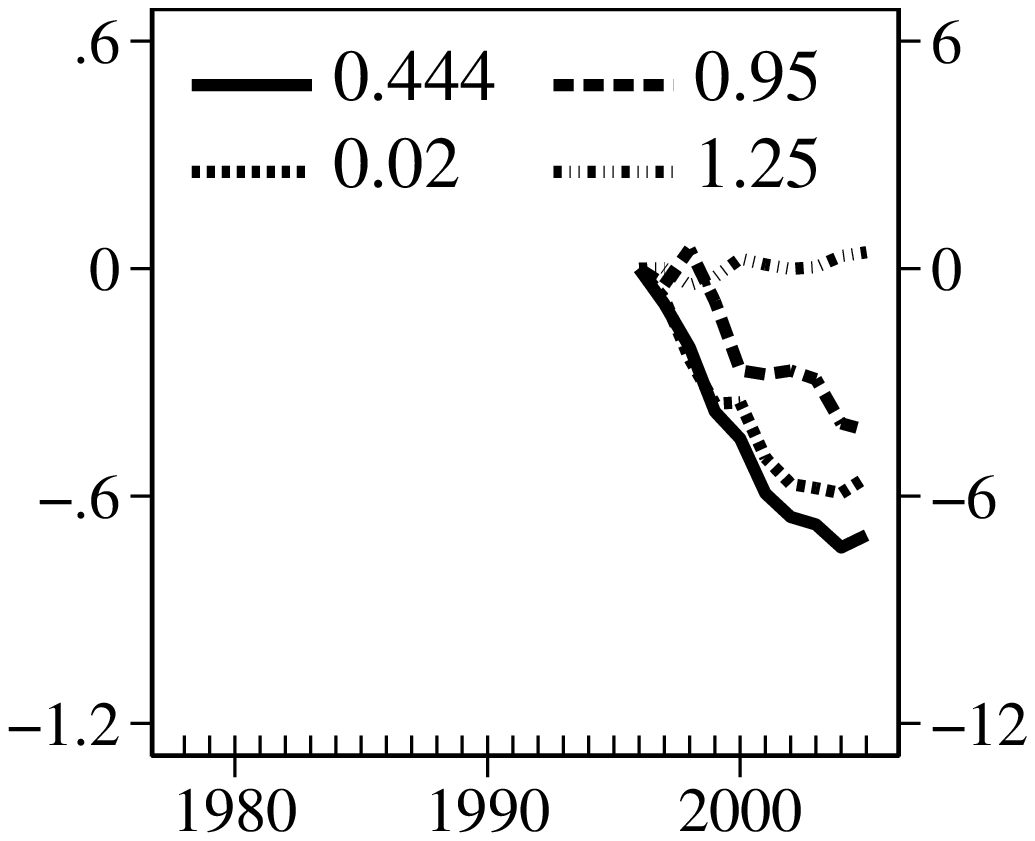}}\subfloat[Sweden]{
\centering{}\includegraphics[scale=0.4]{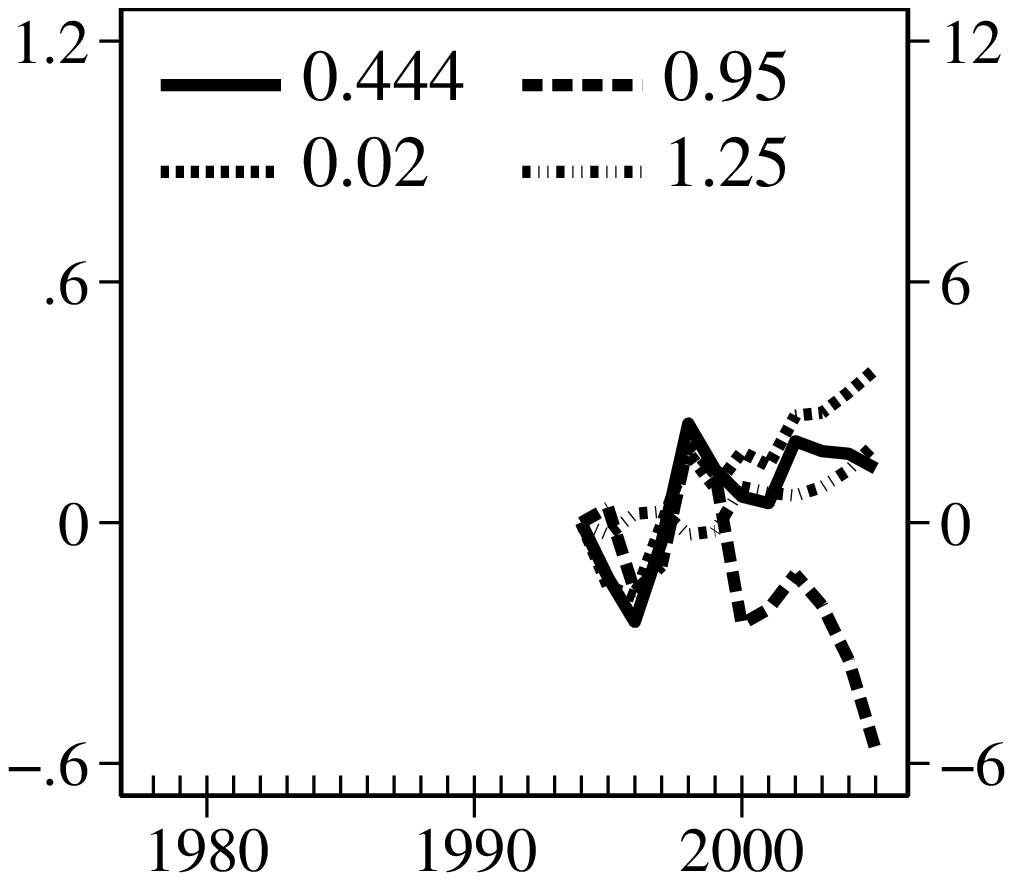}}\subfloat[United Kingdom]{
\centering{}\includegraphics[scale=0.4]{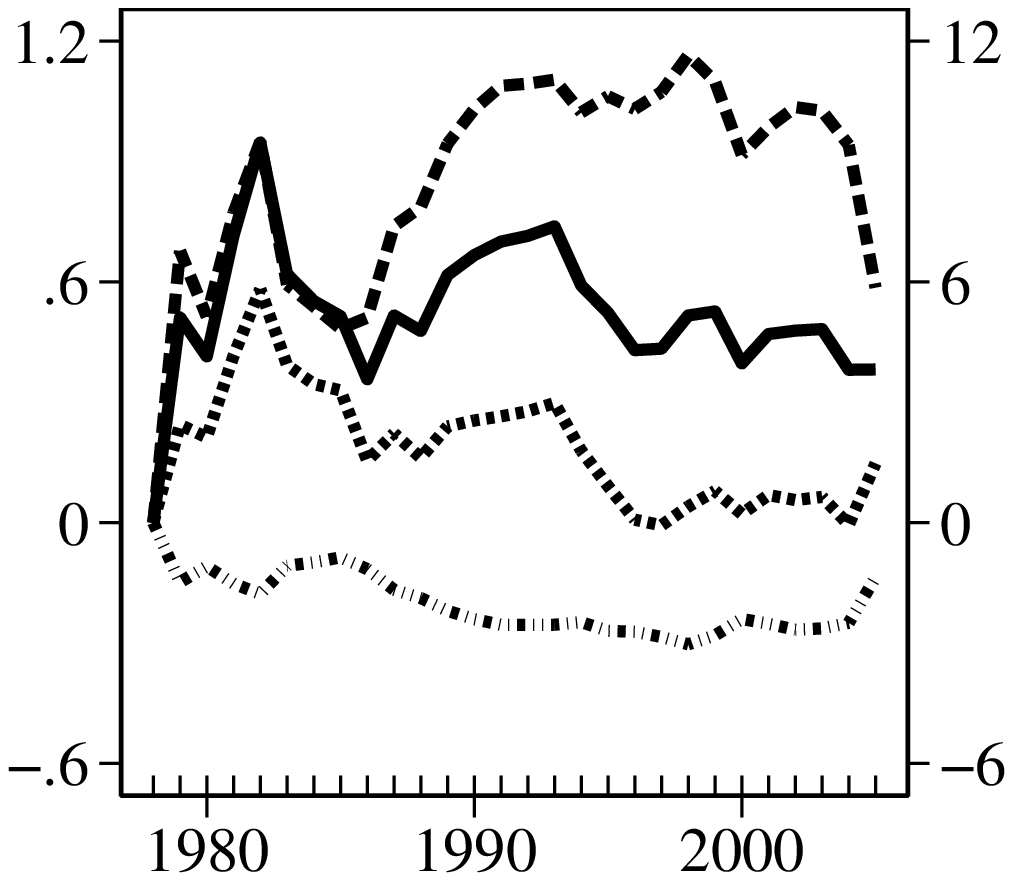}}\subfloat[United States]{
\centering{}\includegraphics[scale=0.4]{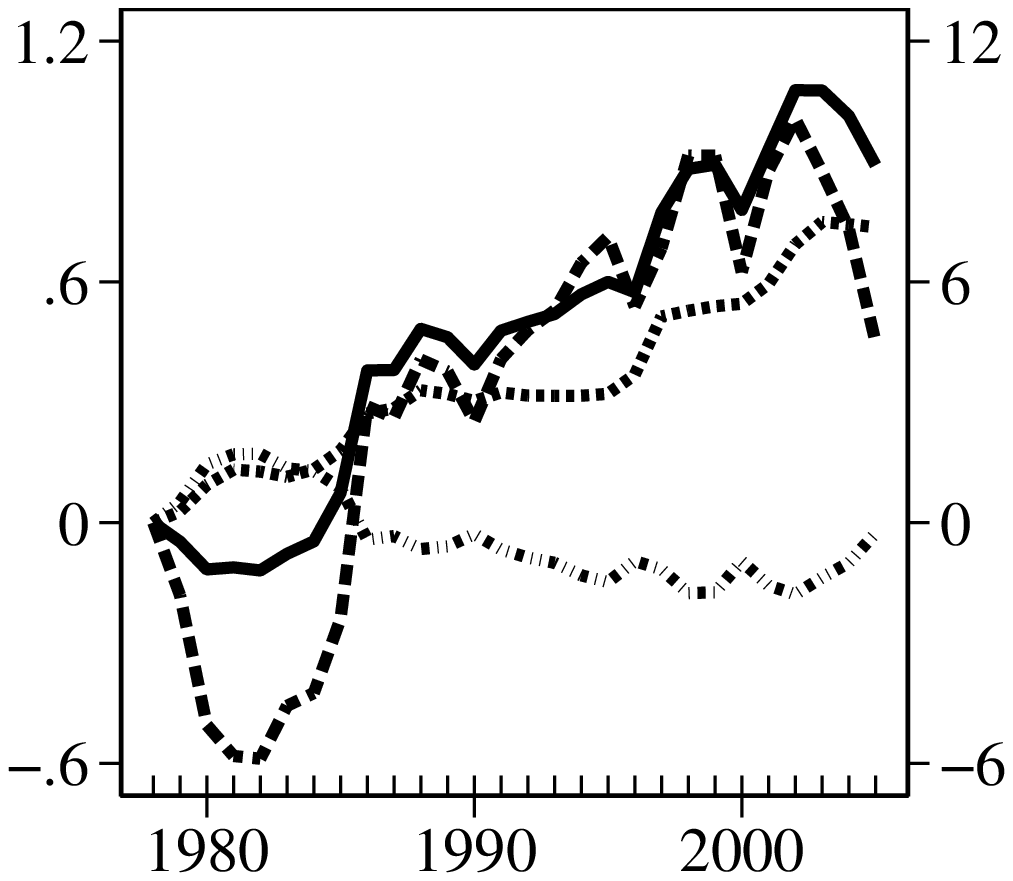}}
\par\end{centering}
\textit{\footnotesize{}Notes}{\footnotesize{}: Each line represents
energy-saving technology ($a_{e}$) when the elasticity of substitution
($\epsilon_{\sigma}$) is 0.444, 0.02, 0.95, or 1.25. The left and
right vertical axes indicate the scale of $a_{e}$ when $\epsilon_{\sigma}$
is 0.444 or 0.02 and when $\epsilon_{\sigma}$ is 0.95 or 1.25, respectively.
All series are expressed as log differences relative to the first
year of observations.}{\footnotesize\par}
\end{figure}

\begin{figure}[H]
\caption{Energy-saving technological change for different values of the elasticity
of substitution in the service sector\label{fig: Ae1_sigma_service}}

\begin{centering}
\subfloat[Austria]{
\centering{}\includegraphics[scale=0.4]{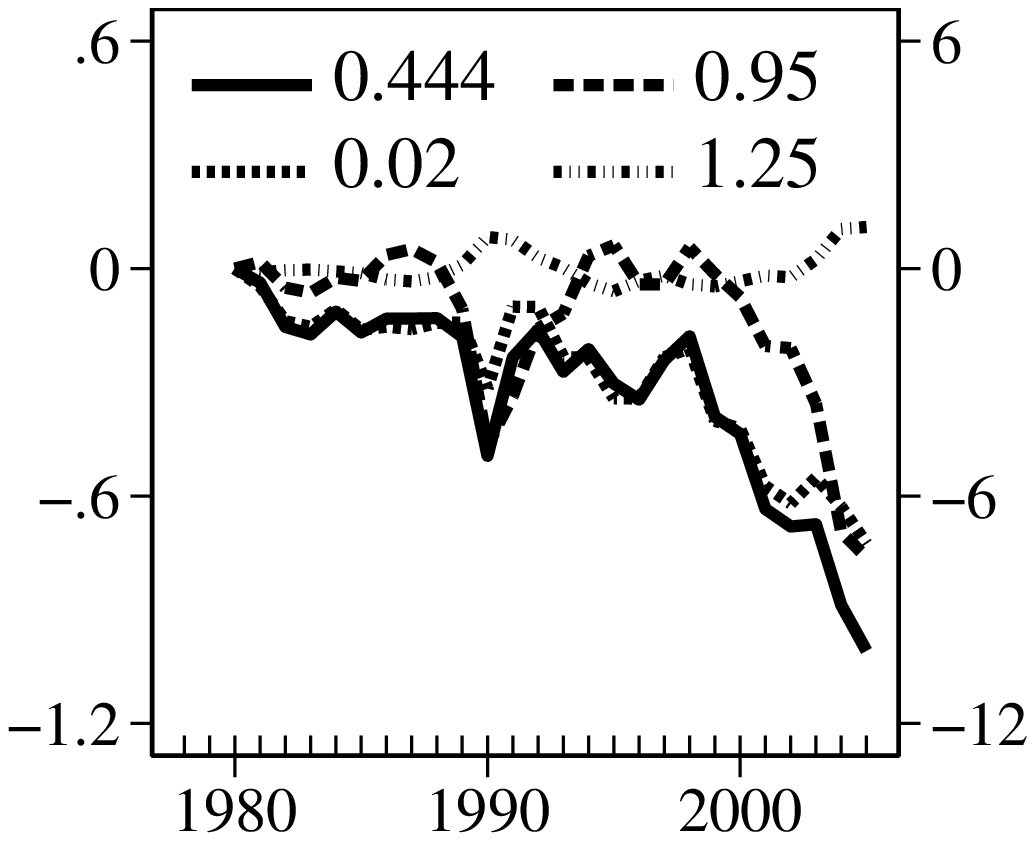}}\subfloat[Czech Republic]{
\centering{}\includegraphics[scale=0.4]{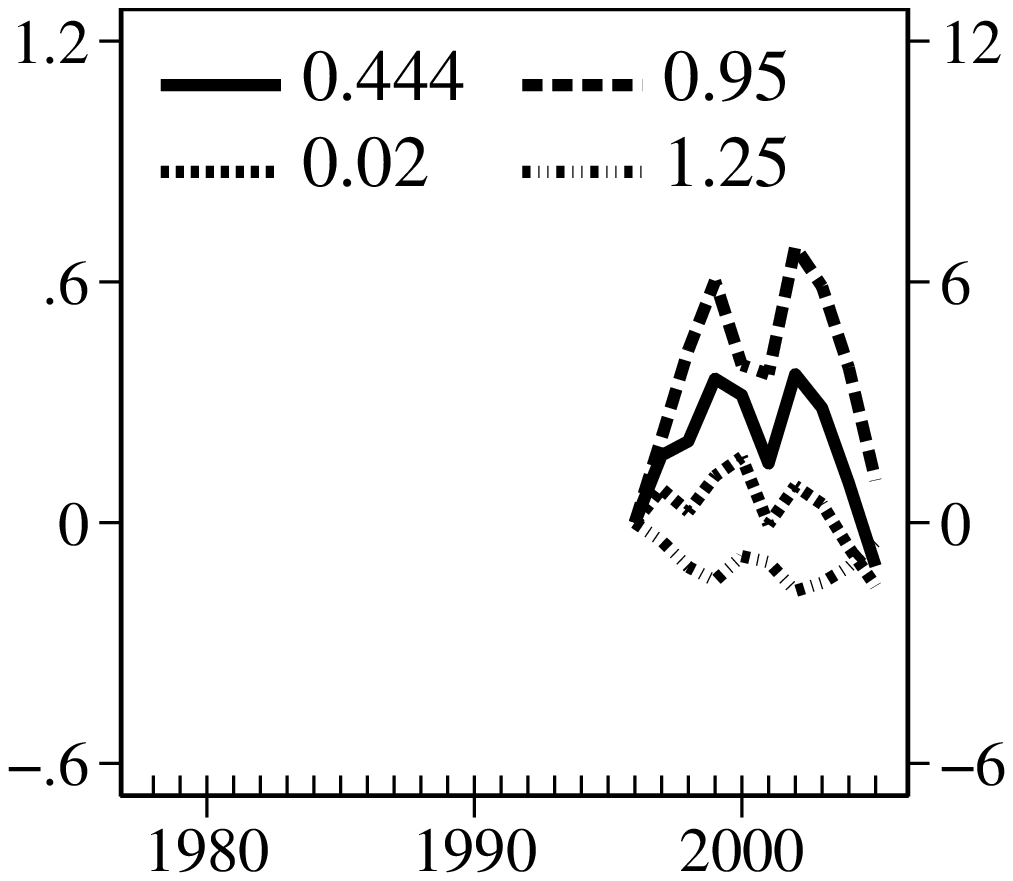}}\subfloat[Denmark]{
\centering{}\includegraphics[scale=0.4]{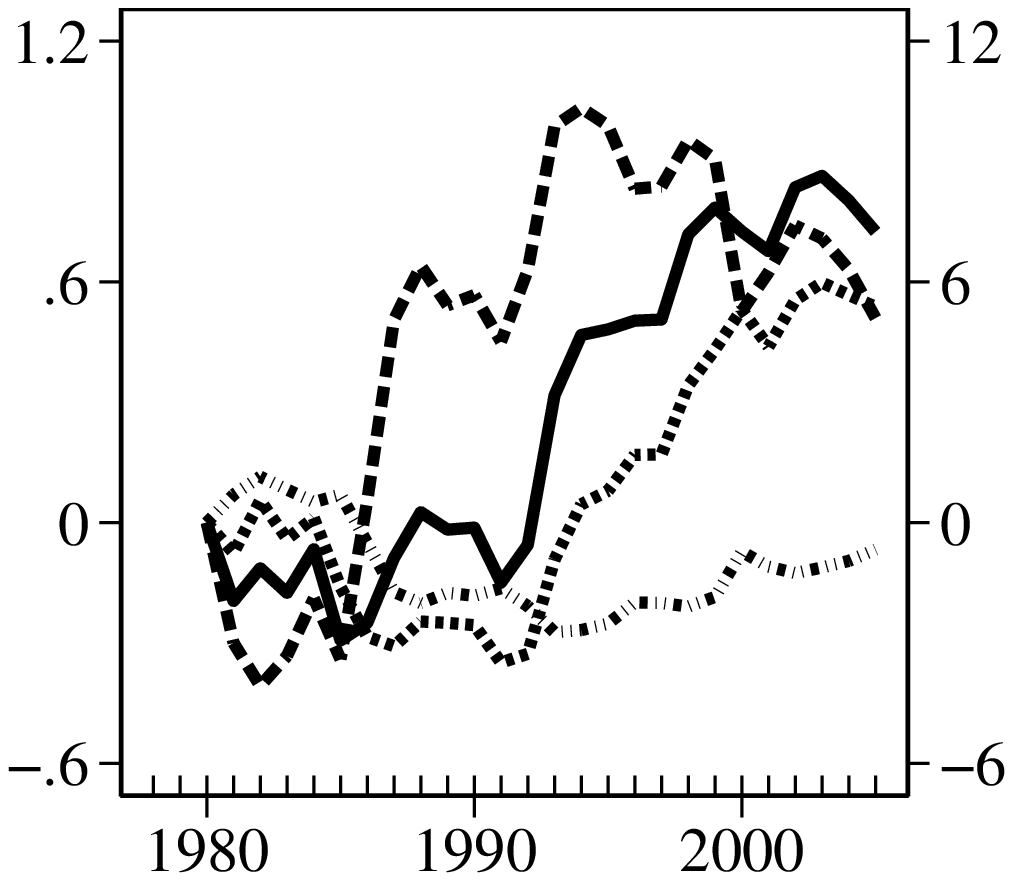}}\subfloat[Finland]{
\centering{}\includegraphics[scale=0.4]{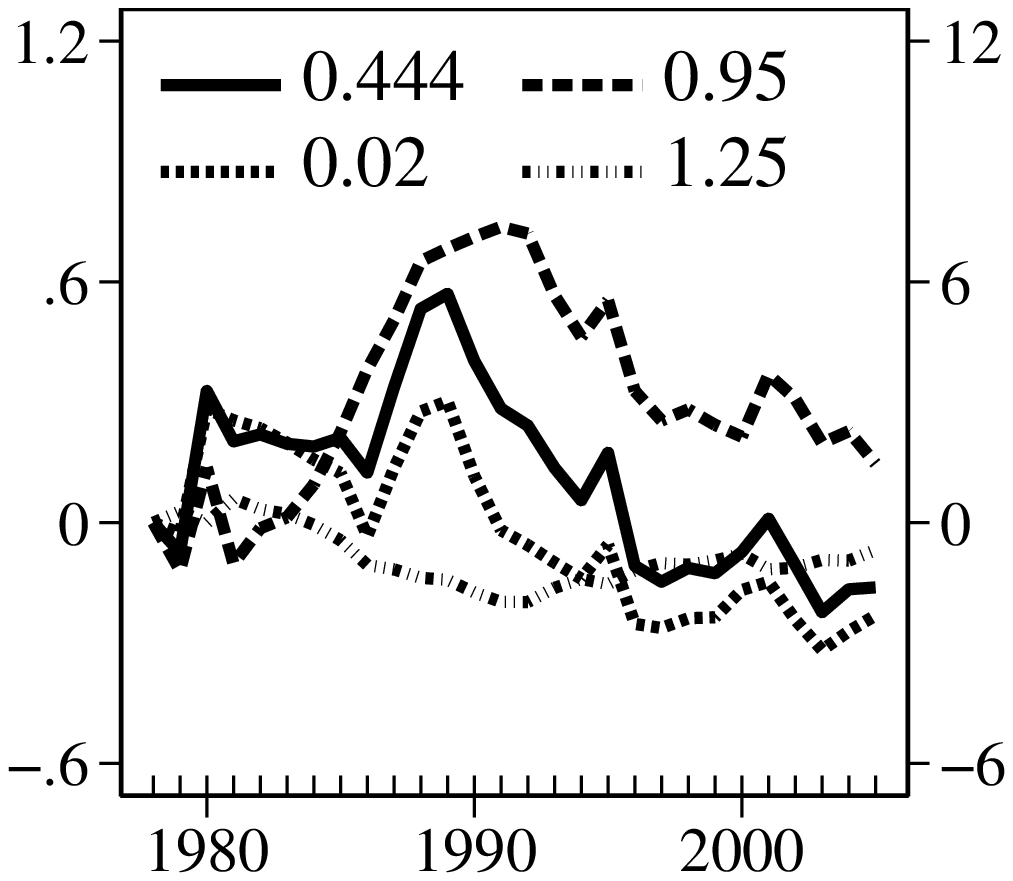}}
\par\end{centering}
\begin{centering}
\subfloat[Germany]{
\centering{}\includegraphics[scale=0.4]{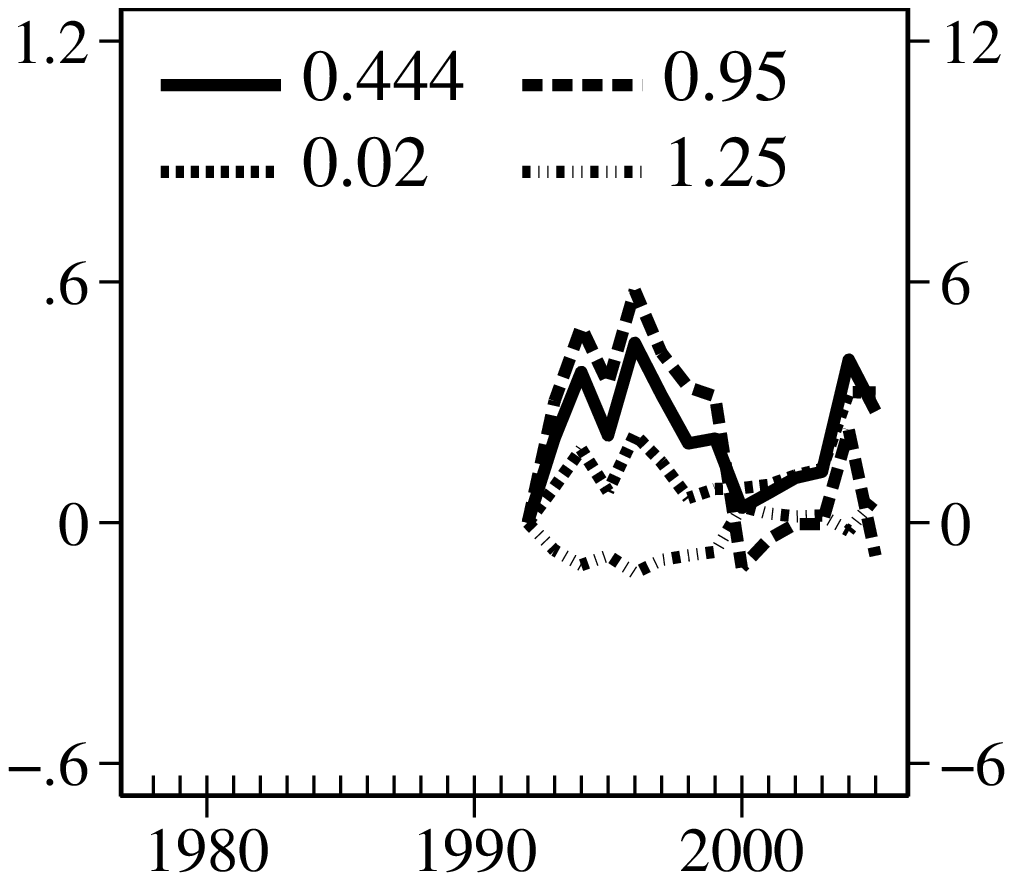}}\subfloat[Italy]{
\centering{}\includegraphics[scale=0.4]{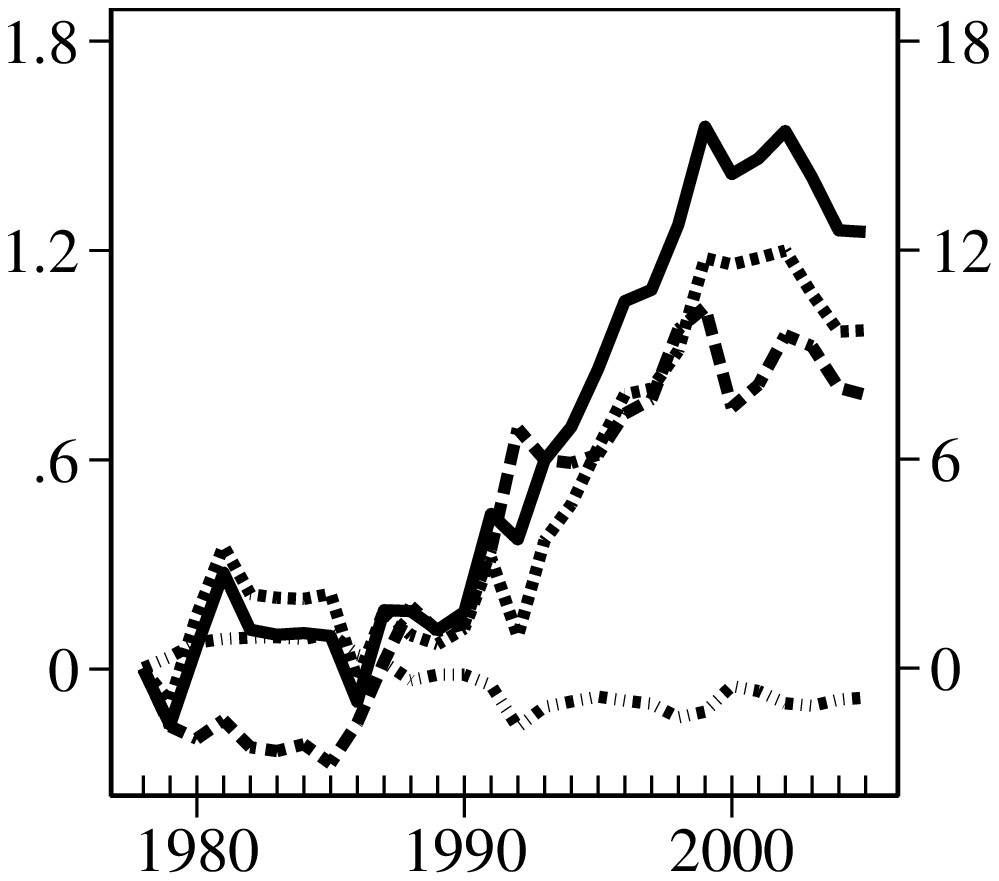}}\subfloat[Japan]{
\centering{}\includegraphics[scale=0.4]{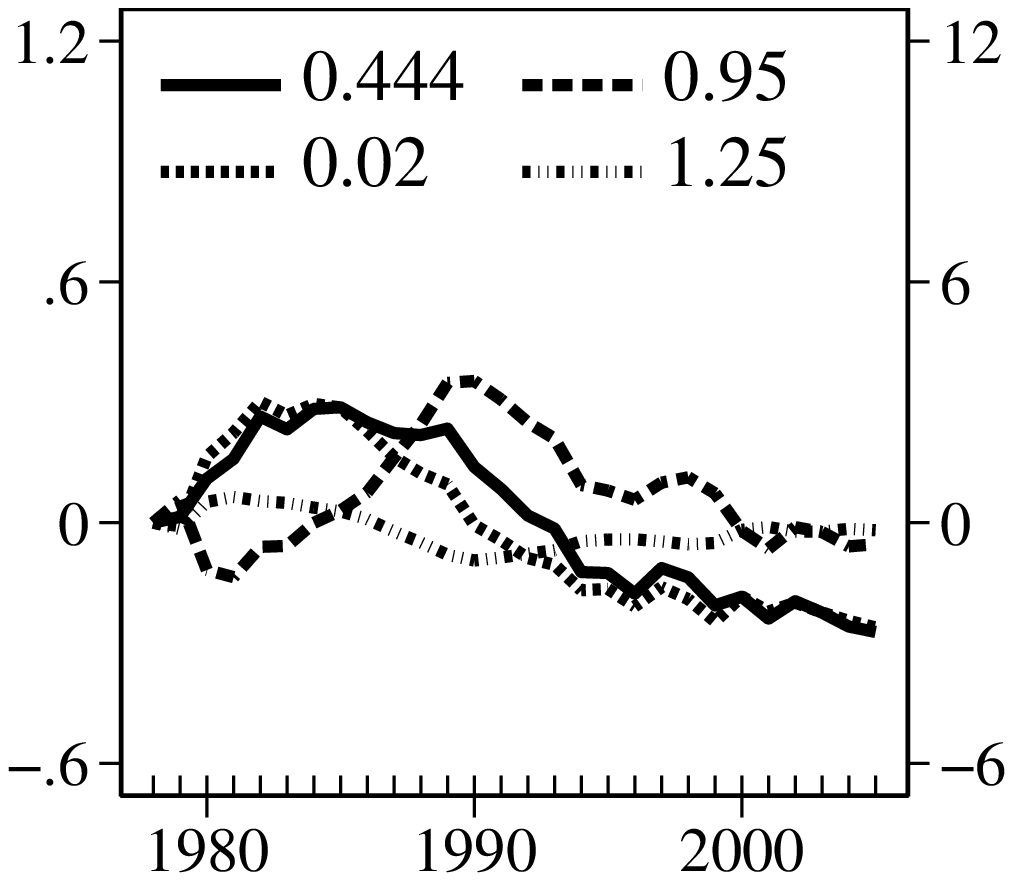}}\subfloat[Netherlands]{
\centering{}\includegraphics[scale=0.4]{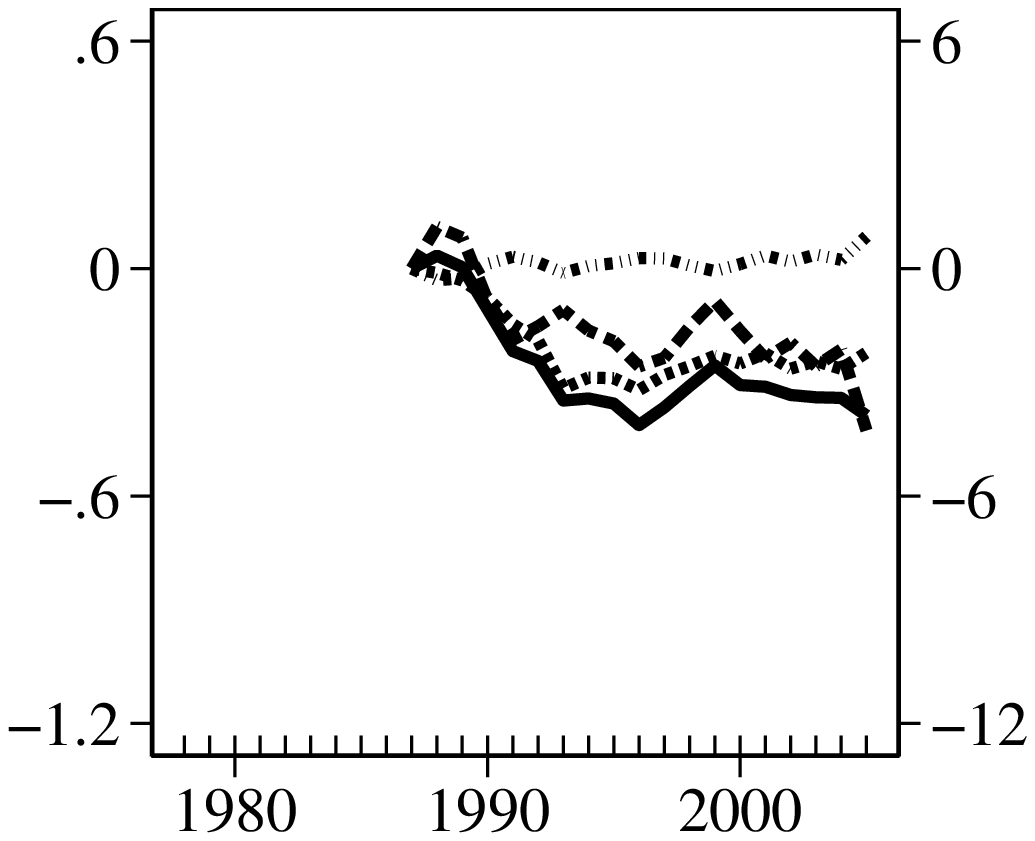}}
\par\end{centering}
\begin{centering}
\subfloat[Portugal]{
\centering{}\includegraphics[scale=0.4]{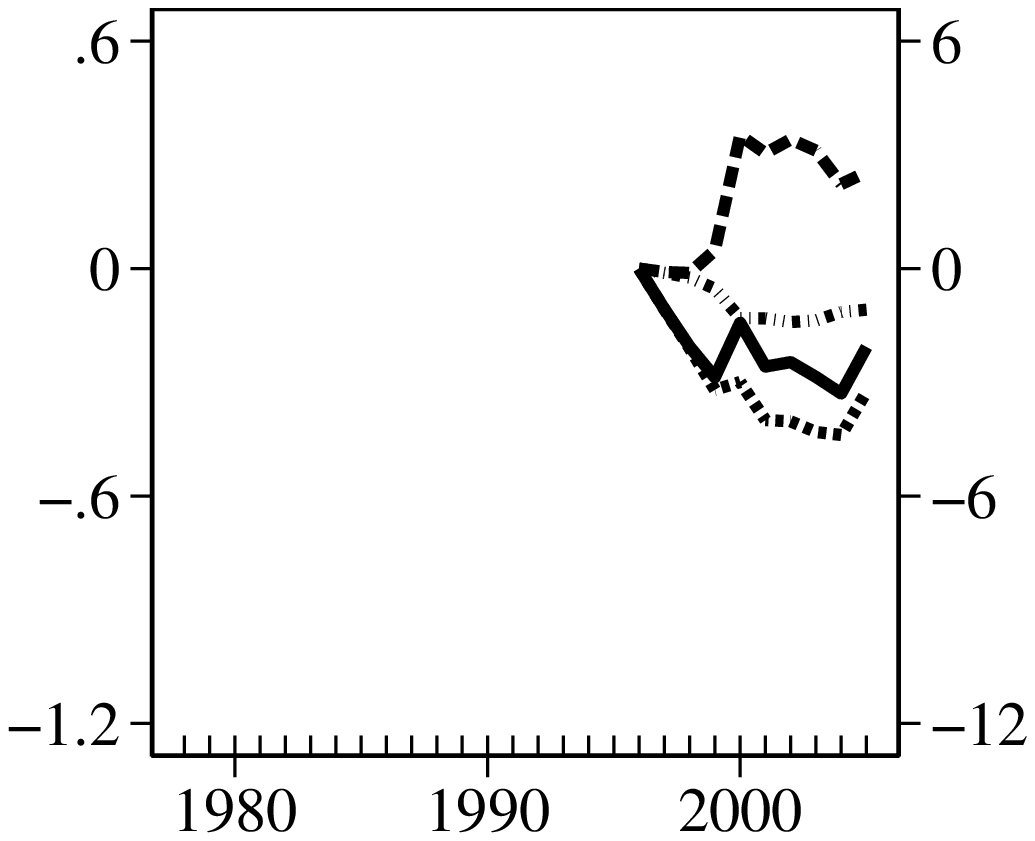}}\subfloat[Sweden]{
\centering{}\includegraphics[scale=0.4]{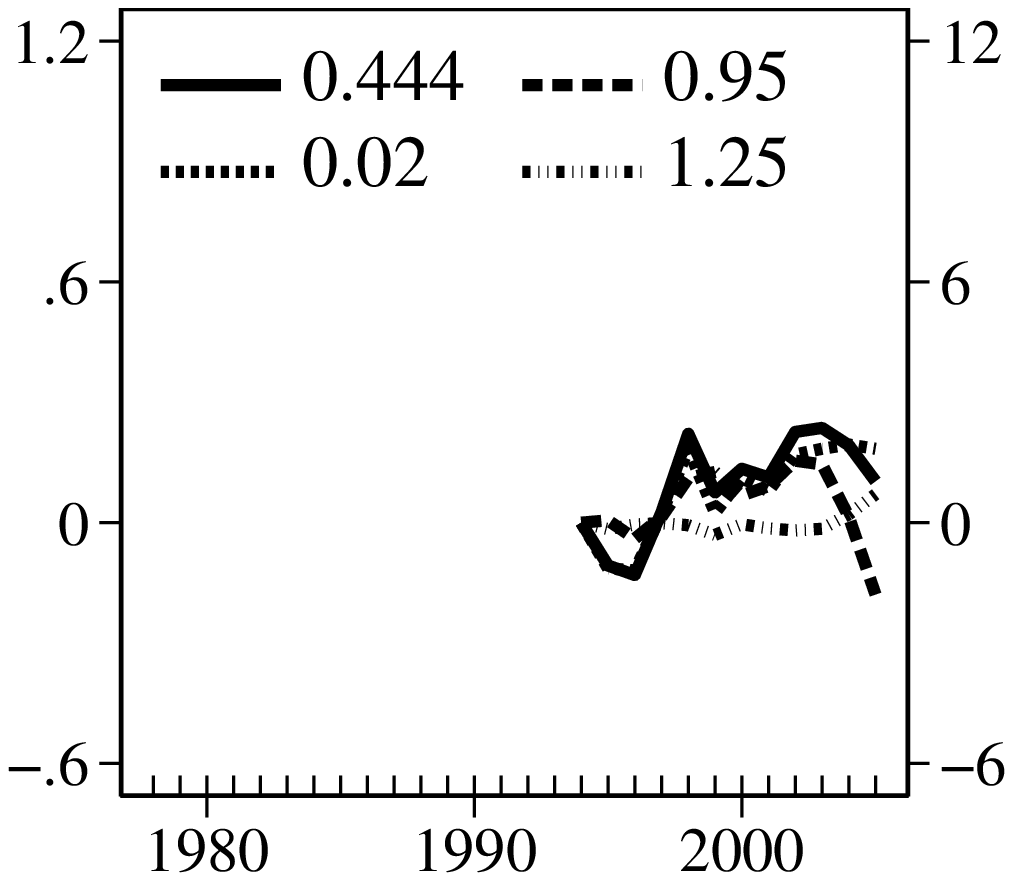}}\subfloat[United Kingdom]{
\centering{}\includegraphics[scale=0.4]{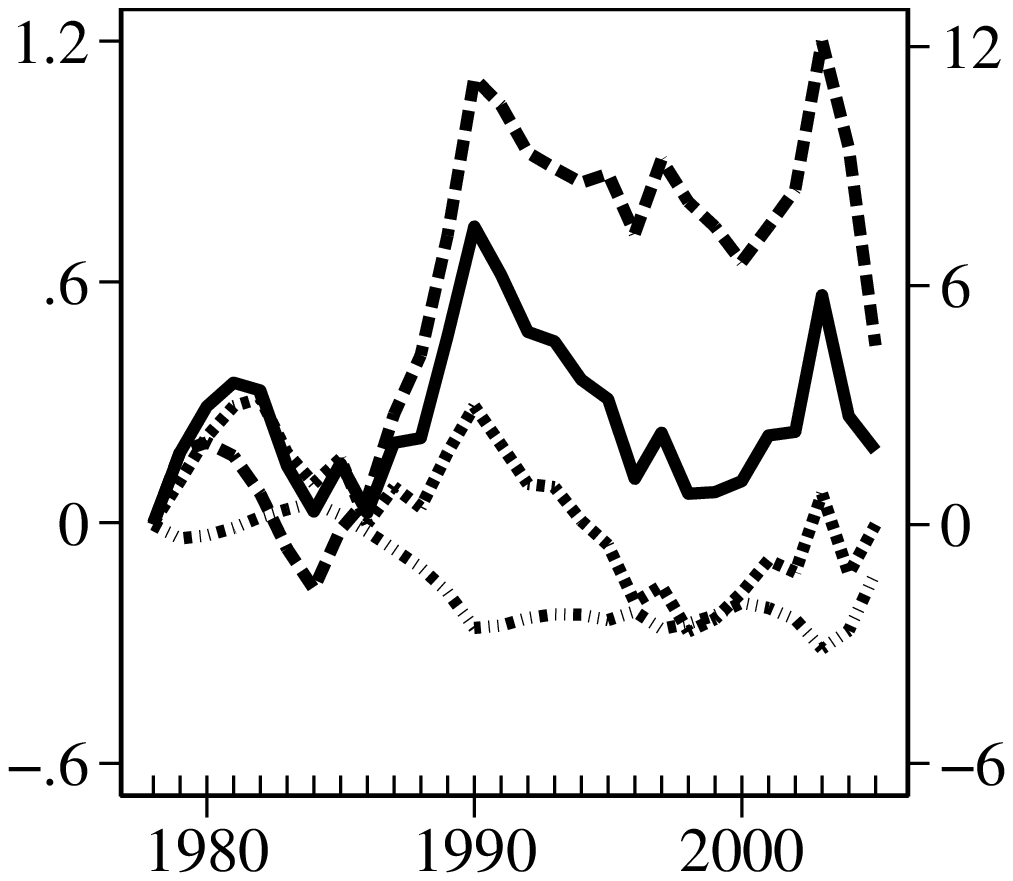}}\subfloat[United States]{
\centering{}\includegraphics[scale=0.4]{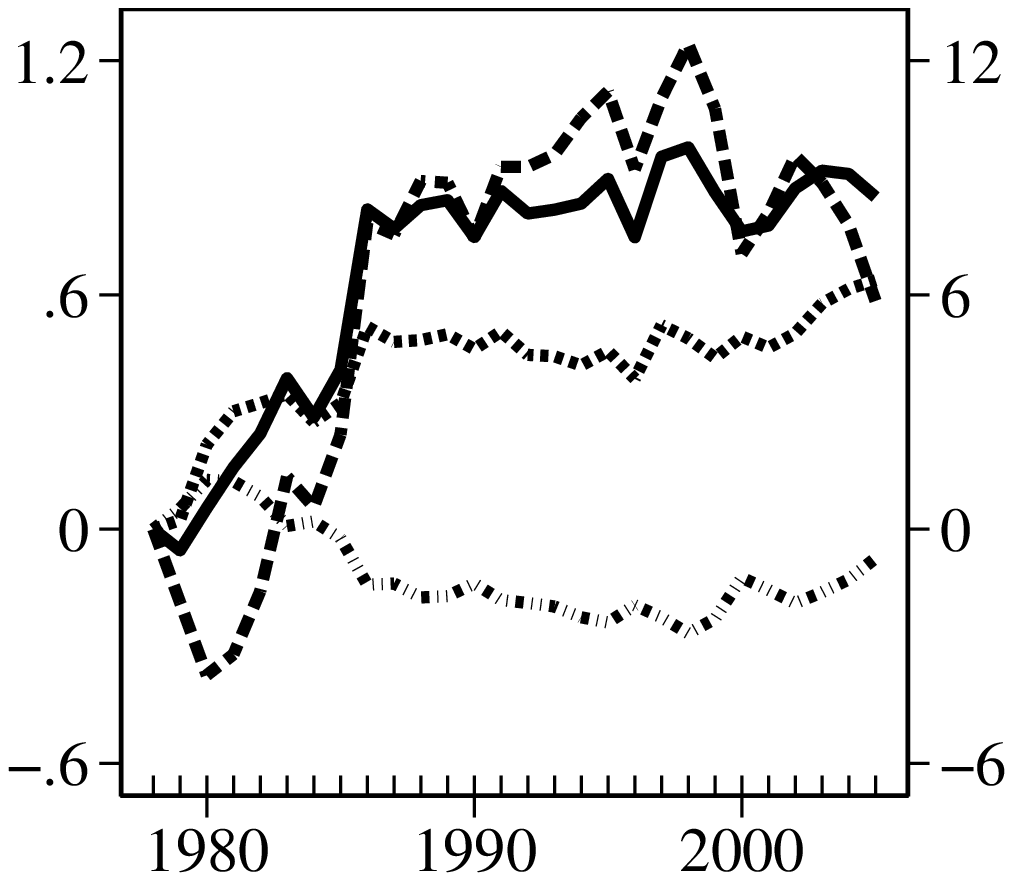}}
\par\end{centering}
\textit{\footnotesize{}Notes}{\footnotesize{}: Each line represents
energy-saving technology ($a_{e}$) when the elasticity of substitution
($\epsilon_{\sigma}$) is 0.444, 0.02, 0.95, or 1.25. The left and
right vertical axes indicate the scale of $a_{e}$ when $\epsilon_{\sigma}$
is 0.444 or 0.02 and when $\epsilon_{\sigma}$ is 0.95 or 1.25, respectively.
All series are expressed as log differences relative to the first
year of observations.}{\footnotesize\par}
\end{figure}

\begin{figure}[H]
\caption{Energy-saving technological change in the goods sector when the elasticity
of substitution varies across sectors\label{fig: Ae1_sigma'_goods}}

\begin{centering}
\subfloat[Austria]{
\centering{}\includegraphics[scale=0.4]{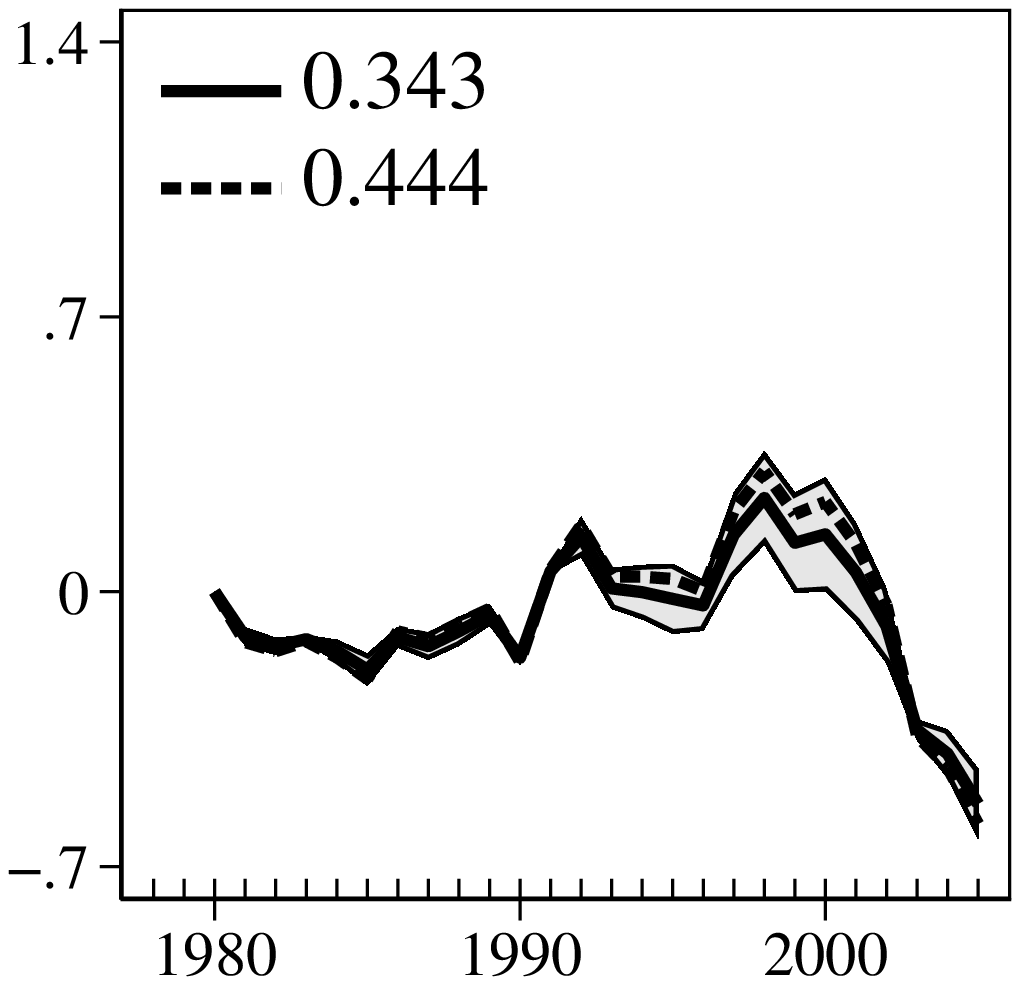}}\subfloat[Czech Republic]{
\centering{}\includegraphics[scale=0.4]{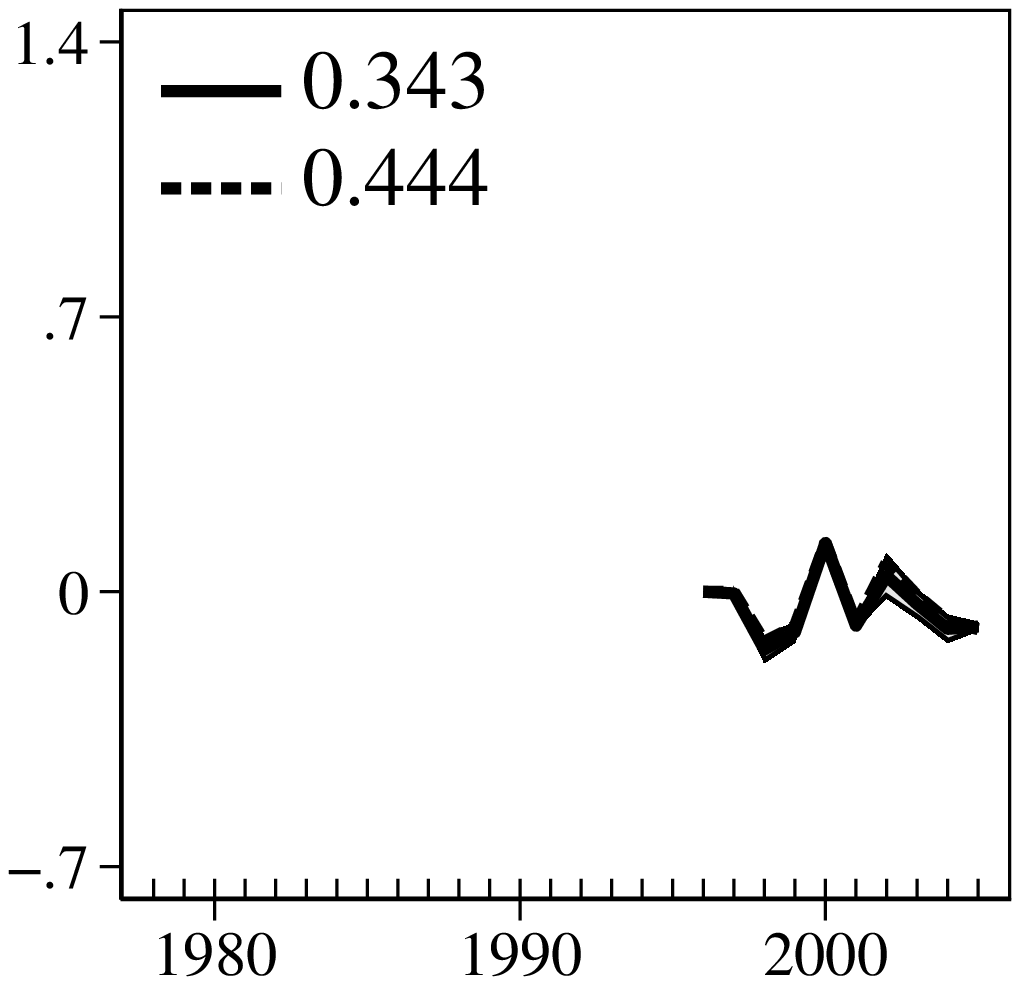}}\subfloat[Denmark]{
\centering{}\includegraphics[scale=0.4]{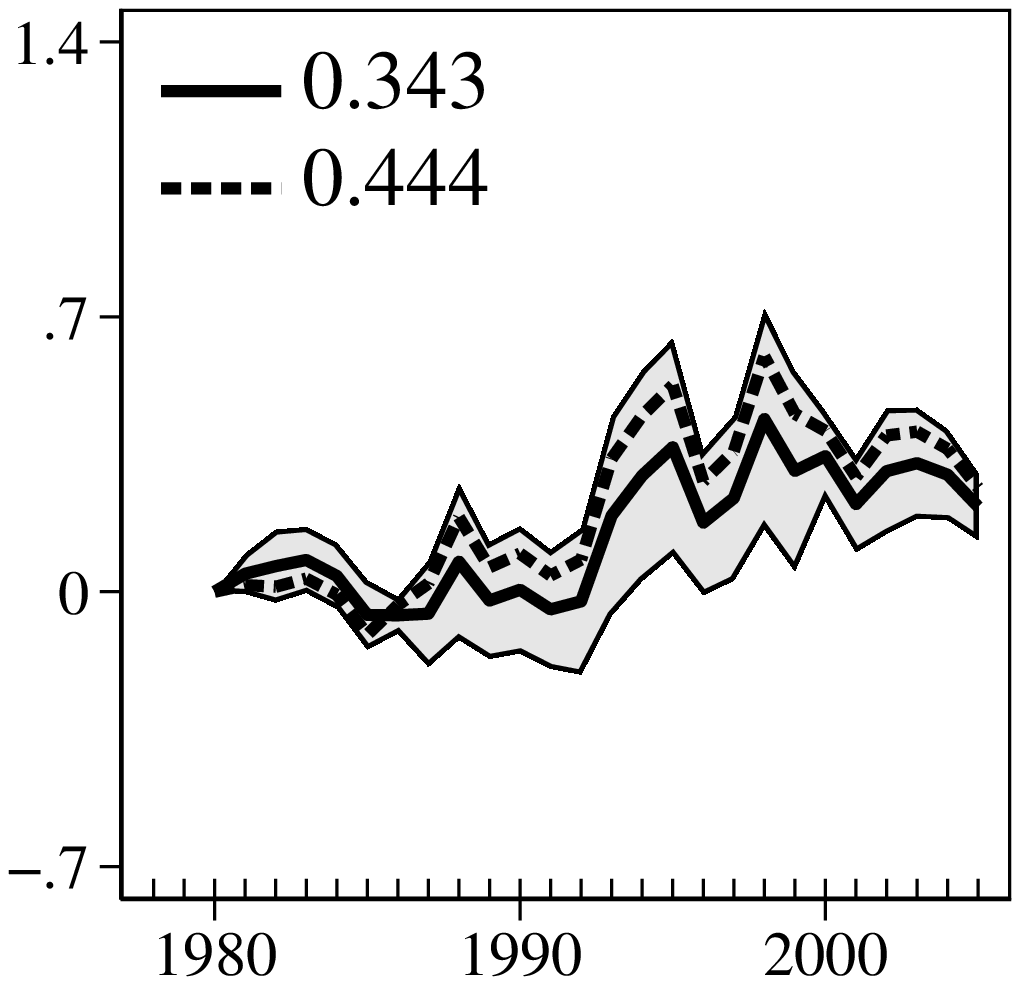}}\subfloat[Finland]{
\centering{}\includegraphics[scale=0.4]{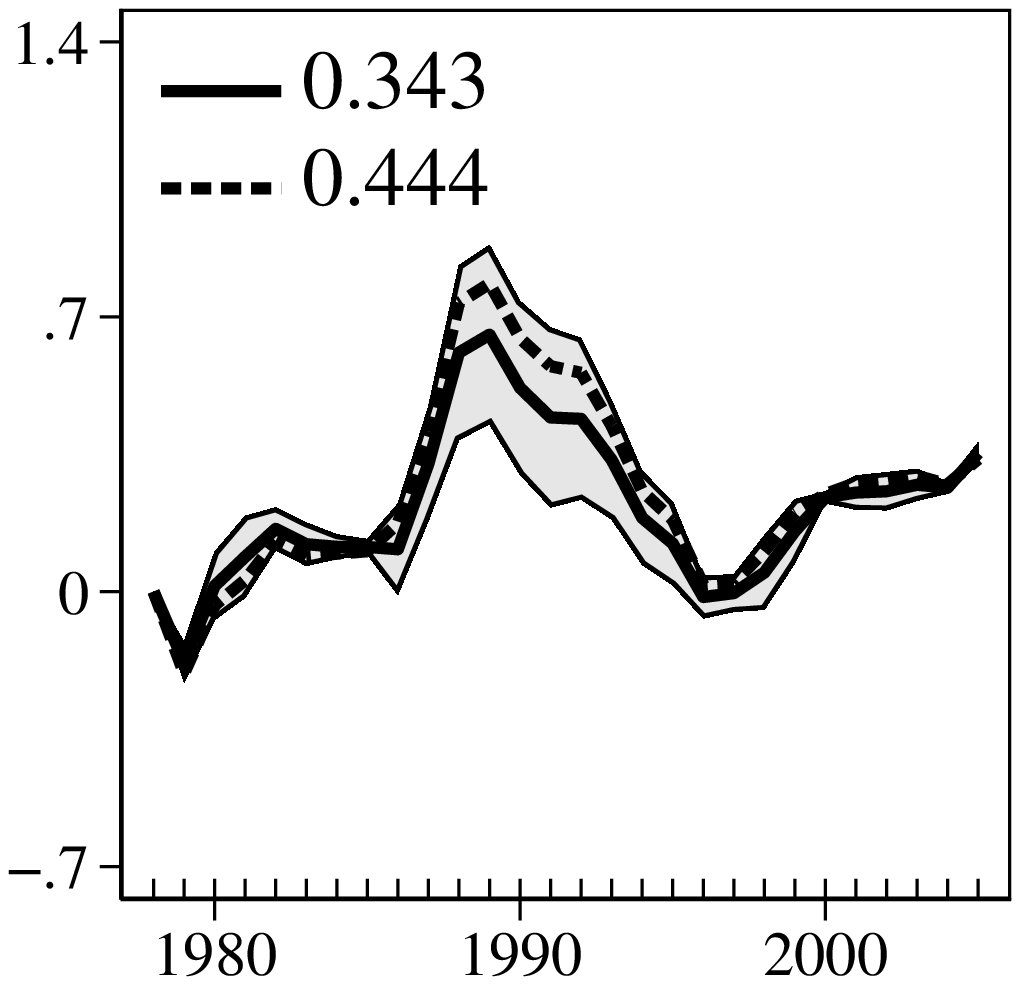}}
\par\end{centering}
\begin{centering}
\subfloat[Germany]{
\centering{}\includegraphics[scale=0.4]{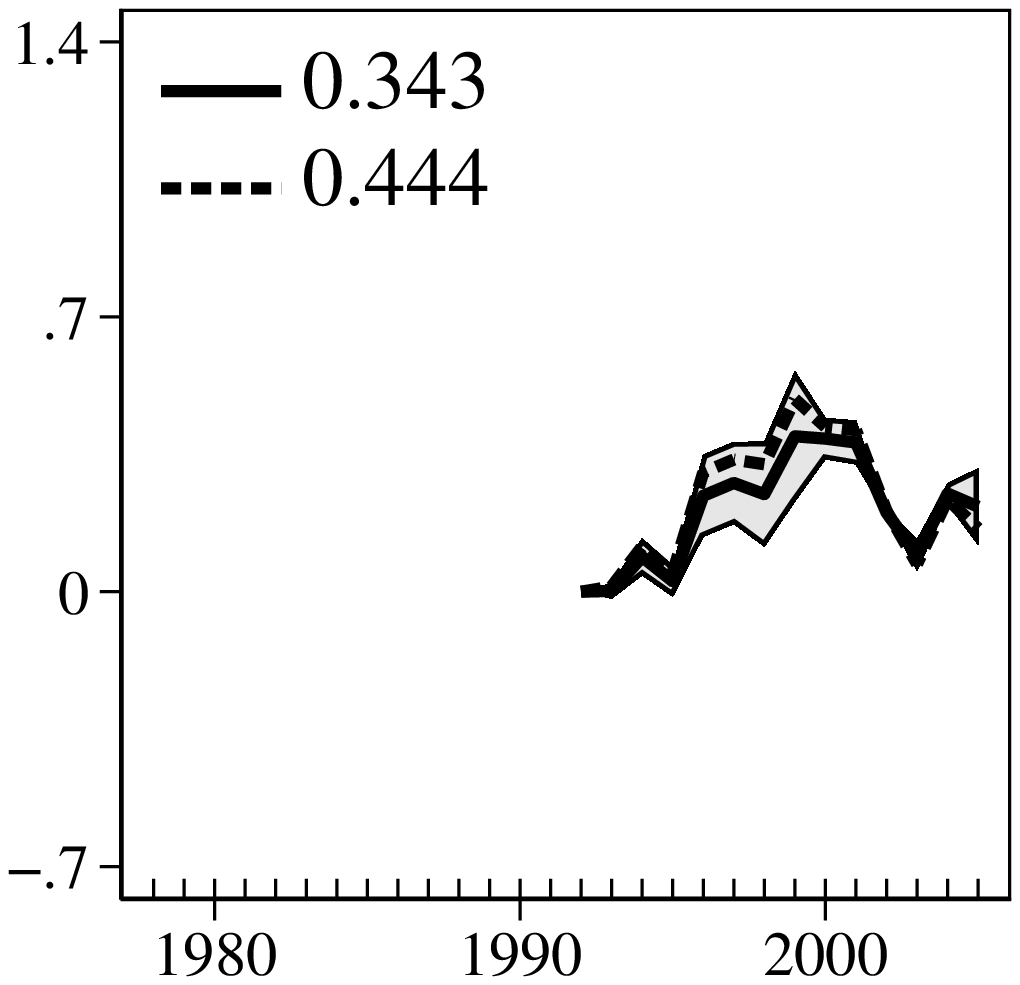}}\subfloat[Italy]{
\centering{}\includegraphics[scale=0.4]{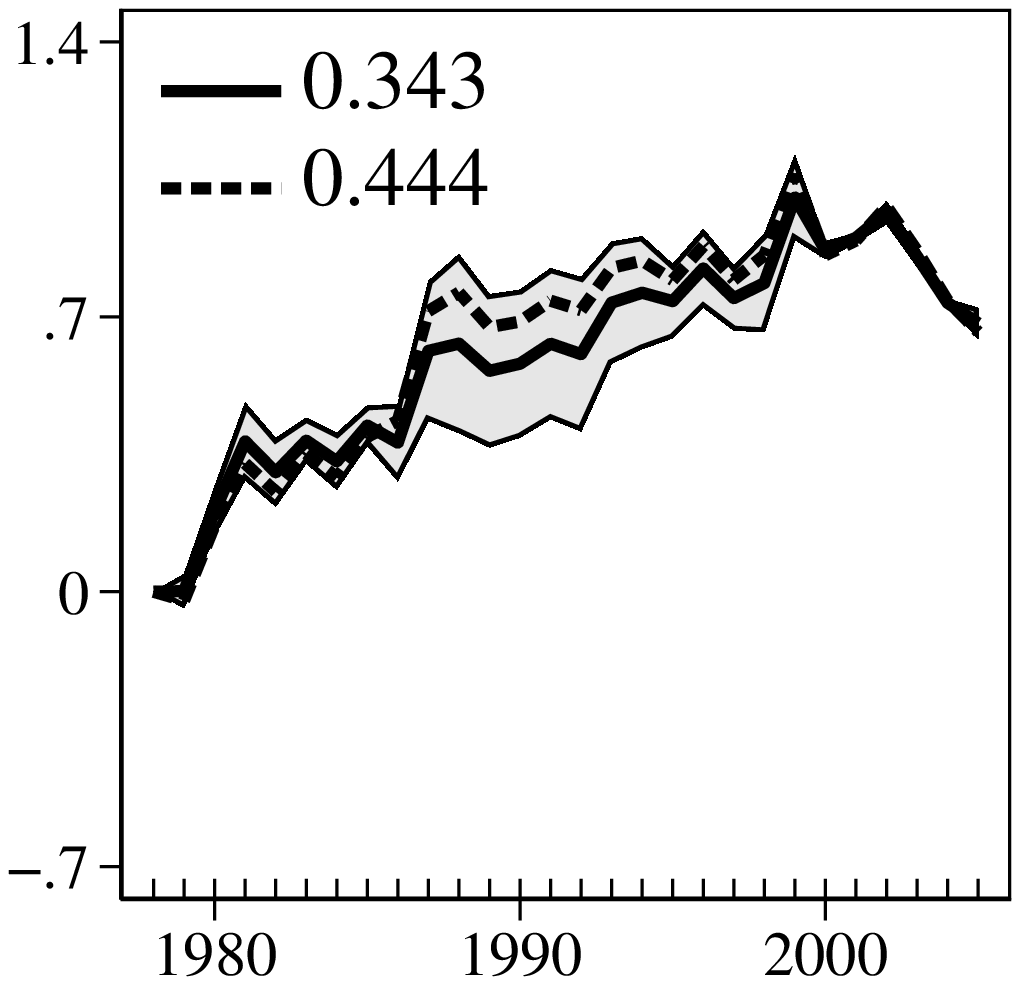}}\subfloat[Japan]{
\centering{}\includegraphics[scale=0.4]{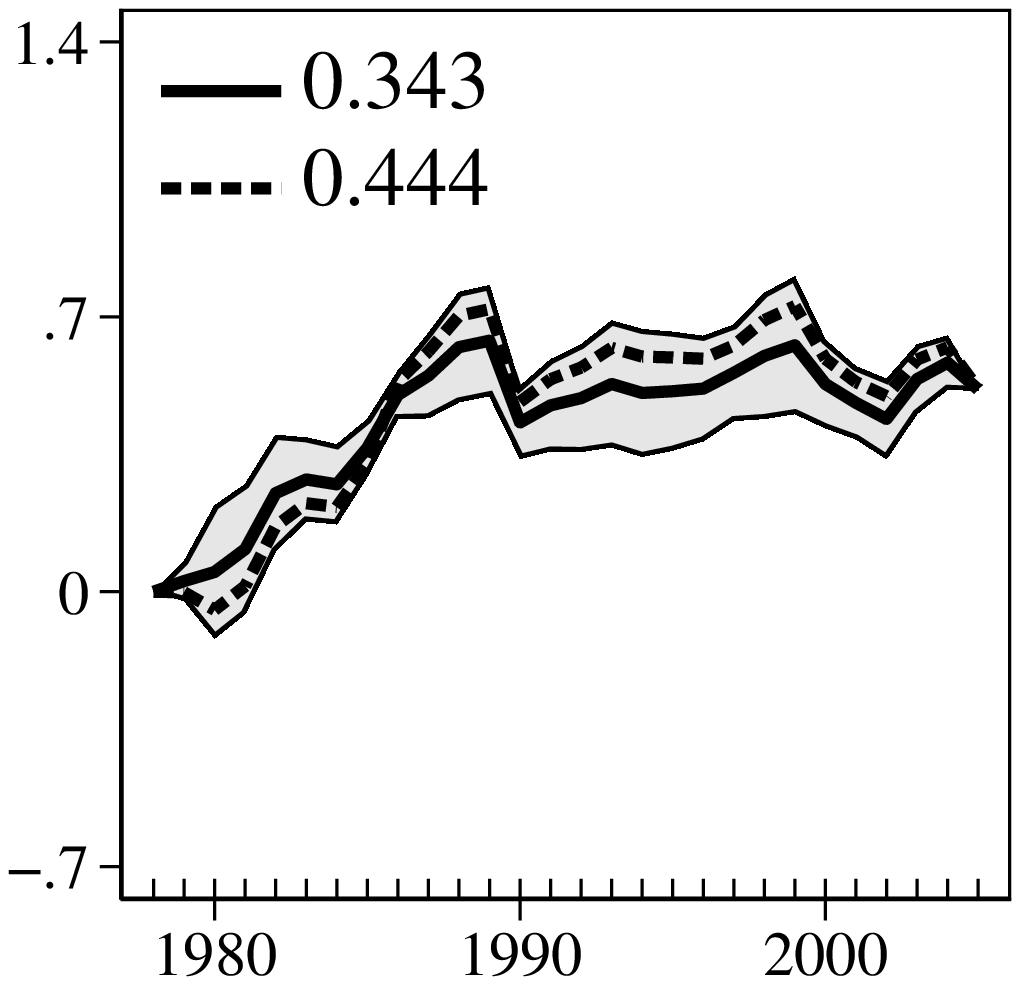}}\subfloat[Netherlands]{
\centering{}\includegraphics[scale=0.4]{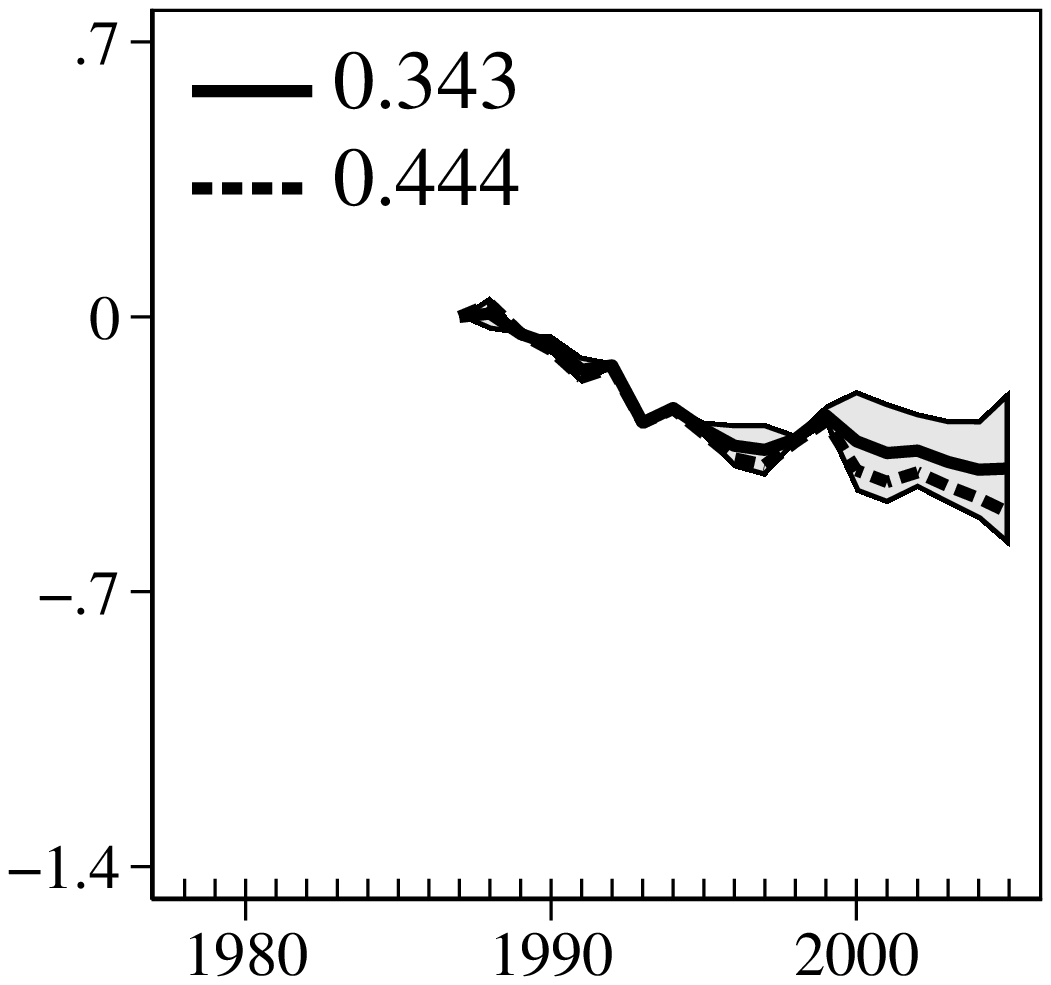}}
\par\end{centering}
\begin{centering}
\subfloat[Portugal]{
\centering{}\includegraphics[scale=0.4]{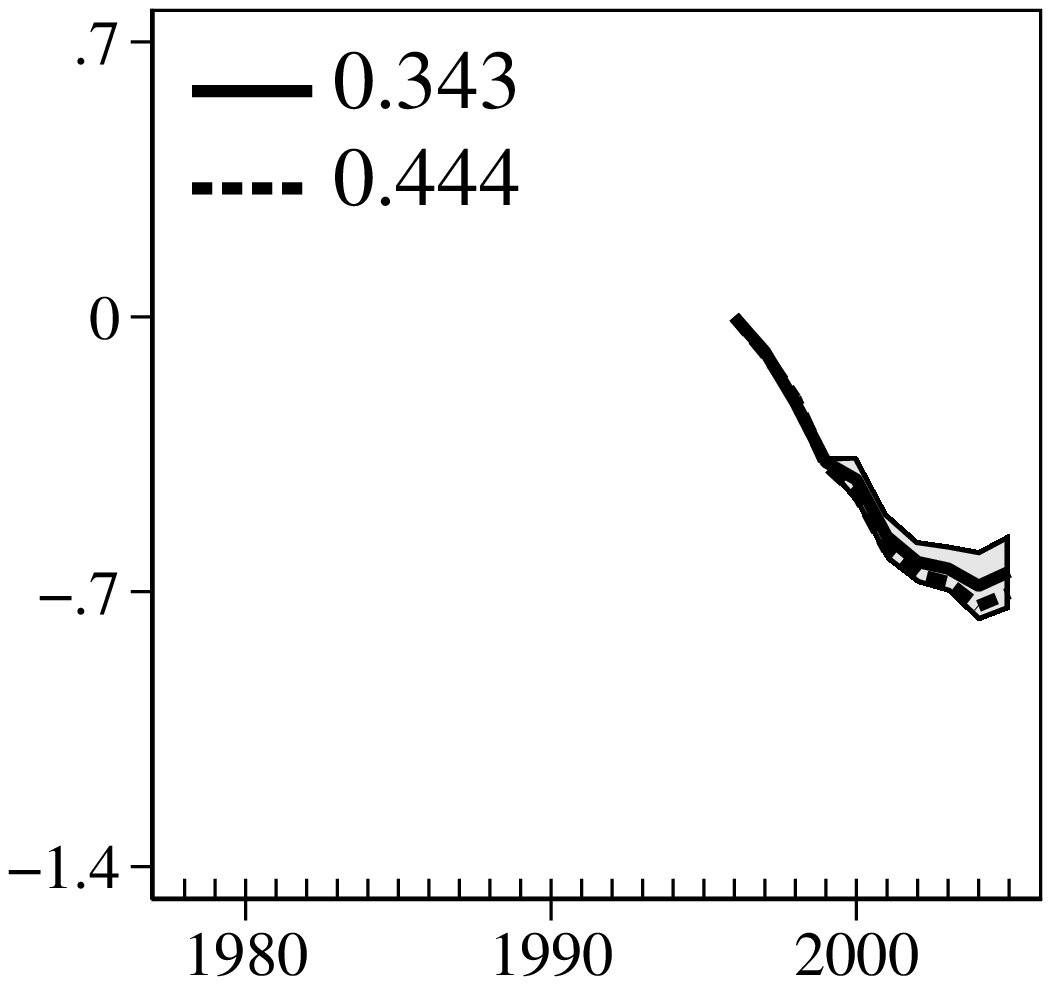}}\subfloat[Sweden]{
\centering{}\includegraphics[scale=0.4]{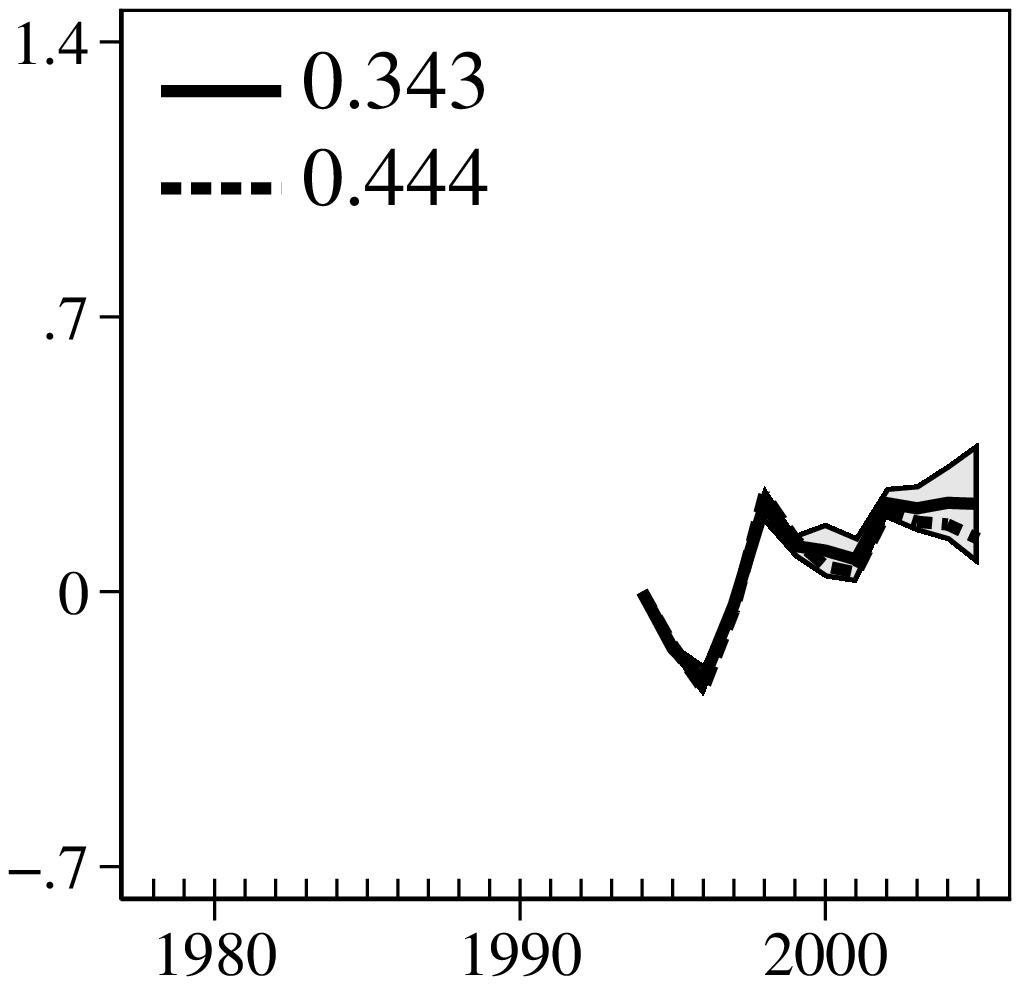}}\subfloat[United Kingdom]{
\centering{}\includegraphics[scale=0.4]{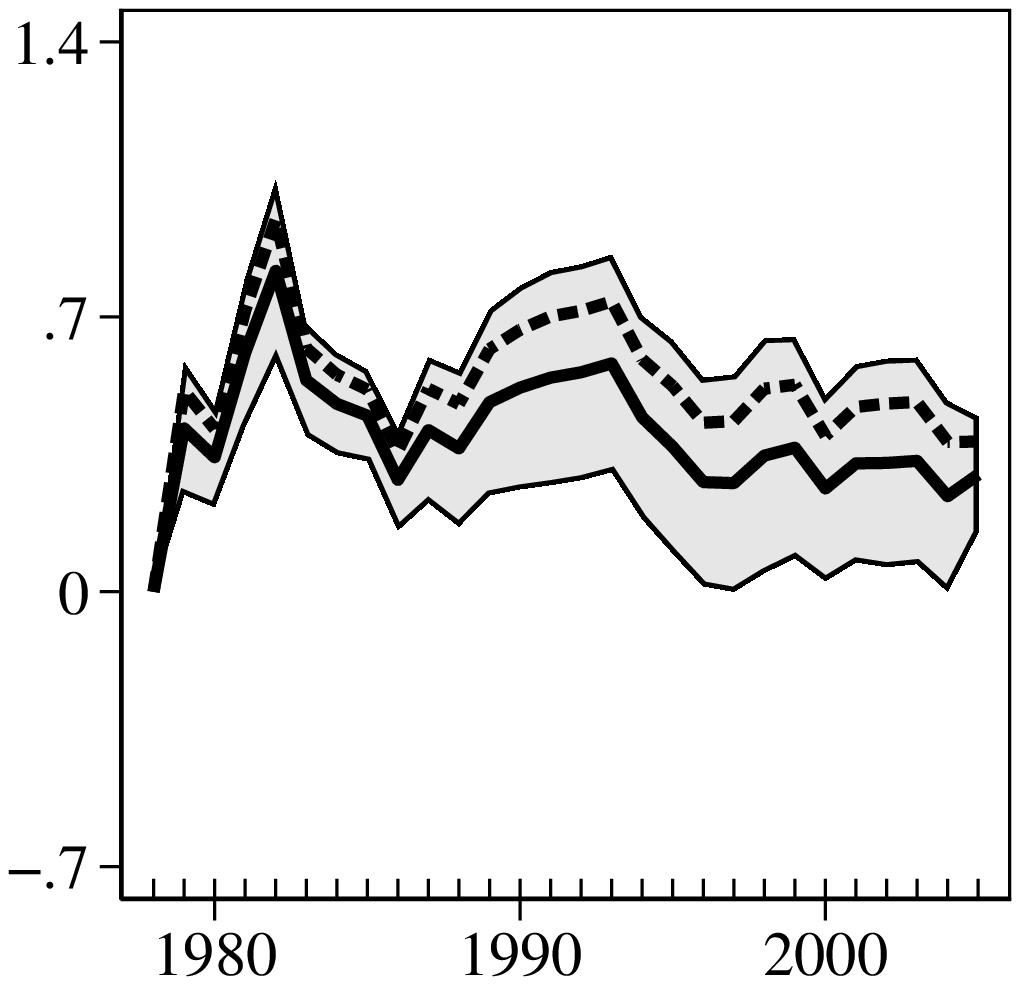}}\subfloat[United States]{
\centering{}\includegraphics[scale=0.4]{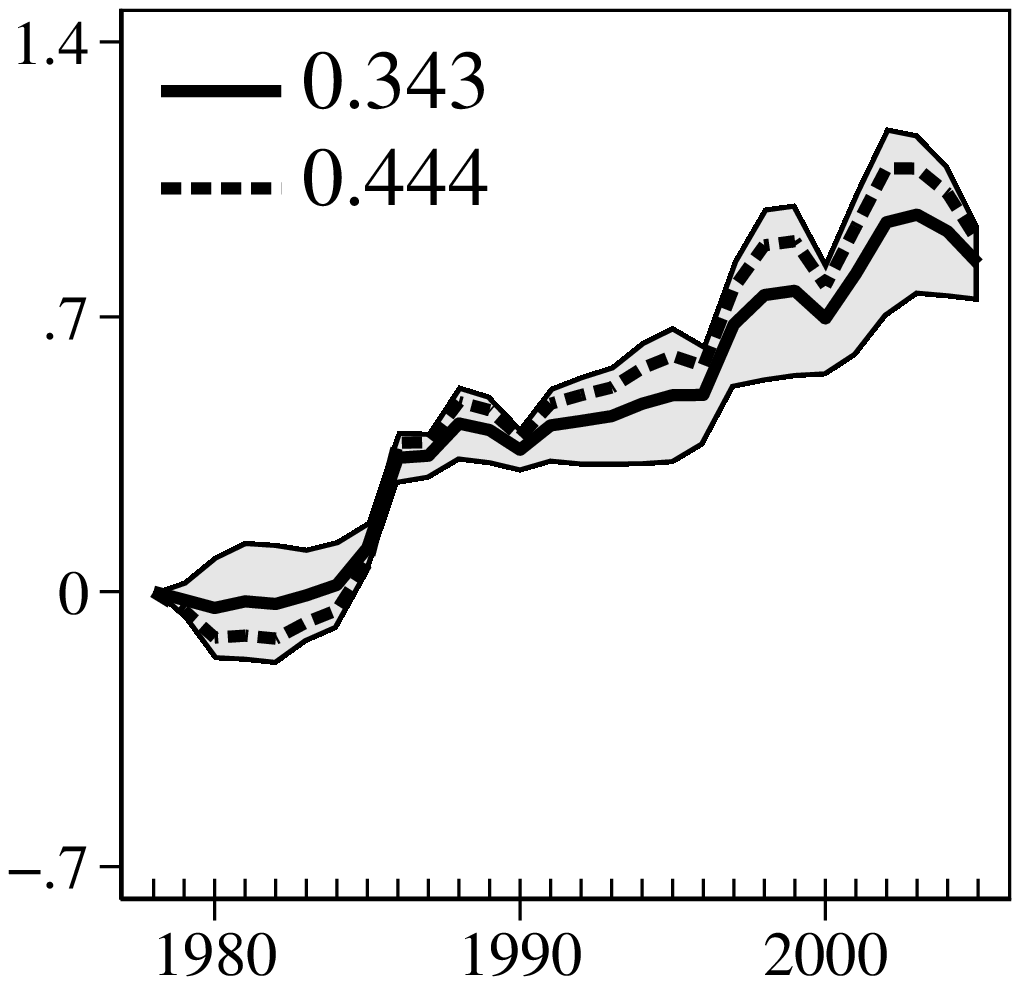}}
\par\end{centering}
\textit{\footnotesize{}Notes}{\footnotesize{}: Each line represents
energy-saving technology ($a_{e}$) when the elasticity of substitution
($\epsilon_{\sigma}$) is 0.343 or 0.444. The shaded area represents
the 90 percent confidence interval. All series are expressed as log
differences relative to the first year of observations.}{\footnotesize\par}
\end{figure}

\begin{figure}[H]
\caption{Energy-saving technological change in the service sector when the
elasticity of substitution varies across sectors\label{fig: Ae1_sigma'_service}}

\begin{centering}
\subfloat[Austria]{
\centering{}\includegraphics[scale=0.4]{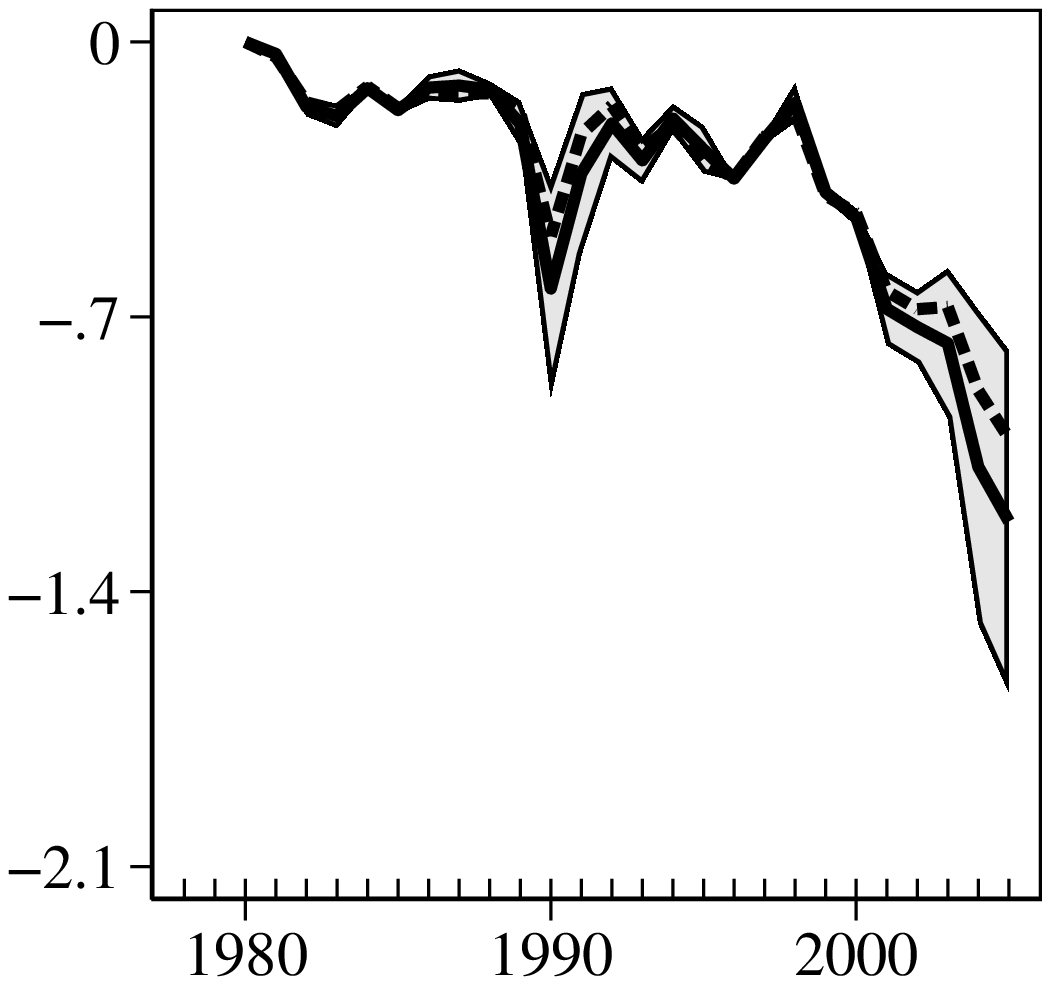}}\subfloat[Czech Republic]{
\centering{}\includegraphics[scale=0.4]{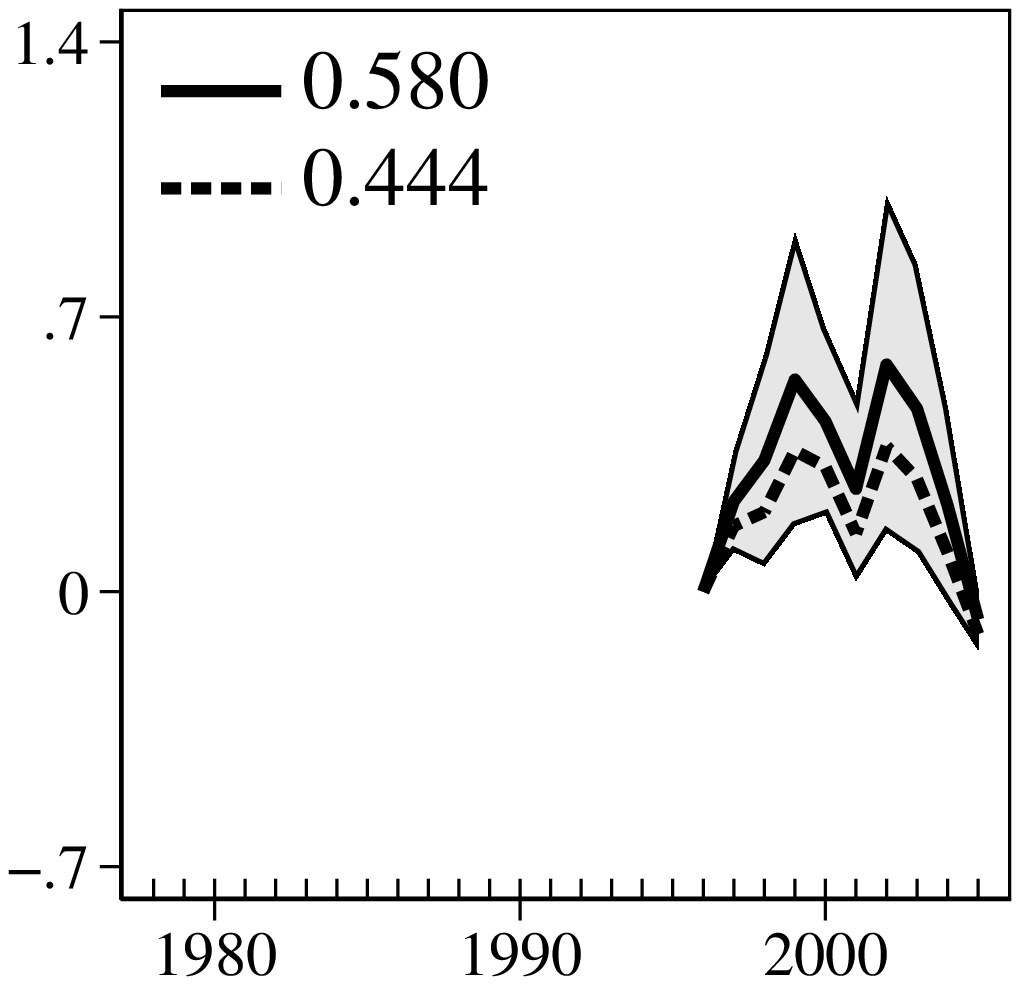}}\subfloat[Denmark]{
\centering{}\includegraphics[scale=0.4]{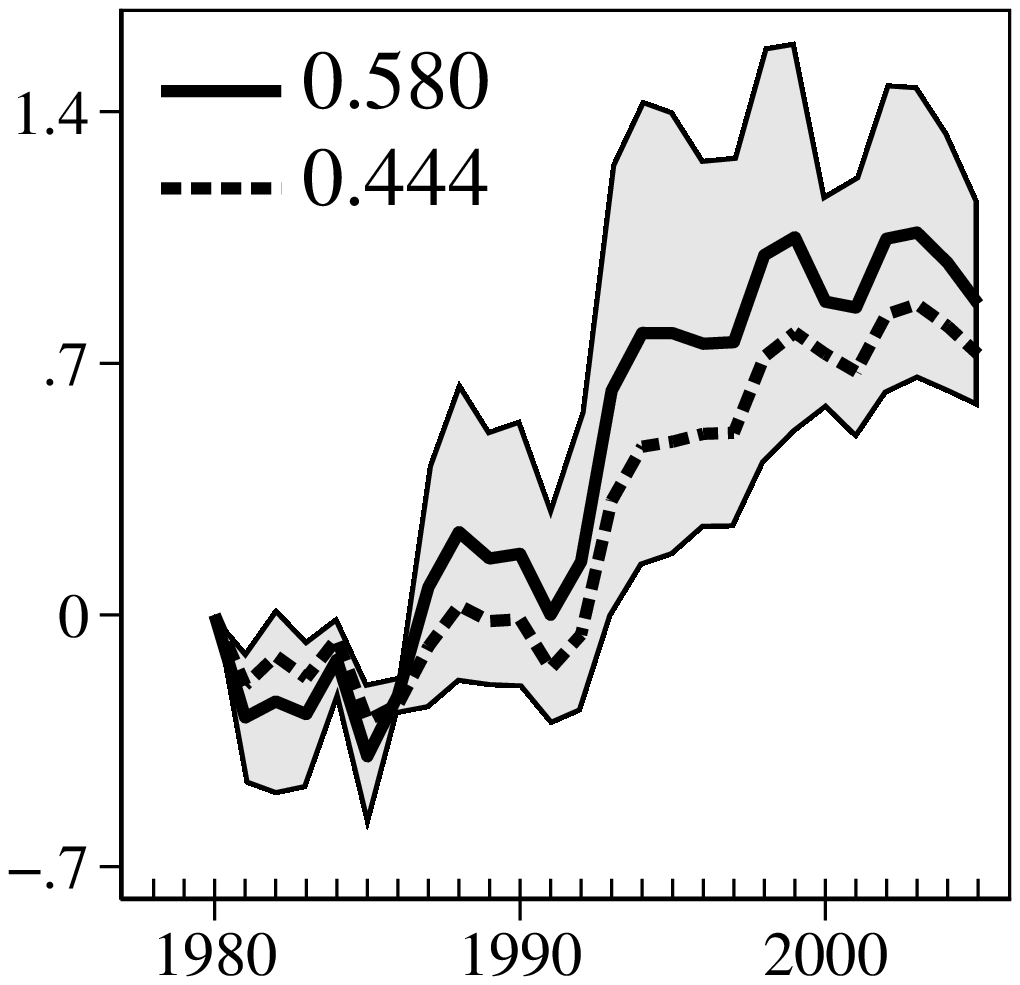}}\subfloat[Finland]{
\centering{}\includegraphics[scale=0.4]{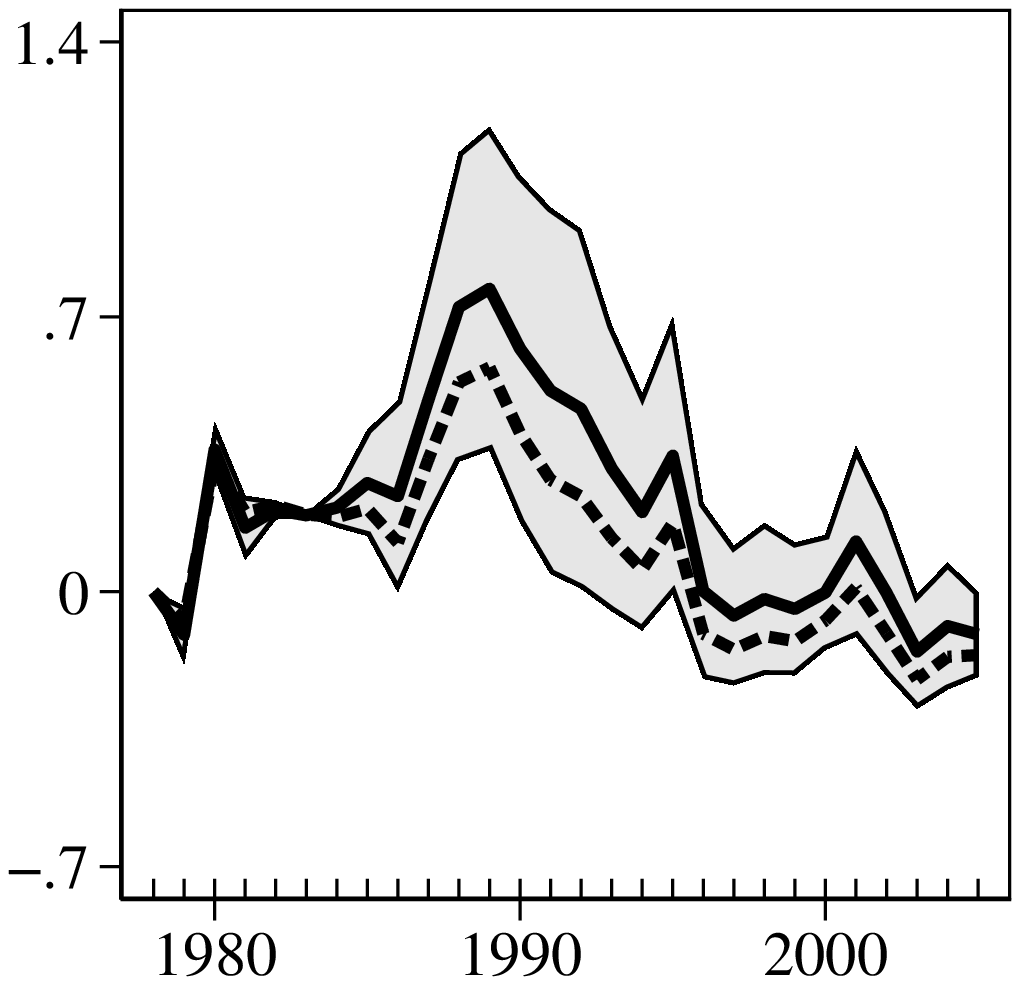}}
\par\end{centering}
\begin{centering}
\subfloat[Germany]{
\centering{}\includegraphics[scale=0.4]{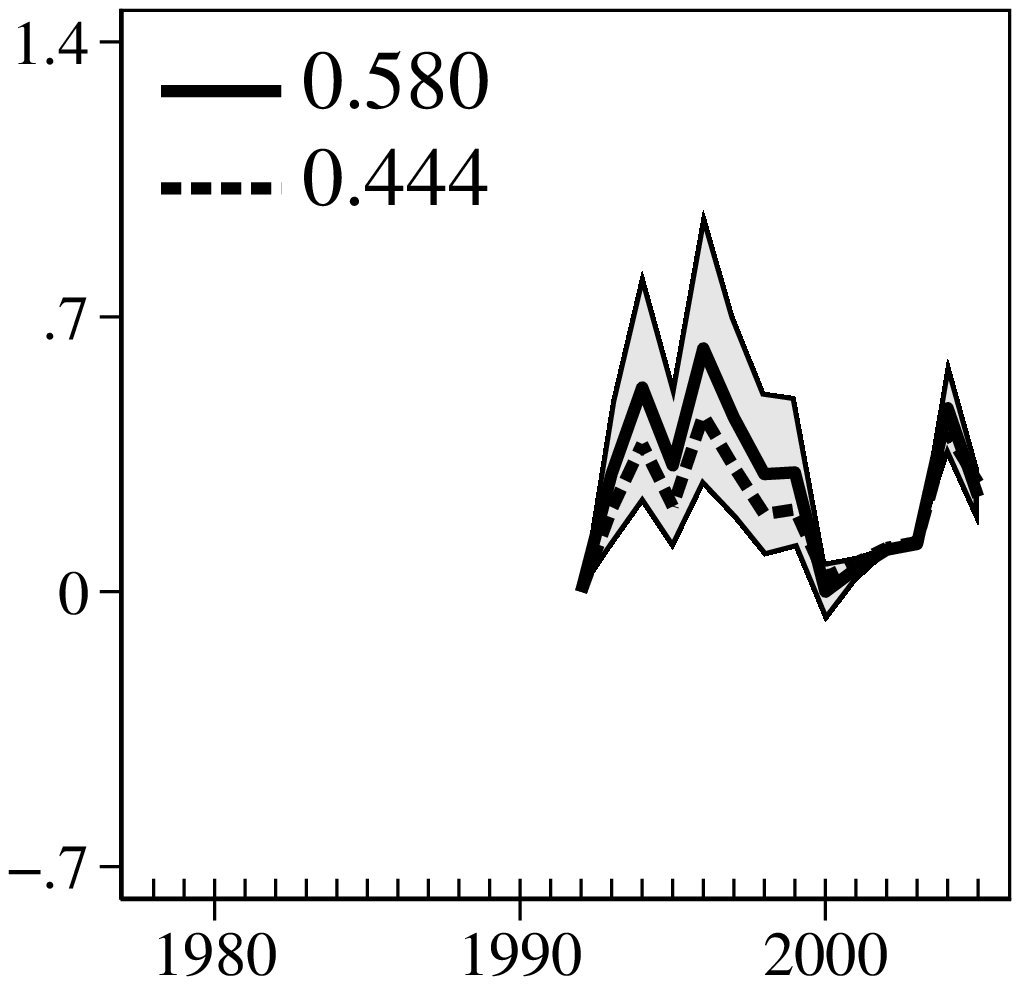}}\subfloat[Italy]{
\centering{}\includegraphics[scale=0.4]{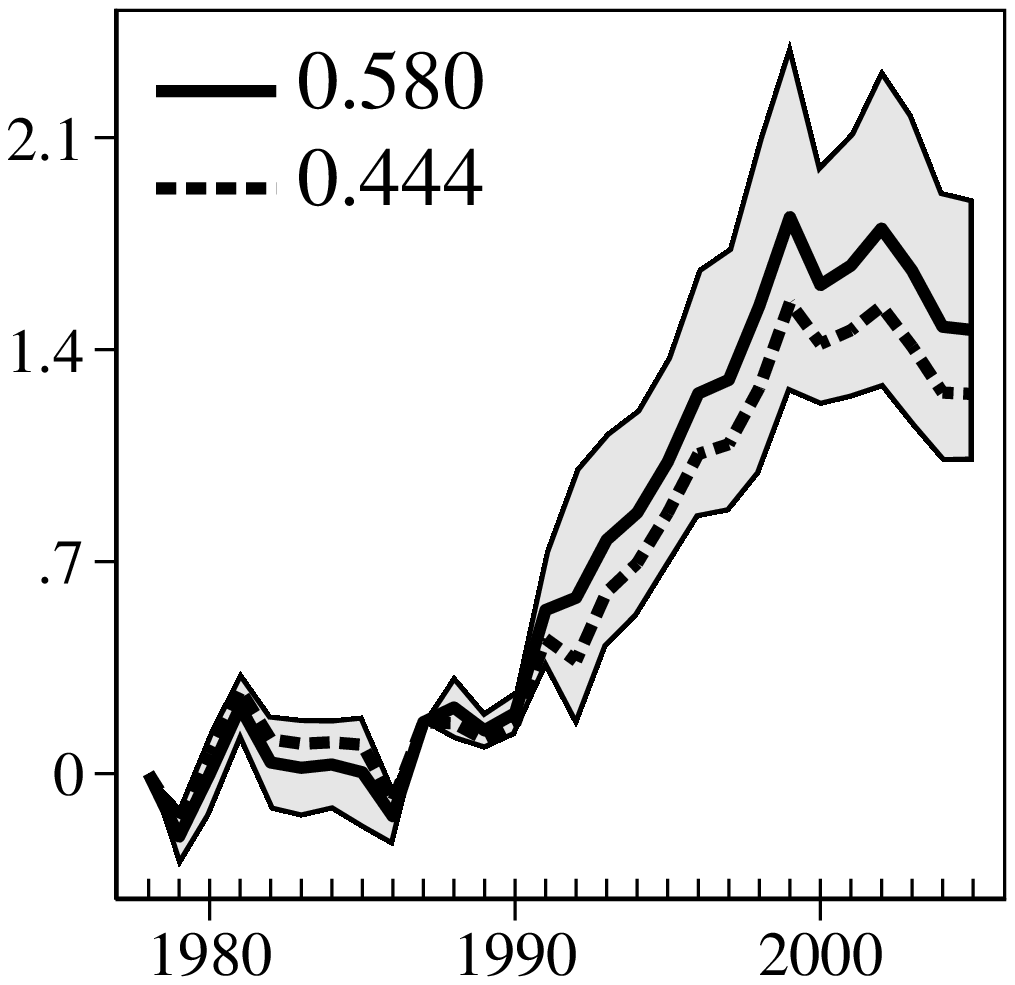}}\subfloat[Japan]{
\centering{}\includegraphics[scale=0.4]{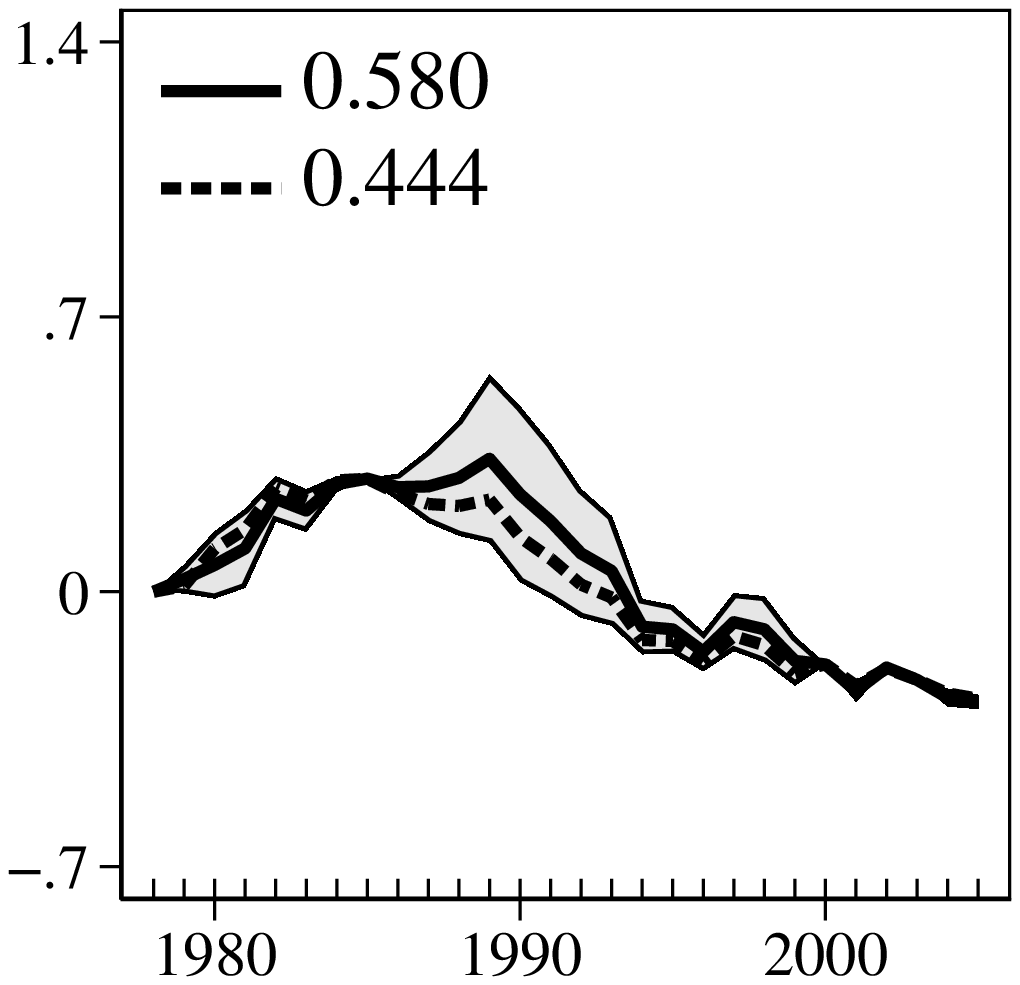}}\subfloat[Netherlands]{
\centering{}\includegraphics[scale=0.4]{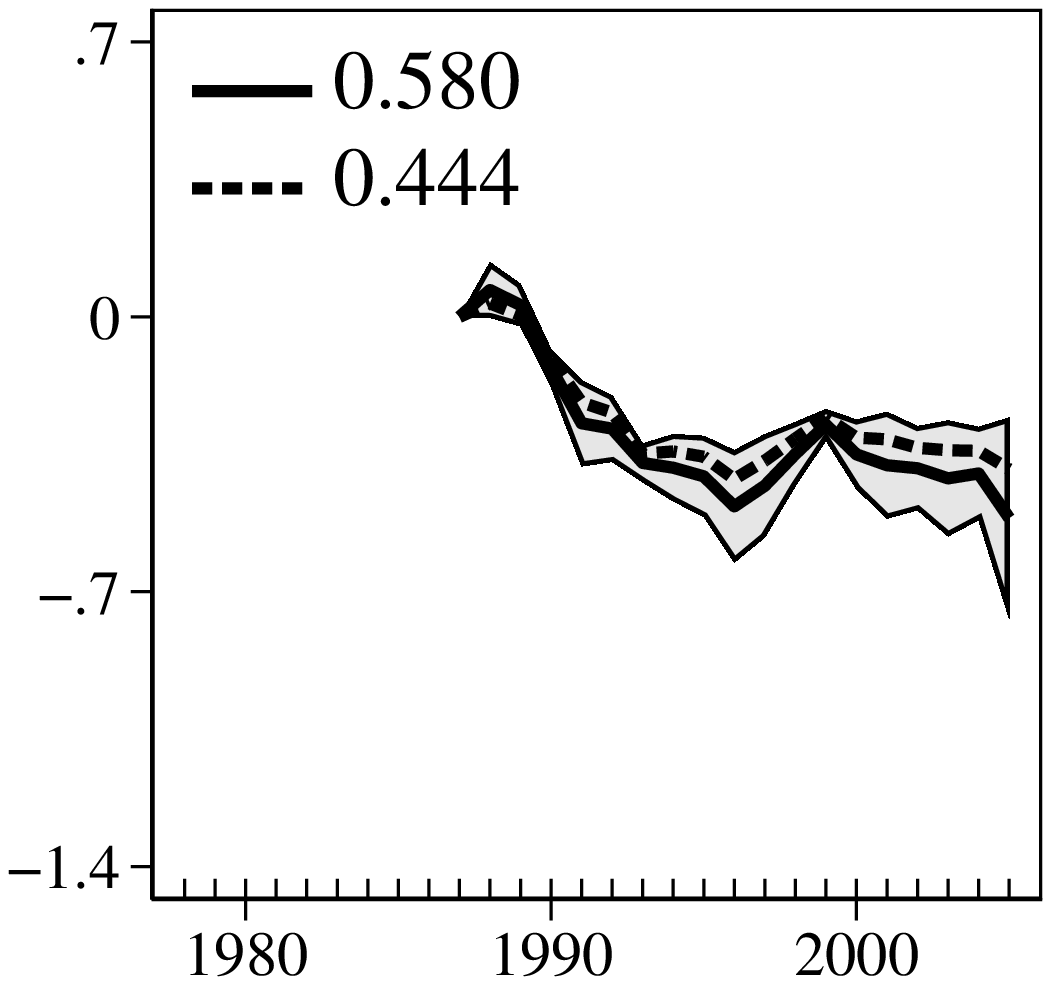}}
\par\end{centering}
\begin{centering}
\subfloat[Portugal]{
\centering{}\includegraphics[scale=0.4]{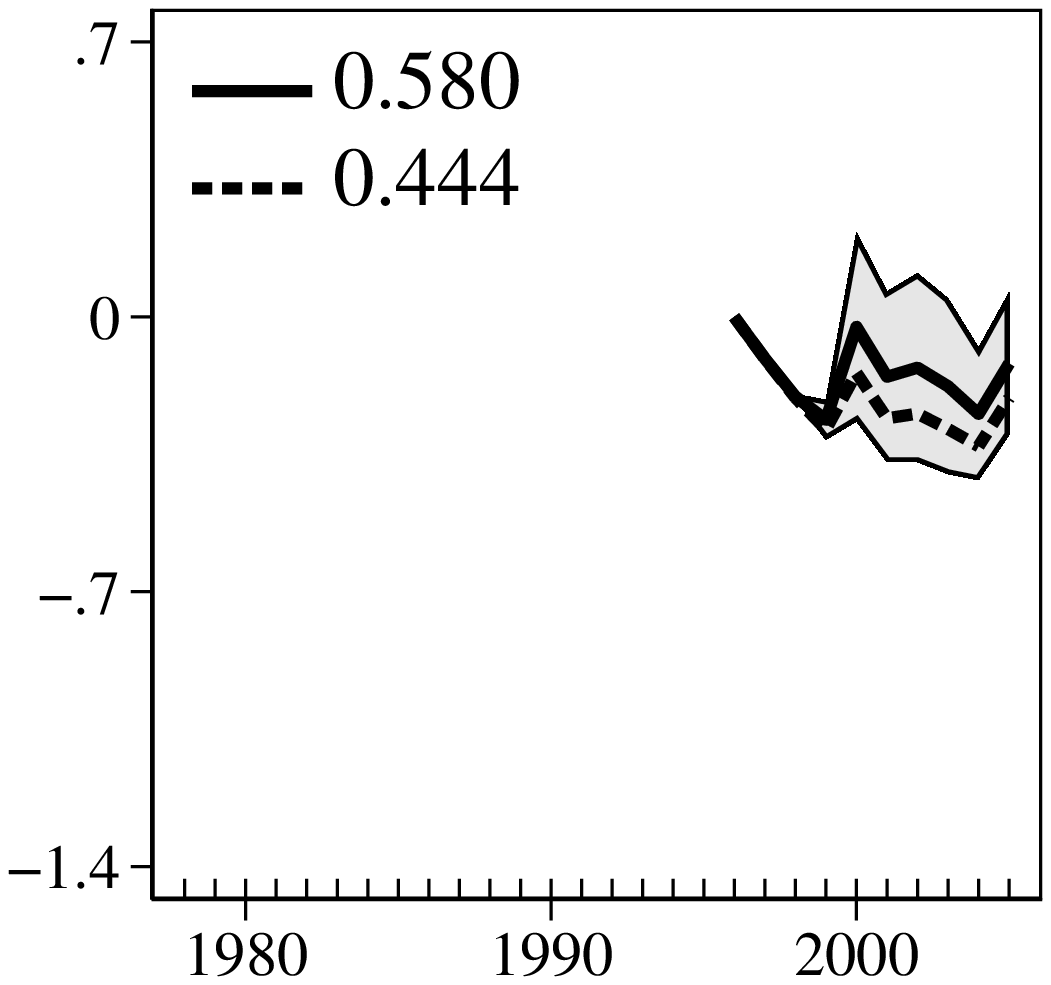}}\subfloat[Sweden]{
\centering{}\includegraphics[scale=0.4]{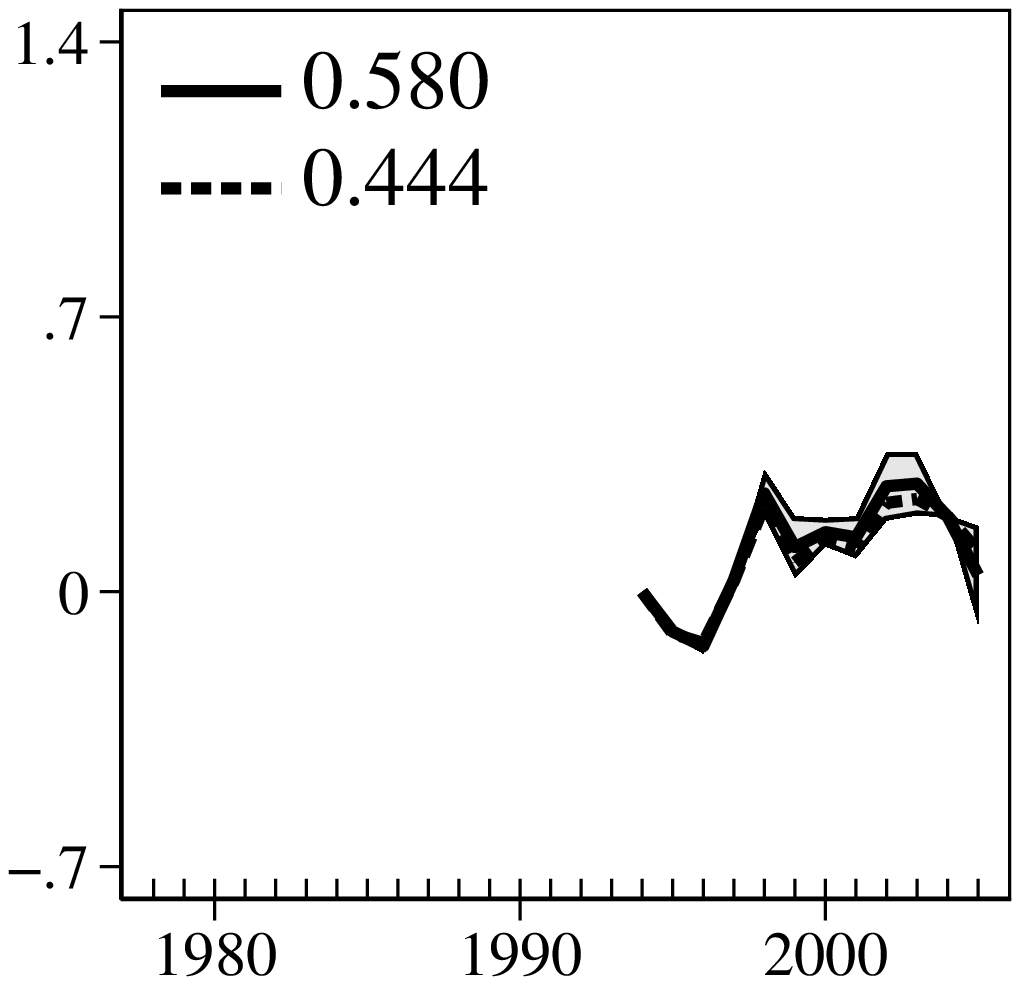}}\subfloat[United Kingdom]{
\centering{}\includegraphics[scale=0.4]{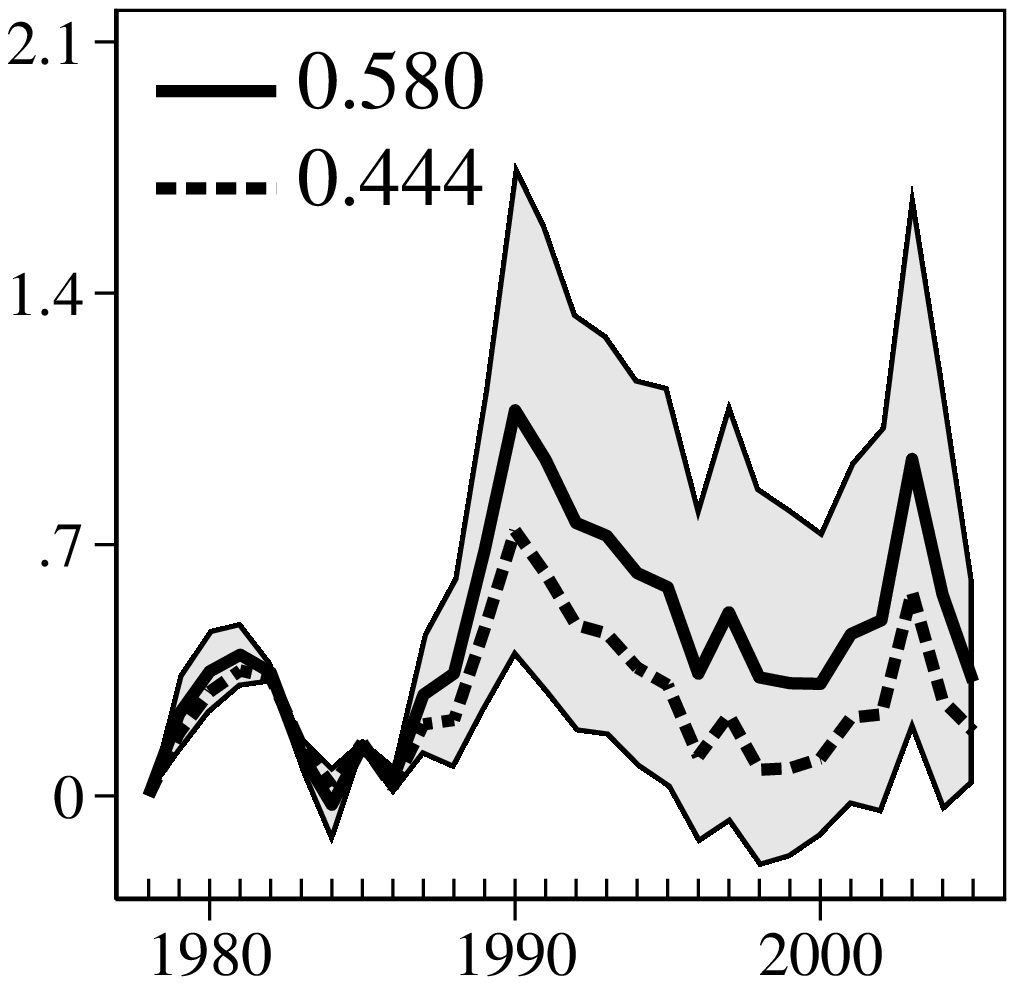}}\subfloat[United States]{
\centering{}\includegraphics[scale=0.4]{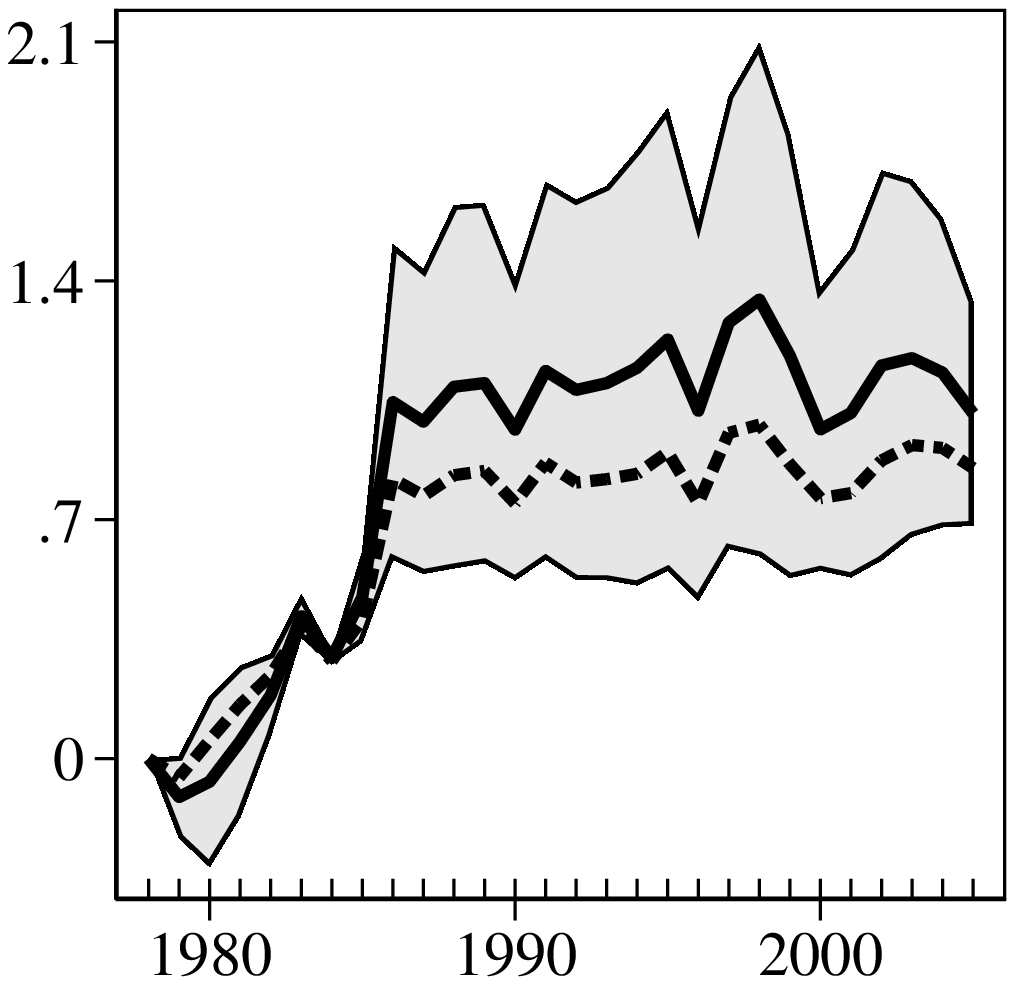}}
\par\end{centering}
\textit{\footnotesize{}Notes}{\footnotesize{}: Each line represents
energy-saving technology ($a_{e}$) when the elasticity of substitution
($\epsilon_{\sigma}$) is 0.580 or 0.444. The shaded area represents
the 90 percent confidence interval. All series are expressed as log
differences relative to the first year of observations.}{\footnotesize\par}
\end{figure}

We present four additional sets of results. First, we show the extent
to which energy-saving technological change can vary according to
the value of the elasticity of substitution. We consider six values
\{0.444, 0.02, 0.95, 1.25, 0.343, 0.580\}, the first of which is used
to measure energy-saving technological change in Figures \ref{fig: AkAlAe1_goods}
and \ref{fig: AkAlAe1_service}. The second and third values are close
to the limit cases when the CES production function converges to Leontief
and Cobb-Douglas, respectively. The second value is also \citeauthor{Hassler_Krusell_Olovsson_JPE21}'s
(2021) estimate of the elasticity of substitution between capital
and fossil energy. \citet{Hassler_Krusell_Olovsson_JPE21} obtain
the elasticity estimate using time-series data on the price and quantity
of fossil energy in the United States. The fourth value is \citeauthor{Karabarbounis_Neiman_QJE14}'s
(2014) estimate of the elasticity of substitution between capital
and labor. \citet{Karabarbounis_Neiman_QJE14} obtain the elasticity
estimate using long-term changes in the labor income of value added
over 15 to 37 years in approximately 50 OECD and non-OECD countries.
The second and fourth values are the lower and upper bounds of the
existing estimates, respectively. The fifth and sixth values are the
estimates obtained separately for the goods and service sectors using
the specification in the second column of Table \ref{tab: elasticity2}.
Figures \ref{fig: Ae1_sigma_goods} and \ref{fig: Ae1_sigma_service}
show how energy-saving technological change varies according to the
first four values in the goods and service sectors, respectively.
Note that the scales of the right and left vertical axes are different.
Energy-saving technological change grows tenfold when the elasticity
of substitution increases from 0.444 to 0.95. Not only the magnitude
but also the direction of energy-saving technological change can vary
when the elasticity of substitution changes from 0.444 to 0.02 or
1.25. Figures \ref{fig: Ae1_sigma'_goods} and \ref{fig: Ae1_sigma'_service}
show that the results in Figures \ref{fig: AkAlAe1_goods} and \ref{fig: AkAlAe1_service}
remain almost unchanged regardless of whether we allow the elasticity
of substitution to vary across sectors.

\begin{figure}[H]
\caption{Energy-saving technological change for different degrees of returns
to scale in the goods sector\label{fig: Ae1_mu_goods}}

\begin{centering}
\subfloat[Austria]{
\centering{}\includegraphics[scale=0.4]{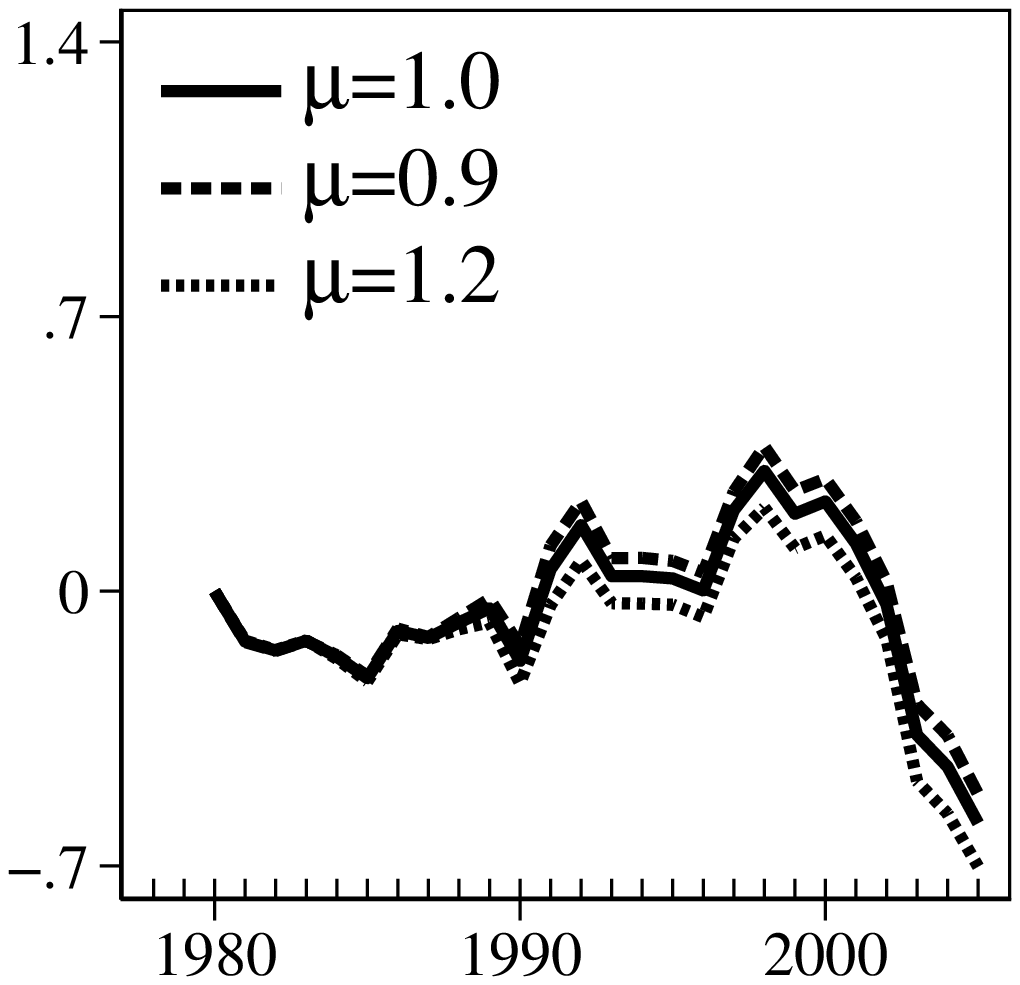}}\subfloat[Czech Republic]{
\centering{}\includegraphics[scale=0.4]{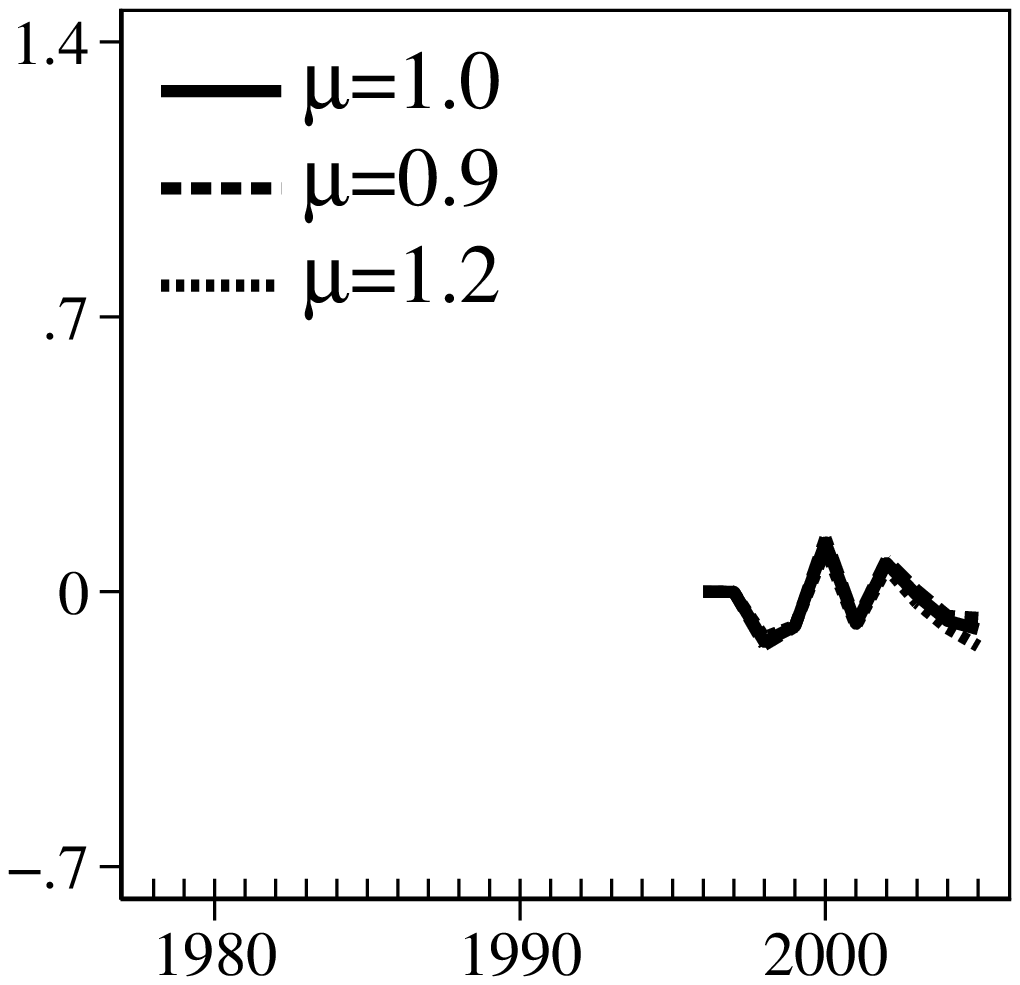}}\subfloat[Denmark]{
\centering{}\includegraphics[scale=0.4]{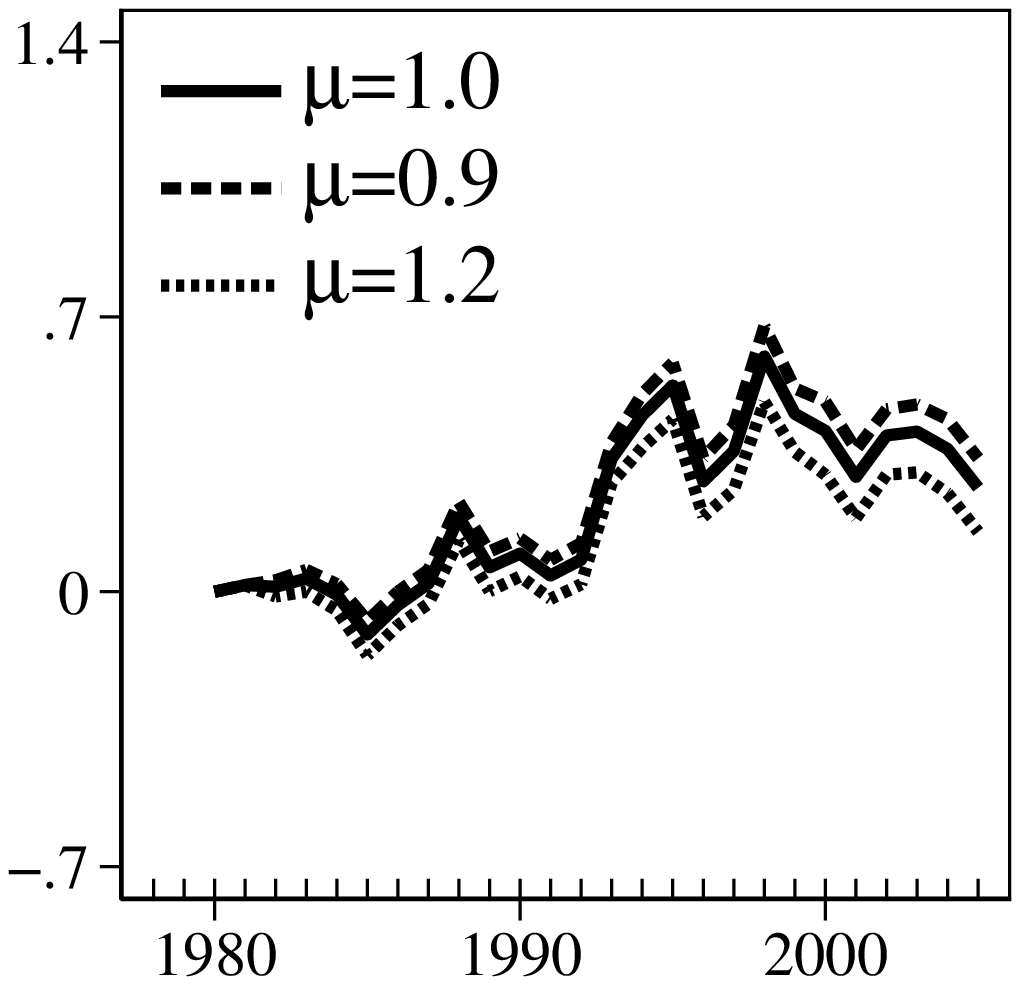}}\subfloat[Finland]{
\centering{}\includegraphics[scale=0.4]{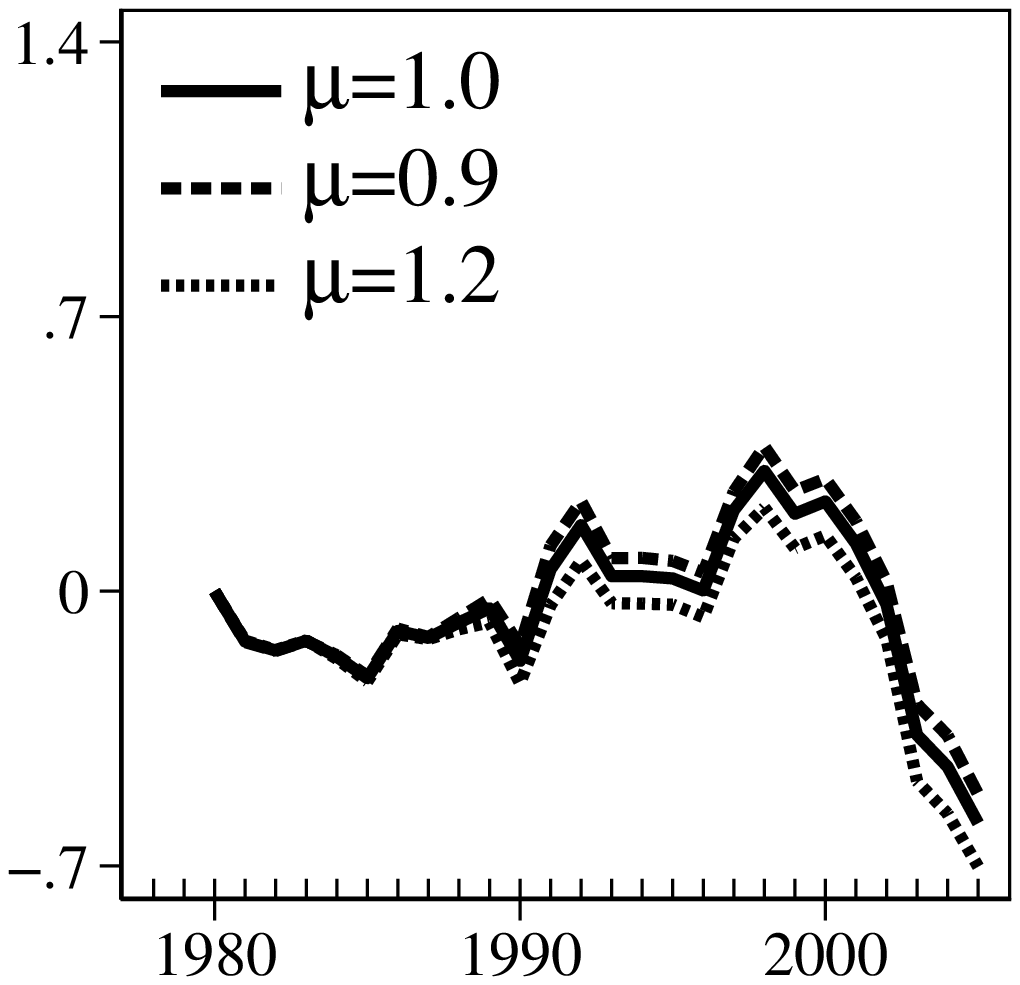}}
\par\end{centering}
\begin{centering}
\subfloat[Germany]{
\centering{}\includegraphics[scale=0.4]{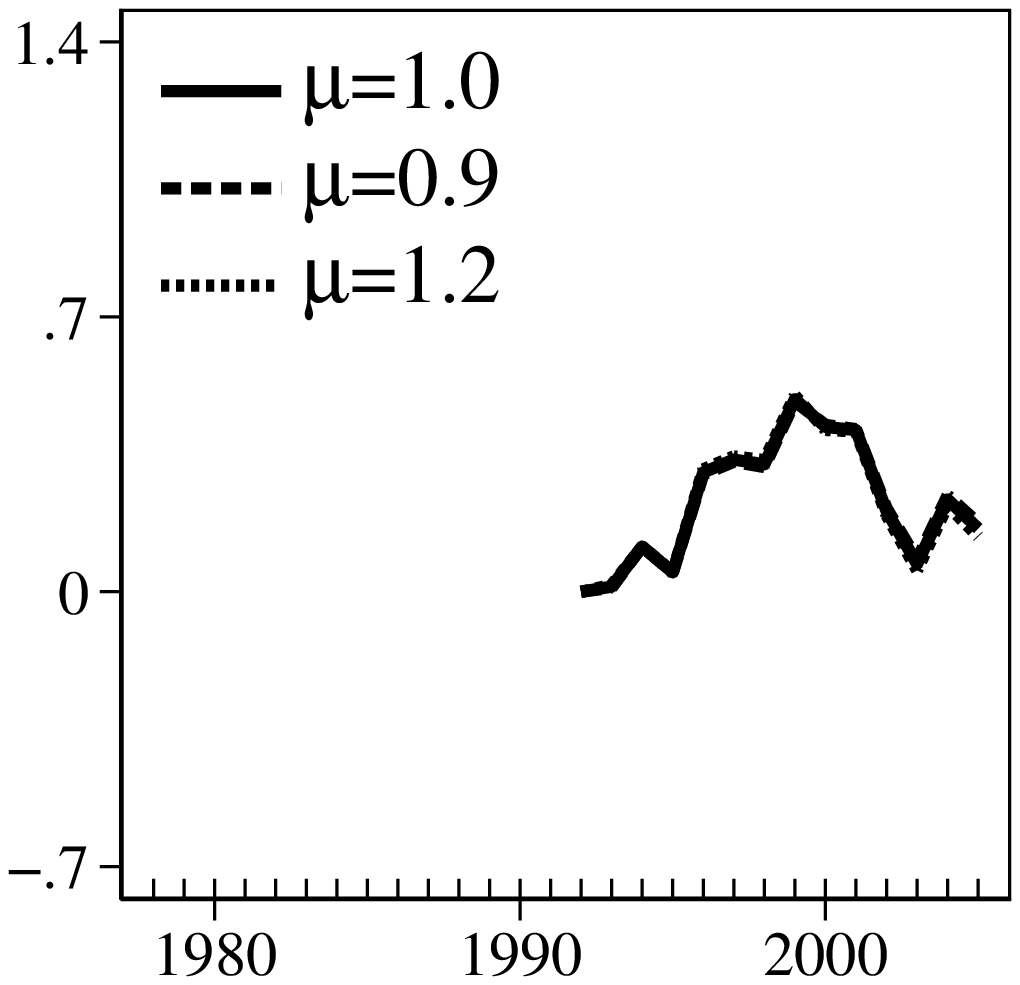}}\subfloat[Italy]{
\centering{}\includegraphics[scale=0.4]{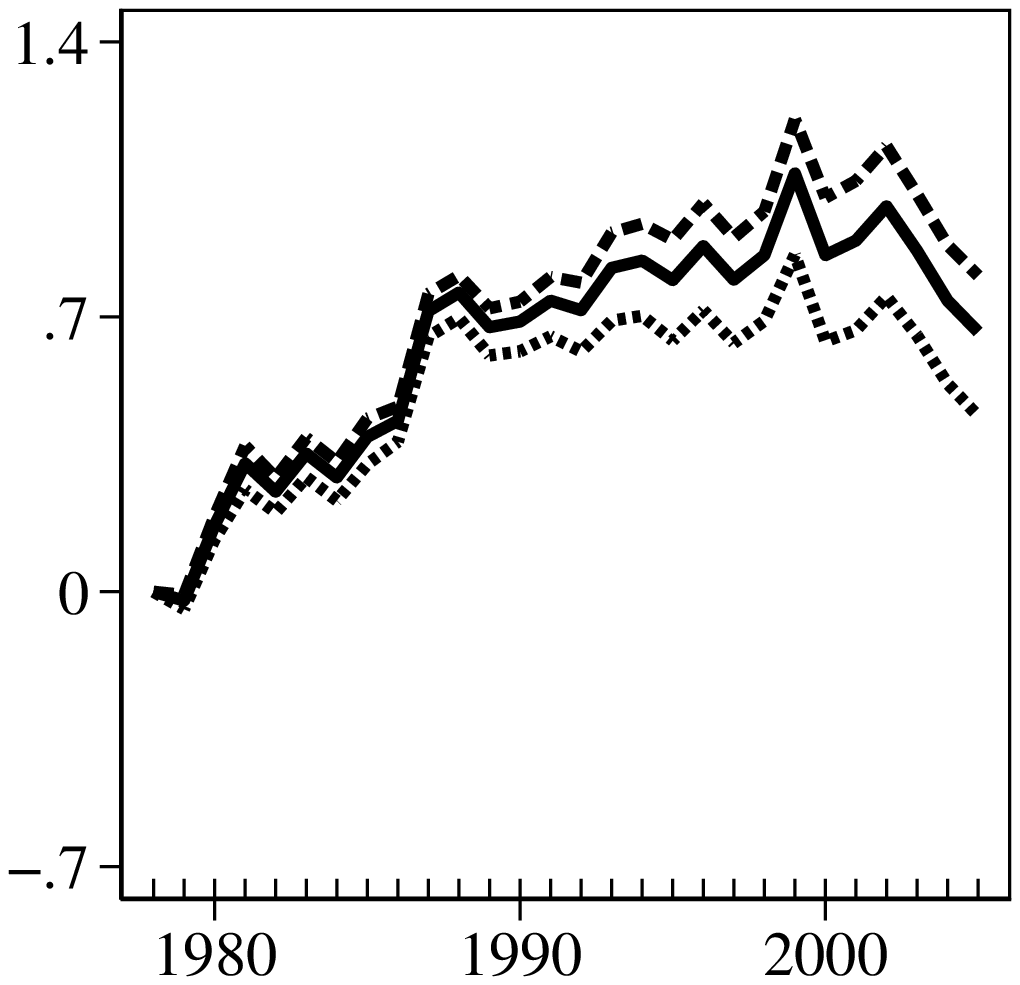}}\subfloat[Japan]{
\centering{}\includegraphics[scale=0.4]{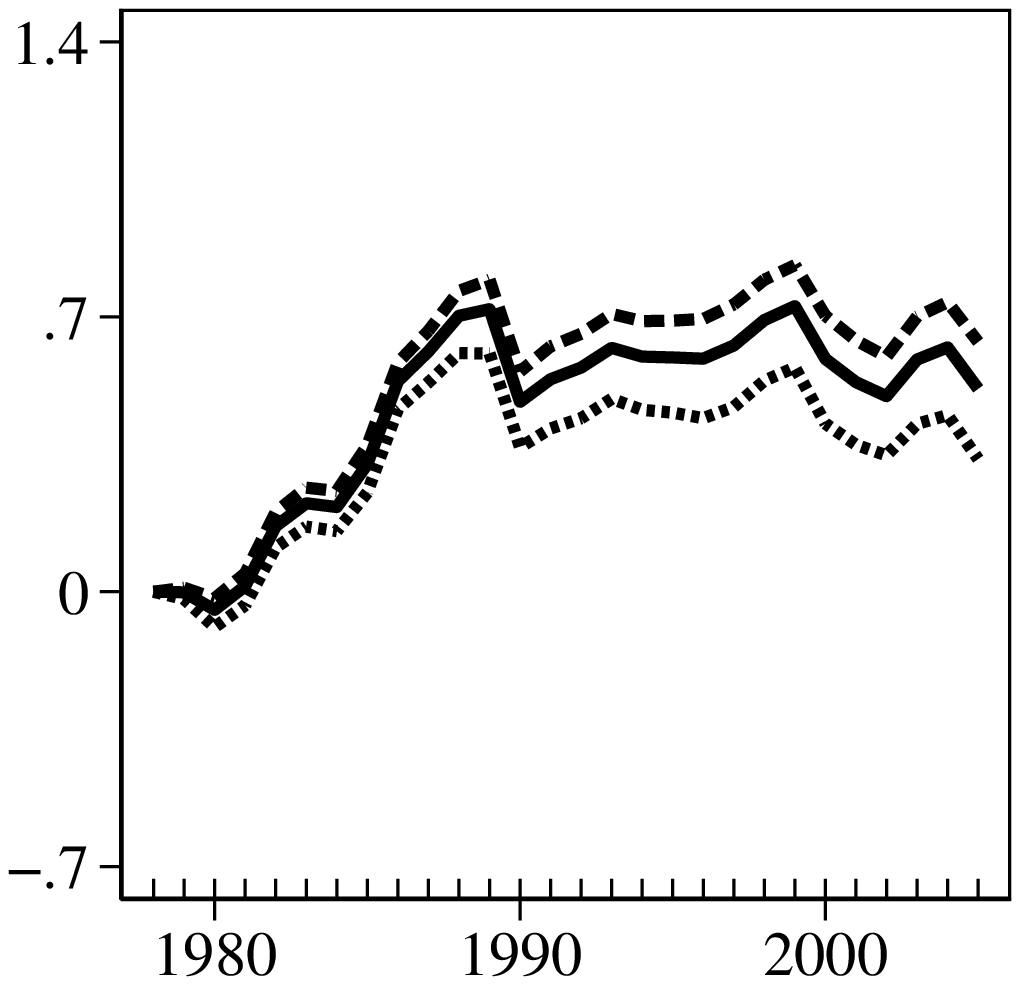}}\subfloat[Netherlands]{
\centering{}\includegraphics[scale=0.4]{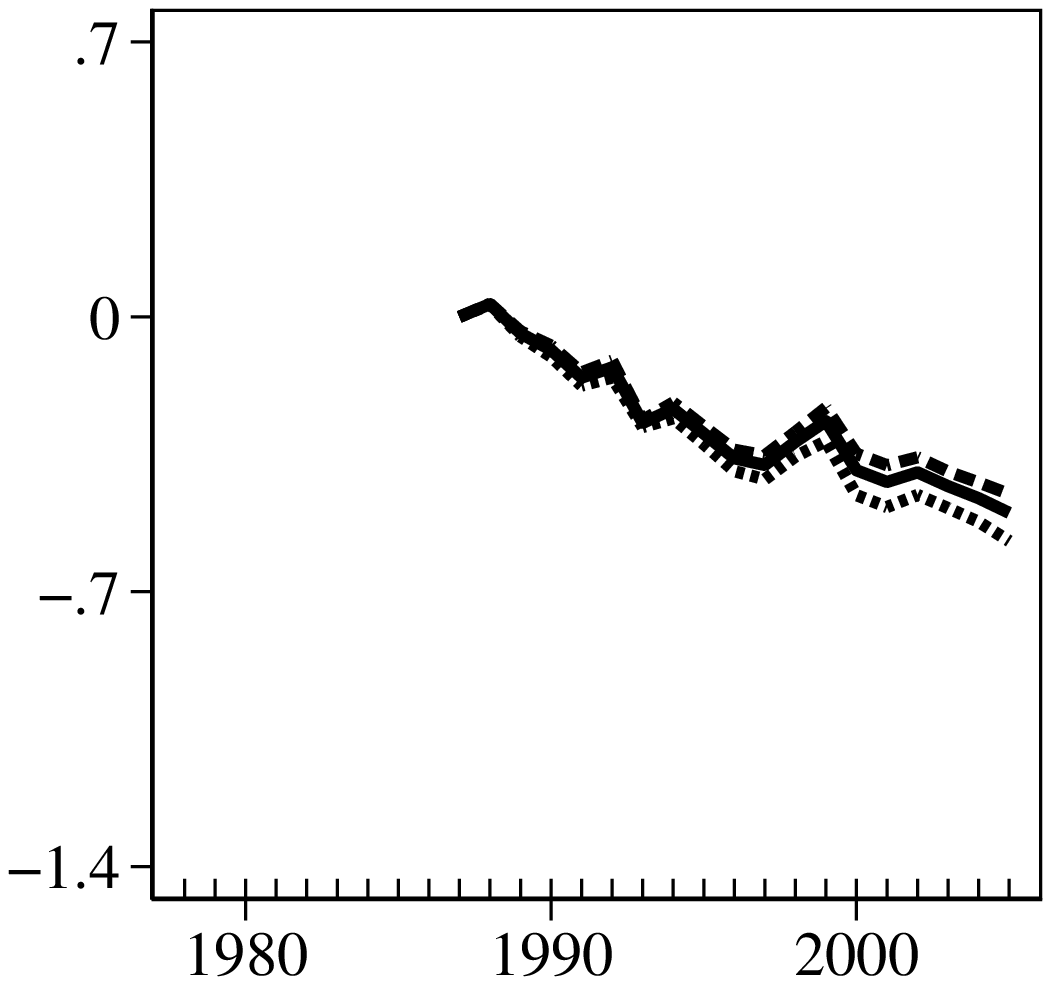}}
\par\end{centering}
\begin{centering}
\subfloat[Portugal]{
\centering{}\includegraphics[scale=0.4]{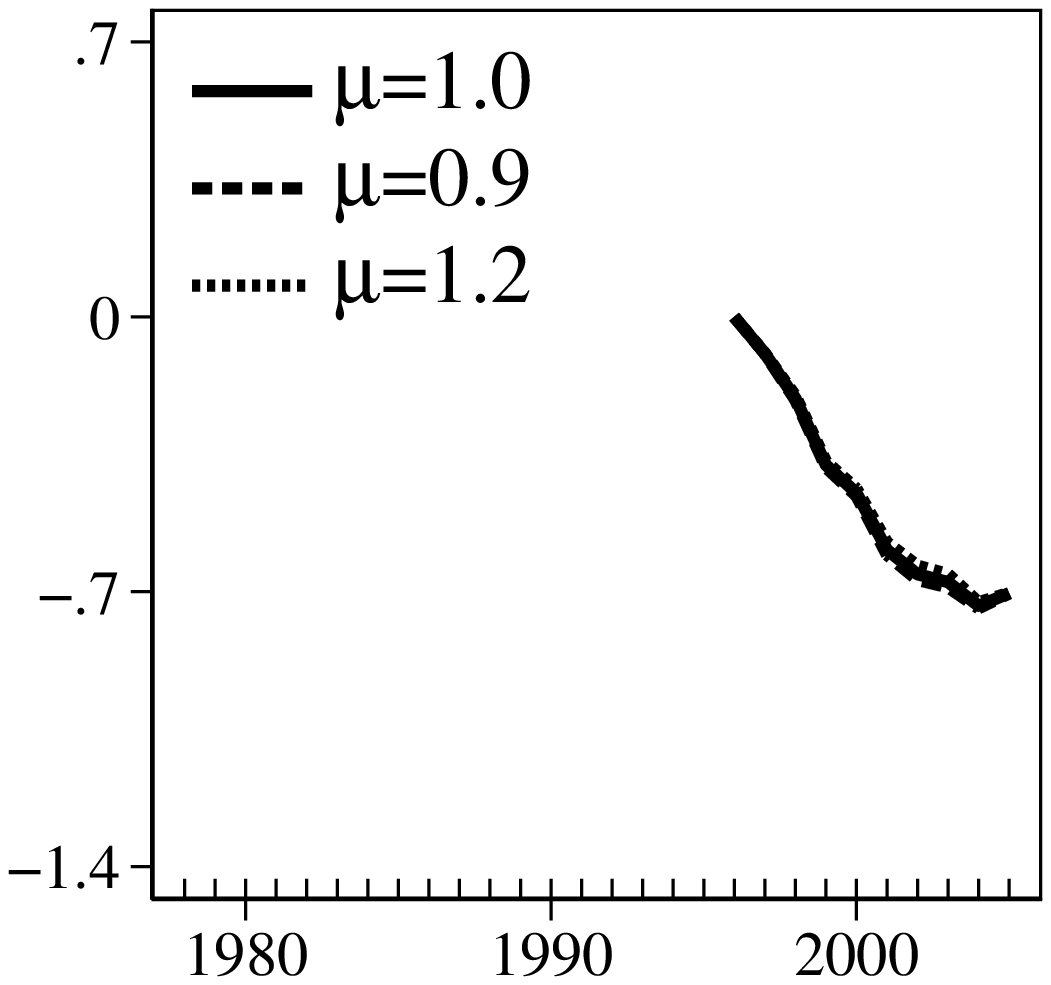}}\subfloat[Sweden]{
\centering{}\includegraphics[scale=0.4]{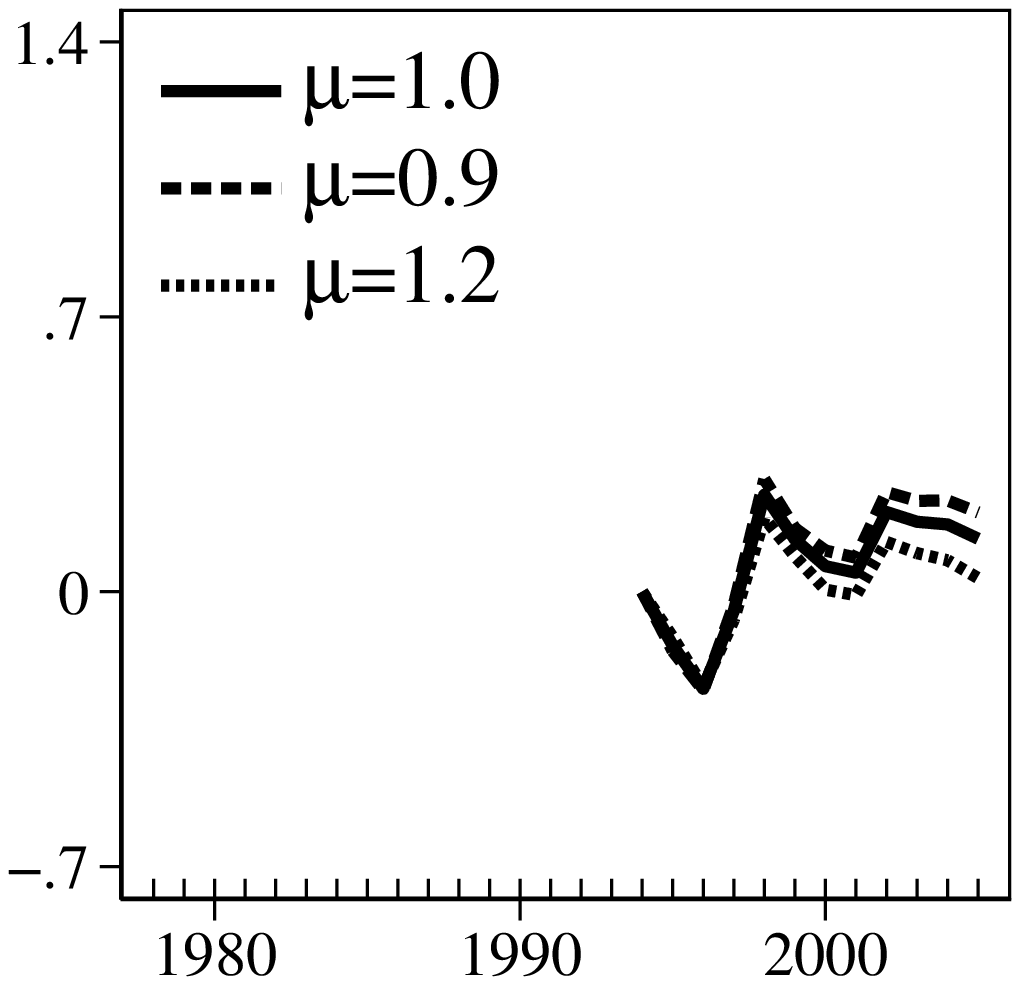}}\subfloat[United Kingdom]{
\centering{}\includegraphics[scale=0.4]{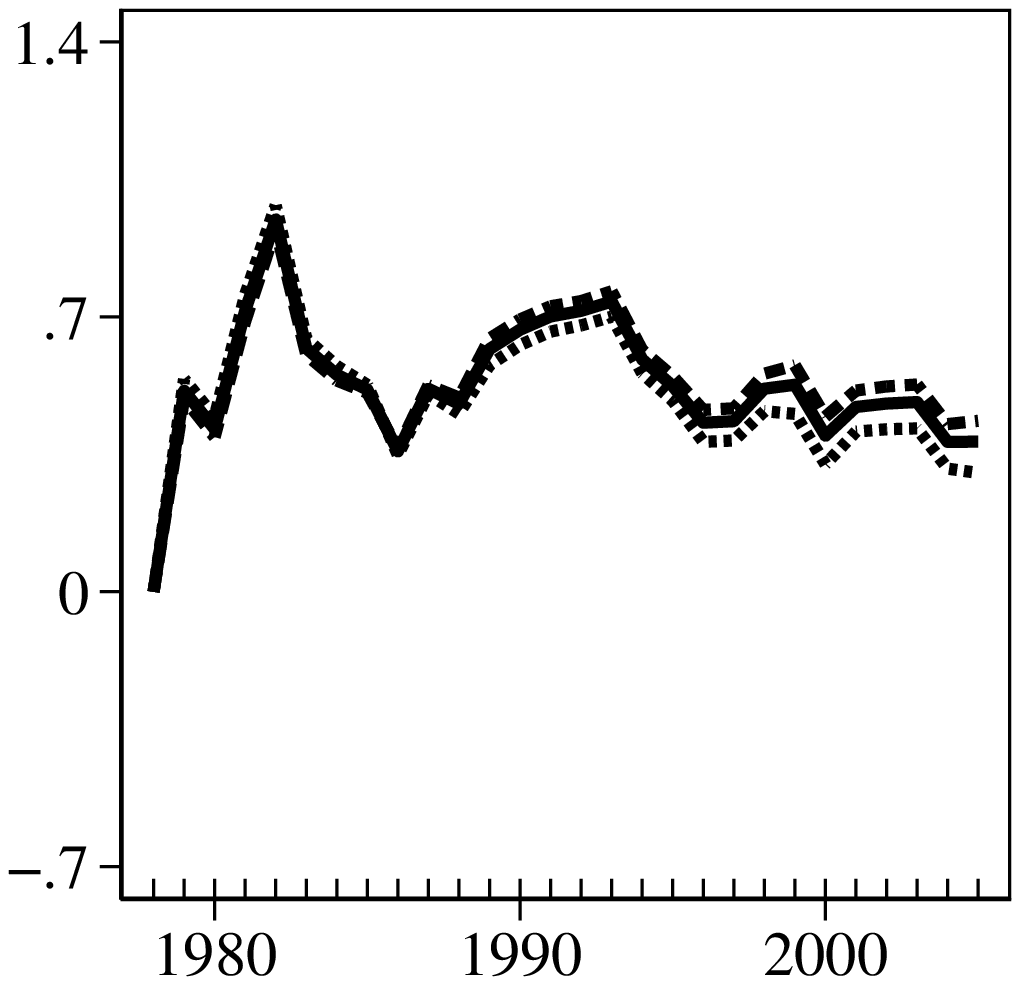}}\subfloat[United States]{
\centering{}\includegraphics[scale=0.4]{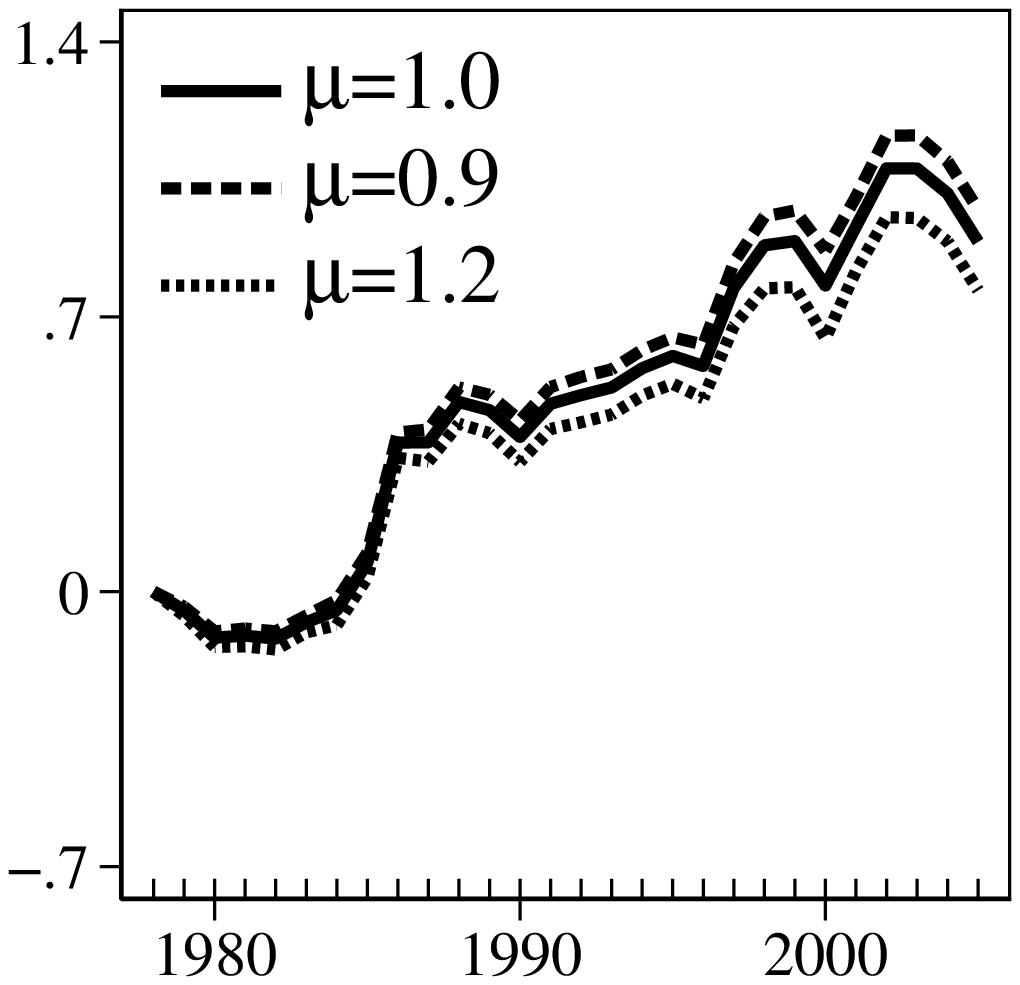}}
\par\end{centering}
\textit{\footnotesize{}Notes}{\footnotesize{}: Each line represents
energy-saving technology ($a_{e}$) when the scale parameter ($\mu$)
is 1, 0.9, or 1.2. All series are expressed as log differences relative
to the first year of observations.}{\footnotesize\par}
\end{figure}

\begin{figure}[H]
\caption{Energy-saving technological change for different degrees of returns
to scale in the service sector\label{fig: Ae1_mu_service}}

\begin{centering}
\subfloat[Austria]{
\centering{}\includegraphics[scale=0.4]{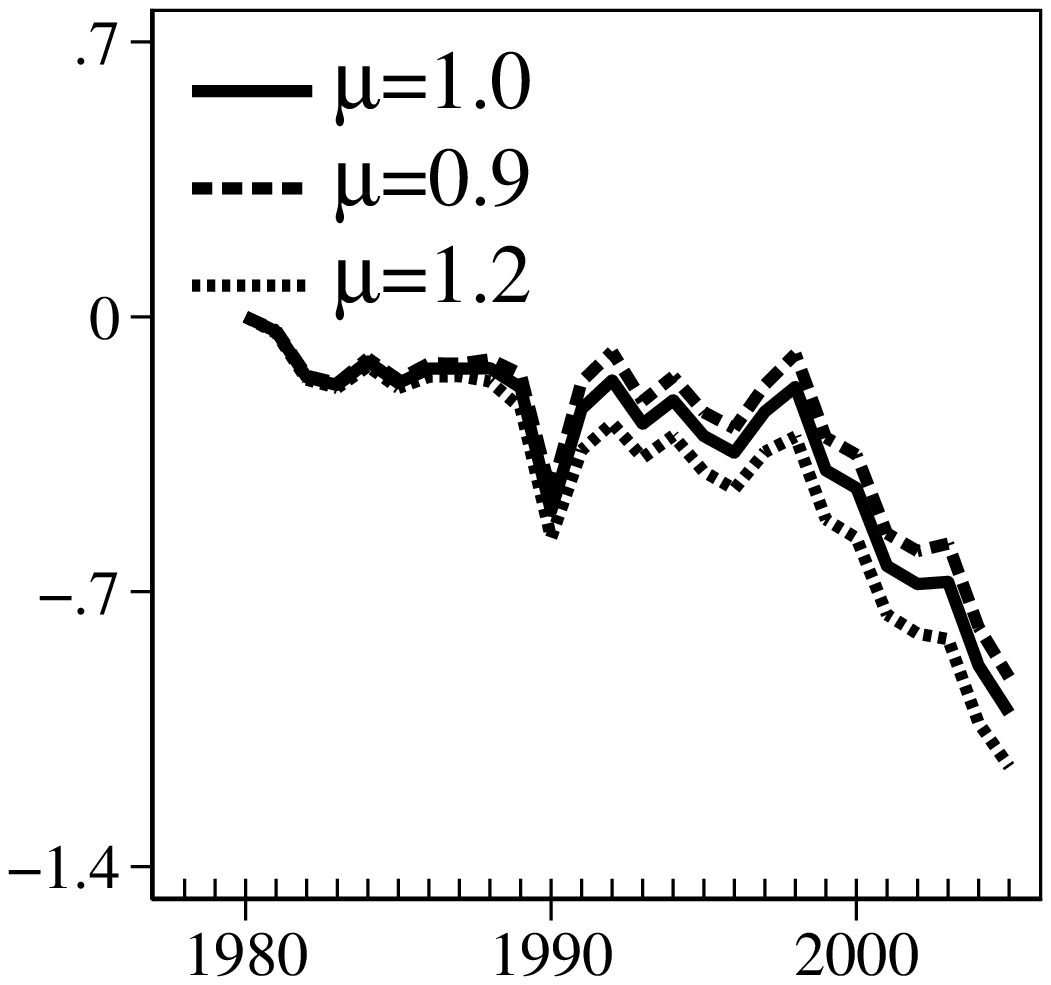}}\subfloat[Czech Republic]{
\centering{}\includegraphics[scale=0.4]{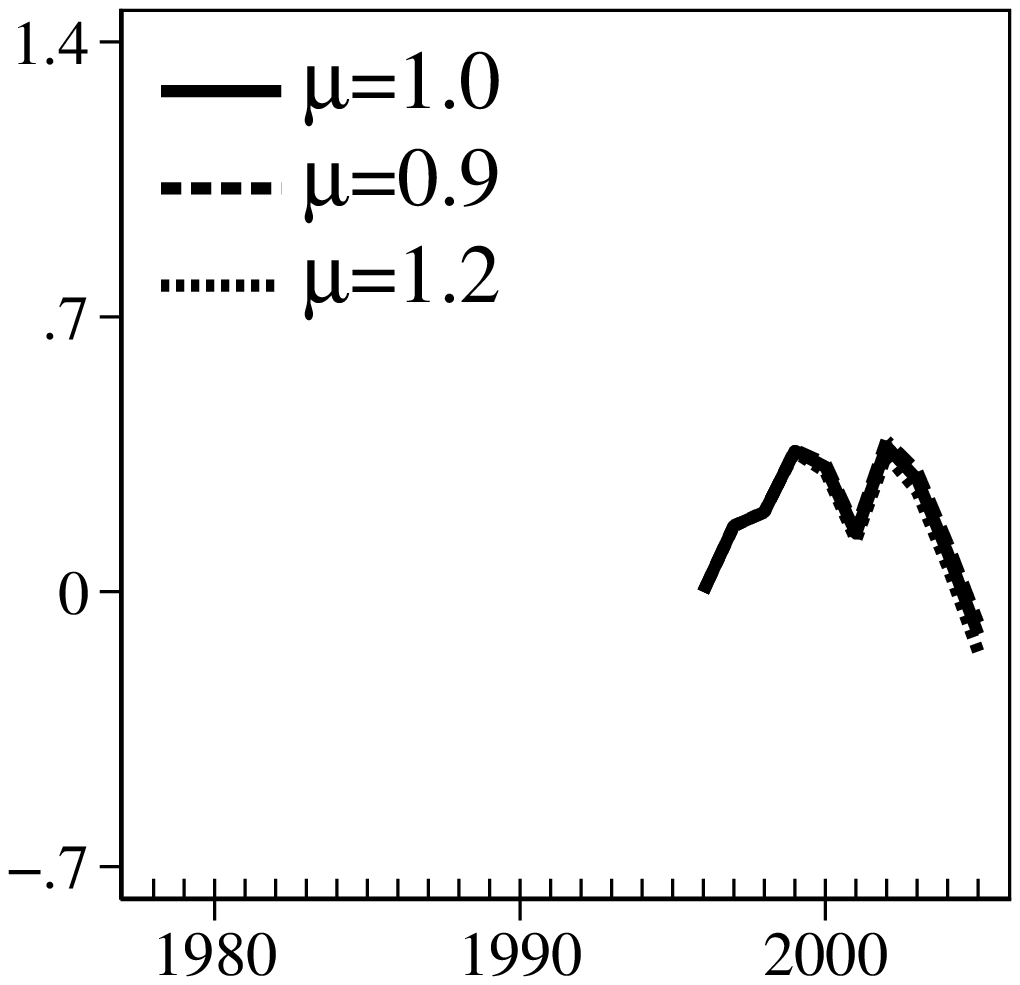}}\subfloat[Denmark]{
\centering{}\includegraphics[scale=0.4]{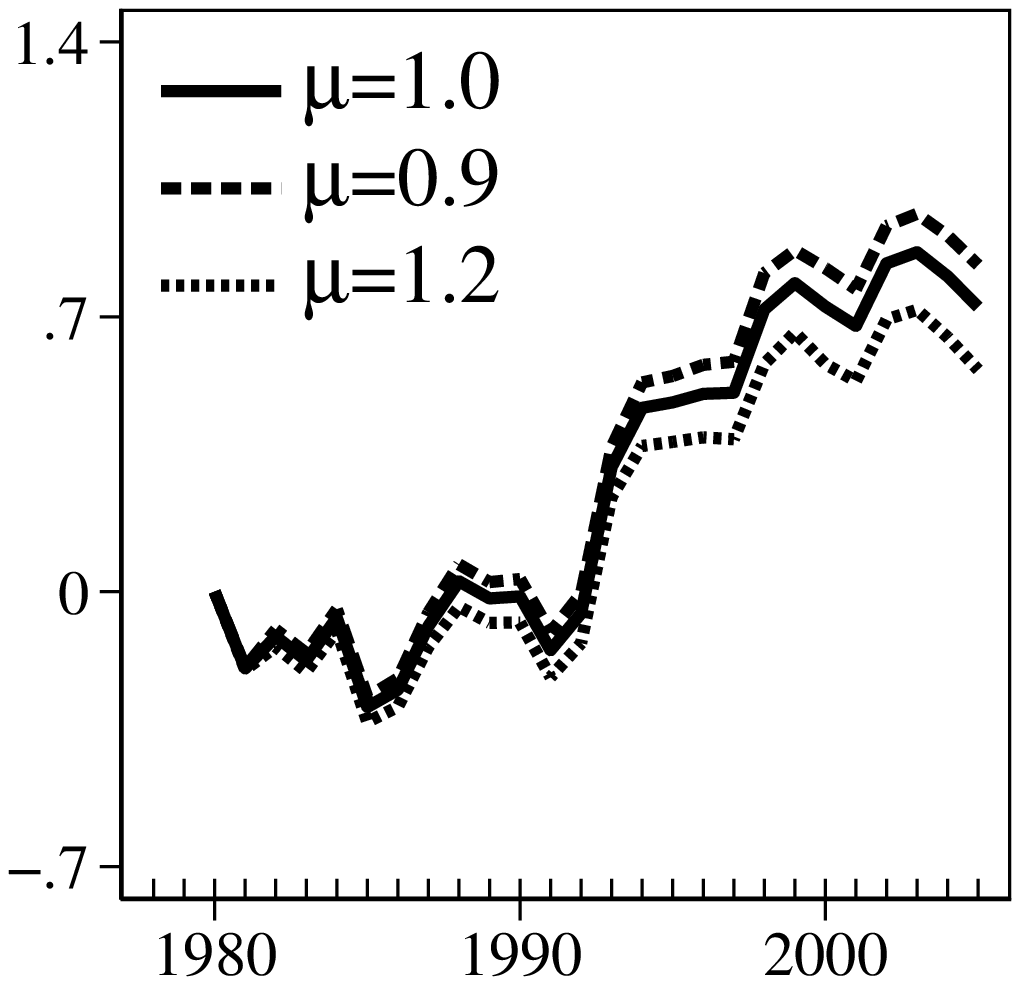}}\subfloat[Finland]{
\centering{}\includegraphics[scale=0.4]{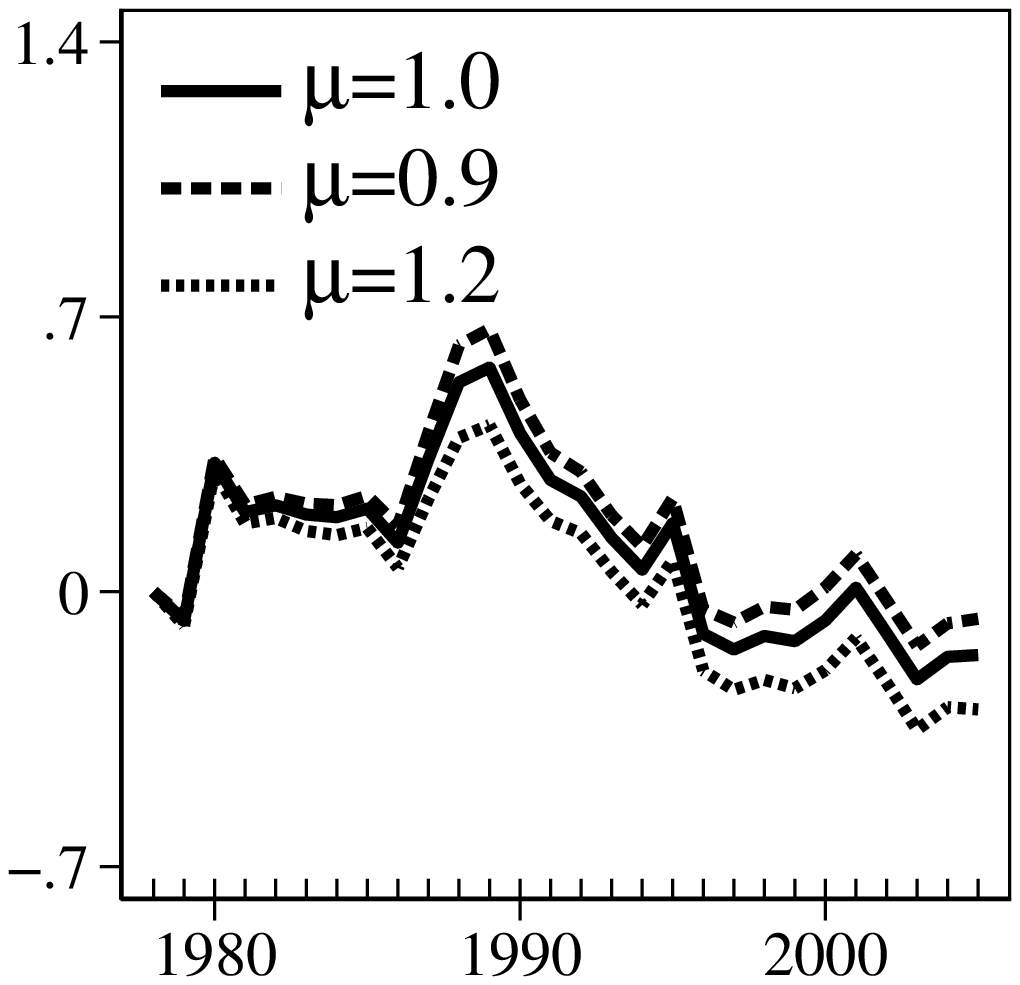}}
\par\end{centering}
\begin{centering}
\subfloat[Germany]{
\centering{}\includegraphics[scale=0.4]{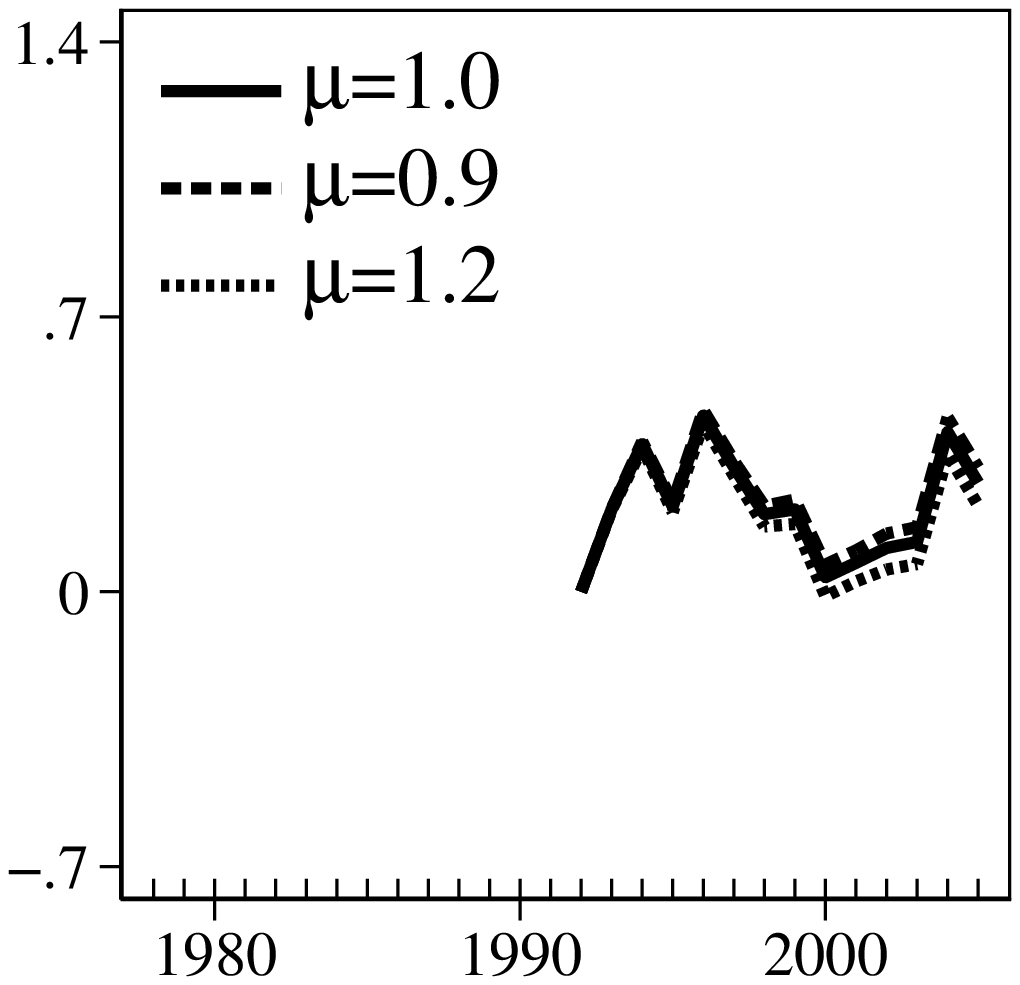}}\subfloat[Italy]{
\centering{}\includegraphics[scale=0.4]{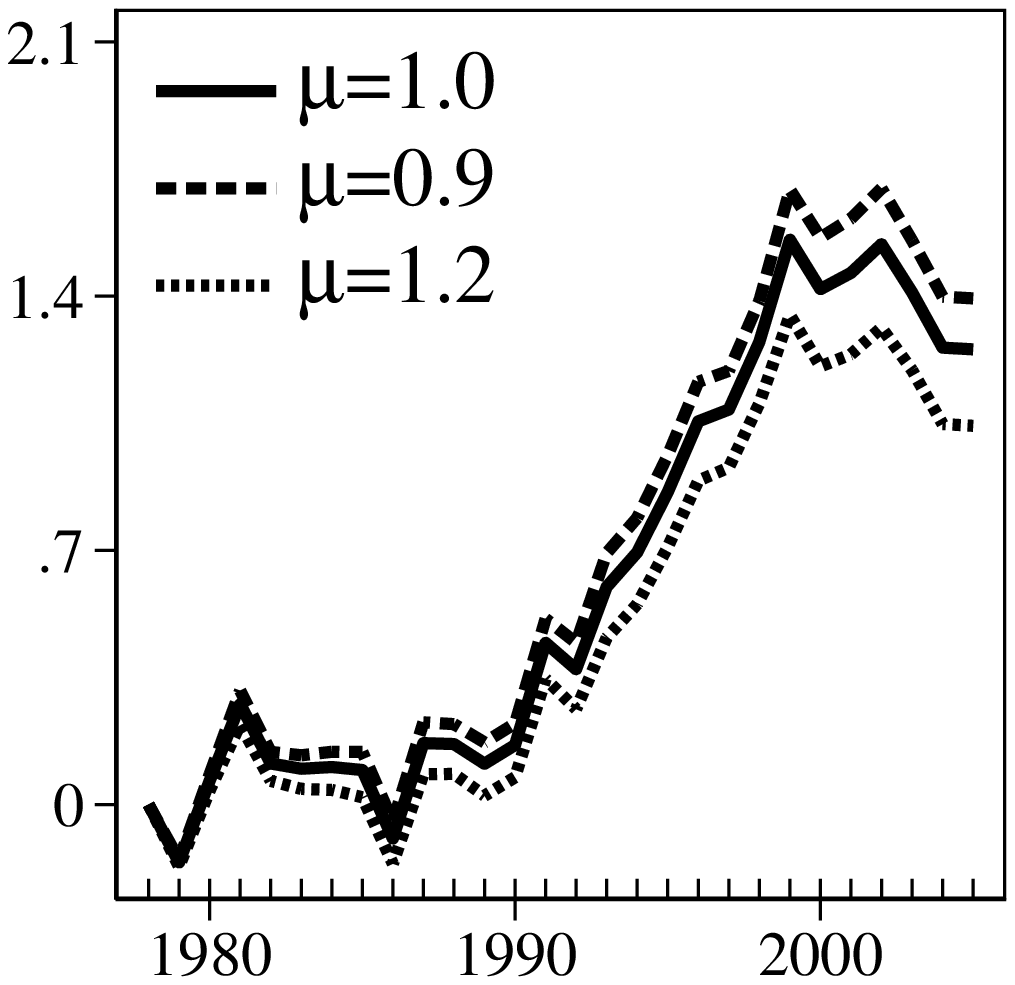}}\subfloat[Japan]{
\centering{}\includegraphics[scale=0.4]{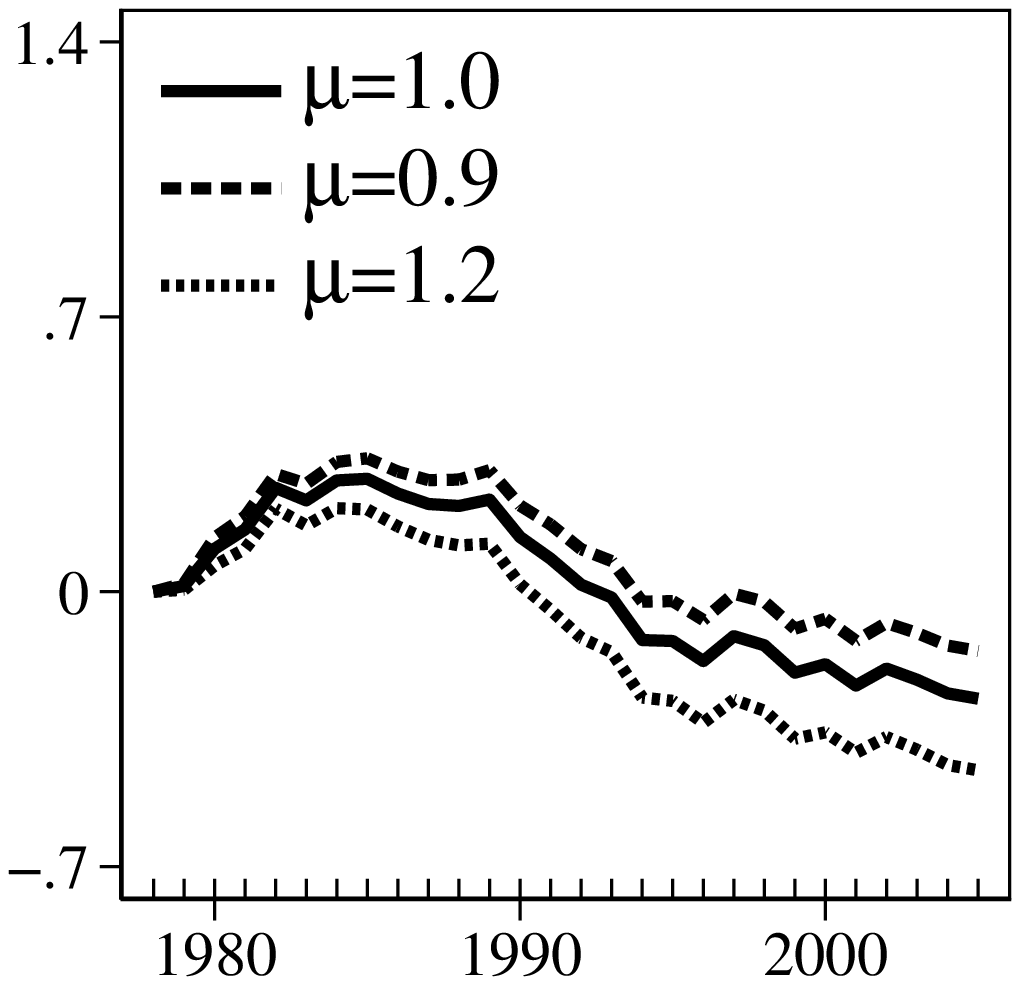}}\subfloat[Netherlands]{
\centering{}\includegraphics[scale=0.4]{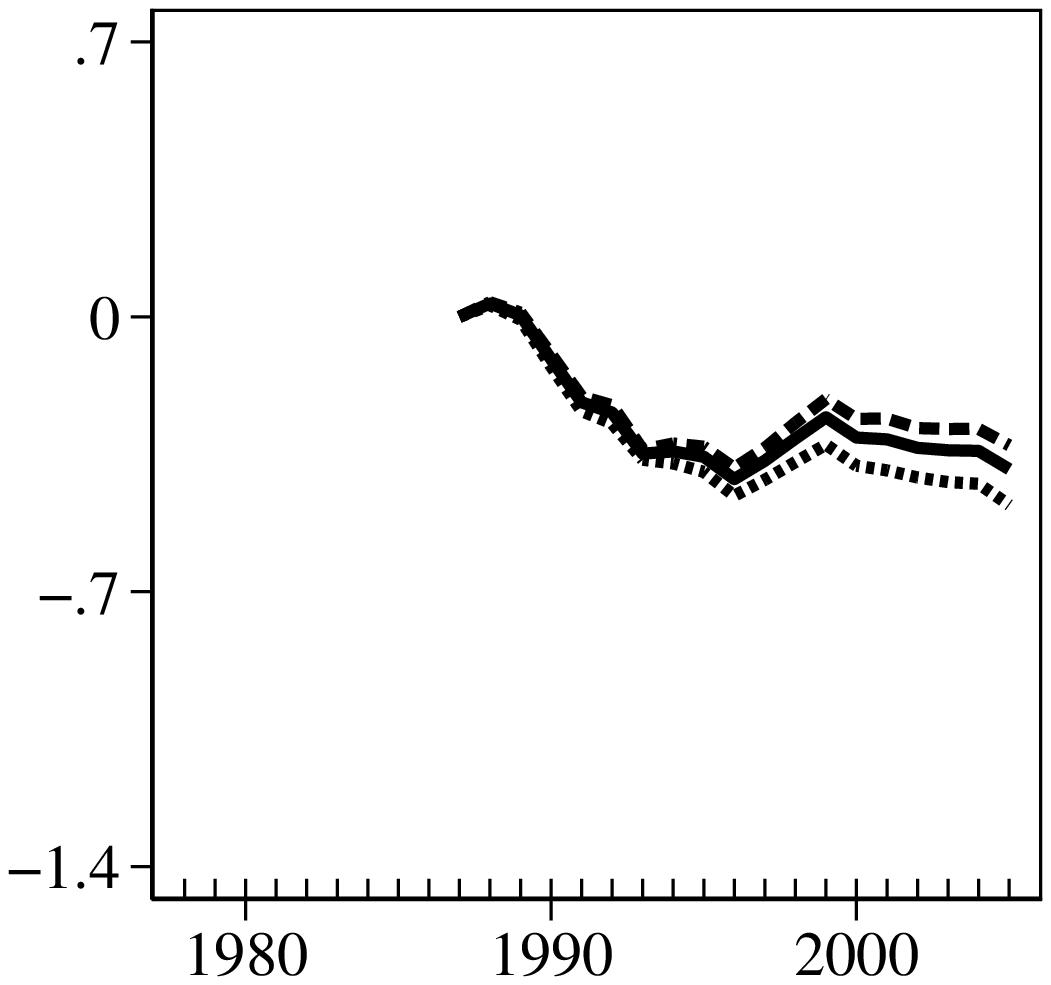}}
\par\end{centering}
\begin{centering}
\subfloat[Portugal]{
\centering{}\includegraphics[scale=0.4]{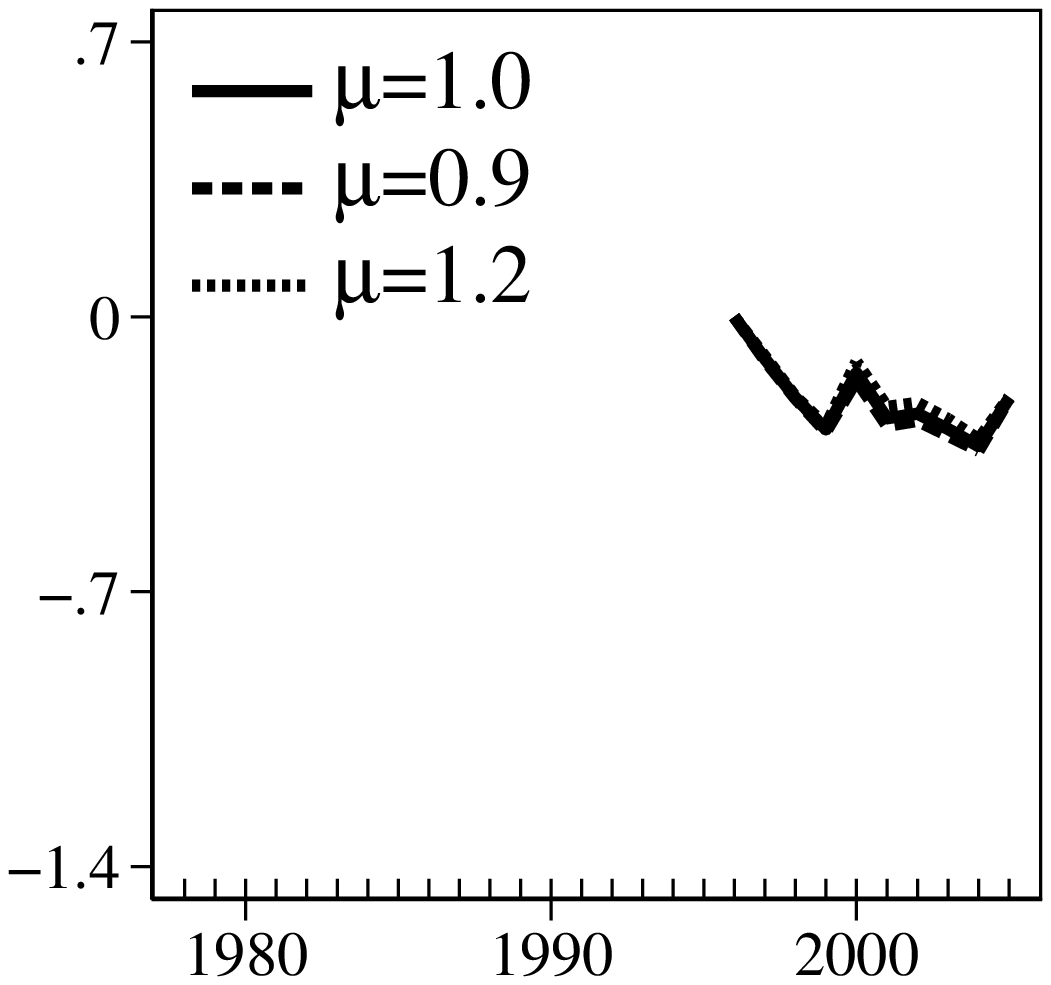}}\subfloat[Sweden]{
\centering{}\includegraphics[scale=0.4]{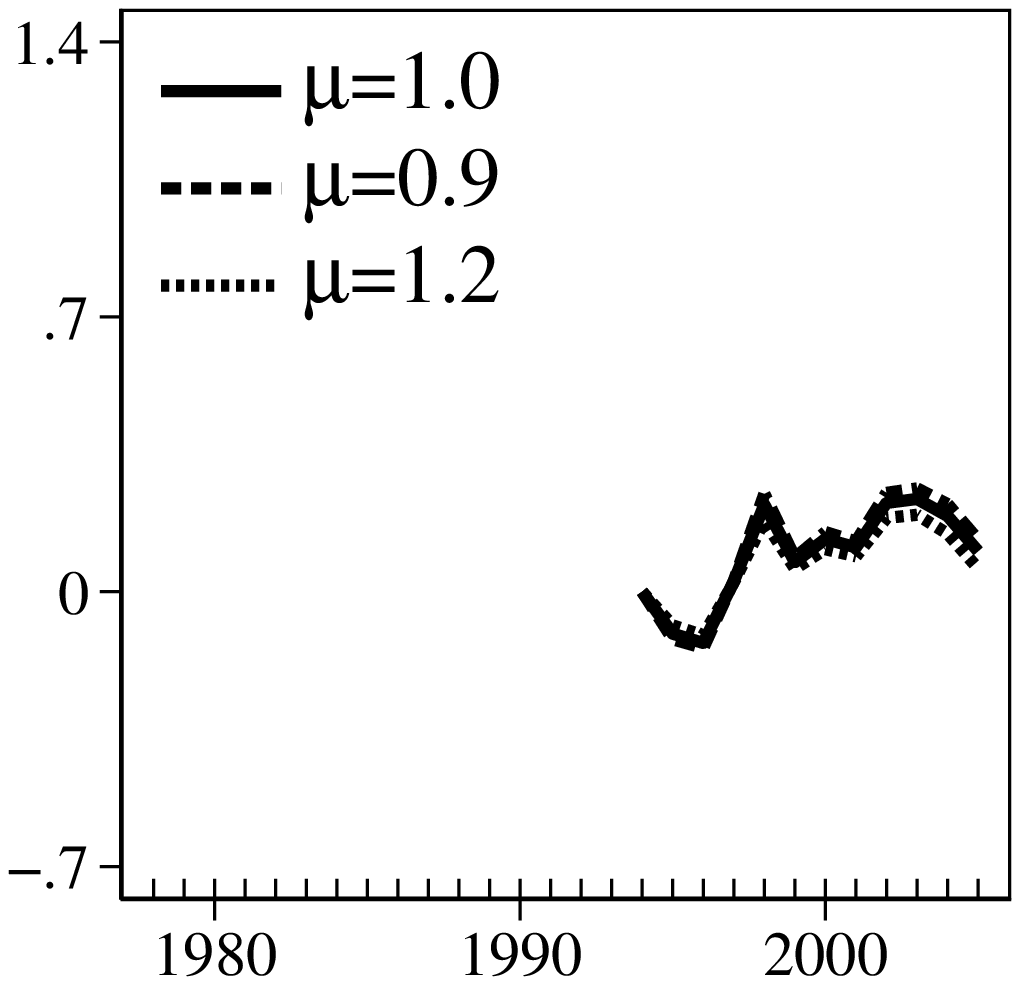}}\subfloat[United Kingdom]{
\centering{}\includegraphics[scale=0.4]{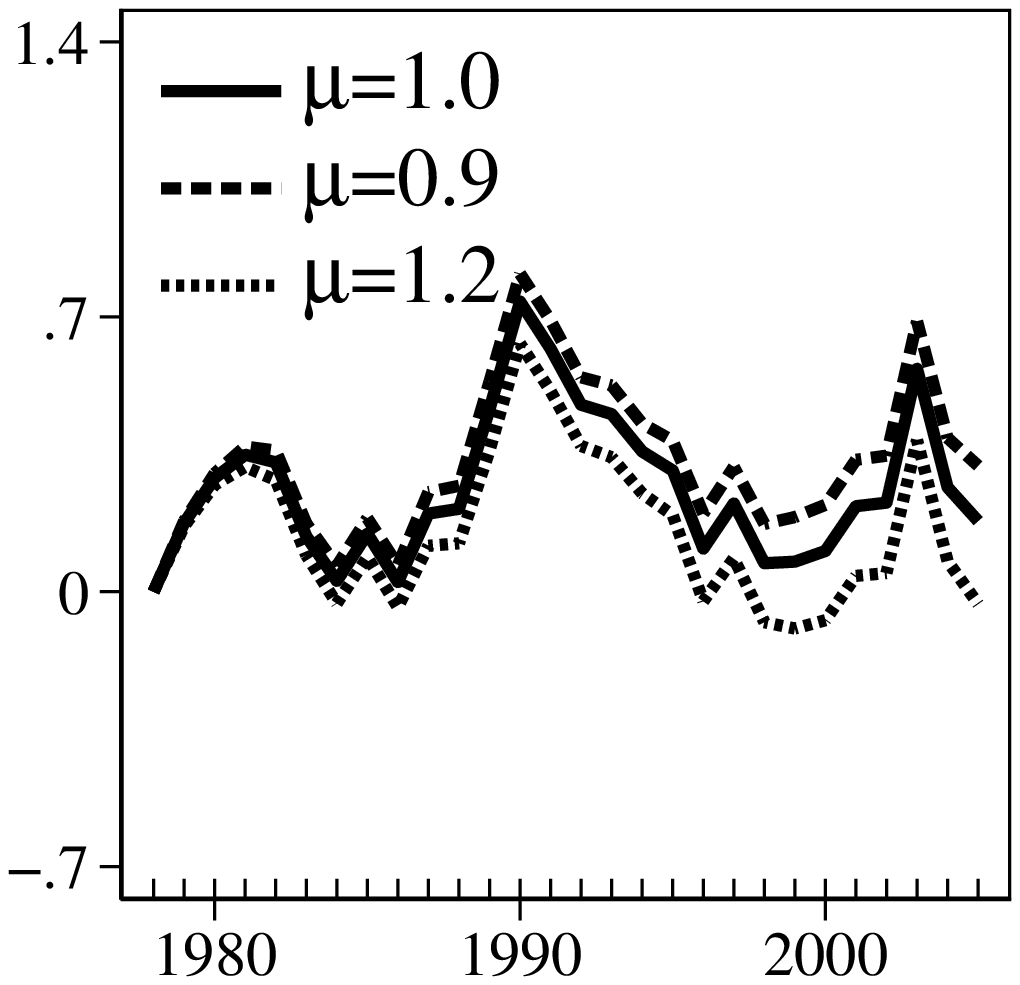}}\subfloat[United States]{
\centering{}\includegraphics[scale=0.4]{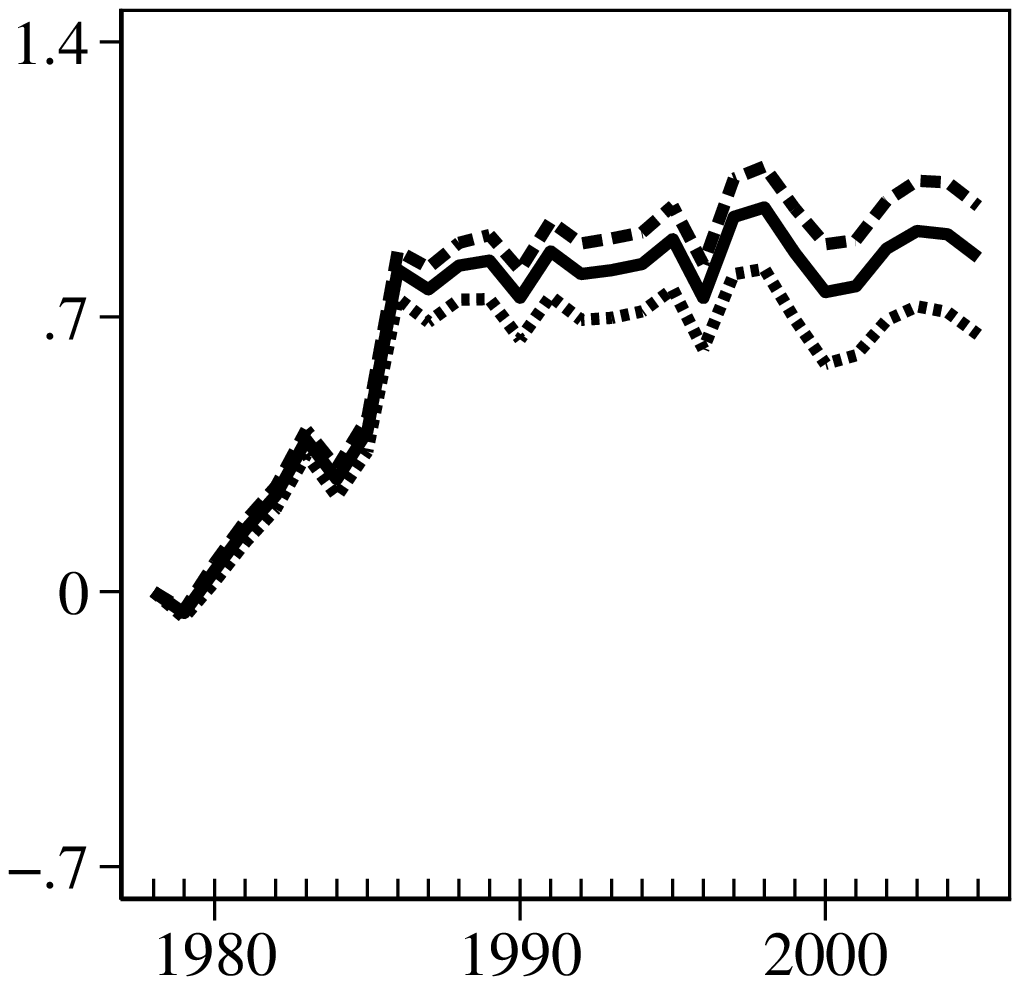}}
\par\end{centering}
\textit{\footnotesize{}Notes}{\footnotesize{}: Each line represents
energy-saving technology ($a_{e}$) when the scale parameter ($\mu$)
is 1, 0.9, or 1.2. All series are expressed as log differences relative
to the first year of observations.}{\footnotesize\par}
\end{figure}

Second, we show the extent to which energy-saving technological change
can vary according to the degree of returns to scale. We consider
the CES production function extended to allow for increasing or decreasing
returns to scale.
\begin{equation}
y=\left[\left(a_{k}k\right)^{\sigma}+\left(a_{\ell}\ell\right)^{\sigma}+\left(a_{e}e\right)^{\sigma}\right]^{\frac{\mu}{\sigma}}\qquad\text{for}\quad\sigma<1.
\end{equation}
The parameter $\mu$ governs the degree of returns to scale. When
the returns to scale are not constant, factor-augmenting technologies
can be rewritten as:
\begin{eqnarray}
a_{k} & = & \left(\frac{rk}{rk+w\ell+ve}\right)^{\frac{\epsilon_{\sigma}}{\epsilon_{\sigma}-1}}\left(\frac{y^{\frac{1}{\mu}}}{k}\right),\label{eq: ak1_mu}\\
a_{\ell} & = & \left(\frac{w\ell}{rk+w\ell+ve}\right)^{\frac{\epsilon_{\sigma}}{\epsilon_{\sigma}-1}}\left(\frac{y^{\frac{1}{\mu}}}{\ell}\right),\label{eq: al1_mu}\\
a_{e} & = & \left(\frac{ve}{rk+w\ell+ve}\right)^{\frac{\epsilon_{\sigma}}{\epsilon_{\sigma}-1}}\left(\frac{y^{\frac{1}{\mu}}}{e}\right),\label{eq: ae1_mu}
\end{eqnarray}
where gross output is $y=\left(rk+w\ell+ve\right)(\omega/\mu)$. \citet*{DeLoecker_Eeckhout_Unger_QJE20}
estimate the returns to scale in the United States from the year 1955
to 2016. They find that the estimates of the returns to scale vary
over time and across specifications, ranging from 0.95 to 1.2. We
consider a wider range of values for the returns-to-scale parameter
$\mu$ from 0.9 to 1.2. Figures \ref{fig: Ae1_mu_goods} and \ref{fig: Ae1_mu_service}
show how energy-saving technological change varies according to these
values in the goods and service sectors, respectively. Energy-saving
technological change varies only marginally according to the degree
of returns to scale.

\begin{figure}[H]
\caption{Capital-, labor-, energy-, and material-augmenting technological change
in the goods sector\label{fig: AkAlAeAm_goods}}

\begin{centering}
\subfloat[Austria]{
\centering{}\includegraphics[scale=0.4]{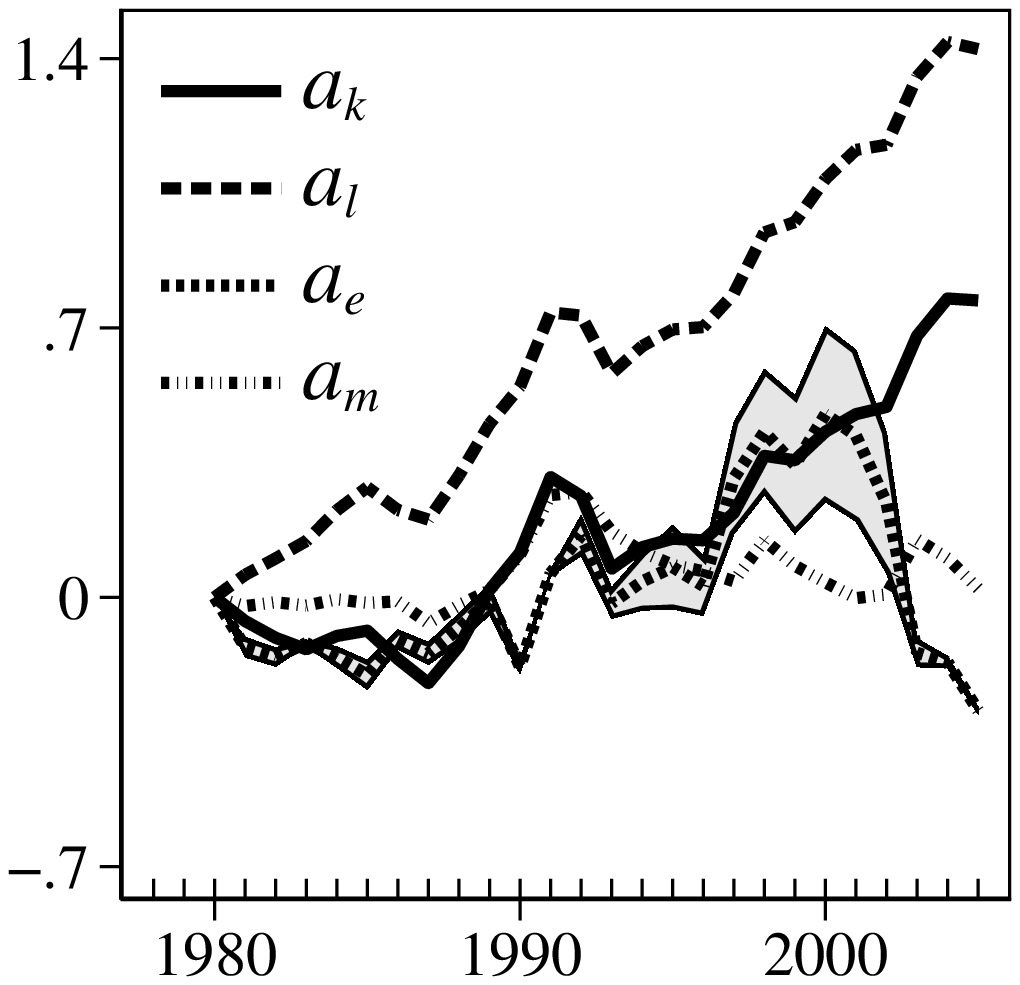}}\subfloat[Czech Republic]{
\centering{}\includegraphics[scale=0.4]{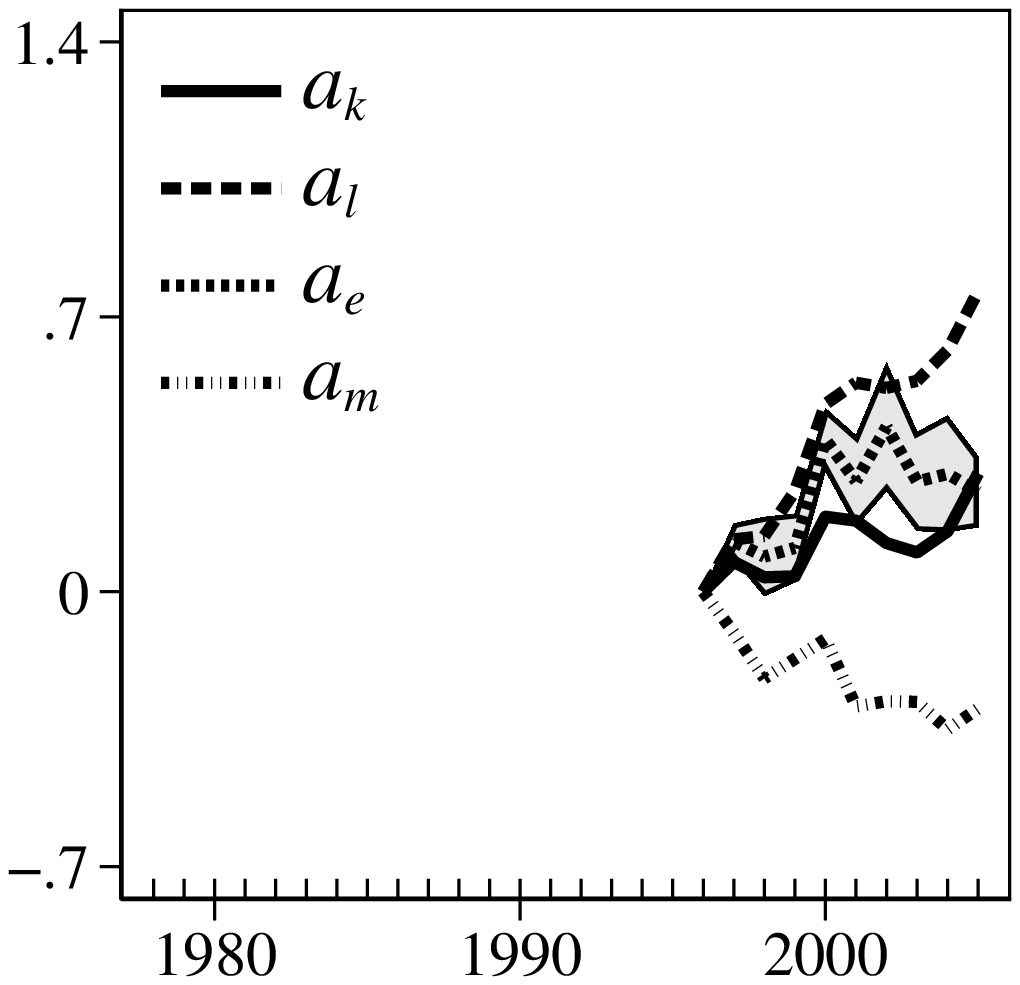}}\subfloat[Denmark]{
\centering{}\includegraphics[scale=0.4]{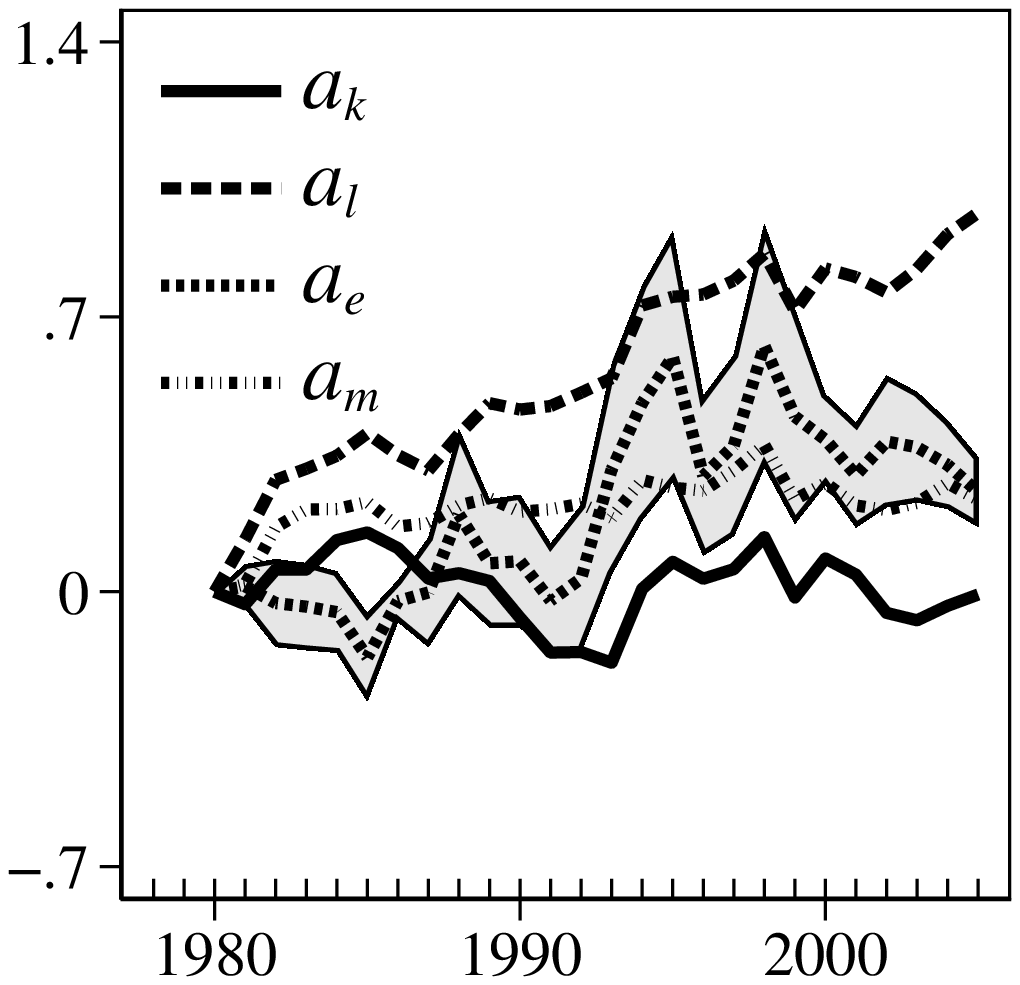}}\subfloat[Finland]{
\centering{}\includegraphics[scale=0.4]{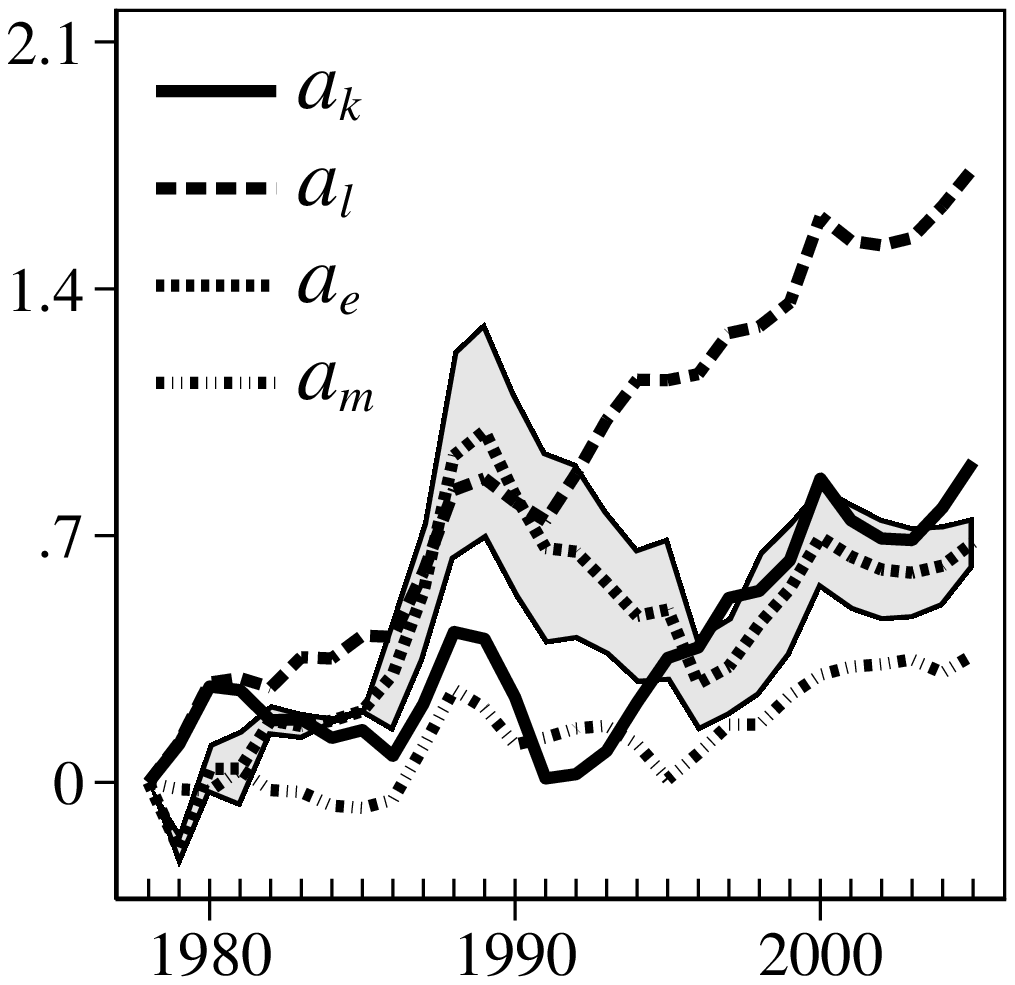}}
\par\end{centering}
\begin{centering}
\subfloat[Germany]{
\centering{}\includegraphics[scale=0.4]{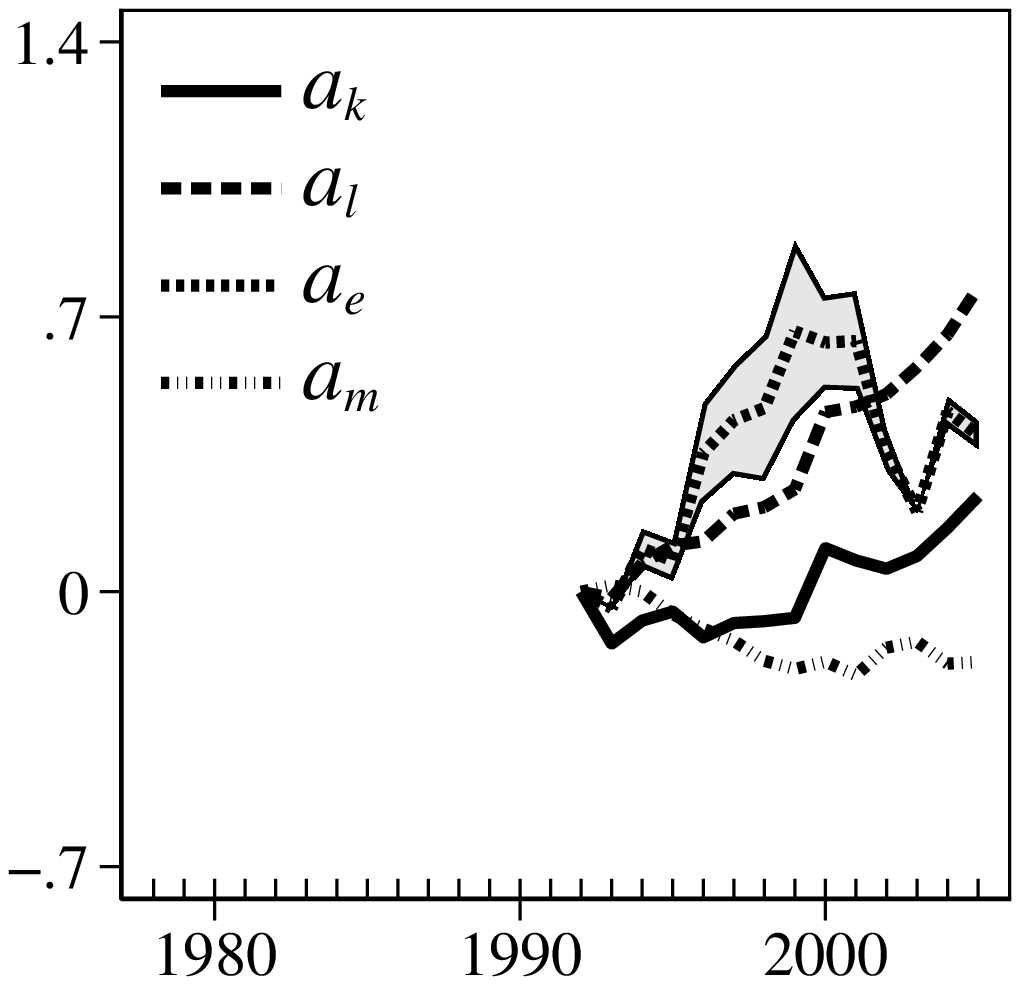}}\subfloat[Italy]{
\centering{}\includegraphics[scale=0.4]{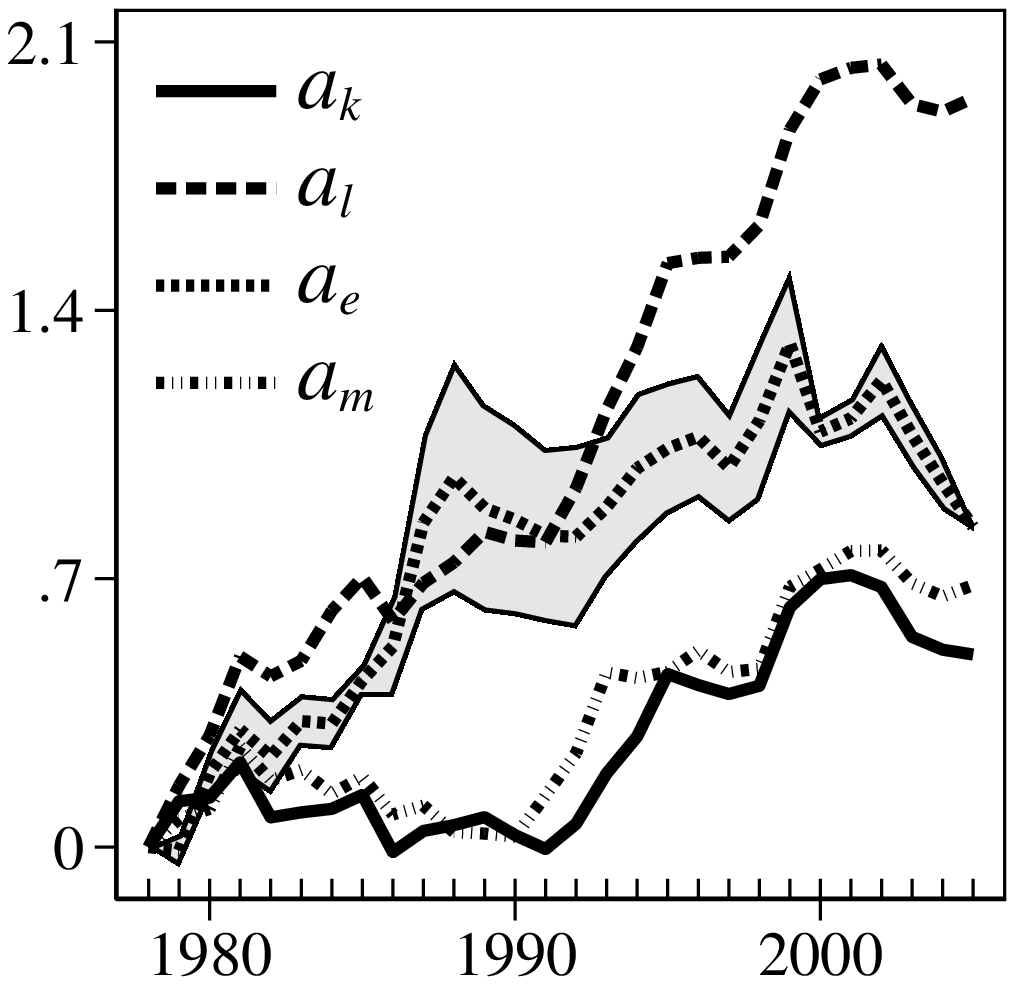}}\subfloat[Japan]{
\centering{}\includegraphics[scale=0.4]{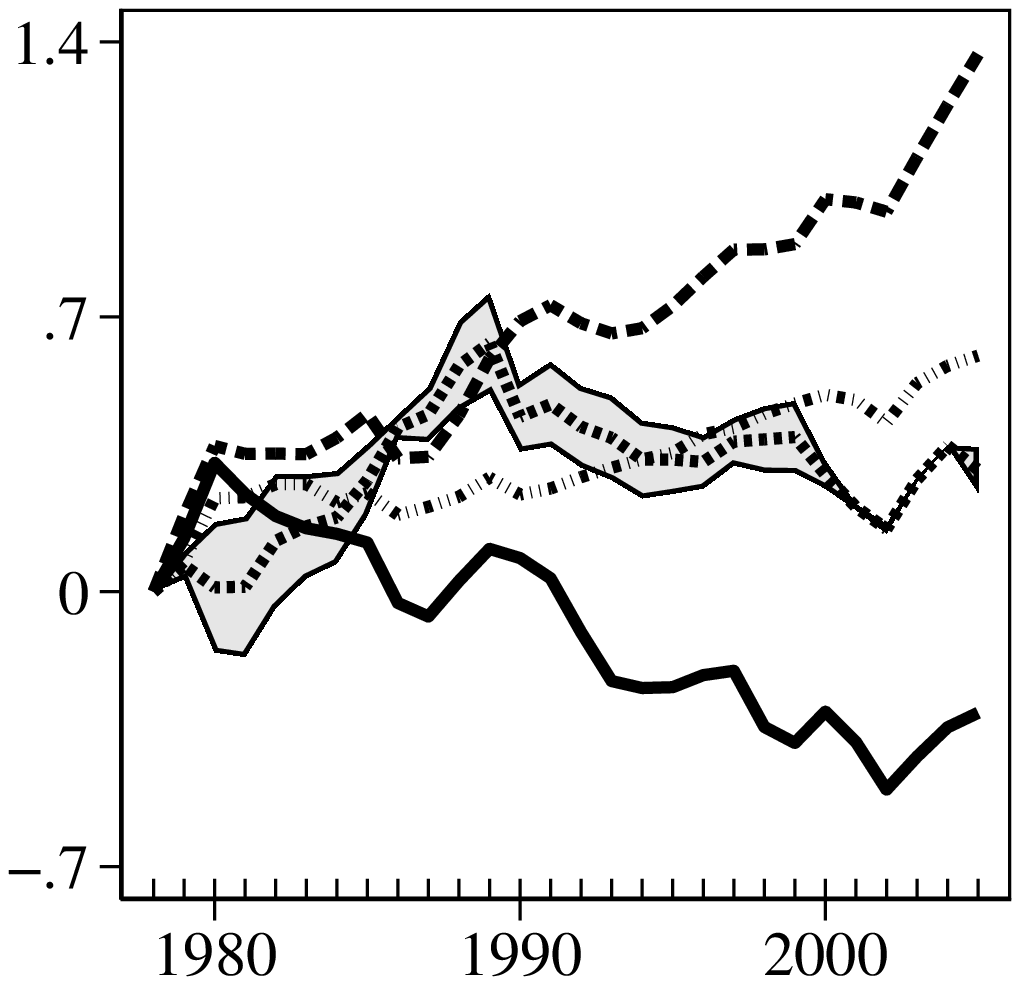}}\subfloat[Netherlands]{
\centering{}\includegraphics[scale=0.4]{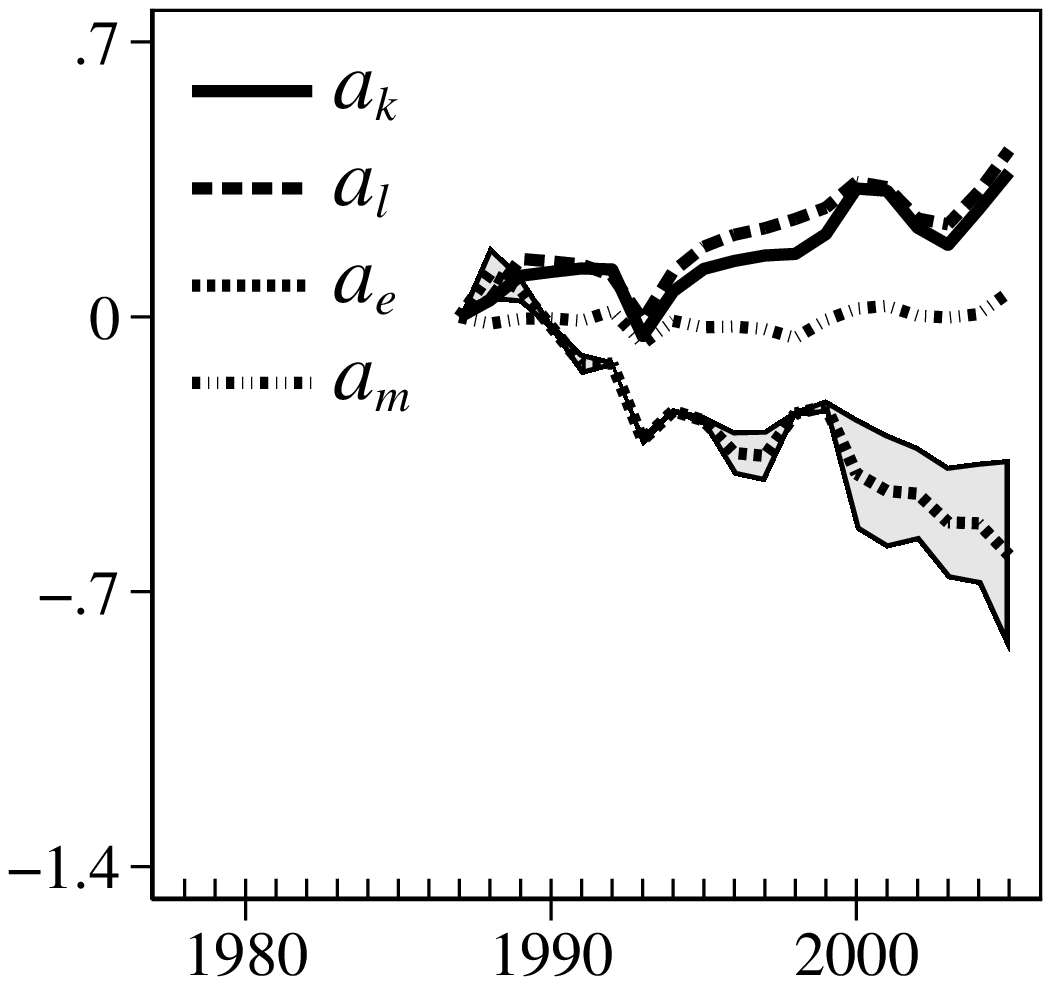}}
\par\end{centering}
\begin{centering}
\subfloat[Portugal]{
\centering{}\includegraphics[scale=0.4]{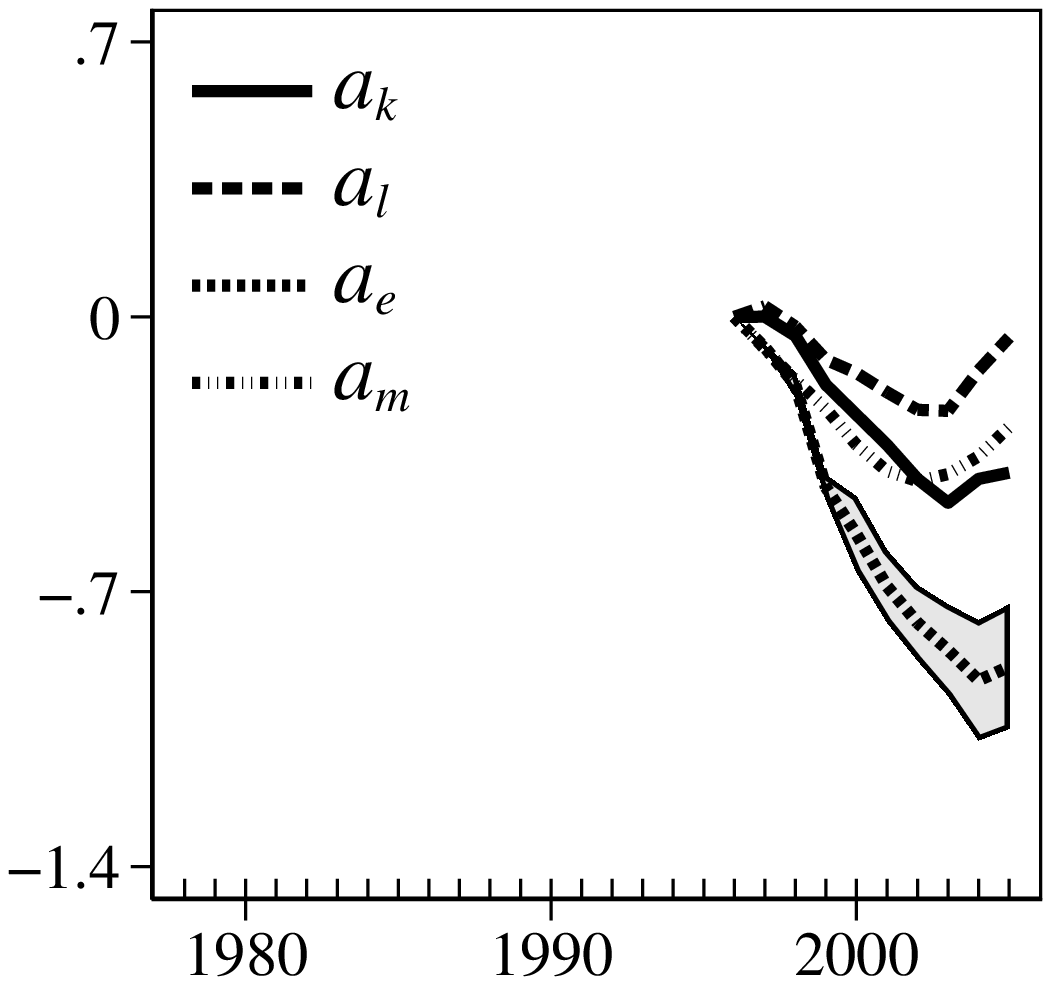}}\subfloat[Sweden]{
\centering{}\includegraphics[scale=0.4]{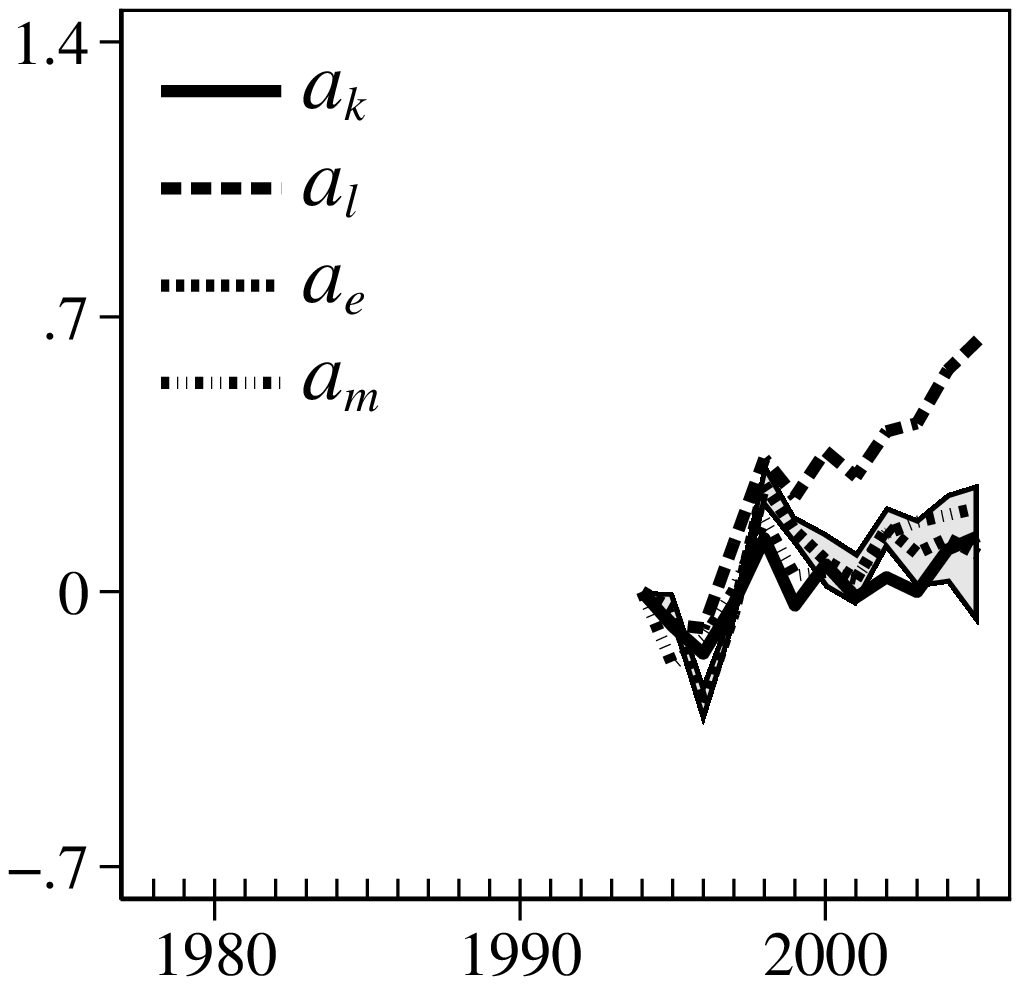}}\subfloat[United Kingdom]{
\centering{}\includegraphics[scale=0.4]{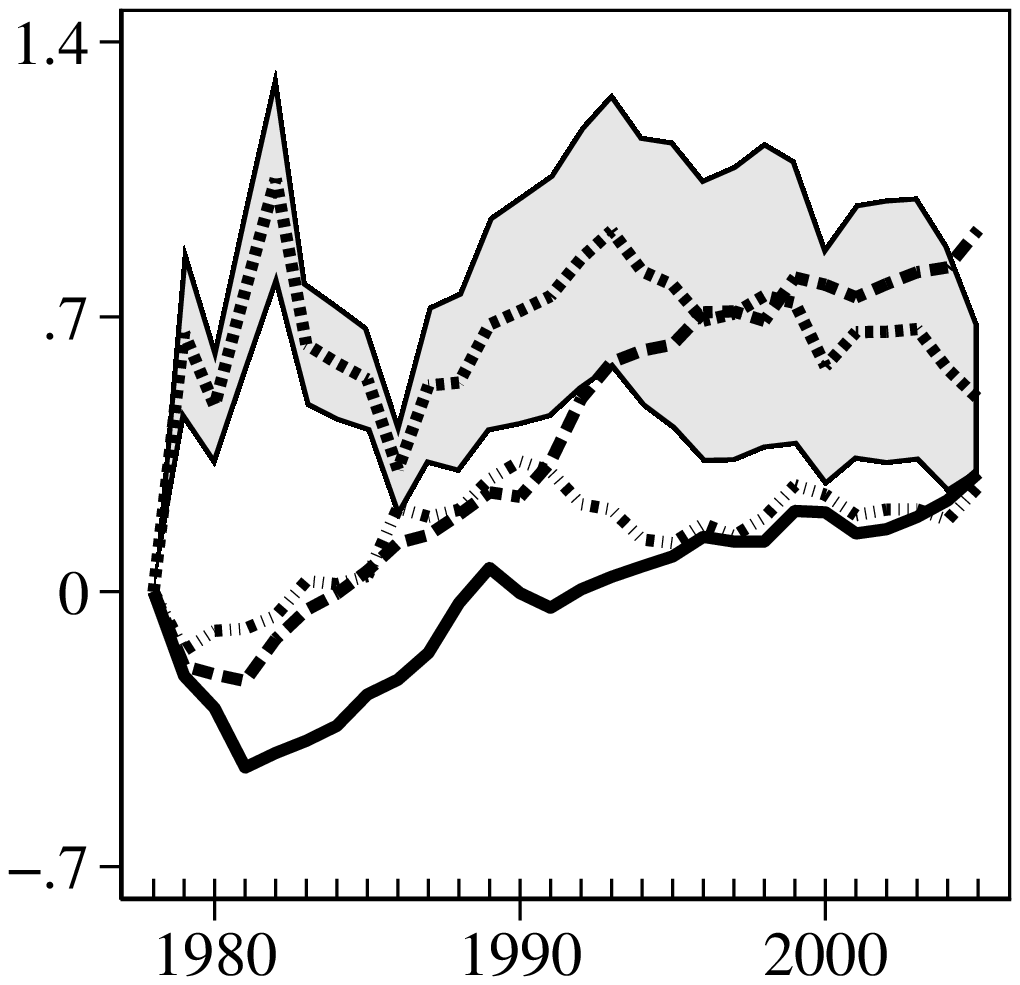}}\subfloat[United States]{
\centering{}\includegraphics[scale=0.4]{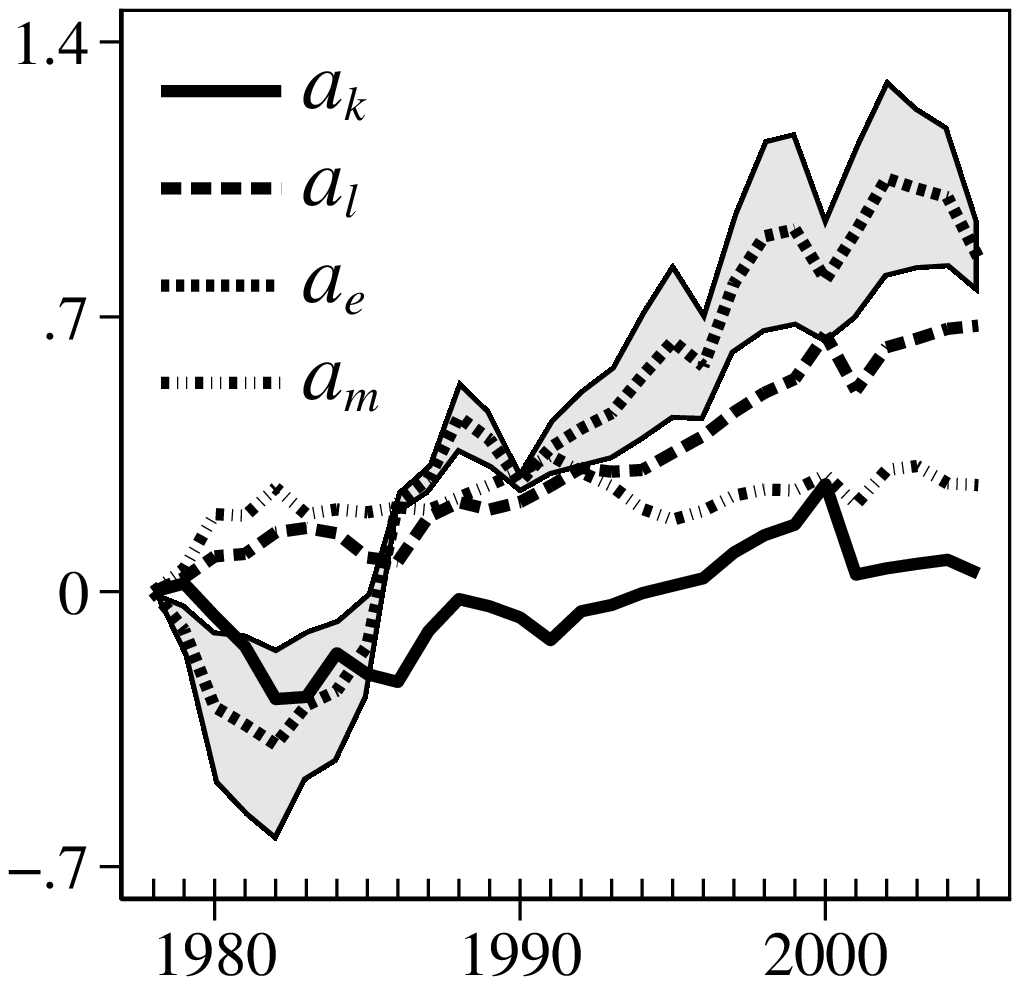}}
\par\end{centering}
\textit{\footnotesize{}Notes}{\footnotesize{}: The solid, dashed,
dotted, and dashed-dotted lines are capital-, labor-, energy-, and
material-augmenting technologies ($a_{k}$, $a_{\ell}$, $a_{e}$,
$a_{m}$), respectively. The shaded area represents the 90 percent
confidence interval for $a_{e}$. All series are expressed as log
differences relative to the first year of observations.}{\footnotesize\par}
\end{figure}

\begin{figure}[H]
\caption{Capital-, labor-, energy-, and material-augmenting technological change
in the service sector\label{fig: AkAlAeAm_service}}

\begin{centering}
\subfloat[Austria]{
\centering{}\includegraphics[scale=0.4]{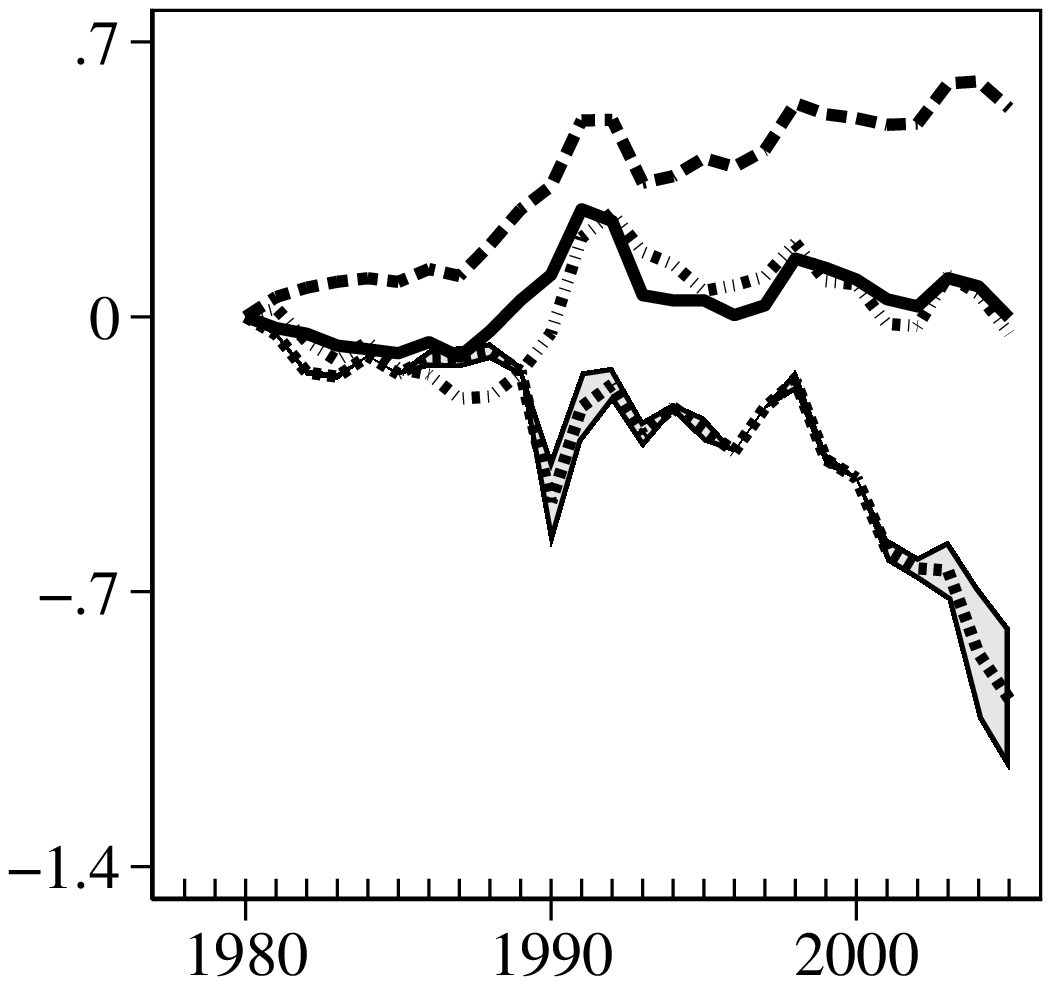}}\subfloat[Czech Republic]{
\centering{}\includegraphics[scale=0.4]{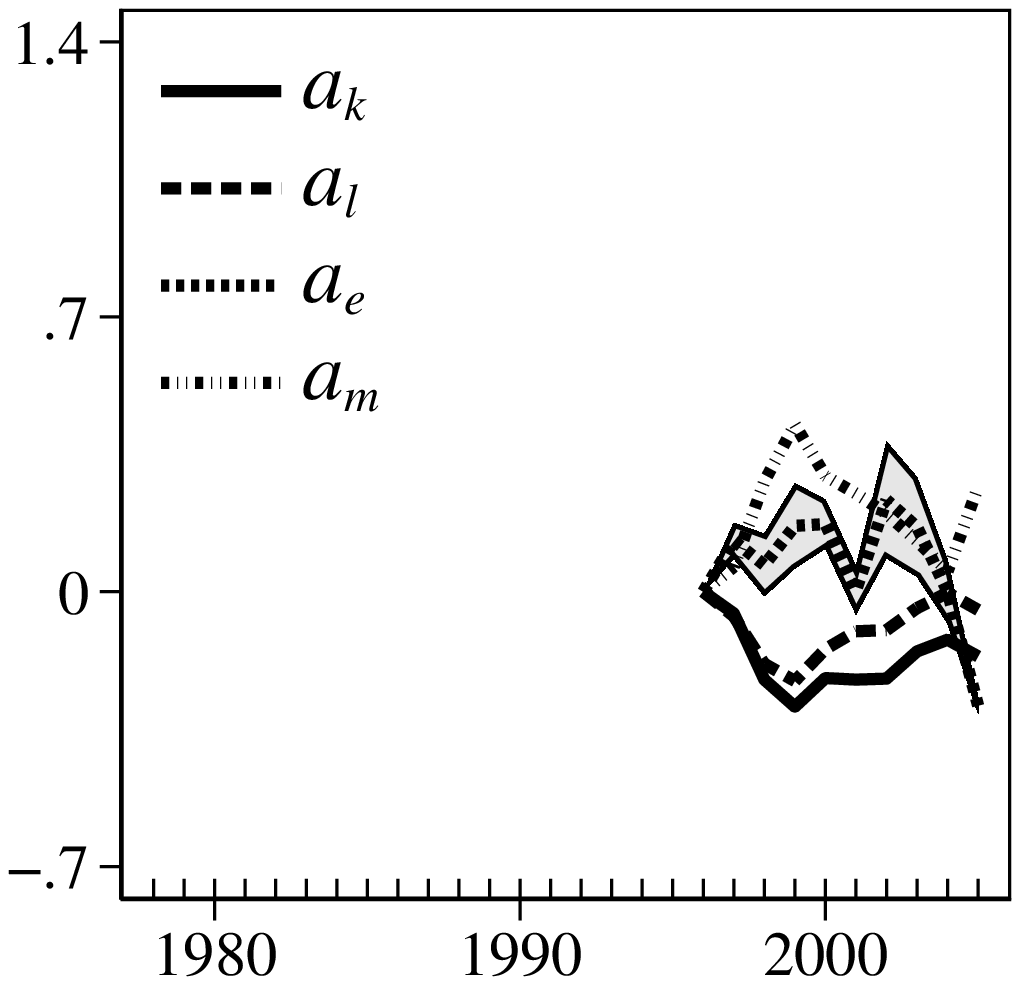}}\subfloat[Denmark]{
\centering{}\includegraphics[scale=0.4]{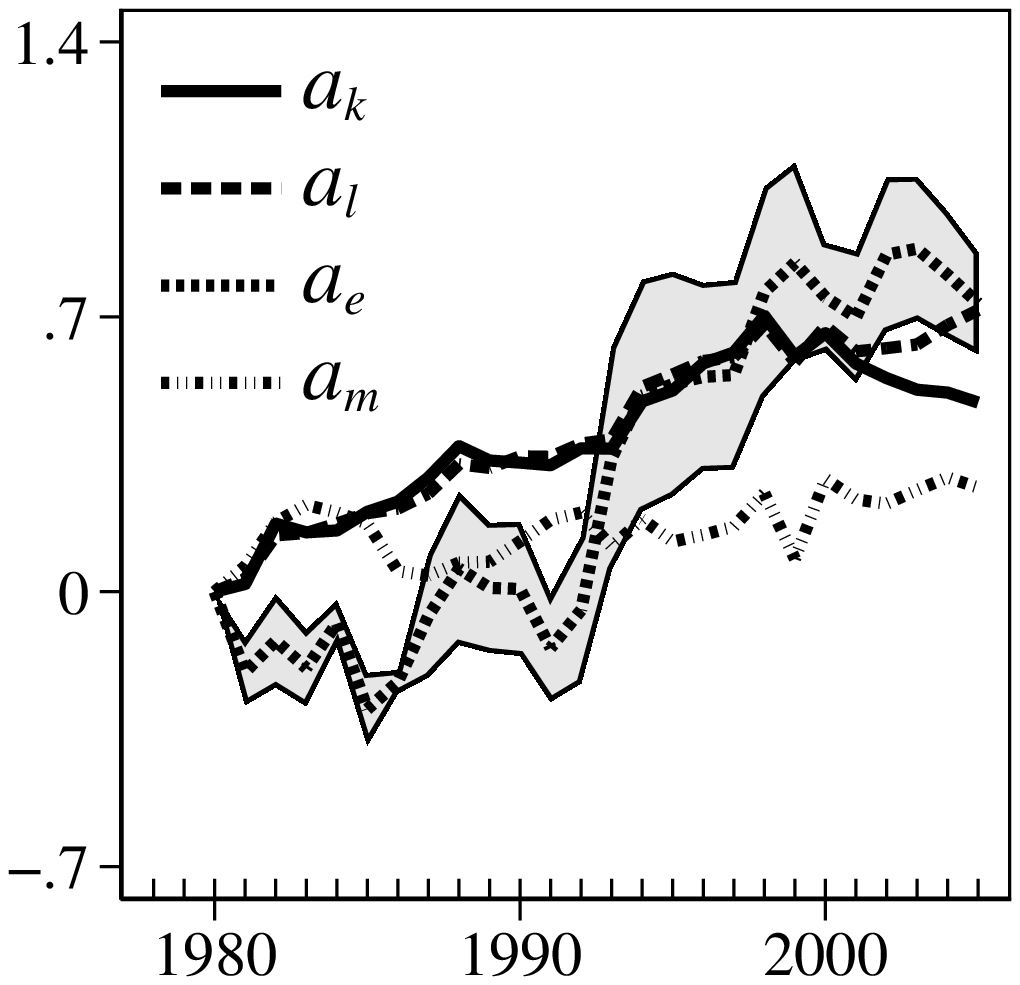}}\subfloat[Finland]{
\centering{}\includegraphics[scale=0.4]{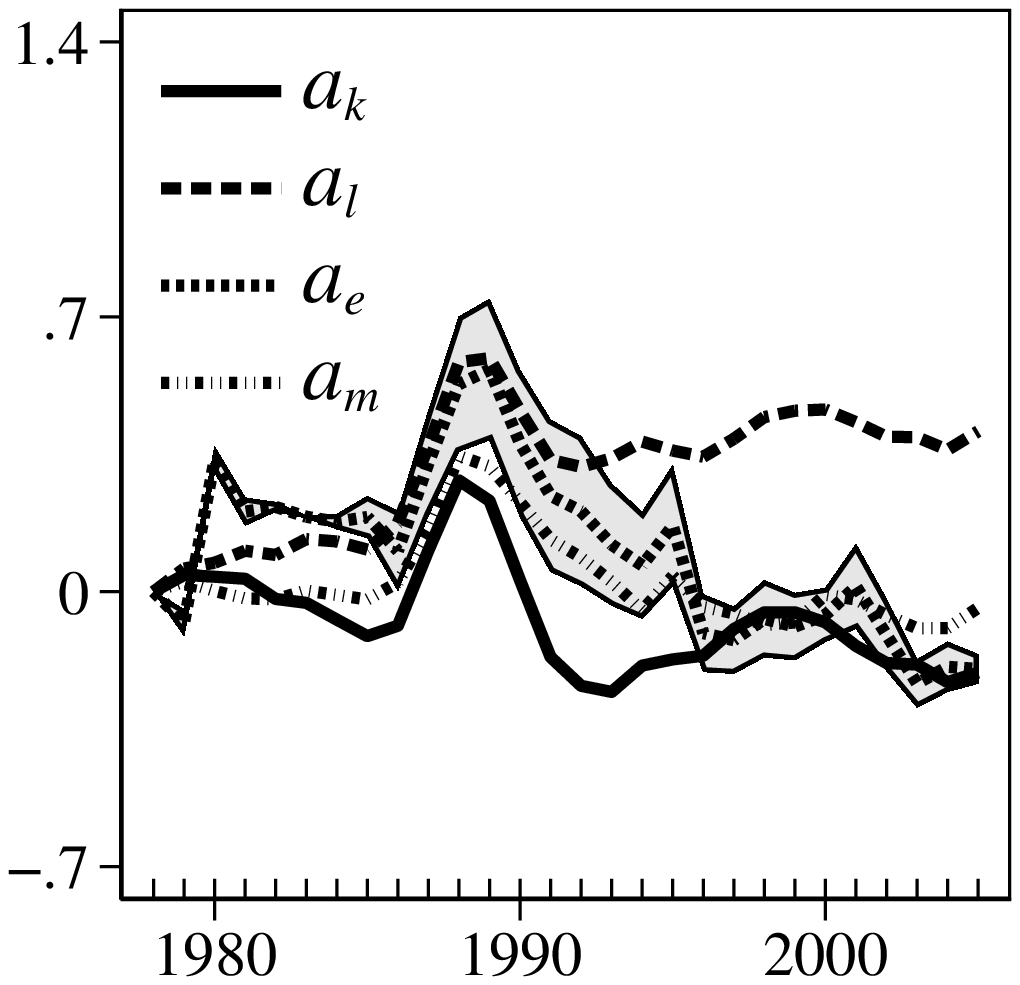}}
\par\end{centering}
\begin{centering}
\subfloat[Germany]{
\centering{}\includegraphics[scale=0.4]{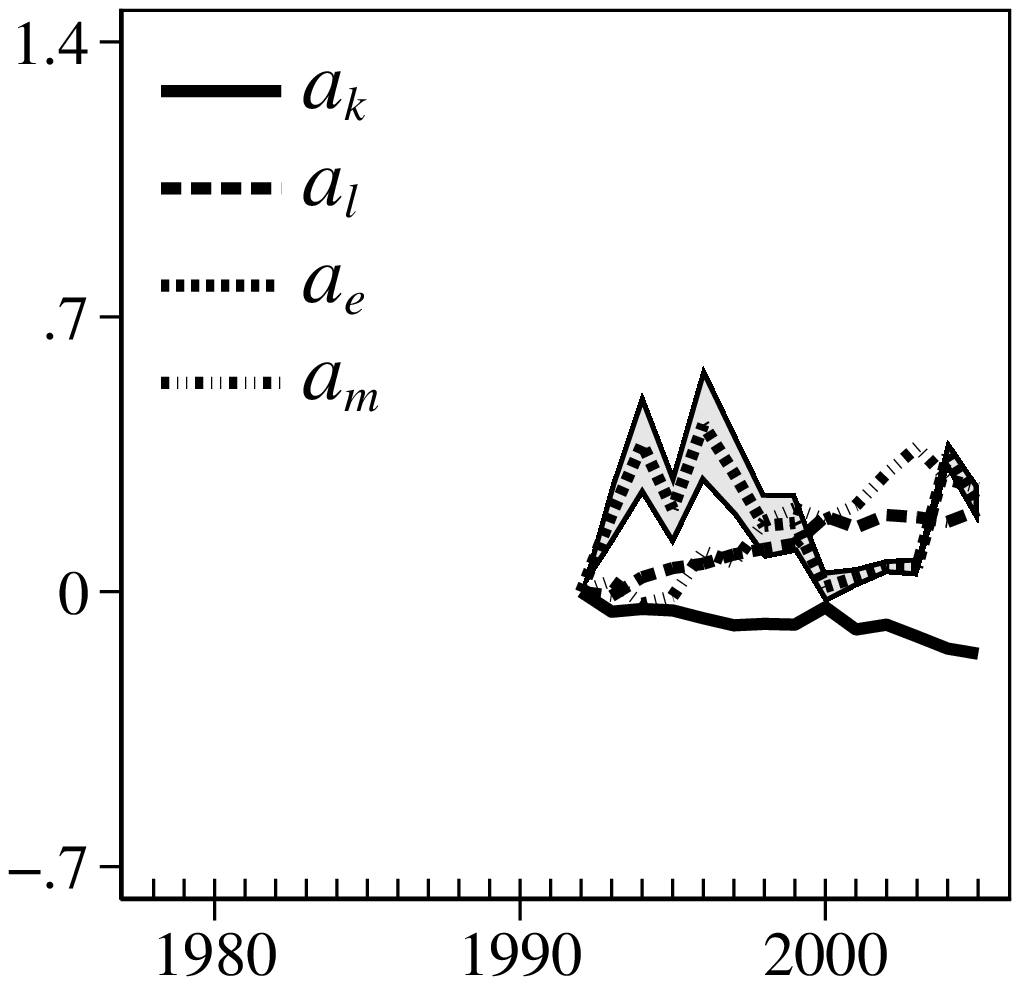}}\subfloat[Italy]{
\centering{}\includegraphics[scale=0.4]{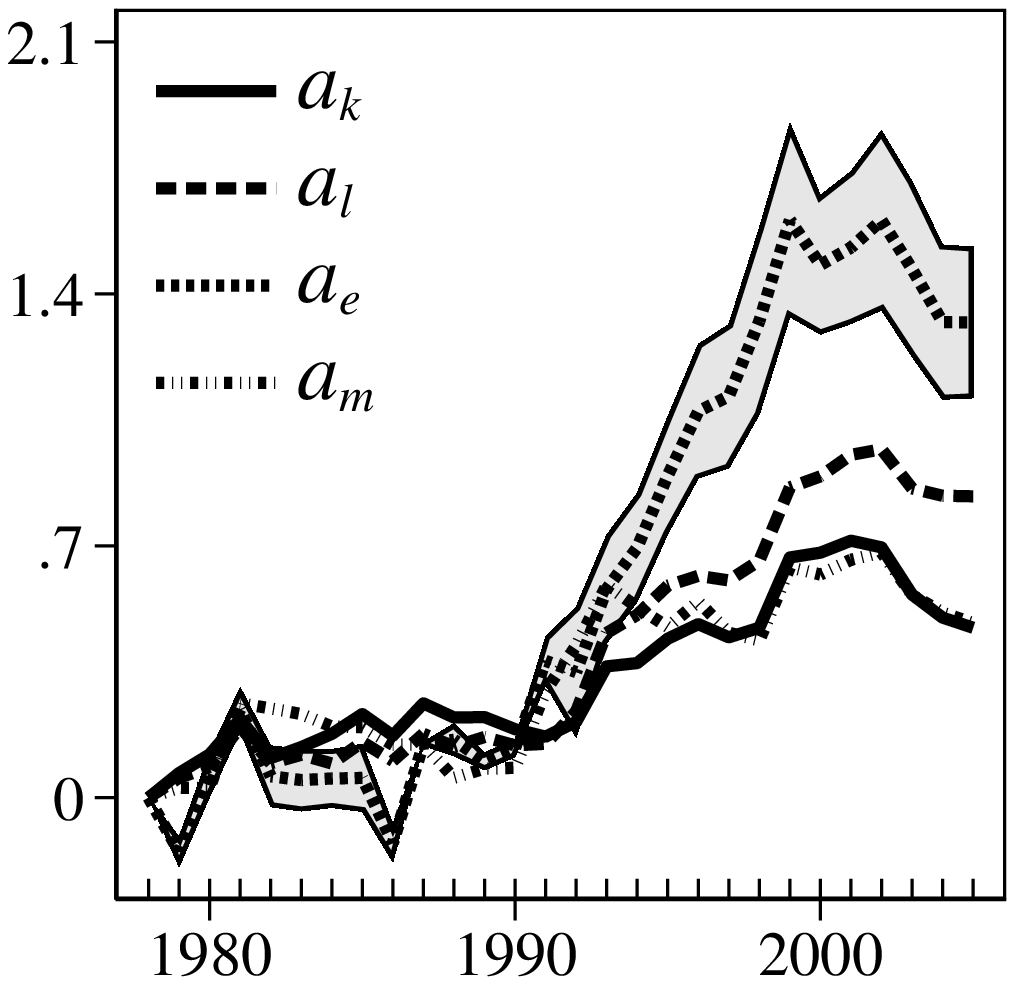}}\subfloat[Japan]{
\centering{}\includegraphics[scale=0.4]{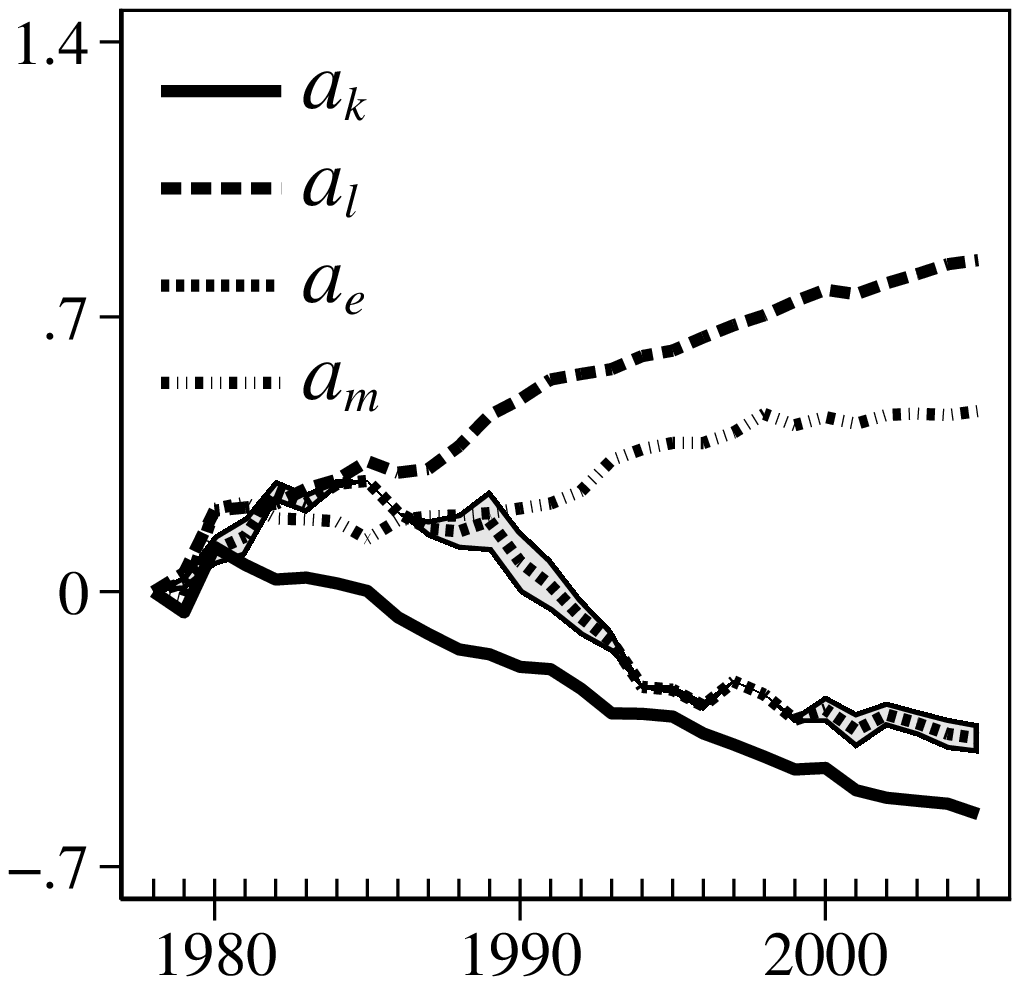}}\subfloat[Netherlands]{
\centering{}\includegraphics[scale=0.4]{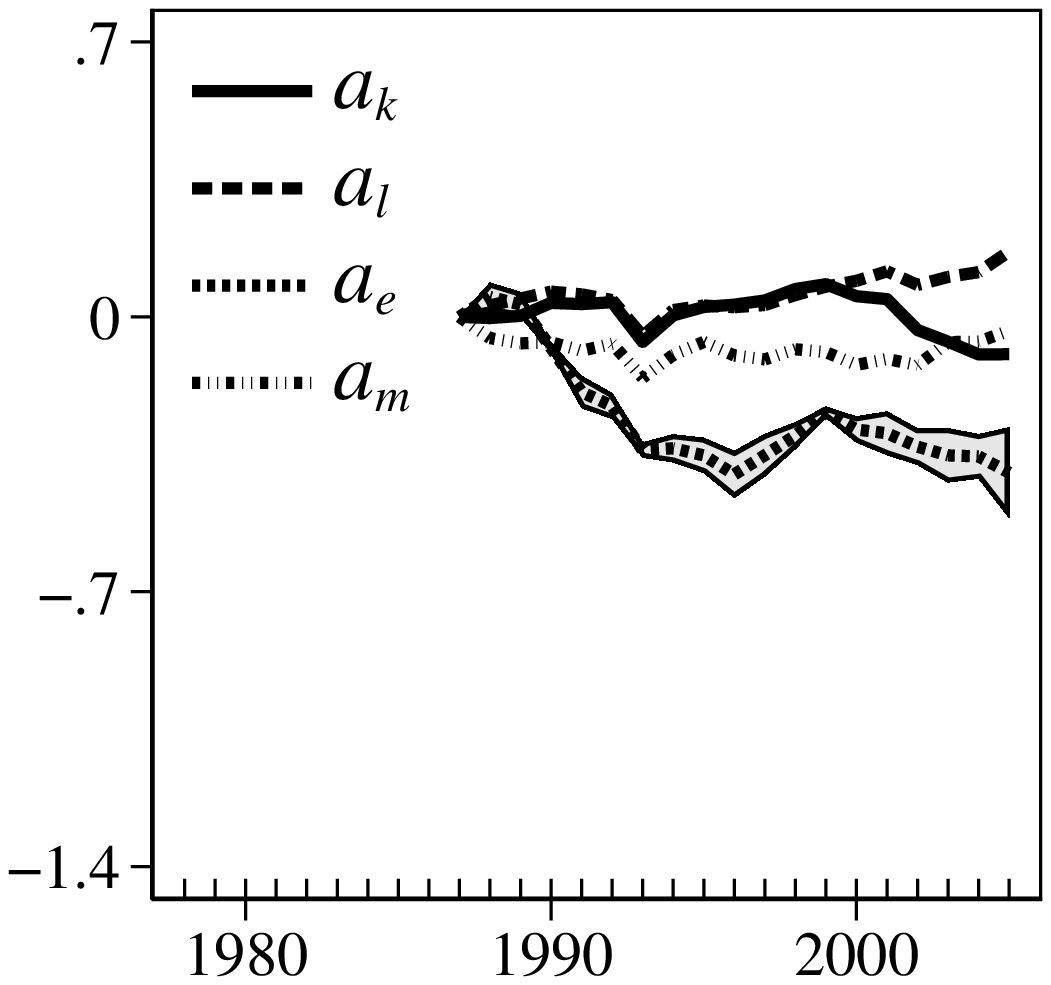}}
\par\end{centering}
\begin{centering}
\subfloat[Portugal]{
\centering{}\includegraphics[scale=0.4]{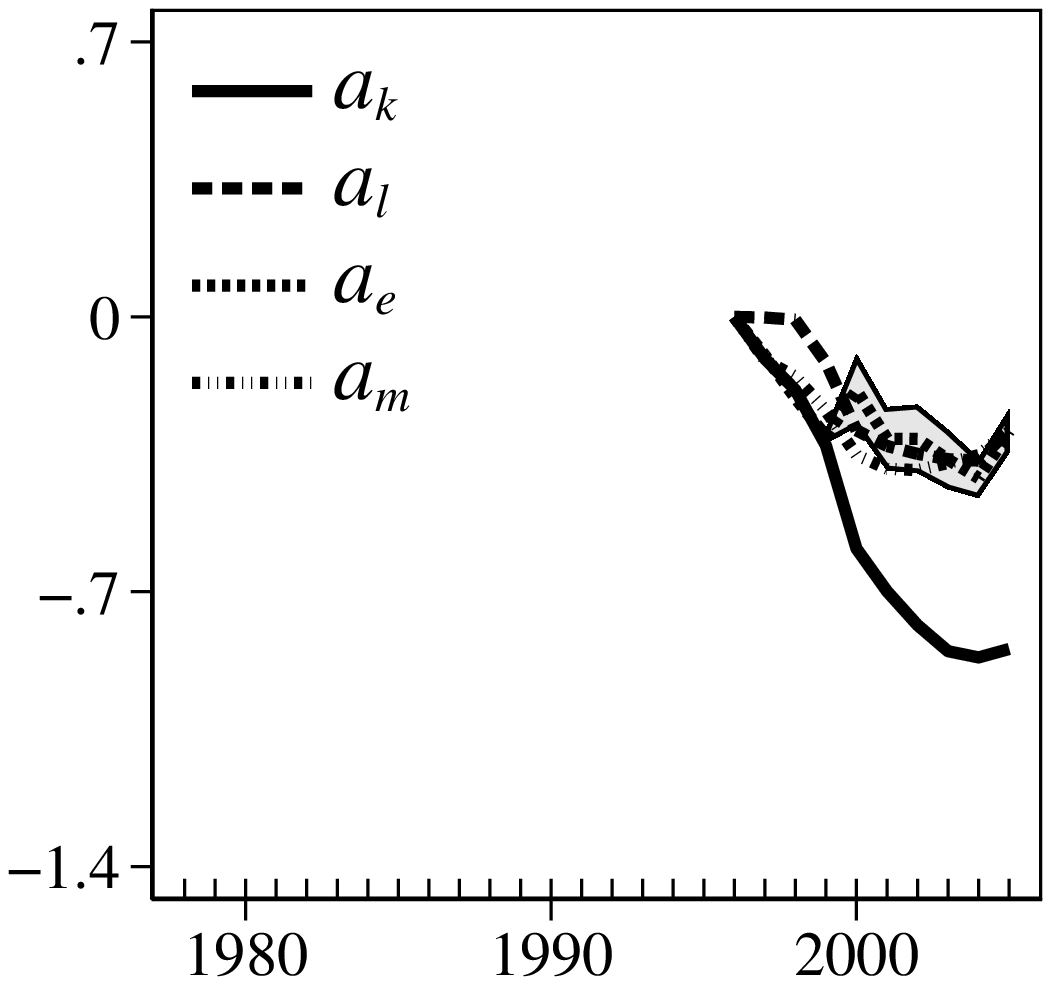}}\subfloat[Sweden]{
\centering{}\includegraphics[scale=0.4]{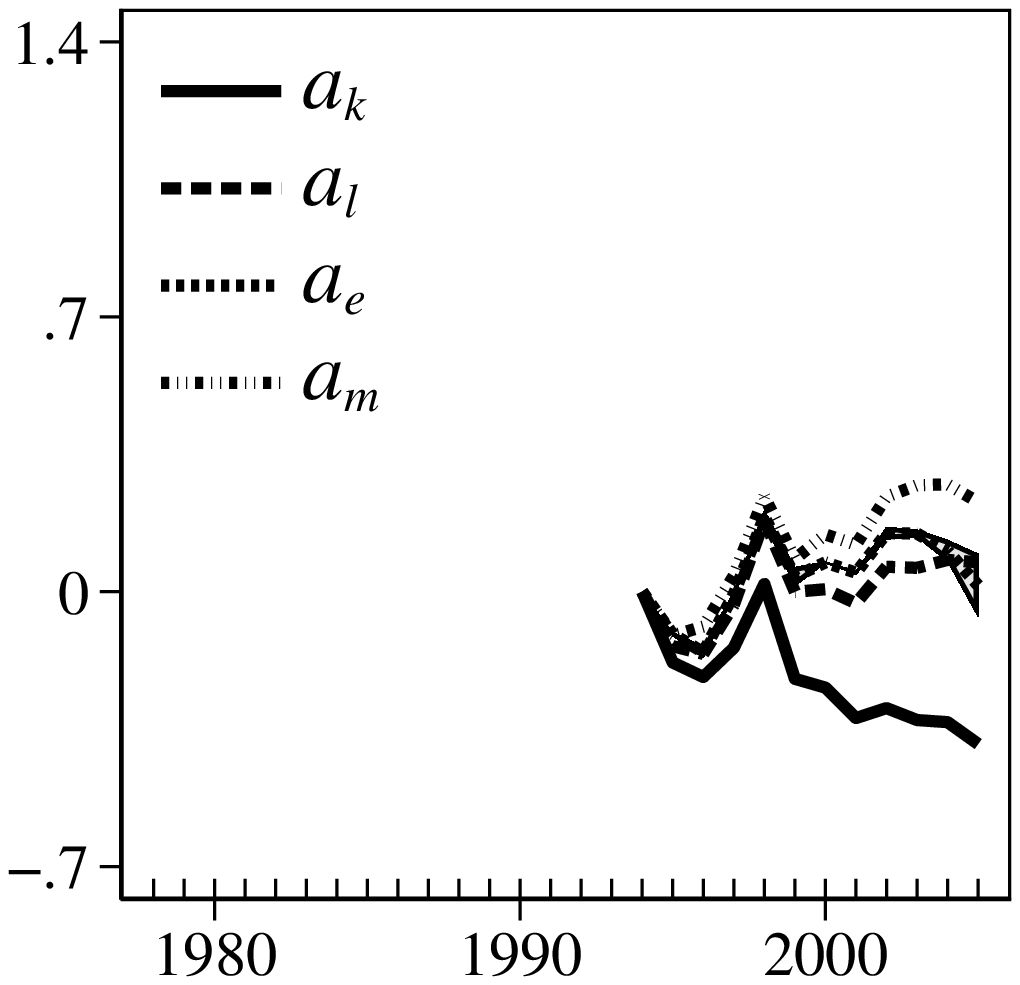}}\subfloat[United Kingdom]{
\centering{}\includegraphics[scale=0.4]{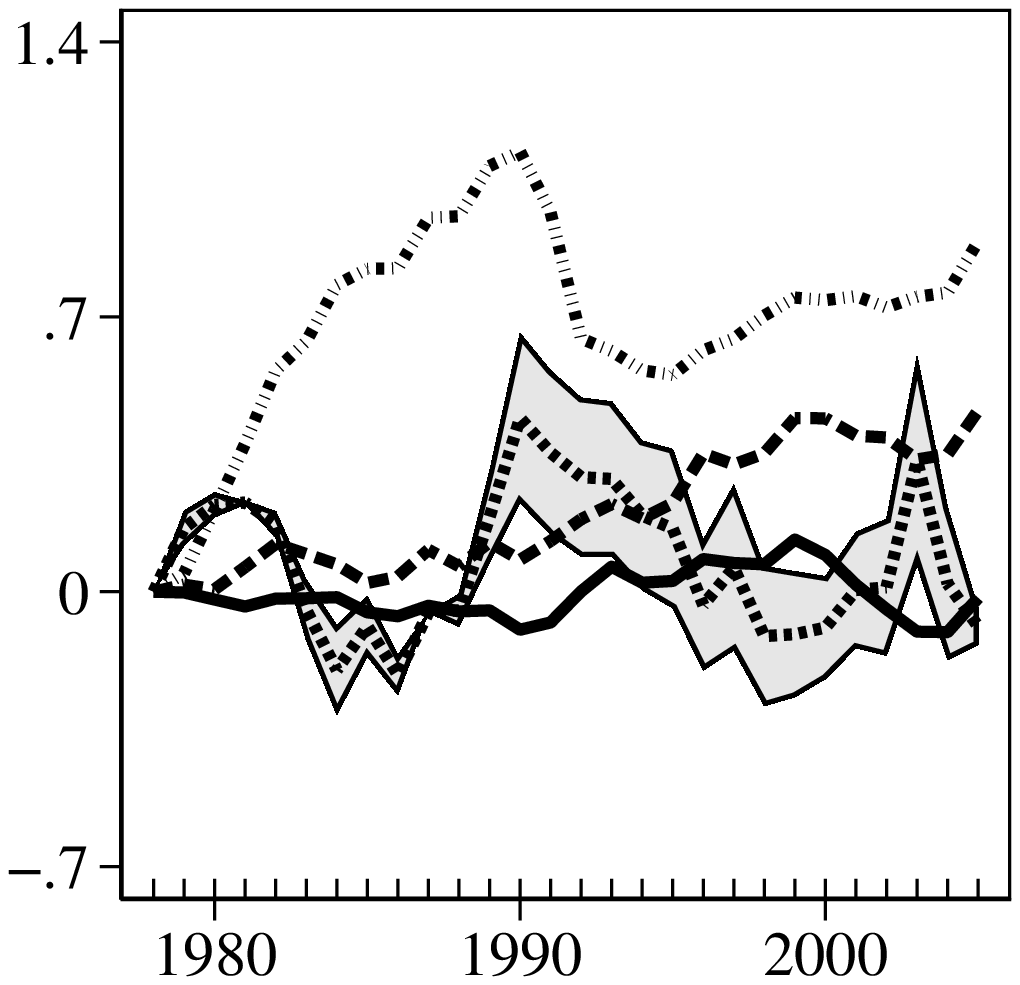}}\subfloat[United States]{
\centering{}\includegraphics[scale=0.4]{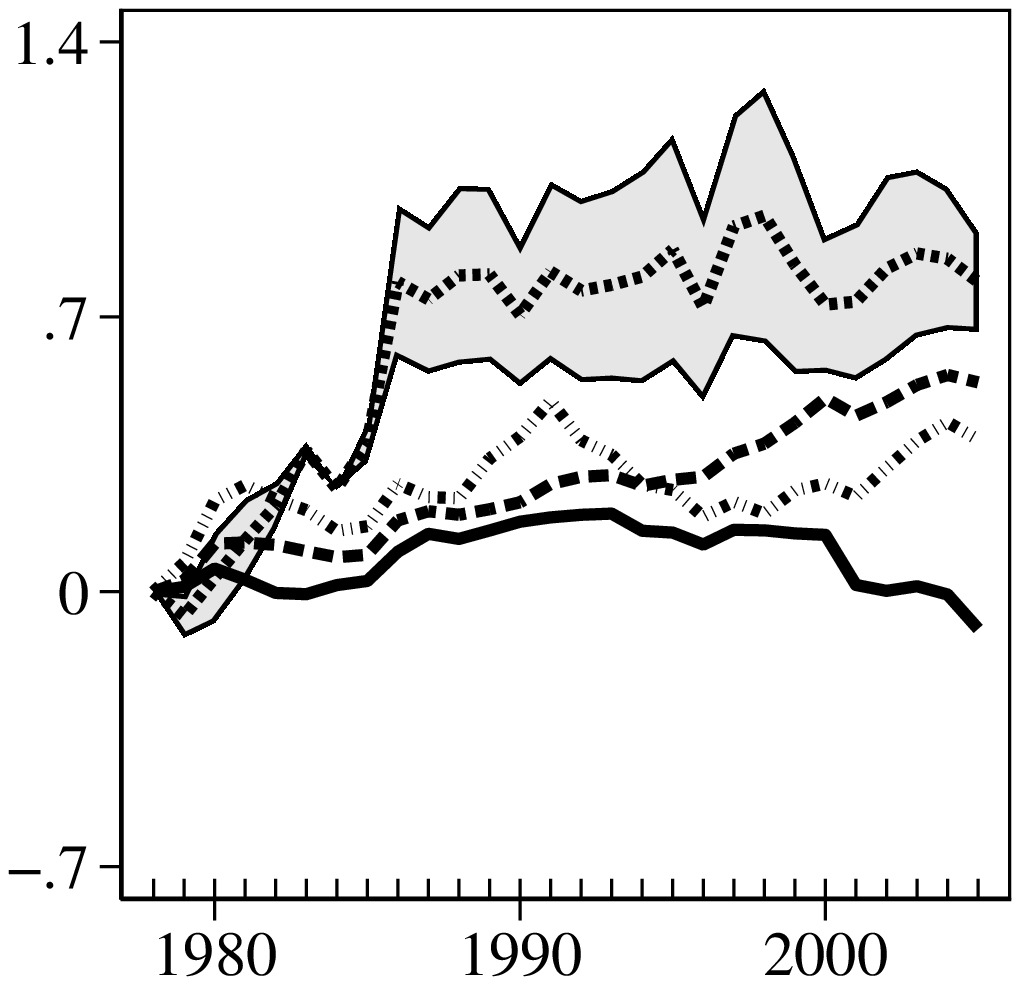}}
\par\end{centering}
\textit{\footnotesize{}Notes}{\footnotesize{}: The solid, dashed,
dotted, and dashed-dotted lines are capital-, labor-, energy-, and
material-augmenting technologies ($a_{k}$, $a_{\ell}$, $a_{e}$,
$a_{m}$), respectively. The shaded area represents the 90 percent
confidence interval for $a_{e}$. All series are expressed as log
differences relative to the first year of observations.}{\footnotesize\par}
\end{figure}

Third, we show the extent to which factor-augmenting technological
change can vary after taking into account material inputs. Let $m$
denote the material quantity, $\tau$ the material price, and $a_{m}$
material-augmenting technology. We consider the CES production function
with capital, labor, energy, and material inputs:
\begin{equation}
y=\left[\left(a_{k}k\right)^{\sigma}+\left(a_{\ell}\ell\right)^{\sigma}+\left(a_{e}e\right)^{\sigma}+\left(a_{m}m\right)^{\sigma}\right]^{\frac{1}{\sigma}}\qquad\text{for}\quad\sigma<1.
\end{equation}
Factor-augmenting technologies retain the same form as equations \eqref{eq: ak1}\textendash \eqref{eq: ae1}.
In this case, gross output is $y=(rk+w\ell+ve+\tau m)\omega$. Figures
\ref{fig: AkAlAeAm_goods} and \ref{fig: AkAlAeAm_service} show that
both the magnitude and the direction of capital-, labor-, and energy-augmenting
technological change remain essentially unchanged. In addition, there
was not much change in material-augmenting technology in most of the
countries.

Finally, we show the extent to which the results of growth accounting
can change when material inputs are included. Table \ref{tab: growth_material}
reports the quantitative contribution of capital-, labor-, energy-,
and material-augmenting technologies in addition to that of capital,
labor, energy, and material inputs to economic growth for each country.
Since the material share of income is large, the contribution of capital,
labor, and energy inputs and of capital-, labor-, and energy-augmenting
technologies to economic growth becomes smaller when material inputs
are included than when they are not. However, the contribution of
energy-saving technology is not negligible in many countries.

\begin{sidewaystable}[ph]
\caption{Sources of economic growth when material inputs are included\label{tab: growth_material}}

\begin{centering}
\subfloat[Goods sector]{
\centering{}\resizebox{1\textwidth}{!}{%
\begin{tabular}{llr@{\extracolsep{0pt}.}lr@{\extracolsep{0pt}.}lr@{\extracolsep{0pt}.}lr@{\extracolsep{0pt}.}lr@{\extracolsep{0pt}.}lr@{\extracolsep{0pt}.}lr@{\extracolsep{0pt}.}lr@{\extracolsep{0pt}.}lr@{\extracolsep{0pt}.}lr@{\extracolsep{0pt}.}lr@{\extracolsep{0pt}.}lr@{\extracolsep{0pt}.}lr@{\extracolsep{0pt}.}lr@{\extracolsep{0pt}.}lr@{\extracolsep{0pt}.}lr@{\extracolsep{0pt}.}lr@{\extracolsep{0pt}.}l}
\hline 
 &  & \multicolumn{2}{c}{$y$} & \multicolumn{2}{c}{} & \multicolumn{2}{c}{$a_{k}k$} & \multicolumn{2}{c}{$a_{\ell}\ell$} & \multicolumn{2}{c}{$a_{e}e$} & \multicolumn{2}{c}{$a_{m}m$} & \multicolumn{2}{c}{} & \multicolumn{2}{c}{$k$} & \multicolumn{2}{c}{$\ell$} & \multicolumn{2}{c}{$e$} & \multicolumn{2}{c}{$m$} & \multicolumn{2}{c}{$a_{k}$} & \multicolumn{2}{c}{$a_{\ell}$} & \multicolumn{2}{c}{$a_{e}$} & \multicolumn{2}{c}{$a_{m}$} & \multicolumn{2}{c}{} & \multicolumn{2}{c}{$a$}\tabularnewline
\cline{3-4} \cline{7-14} \cline{9-14} \cline{11-14} \cline{13-14} \cline{17-32} \cline{19-32} \cline{21-32} \cline{23-32} \cline{25-32} \cline{27-32} \cline{29-32} \cline{31-32} \cline{35-36} 
Sweden & 1994\textendash 2005 & 5&27 & \multicolumn{2}{c}{} & 0&57 & 1&47 & 0&27 & 2&96 & \multicolumn{2}{c}{} & 0&43 & 0&03 & 0&19 & 1&90 & 0&14 & 1&43 & 0&08 & 1&06 & \multicolumn{2}{c}{} & 2&72\tabularnewline
Italy & 1978\textendash 2005 & 5&12 & \multicolumn{2}{c}{} & 0&45 & 1&64 & 0&63 & 2&40 & \multicolumn{2}{c}{} & 0&27 & \textendash 0&38 & 0&25 & 1&15 & 0&19 & 2&02 & 0&37 & 1&25 & \multicolumn{2}{c}{} & 3&83\tabularnewline
Czech Republic & 1996\textendash 2005 & 5&02 & \multicolumn{2}{c}{} & 0&69 & 0&90 & 0&98 & 2&45 & \multicolumn{2}{c}{} & 0&37 & \textendash 0&25 & 0&55 & 4&49 & 0&32 & 1&15 & 0&42 & \textendash 2&04 & \multicolumn{2}{c}{} & \textendash 0&14\tabularnewline
Finland & 1978\textendash 2005 & 4&49 & \multicolumn{2}{c}{} & 0&45 & 1&39 & 0&68 & 1&96 & \multicolumn{2}{c}{} & 0&16 & \textendash 0&38 & 0&33 & 1&30 & 0&29 & 1&78 & 0&35 & 0&66 & \multicolumn{2}{c}{} & 3&08\tabularnewline
Japan & 1978\textendash 2005 & 3&55 & \multicolumn{2}{c}{} & 0&25 & 0&95 & 0&17 & 2&19 & \multicolumn{2}{c}{} & 0&37 & \textendash 0&37 & 0&11 & 0&90 & \textendash 0&12 & 1&31 & 0&06 & 1&29 & \multicolumn{2}{c}{} & 2&54\tabularnewline
Austria & 1980\textendash 2005 & 3&32 & \multicolumn{2}{c}{} & 0&39 & 0&99 & 0&43 & 1&51 & \multicolumn{2}{c}{} & 0&08 & \textendash 0&37 & 0&58 & 1&46 & 0&32 & 1&36 & \textendash 0&15 & 0&05 & \multicolumn{2}{c}{} & 1&57\tabularnewline
Denmark & 1980\textendash 2005 & 2&67 & \multicolumn{2}{c}{} & 0&20 & 0&65 & 0&31 & 1&52 & \multicolumn{2}{c}{} & 0&20 & \textendash 0&29 & 0&21 & 0&99 & 0&00 & 0&94 & 0&10 & 0&53 & \multicolumn{2}{c}{} & 1&56\tabularnewline
Unites States & 1978\textendash 2005 & 2&66 & \multicolumn{2}{c}{} & 0&18 & 0&72 & 0&44 & 1&32 & \multicolumn{2}{c}{} & 0&16 & 0&00 & 0&00 & 0&83 & 0&02 & 0&73 & 0&44 & 0&49 & \multicolumn{2}{c}{} & 1&67\tabularnewline
Netherlands & 1987\textendash 2005 & 2&06 & \multicolumn{2}{c}{} & 0&28 & 0&46 & 0&02 & 1&30 & \multicolumn{2}{c}{} & 0&11 & 0&00 & 0&48 & 1&10 & 0&17 & 0&46 & \textendash 0&46 & 0&20 & \multicolumn{2}{c}{} & 0&37\tabularnewline
United Kingdom & 1978\textendash 2005 & 1&99 & \multicolumn{2}{c}{} & 0&17 & 0&34 & 0&59 & 0&89 & \multicolumn{2}{c}{} & 0&09 & \textendash 0&47 & 0&24 & 0&40 & 0&09 & 0&82 & 0&35 & 0&49 & \multicolumn{2}{c}{} & 1&74\tabularnewline
Germany & 1992\textendash 2005 & 1&97 & \multicolumn{2}{c}{} & 0&25 & 0&86 & 0&18 & 0&67 & \multicolumn{2}{c}{} & 0&08 & \textendash 0&70 & \textendash 0&18 & 1&40 & 0&17 & 1&56 & 0&36 & \textendash 0&73 & \multicolumn{2}{c}{} & 1&36\tabularnewline
Portugal & 1996\textendash 2005 & \textendash 1&22 & \multicolumn{2}{c}{} & \textendash 0&08 & \textendash 0&13 & \textendash 0&60 & \textendash 0&41 & \multicolumn{2}{c}{} & 0&31 & \textendash 0&05 & 1&00 & 1&51 & \textendash 0&38 & \textendash 0&08 & \textendash 1&60 & \textendash 1&92 & \multicolumn{2}{c}{} & \textendash 3&98\tabularnewline
\hline 
\end{tabular}}}
\par\end{centering}
\begin{centering}
\subfloat[Service sector]{
\centering{}\resizebox{1\textwidth}{!}{%
\begin{tabular}{llr@{\extracolsep{0pt}.}lr@{\extracolsep{0pt}.}lr@{\extracolsep{0pt}.}lr@{\extracolsep{0pt}.}lr@{\extracolsep{0pt}.}lr@{\extracolsep{0pt}.}lr@{\extracolsep{0pt}.}lr@{\extracolsep{0pt}.}lr@{\extracolsep{0pt}.}lr@{\extracolsep{0pt}.}lr@{\extracolsep{0pt}.}lr@{\extracolsep{0pt}.}lr@{\extracolsep{0pt}.}lr@{\extracolsep{0pt}.}lr@{\extracolsep{0pt}.}lr@{\extracolsep{0pt}.}lr@{\extracolsep{0pt}.}l}
\hline 
 &  & \multicolumn{2}{c}{$y$} & \multicolumn{2}{c}{} & \multicolumn{2}{c}{$a_{k}k$} & \multicolumn{2}{c}{$a_{\ell}\ell$} & \multicolumn{2}{c}{$a_{e}e$} & \multicolumn{2}{c}{$a_{m}m$} & \multicolumn{2}{c}{} & \multicolumn{2}{c}{$k$} & \multicolumn{2}{c}{$\ell$} & \multicolumn{2}{c}{$e$} & \multicolumn{2}{c}{$m$} & \multicolumn{2}{c}{$a_{k}$} & \multicolumn{2}{c}{$a_{\ell}$} & \multicolumn{2}{c}{$a_{e}$} & \multicolumn{2}{c}{$a_{m}$} & \multicolumn{2}{c}{} & \multicolumn{2}{c}{$a$}\tabularnewline
\cline{3-4} \cline{7-14} \cline{9-14} \cline{11-14} \cline{13-14} \cline{17-32} \cline{19-32} \cline{21-32} \cline{23-32} \cline{25-32} \cline{27-32} \cline{29-32} \cline{31-32} \cline{35-36} 
Italy & 1978\textendash 2005 & 4&82 & \multicolumn{2}{c}{} & 0&78 & 2&69 & 0&41 & 0&93 & \multicolumn{2}{c}{} & 0&49 & 0&99 & 0&08 & 0&55 & 0&30 & 1&70 & 0&33 & 0&38 & \multicolumn{2}{c}{} & 2&71\tabularnewline
Unites States & 1978\textendash 2005 & 4&24 & \multicolumn{2}{c}{} & 0&68 & 2&65 & 0&27 & 0&64 & \multicolumn{2}{c}{} & 0&74 & 1&37 & 0&11 & 0&46 & \textendash 0&06 & 1&27 & 0&16 & 0&19 & \multicolumn{2}{c}{} & 1&56\tabularnewline
United Kingdom & 1978\textendash 2005 & 4&14 & \multicolumn{2}{c}{} & 0&64 & 1&83 & 0&28 & 1&40 & \multicolumn{2}{c}{} & 0&65 & 0&95 & 0&30 & 0&55 & \textendash 0&01 & 0&88 & \textendash 0&02 & 0&85 & \multicolumn{2}{c}{} & 1&70\tabularnewline
Denmark & 1980\textendash 2005 & 3&86 & \multicolumn{2}{c}{} & 0&93 & 2&01 & 0&29 & 0&64 & \multicolumn{2}{c}{} & 0&50 & 0&45 & 0&11 & 0&46 & 0&43 & 1&56 & 0&18 & 0&18 & \multicolumn{2}{c}{} & 2&35\tabularnewline
Japan & 1978\textendash 2005 & 3&84 & \multicolumn{2}{c}{} & 0&58 & 2&11 & 0&13 & 1&03 & \multicolumn{2}{c}{} & 0&98 & 0&43 & 0&18 & 0&63 & \textendash 0&40 & 1&67 & \textendash 0&05 & 0&40 & \multicolumn{2}{c}{} & 1&62\tabularnewline
Austria & 1980\textendash 2005 & 3&35 & \multicolumn{2}{c}{} & 0&61 & 1&89 & 0&21 & 0&64 & \multicolumn{2}{c}{} & 0&61 & 0&78 & 0&52 & 0&68 & 0&00 & 1&12 & \textendash 0&31 & \textendash 0&03 & \multicolumn{2}{c}{} & 0&77\tabularnewline
Netherlands & 1987\textendash 2005 & 3&02 & \multicolumn{2}{c}{} & 0&57 & 1&73 & 0&04 & 0&68 & \multicolumn{2}{c}{} & 0&68 & 1&20 & 0&08 & 0&72 & \textendash 0&10 & 0&53 & \textendash 0&04 & \textendash 0&04 & \multicolumn{2}{c}{} & 0&34\tabularnewline
Finland & 1978\textendash 2005 & 3&00 & \multicolumn{2}{c}{} & 0&39 & 1&54 & 0&36 & 0&71 & \multicolumn{2}{c}{} & 0&50 & 0&75 & 0&45 & 0&74 & \textendash 0&11 & 0&79 & \textendash 0&08 & \textendash 0&03 & \multicolumn{2}{c}{} & 0&56\tabularnewline
Germany & 1992\textendash 2005 & 2&45 & \multicolumn{2}{c}{} & 0&48 & 1&35 & 0&09 & 0&54 & \multicolumn{2}{c}{} & 0&75 & 0&43 & 0&01 & 0&22 & \textendash 0&27 & 0&92 & 0&08 & 0&31 & \multicolumn{2}{c}{} & 1&04\tabularnewline
Sweden & 1994\textendash 2005 & 1&97 & \multicolumn{2}{c}{} & 0&20 & 1&04 & 0&05 & 0&68 & \multicolumn{2}{c}{} & 0&76 & 0&64 & 0&04 & 0&26 & \textendash 0&56 & 0&40 & 0&01 & 0&43 & \multicolumn{2}{c}{} & 0&27\tabularnewline
Czech Republic & 1996\textendash 2005 & 1&80 & \multicolumn{2}{c}{} & 0&45 & 0&07 & 0&13 & 1&14 & \multicolumn{2}{c}{} & 0&96 & 0&26 & 0&42 & 0&34 & \textendash 0&50 & \textendash 0&19 & \textendash 0&29 & 0&80 & \multicolumn{2}{c}{} & \textendash 0&18\tabularnewline
Portugal & 1996\textendash 2005 & \textendash 0&68 & \multicolumn{2}{c}{} & \textendash 0&42 & \textendash 0&45 & 0&02 & 0&16 & \multicolumn{2}{c}{} & 1&60 & 1&01 & 0&37 & 0&88 & \textendash 2&01 & \textendash 1&45 & \textendash 0&34 & \textendash 0&72 & \multicolumn{2}{c}{} & \textendash 4&53\tabularnewline
\hline 
\end{tabular}}}
\par\end{centering}
\textit{\footnotesize{}Notes}{\footnotesize{}: The first column reports
the annual rate of growth in gross output ($y$) from the first year
of observation ($t_{0}$) to 2005 (i.e., $100\times(\ln y_{2005}-\ln y_{t_{0}})/(2005-t_{0})$).
The sixth to thirteenth columns report the results of the Shapley
decomposition based on the one-level CES production function. The
second, third, fourth, and fifth columns report the sum of the numbers
in the sixth and tenth columns, the seventh and eleventh columns,
the eighth and twelfth columns, and the ninth and thirteenth columns,
respectively. The last column reports the sum of the numbers from
the tenth to thirteenth columns. Countries are arranged in descending
order of growth rate of gross output by sector.}{\footnotesize\par}
\end{sidewaystable}

\end{document}